\documentclass{emulateapj}

\usepackage{xspace}
\usepackage{graphics,graphicx}
\usepackage{natbib}
\usepackage{multirow}
\usepackage{longtable}
\usepackage{rotating}
\usepackage[section]{placeins}
\usepackage[percent]{overpic}

\newcommand{\as}{\mbox{\ensuremath{.\!\!^{\prime\prime}}}}

\newcommand{\asn}{$^{\prime\prime}$\xspace}
\newcommand{\am}{$^{\prime}$\xspace}
\newcommand{\nH}{$N_{\rm H}$\xspace}
\newcommand{\PL}{$\Gamma$\xspace}
\newcommand{\Msun}{$M_{\odot}$\xspace}

\newcommand{\Chandra}{{\it Chandra}\xspace}
\newcommand{\HST}{{\it HST}\xspace}
\newcommand{\XMM}{{\it XMM-Newton}\xspace}
\newcommand{\lum}{erg s$^{-1}$\xspace}
\newcommand{\flux}{erg s$^{-1}$ cm$^{-2}$\xspace}

\newcommand{\wavdetect}{\texttt{wavdetect}\xspace}
\newcommand{\AEx}{\texttt{AE}\xspace}
\newcommand{\pns}{{\it pns}\xspace}

\newcommand{\lognlogs}{log$N$-log$S$\xspace}

\shorttitle{CLVS Archival X-ray Source Catalogs}
\shortauthors{Binder et al.}

\begin{document}

\title{The {\it Chandra} Local Volume Survey I: The X-ray Point Source Populations of NGC~55, NGC~2403, and NGC~4214}
\author{B. Binder\altaffilmark{1}, 
B. F. Williams\altaffilmark{1}, 
M. Eracleous\altaffilmark{2},
P. P. Plucinsky\altaffilmark{3},
T. J. Gaetz\altaffilmark{3},
S. F. Anderson\altaffilmark{1},
E. D. Skillman\altaffilmark{5},
J. J. Dalcanton\altaffilmark{1},
A. K. H. Kong\altaffilmark{4},
D. R. Weisz\altaffilmark{1}
}
\altaffiltext{1}{University of Washington, Department of Astronomy, Box 351580, Seattle, WA 98195}
\altaffiltext{2}{Department of Astronomy \& Astrophysics and Center for Gravitational Wave Physics, The Pennsylvania State University, 525 Davey Lab, University Park, PA 16802}
\altaffiltext{3}{Harvard-Smithsonian Center for Astrophysics, 60 Garden Street, Cambridge, MA 02138, USA}
\altaffiltext{4}{Institute of Astronomy and Department of Physics, National Tsing Hua University, Hsinchu 30013, Taiwan}
\altaffiltext{5}{University of Minnesota, Astronomy Department, 116 Church St. SE, Minneapolis, MN 55455}

\begin{abstract}
We present comprehensive X-ray point source catalogs of NGC~55, NGC~2403, and NGC~4214 as part of the \Chandra Local Volume Survey. The combined archival observations have effective exposure times of 56.5 ks, 190 ks, and 79 ks for NGC~55, NGC~2403, and NGC~4214, respectively. When combined with our published catalogs for NGC 300 and NGC 404, our survey contains 629 X-ray sources total down to a limiting unabsorbed luminosity of $\sim5\times10^{35}$ \lum in the 0.35-8 keV band in each of the five galaxies. We present X-ray hardness ratios, spectral analysis, radial source distributions, and an analysis of the temporal variability for the X-ray sources detected at high significance.  To constrain the nature of each X-ray source, we carried out cross-correlations with multi-wavelength data sets. We searched overlapping {\it Hubble Space Telescope} observations for optical counterparts to our X-ray detections to provide preliminary classifications for each X-ray source as a likely X-ray binary, background AGN, supernova remnant, or foreground star. 
\end{abstract}

\section{Introduction}
The luminous, discrete X-ray source populations ($L_X>10^{37}$ \lum) of inactive galaxies are dominated by X-ray binaries \citep[XRBs;][]{Fabbiano06,FabWhite06}. These are stellar systems containing an accreting compact object (either a neutron star, NS, or a black hole, BH) and a donor stellar companion, and are divided into low-mass XRBs (LMXBs) and high-mass XRBs (HMXBs) depending on the mass of the stellar companion. The donor stars of HMXBs are OB stars, while LMXBs have later type companions (usually with stellar masses $<$2 \Msun). Other sources of X-ray emission are also present, such as hot gas, and young supernova remnants (SNRs). Some galaxies additionally have ultraluminous X-ray sources (ULXs, $L_X>10^{39}$ \lum) associated with them, which are believed to be either binaries accreting at super-Eddington rates or intermediate-mass black holes.  Finally, background AGN are sometimes incorrectly identified as belonging to the host galaxy \citep[][and references therein]{Heida+13}.

Studies of XRBs and X-ray emitting SNRs in the Milky Way often suffer from significant distance uncertainties and considerable extinction along Galactic lines of sight. These difficulties are minimized, however, when studying nearby galaxies. The effects of Galactic extinction become less  problematic when observing galaxies at higher Galactic latitudes, and the impact of distance uncertainties is reduced since all X-ray sources located within a particular galaxy are essentially equidistant from the observer. Therefore, observing a large number of X-ray sources in nearby galaxies can provide new insights into the characteristics of both individual X-ray sources and the global properties of the X-ray point source population.
 
With these considerations in mind, we undertook the \Chandra Local Volume Survey (CLVS), a deep, volume-limited survey of five nearby galaxies (NGC~55, NGC~300, NGC~404, NGC~2403, and NGC~4214) with overlapping {\it Hubble Space Telescope} (\HST) observations \citep[down to $M_V\sim0$;][]{Dalcanton+09}. With its excellent angular resolution ($\sim$0\as5) and positional accuracy, \Chandra is the best X-ray telescope for separating X-ray point source populations in nearby galaxies from the bulk of their diffuse emission. When combined with deep, optical \HST imaging, reliable optical counterpart identification and source classification may be carried out even at distances of a few Mpc. Our collaboration has already examined the X-ray point source populations of NGC~300 \citep[][hereafter Paper~I]{Binder+12} and NGC~404 \citep[][hereafter Paper~II]{Binder+13}. 

The goal of the CLVS is to identify strong XRB candidates in nearby galaxies, and to compare the properties of these populations to the star formation histories (SFHs) of their host galaxies. Previous studies have shown that the fast evolutionary timescales of massive stars ($\sim$10$^7$ years) makes X-ray emission from HMXBs nearly simultaneous with their formation \citep{Grimm+03,Gilfanov+04,Shty+07,Mineo+11,Mineo+12}, while the longer-lived LMXB systems trace older underlying stellar populations \citep{Kong+02,Soria+02,Trudolyubov+02}; thus, the shape of the X-ray luminosity function (XLF) can be correlated with the age and SFH of the host galaxy \citep{Grimm+03, Belczynski+04,Eracleous+06}. In many of these early works, however, the current star formation rate (SFR) of the host galaxy was calculated through indirect means (i.e., H$\alpha$ luminosity, UV flux, radio flux, etc.), and the contribution of background AGN to the XLFs could only be corrected for in a statistical sense. 

By using \Chandra to study nearby galaxies with available deep \HST imaging, we are able to (1) directly measure both the recent and long-term SFH of each galaxy using resolved stellar populations, and (2) identify strong HMXB candidates, thereby greatly reducing the degree of contamination (from AGN, SNRs, or foreground stars) present in our XLF analysis. 

In this paper, we present the X-ray point source catalogs from archival \Chandra observations of NGC~55, NGC~2403, and NGC~4214 and compare these catalogs to those of NGC~300 and NGC~404. We describe the galaxy sample (Section~\ref{sample}), the archival observations and data reduction procedures (Section~\ref{data}), and the methodology used to generate the X-ray source catalogs (Section~\ref{catalogs}). We then present a spectral and timing analysis of the sources in each galaxy (Section~\ref{xray}) and perform optical counterpart identification and X-ray source classification (Section~\ref{optical}).

\section{Galaxy Sample}\label{sample}
The five target galaxies of the \Chandra Local Volume Survey (CLVS) span a representative sample of disk galaxies with a range of masses, metallicities, and morphologies. When combined with the already well-studied disks of M~31 \citep{Kong+03,Williams+04,Stiele+11} and M~33 \citep{Williams+08,Plucinsky+08,Tullmann+11}, these galaxies contain $\sim$99\% of the stellar mass and $\sim$90\% of the recent star formation out to a distance of $\sim$3.3 Mpc \citep{Freedman+88,Maiz+02,Tikhonov+05}. Table~\ref{galbasic} summarizes some basic properties of each galaxy included in our sample. Row (1) lists the distance (in Mpc) to each galaxy. Row (2) provides the morphological type. Rows (3)-(4) list the right ascension (RA) and declination (Decl.), respectively (J2000.0). Row (5) provides the distance modulus to each galaxy. Row (6) provides the absolute $B$-magnitude. Row (7) provides the inclination angle of each galaxy, and row (8) lists the position angle (both given in degrees). Row (9) lists the major linear diameter of each galaxy, in kpc. Row (10) provides the log of the stellar mass of each galaxy, in units of $M_*$/\Msun. Row (11) provides the Galactic absorbing column. Row (12) lists 12+log(O/H) to give an indication of the metallicity of each galaxy. Row (13) provides the integrated (unabsorbed) 0.35-8 keV luminosity of each galaxy.

\subsection{NGC~55}
Along with NGC~300, NGC~55 is a member of the nearby Sculptor group of galaxies. Classified as SB(s)m type galaxy \citep{RC3}, NGC~55 is an edge-on analog of the Large Magellanic Cloud \citep{Schmidt+93} and is the nearest bright edge-on galaxy \citep{Kara+04}. Its optical morphology is asymmetric, with a bright region displaced from the geometrical center of the galaxy \citep{Robinson+66} interpreted as a bar viewed near- to end-on \citep{deV61}. The edge-on orientation of NGC~55 offers excellent insight into the effect of disk-based star formation activity on extra-planar regions, and as such, NGC~55 has been a popular target for ISM studies \citep{Puche+91,Ferguson+96, Hoopes+96, Otte+99}. Ionized gas shells are observed to protrude well above the plane of the galaxy, and at least some of this gas appears to cool sufficiently to form new stars in the halo \citep{Tullmann+03,Tullmann+04}. The star formation rate derived from the total infrared luminosity of NGC~55 is 0.22 \Msun yr$^{-1}$ \citep{Engelbracht+04}.

NGC~55 was first observed at X-ray wavelengths by {\it ROSAT}, revealing 15 discrete X-ray sources and localized diffuse emission within the optical confines of the galaxy \citep{Read+97,Schlegel+97,Dahlem+98}. One of these sources has been established as a ULX and likely BH-XRB \citep{Stobbart+04}. A detailed X-ray view of NGC~55 by \XMM was presented by \cite{Stobbart+06a}, with 137 X-ray sources detected within the field of view down to a flux of $\sim5\times10^{-15}$ \flux in the 0.3-6 keV energy band. Of these sources, 42 are within the $D_{\rm 25}$ ellipse, and the authors find $\sim$20 XRB candidates, five SNRs, and seven candidate super soft sources (SSSs).

\subsection{NGC~2403}
NGC~2403 is an outlying member of the M81 group. Although comparable in $M_B$, morphology, inclination, and dynamical mass to M~33, it lies more than four times farther away from M81 (the nearest large disk galaxy) than M33 is from the Milky Way \citep{Kara+02}, making it a far more isolated system. NGC~2403 lacks a central bulge, but like M33 hosts a luminous compact nuclear star cluster \citep{Davidge+02} containing a recently discovered X-ray transient source \citep{Yukita+07}. Additionally, NGC~2403 hosts a ULX \citep{Swartz+04}, a recent SN Type IIp \citep[SN 2004dj,][]{Nakano+04,Beswick+05}, 35 spectroscopically confirmed or suspected SNRs \citep{Matonick+97}, and 52 Wolf-Rayet candidates \citep{Drissen+99}. NGC~2403 has been observed multiple times with \XMM \citep{Feng+05,Stobbart+06b}.  

The dominant age of stars within the nuclear star cluster is $\sim1.4$ Gyr \citep{Yukita+07}, while numerous stars within the inner disk show stellar ages of $\sim$100 Myr \citep{Davidge+02}. The youngest ($\sim2-10$ Myr) and most massive \ion{H}{2} regions are 0.7-1.6 kpc from the center \citep{Drissen+99}. The global star formation rate of NGC~2403 is $\sim1.2$ \Msun yr$^{-1}$ \citep{Kennicutt98}, and the star formation history of NGC~2403 is also well measured from \HST stellar photometry \citep{Williams+13}.

\subsection{NGC~4214} 
Classified as an IAB(s)m similar to the Large Magellanic Cloud \citep{RC3}, NGC~4214 is one of the nearest examples of a starburst galaxy \citep{Huchra+83, Sargent+91} with a substantial population of Wolf-Rayet stars. Despite its active star formation \citep[$\sim$0.1 \Msun yr$^{-1}$ over the past 100 Myr;][]{Williams+11}, only a small percentage of the stellar population is young ($<$1\% formed in the past 50 Myr). Additionally, evidence from \HST imaging and spectroscopy suggests the stellar initial mass function (IMF) of NGC~4214 is steeper than that of \cite{Salpeter55} at high masses \citep[$>$20 \Msun,][]{Ubeda+07a,Ubeda+07b}. Despite the elevated levels of recent star formation, however, only a small fraction of the galaxy's stars have formed since $z\sim1$. The overall stellar populations of NGC~4214 are surprisingly similar to that of the nearby early-type dwarf galaxy NGC~404 \citep{Williams+10}.

NGC~4214 was the host of a Type Ia supernova \citep{Wellmann55} and a recent nova \citep{Humphreys+10}. The most luminous X-ray source in NGC~4214, CXOU J121538.2+361921, is an XRB with a 3.62 hour period and is likely composed of a slightly evolved He-burning donor (of $\sim2-3$ \Msun) and a NS \citep{Ghosh+06, Dewi06}. The X-ray point source population of NGC~4214 was investigated using both \Chandra and \XMM observations by \cite{Hartwell+04}, who demonstrated that the high energy properties of NGC~4214 were comparable to other dwarf starbursts down to a limiting luminosity of $\sim10^{36}$ \lum.

\section{Observations and Data Reduction}\label{data}
\subsection{Observations and Image Alignment}
We constructed the X-ray point source catalogs of NGC~55, NGC~2403, and NGC~4214 using archival \Chandra ACIS data. All data reduction used CIAO v4.3 and CALDB v4.4.2 using standard reduction procedures. All observations were reprocessed from \texttt{evt1} level files using the CIAO routine \texttt{acis\_process\_events}. The data were corrected for bad pixels/columns, charge transfer inefficiency, time-dependent gain variations, and the pixel randomization was removed. The event list was restricted to the pipeline-provided good time intervals, grades = 0, 2, 3, 4, 6, and status = 0. Background light curves were generated from source-free event lists for each observation, and a sigma-clipping algorithm was applied to remove time intervals with count rates more than 5$\sigma$ from the mean. Exposure maps and exposure-corrected images were made using \texttt{fluximage} and observations were then merged using \texttt{reproject\_image}. 

Spectral weights used for the instrument maps were generated assuming a power law spectrum with $\Gamma$ =1.9 (appropriate for both XRBs and background AGN) and the foreground hydrogen column density provided in Table~\ref{galbasic}. Since many background AGN (and XRBs) show evidence for absorption beyond that of the Galactic column (see Section~\ref{spectral_analysis} for further discussion of these sources), we generated spectral weights assuming a highly absorbed power law (with $\Gamma=1.9$ but \nH an order of magnitude larger than the Galactic value). From the resulting instrument and exposure maps, we found that the assumption of Galactic absorption will lead to a flux overestimation  in sources that are intrinsically more absorbed: the predicted unabsorbed fluxes differ by $\sim$35\% in the 0.5-2 keV band and $\sim$5\% in the 2-8 keV band.

Table~\ref{obs} summarizes the archival observations used in this work. Column (1) lists the galaxy. Column (2) provides the \Chandra observation IDs. Column (3) lists the date on which each observation was performed. Column (4) gives the usable exposure time acquired during each observation from the sigma-clipping algorithm, and column (5) lists this time as a fraction of the total exposure. Columns (6)-(7) provide the RA and Decl. (J2000.0) of the \Chandra aim-point. Column (8) lists the detector used during each observation.

\subsection{Image Alignment}
A total of 13 \HST fields were used to search for optical counterparts for each X-ray source: six fields in NGC~55, five in NGC~2403, and two in NGC~4214. The \HST fields were selected from the ACS Nearby Galaxy Survey Treasury; details of the \HST data acquisition and reduction are provided in \cite{Dalcanton+09}. In the case of NGC~2403, three of the five \HST fields targeted the halo and were far off the nominal \Chandra aim-points.

To identify candidate optical counterparts to the \Chandra X-ray sources, both the \Chandra and \HST fields need to be placed onto a common, reference coordinate frame. We place all observations (X-ray and optical) onto the International Celestial Reference System (ICRS) by finding matches (within $\sim$5\asn) between stars or background galaxies in the Two Micron All Sky Survey (2MASS) Point Source Catalog \citep{Skrutskie+06}. We were able to identify between three and six bright 2MASS sources per field which matched either a \Chandra X-ray source, optical foreground star, or background galaxy. The plate solutions were computed using the IRAF task \texttt{ccmap}, with $r.m.s.$ residuals from the ICRS frame typically less than a few hundredths of an arcsecond in both right ascension and declination. The total alignment error applied to the coordinates of each X-ray source was computed by summing the \Chandra and \HST $r.m.s.$ residuals in quadrature. For both NGC~2403 and NGC~4214, two of the available \HST fields overlapped with one another; we were able to identify one source in NGC~2403 that was suitable for alignment in both the \HST fields, and two sources in the overlap region of NGC~4214. 

\begin{figure*}
\centering
	\begin{overpic}[width=1\linewidth,clip=true,trim=0.25cm 0.25cm 0.25cm 0.25cm]{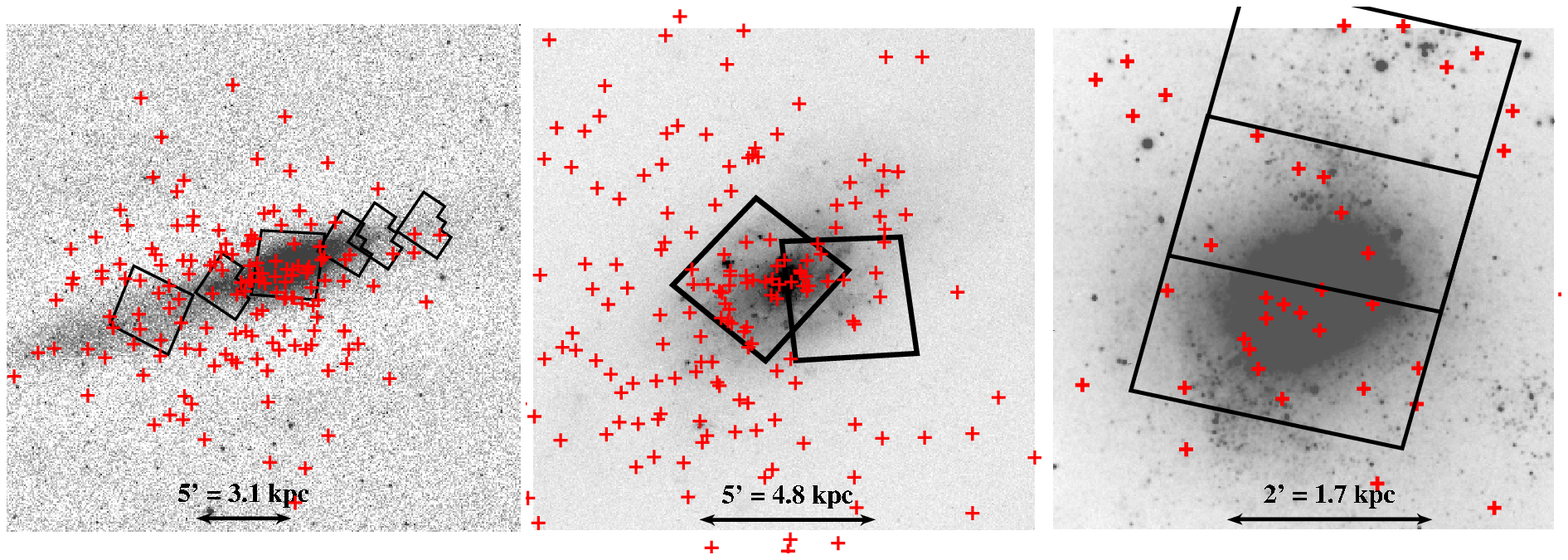} 
 		\put (13,30) {\large NGC 55}
		\put (45,30) {\large NGC 2403}
		\put (80,30) {\large NGC 4214}
	\end{overpic}	\\
\caption{Images of NGC~55 ({\it left}, 2MASS $J$-band), NGC~2403 ({\it middle}, ground-based $r$-band), and NGC~4214 ({\it right}, ground-based $r$-band). Highly significant discrete point sources (i.e., those with \pns$<4\times10^{-6}$) are shown as red crosses. \HST fields are shown in black.}
\label{optical_xray}
\end{figure*}

Figure~\ref{optical_xray} shows the 2MASS $J$-band image of NGC~55 and ground-based $r$-band images of NGC~2403 and NGC~4214. The positions of discrete X-ray sources are shown in red. Only the two \HST fields covering the NGC~2403 disk are shown, as the remaining three fields targeted the galaxy's halo and contain only a small number of X-ray sources.

A summary of our \HST observations and the results of our astrometry are provided in Table~\ref{align}. Column (1) lists the galaxy name. Column (2) provides a description of the field (e.g., ``\Chandra'' for the X-ray image, ``DISK'' for the archival \HST field labeled ``disk'') and column (3) provides the corresponding \HST observation IDs. Column (4) lists the number of X-ray sources contained within each field. Columns (5)-(6) provide the RA and Decl. (J2000.0) of the telescope aim-points. Column (7) lists the number of sources used to align the field to the ICRS. Columns (8)-(9) provide the root-mean-squared uncertainty in the RA and Decl., respectively. Overall, these uncertainties are smaller than the individual source position uncertainties. Thus, our counterpart analysis was limited by the uncertainties in the X-ray source centroids.

\section{X-ray Source Catalogs}\label{catalogs}
\subsection{Catalog Creation}
For consistency with Papers~I and II, we employed the same iterative source detection strategy \citep[developed by][]{Tullmann+11}. A summary of the approach is given here, and the reader is referred to Papers I and II for further details. The CIAO task \wavdetect \citep{Freeman+02} was used to create a list of source candidates; \wavdetect was run on each individual exposure for each galaxy {\it and} on the stacked image. These source candidates were used as input to \texttt{ACIS-Extract} \citep[\AEx;][]{Broos+10}. \AEx is a source extraction and characterization tool which was used to determine various source properties (source and background count rates, detection significances, fluxes, etc.) and to generate light curves and spectra with appropriate response matrices. The Poisson probability of not being a source ({\it prob\_no\_source}; hereafter \pns) provided by \AEx was used to select highly significant sources for inclusion in the X-ray source catalogs. The \pns value provides the Poisson probability that all of the counts in the source extraction region are actually from the background (in the given energy band); therefore, candidate sources with low \pns values are most likely to be genuine.

For each input source, \AEx creates background regions by first removing all sources from the data and then searching for the smallest circular region around each source that encompasses at least 50 counts \citep[changing the number of background counts required does not significantly affect source properties;][]{Tullmann+11}. These background regions are used to extract background spectra. To extract the X-ray spectrum for each source, \AEx constructs a polygon region that approximates the 90\% contour of the \Chandra-ACIS PSF at the location of the source. The CIAO tools \texttt{mkacisrmf} and \texttt{mkarf} are used to create appropriate response and ancillary response matrices, respectively, for each source.

The initial source lists generated by \wavdetect were deliberately constructed to include many more sources than we anticipated being statistically significant. Each source list was then filtered to remove spurious sources and improve source and background regions following the iterative procedure of Papers I, II, and \cite{Tullmann+11}. Both the construction of the initial source list and all subsequent iterations with \AEx were run on the combined images for each galaxy (e.g., using all available exposures, to ensure detection of faint sources). For our final iteration with \AEx, we required that a candidate X-ray source have \pns$<4\times10^{-6}$ \citep{Nandra+05,Georgakakis+08,Tullmann+11} in any of the following energy bands: 0.5-8, 0.5-2, 2-8, 0.5-1, 1-2, 2-4, 4-8, 0.35-1, or 0.35-8 keV.  If we would have considered only the 0.35-8.0 keV band, $\sim$4-8\% of significant sources would have been missed. 

Using a \pns threshold of $4\times10^{-6}$ results in $\sim0.5$ false source detections per megapixel \citep[determined by e.g.,][]{Nandra+05,Georgakakis+08}. There is approximately one megapixel per ACIS CCD, and source detection was performed in nine separate energy bands. Thus, we calculate the number of false positives in our catalogs as $0.5\times9\times$ the number of CCDs used in the source detection process. NGC~55 was observed with ACIS-I (4 CCDs), while NGC~2403 and NGC~4214 were observed with ACIS-S (with source detection performed on S2-4 in both cases). We therefore expect $\sim$18, 14, and 14 false sources to be present in the X-ray source catalogs of NGC~55, NGC~2403, and NGC~4214, respectively. Of the 629 X-ray sources detected at high significance across all five galaxies, $\sim$94\% have fewer than 200 net counts. Half of all X-ray sources have error circles smaller than 2\asn in radius and 54\% have off-axis angles less than 5\am.

Table~\ref{ae_passes} summarizes the number of candidate X-ray sources included in the source lists for each CLVS galaxy at each step in our iterative source detection procedure. Column (1) lists the galaxy. Column (2) lists the number of sources contained in the original \wavdetect source list. Column (3) provides the number of sources remaining after our first pass through \AEx. Column (4) lists the number of sources remaining after the second pass through \AEx, and column (5) lists the number of sources in the final source catalog.

The final source catalogs for NGC~55, NGC~2403, and NGC~4214 contain 154, 190, and 116 X-ray sources, respectively, and 95 and 74 sources for earlier studies of NGC~300 and NGC~404, respectively. A total of 629 highly significant X-ray sources were detected as part of the CLVS\footnote{Complete source catalogs for all five CLVS galaxies are available in FITS format at \url{http://www.astro.washington.edu/users/bbinder/CLVS/}}. The remainder of this work utilizes these highly-significant sources. However, it is possible that some marginally-detected sources (i.e., those sources that failed to meet our final \pns requirement) are in fact genuine X-ray sources, and other observers may wish to use different source selection criteria than we employed here. The locations of the lower-significance sources have been excluded in Figure~\ref{optical_xray}, where there exist a handful of regions showing evidence for X-ray emission but an X-ray point source was not identified. In these regions, either the X-ray emission was too diffuse to be categorized as a point source (and therefore failed to be included in our catalog), or the X-ray source failed to meet our \pns criteria of $<4\times10^{-6}$. For that reason, we additionally release an expanded version of each X-ray source catalog, where X-ray sources are required only to meet a \pns threshold of $10^{-3}$. 

The positions and properties of the sources in the expanded X-ray point source catalogs (for all five CLVS galaxies) are listed in Table~\ref{srclist}. Column (1) provides a unique X-ray ID number, with the highly significant sources listed first according to increasing right ascension (J2000.0), followed by the marginally detected (and mostly spurious) sources (marked with an asterisk). Column (2) provides the source catalog name, using RA and Decl. nomenclature. Columns (3)-(4) provide the RA and Decl., respectively (J2000.0). Column (5) provides the positional uncertainty of the X-ray sources, in arcseconds. Column (6) lists the off-axis angle from the nominal \Chandra aim-point, in arcminutes. For sources that were detected during multiple \Chandra observations, the minimum off-axis angle is reported. Column (7) provides the net counts detected in the 0.35-8 keV energy band for all available \Chandra observations. Column (8) provides the log of 0.35-8 keV \pns value for each source\footnote{Although Table~\ref{srclist} only lists the \pns value in the 0.35-8 keV band, \pns values in all energy bands will be included in the FITS catalogs.}. Column (9) provides the unabsorbed 0.35-8 keV flux predicted for each source, assuming a power law spectrum with \PL=1.9 and absorption due to the Galactic column only.

\subsection{Sensitivity Maps}\label{sens_maps}
The remainder of the analyses in this work require sensitivity maps, which provide the energy flux level at which a source could be detected for each point in our survey area. We convert the \AEx-provided photon flux for each source ({\it flux2}, which is based on the net source counts, exposure time, and the mean auxiliary response file (ARF) in the given energy band) into energy fluxes assuming a power law with $\Gamma$=1.9 that is absorbed by the appropriate Galactic column density for each galaxy. Since the majority of X-ray sources in each catalog are either XRBs or background AGN, we expect this model to be appropriate and not to systematically bias our subsequent analyses.

To create sensitivity maps, we calculated the number of source counts that would be required to meet our \pns criteria ($<4\times10^{-6}$) for each point in the survey area, following \cite{Georgakakis+08}. The limiting fluxes (and corresponding luminosities) in the 0.35-8, 0.5-2, and 2-8 keV bands for each galaxy are summarized in Table~\ref{sens}. Column (1) lists the galaxy and column (2) lists the energy band, in keV, being considered. Columns (3)-(4) provide the fluxes and corresponding luminosities to which 70\% of the survey area is complete, respectively. Columns (5)-(6) provide the fluxes and corresponding luminosities to which 90\% of the survey area is complete, respectively. Columns (7)-(8) provide the fluxes and corresponding luminosities to which 95\% of the survey area is complete, respectively. Columns (9)-(10) list the flux and luminosity, respectively, of the faintest source that met the \pns threshold to be included in our catalogs. All fluxes are reported in units of $10^{-15}$ \flux, and all luminosities are in units of $10^{36}$ \lum.

\section{The X-ray Point Source Populations}\label{xray}
The X-ray properties of individual sources, such as their spectral shape and temporal variability, can be used to constrain the physical origin of the X-ray emission. Additionally, bulk population properties (such as the radial distribution of sources) enable us to identify different X-ray populations statistically. We expect the X-ray point source populations of NGC~55, NGC~2403, and NGC~4214 to consist of HMXBs, LMXBs, SNRs, background AGN, and foreground stars. We use our X-ray point source catalogs to search for time-variable sources, analyze hardness ratios (for faint sources) and perform spectral fitting (for bright sources), and construct radial source distributions. In a follow-up paper, we consider the properties of the source populations in more depth, and discuss the implications of our data for scenarios of HMXB evolution.

\subsection{Hardness Ratios}
The X-ray spectrum of a source can provide a key diagnostic for separating different populations. However, it is not reliably possible to constrain spectral parameters to any degree of accuracy for sources with $\lesssim50$ counts. The majority of the CLVS X-ray sources fall within this low-count regime.

Instead of directly measuring the X-ray spectral shape, we define hardness ratios ($HRs$, also called X-ray colors) to separate different populations. $HRs$ measure the fraction of photons emitted by a source in different energy ranges. We use the following energy bands in our analysis: soft ($S$, 0.35-1.1 keV), medium ($M$, 1.1-2.6 keV), and hard ($H$, 2.6-8 keV). We evaluate two $HRs$ for each source using the same approach as in Papers~I and II. Source and background counts were determined by \AEx in each band, and we define a ``soft'' color,

\begin{equation}
HR1 = \frac{M-S}{H+M+S},
\end{equation}

\noindent and a ``hard'' color,

\begin{equation}
HR2 = \frac{H-M}{H+M+S}.
\end{equation}

We use the Bayesian Estimation of Hardness Ratios\footnote{\url{http://hea-www.harvard.edu/AstroStat/BEHR/}} \citep[{\it BEHR};][]{Park+06} to determine the $HRs$ for each X-ray source in NGC~55, NGC~2403, and NGC~4214. For faint sources, the net counts in a band can be negative due to fluctuations in the source and background estimates, resulting in unphysical $HRs$ when calculated using a traditional, non-Bayesian approach. The {\it BEHR} code accounts for the fact that source and background counts are non-negative, ensuring that all $HRs$ are physically meaningful. In the high-count regime, the results of the Bayesian approach become equivalent to traditionally-computed $HRs$.

While the {\it BEHR} code accounts for the fact that source and background counts are non-negative, it can only compute a hardness ratio for two input bands (e.g., only using $H$ and $S$) and does not directly handle the three-band forms such as those defined above. We therefore use the {\it BEHR} code to construct 50,000 probability distributions for the $S$, $M$, and $H$ counts. Using these distributions, we compute 50,000 values for $HR1$ and $HR2$ for each source. The $HR$ value is defined as the mean of the distribution, and the credible interval is evaluated based on the 68.2\% equal-tail estimates (i.e., 0.682/2 of the samples have values below the lower limit, and 0.682/2 of the samples have values above the upper limit). We note that the median $HR$ values (with a credible interval defined by the 16\% and 84\% values) did not differ significantly from the mean values reported. 

For each source, the inputs to {\it BEHR} are source counts, background counts, the \AEx ``backscale'' parameter (which accounts for the ratio of the source and background extraction areas and efficiencies), and a factor converting from counts to photon flux (i.e., the exposure time multiplied by the mean ARF over the extraction region). We additionally set the ``softeff'' and ``hardeff'' parameters equal to the exposure time multiplied by the mean ARF values computed by \AEx for each observation in which the source was observed. The ``eff'' parameters take into account variations in effective area and exposure times between different observations and different energy bands, so that all $HR$ calculations are normalized to the same instrument (e.g., one could correct for the differences in low-energy response of an ACIS-S detection when comparing to an ACIS-I detection). This is especially important considering the differences in ACIS filter contamination between the observations utilized in this work. For each energy band, we set the {\it BEHR} ``burnin'' parameter to 50,000 and the ``total draws'' parameter to 100,000.

To aid in the interpretation of our $HR$ calculations, we use a version of the X-ray color-color classification used in \cite{Kilgard+05} and \cite{Prestwich+03}, modified here to account for our different definition of the $HRs$. The $HR$ classes are described in Table~\ref{color_class}. Column (1) lists the X-ray source classification category. Column (2) provides the definition of the classification category in terms of $HR1$ and $HR2$. Column (3) lists the number of sources falling within each classification. Six categories of X-ray sources are defined: `XRB/AGN'-like (which contains both XRBs and background AGN, given their similar $HR$s), `SNR'-like (which likely also contains foreground stars and super soft sources), `ABS' for heavily absorbed sources, indeterminate `HARD' and `SOFT' sources, and `INDET' for sources with an indeterminate spectral shape.  We note that these classifications are uncertain - for example, a heavily obscured SNR could exhibit X-ray colors consistent with being an `XRB/AGN'-like source. However, for low-count sources, $HRs$ are the only method for constraining the underlying spectral shape. Additional information (e.g., variability, optical counterpart, etc.) is required to refine each source's classification.

\begin{figure*}
\centering
\includegraphics[width=0.85\linewidth,clip=true]{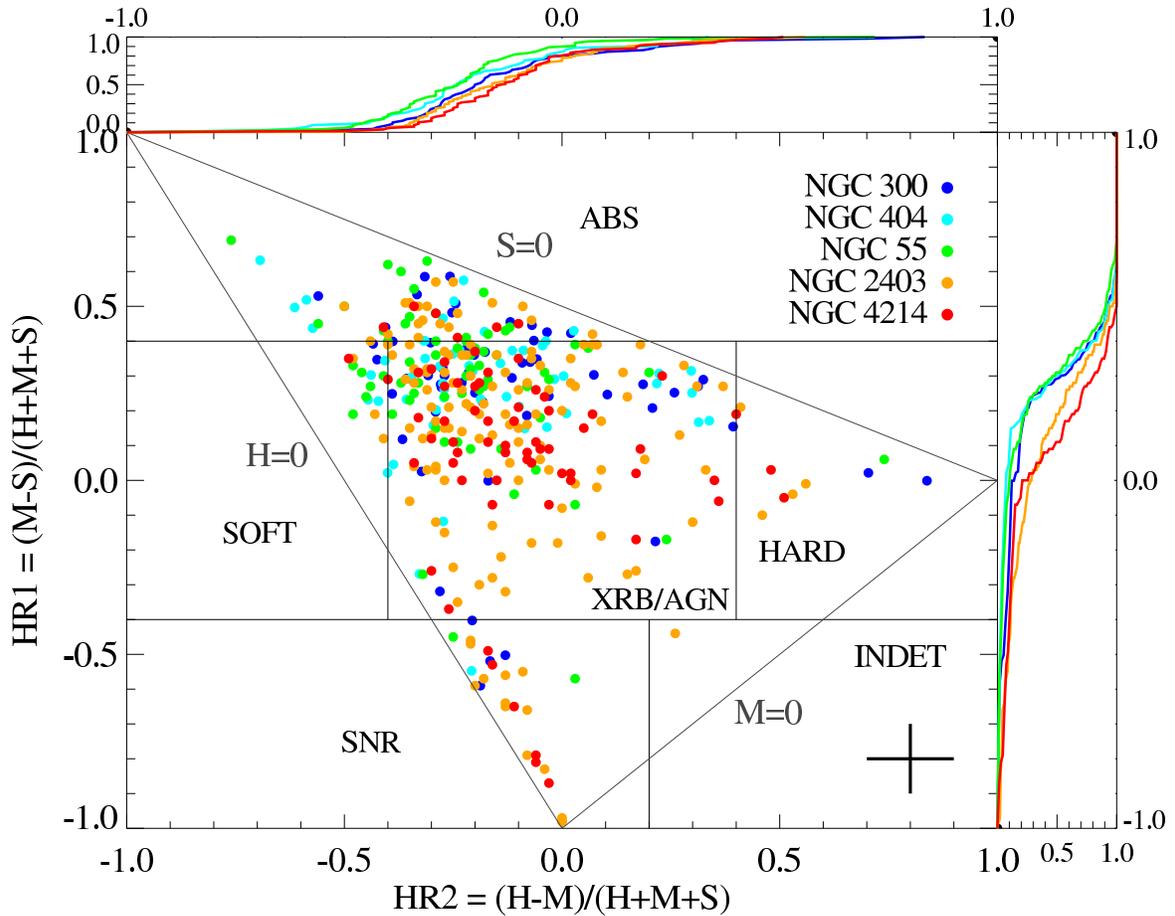} 
\caption{X-ray color-color diagram of all X-ray sources with more than 20 counts detected in the five CLVS galaxies. The thick cross in the lower-right corner shows the typical size of the errors. The diagram has been broken into preliminary source identification regions.  The regions are labeled as: `ABS' for absorbed sources, `XRB/AGN' for X-ray binaries and AGN, `SNR' for supernova remnants, `SOFT' for indeterminate soft sources, `HARD' for indeterminate hard sources,  and `INDET' for sources with an indeterminate spectral shape. The gray lines indicate the zero-count limits in the hard (`H'), medium (`M'), and soft (`S') bands. The top and right panels show the cumulative fraction of sources at each HR for each of the five galaxies.}
\label{HRplot}
\end{figure*}

The $HRs$ and corresponding source classifications for each galaxy are given in Table~\ref{HRtable}. Columns (1)-(4) provide the unique source number, $HR1$, $HR2$, and the classification, respectively, of the sources in NGC~55. Columns (5)-(8) provide the same information for sources in NGC~2403, and Columns (9)-(12) provide the corresponding information for sources in NGC~4214. Figure~\ref{HRplot} shows the $HR$ values for all sources with more than 20 net counts in the 0.35-8 keV band for all five CLVS galaxies, and hints at some differences in the X-ray source populations of the five CLVS galaxies. For example, NGC~404 (cyan) shows significantly fewer sources in the `SNR'-like category than the disk galaxies. This is not surprising, as one would not expect to find many SNRs in the older stellar populations and earlier morphology of NGC~404. The three `SNR'-like sources in NGC~404 are likely to be foreground stars. The top and right panels of Figure~\ref{HRplot} show the cumulative fraction of sources at each HR for all five galaxies. 

To compare the X-ray source populations of the five CLVS galaxies in a quantitative manner, we calculate the fraction of each HR class found for each galaxy and for the total CLVS sample (Table~\ref{HRclass_percent}). Column (1) lists each galaxy, and Columns (2)-(6) provide the percentage (\%) of X-ray sources in that galaxy falling within each classification as determined solely by their hardness ratios. We find that most of the X-ray sources fall within the `XRB/AGN' category ($\sim58-67$\%), with an additional $\sim$12-20\% of sources showing evidence for additional absorption beyond the Galactic column. These two categories make up approximately three quarters of all the X-ray point sources and are likely to contain the majority of XRBs associated with their host galaxies with some contamination from background AGN (whose X-ray spectra are similar to those of many XRBs). 

The softer HR categories (i.e., `SNR' and 'SOFT') likely contain contamination by foreground stars. This is especially true for the $\sim$4\% X-ray sources in NGC~404 designated as `SNR'-like by their hardness ratios. The fraction of `SNR'-type sources in NGC~404 is below the sample average by $\sim$3$\sigma$, as would be expected when comparing an S0-type galaxy with very little recent star formation to the later-type spirals and irregulars. `SOFT' sources are observed in comparable fractions to `SNR'-like sources, and may consist of very soft XRBs, background sources with unusual absorption properties, or foreground objects. A small fraction of X-ray sources, $<8$\%, are classified as `HARD.' These sources are likely background AGN that have experienced a high degree of absorption.

\subsection{X-ray Spectral Analysis}\label{spectral_analysis}
For sufficiently bright sources, the X-ray spectrum can constrain the physical origin of the X-ray emission. Both XRBs and AGN are typically described by power laws, with photon indices ranging from $\Gamma\sim1-2$. If the source is a likely XRB, the photon index can sometimes be used to constrain the nature of the compact object: NS primaries typically exhibit harder X-ray emission, with $\Gamma<1.5$, while BHs produce $\Gamma\sim$2-2.5 (or softer) due to the lack of a solid surface \citep{McClintock+06}. Foreground stars and SNRs exhibit significantly softer X-ray emission with emission lines indicative of a thermal plasma. 

Spectra were extracted by \AEx using the CIAO tool \texttt{dmextract}, and response files were created using \texttt{mkacisrmf} and \texttt{mkarf}. Due to the low number of counts found for the majority of sources, the traditional approach of binning the spectrum and using the $\chi^2$ statistic would result in significant biases in our spectral models (especially at high energies). Instead, we utilize the unbinned spectra and the Cash statistic \citep[see e.g.,][]{Cash79,Nousek+89,Humphrey+09,Arnaud+11}. 

We define the $C$-statistic as our ``fit statistic'' (i.e., to be minimized during the fit) and use the Pearson $\chi^2$ as our ``test statistic'' for the same fit. Additionally, we use the \texttt{goodness} command \citep{Arnaud96} to perform 5,000 Monte Carlo simulations of each spectrum to evaluate the quality of the fit; if the observed spectrum was described well by that model, the ``goodness'' should be $\sim$50\%. In most of the high-count cases (e.g., those sources with $>$500 counts), spectra with a \texttt{goodness} in the range of $\sim$40-60\% correspond with a minimized $C$-statistic and a reduced $\chi^2$ near unity. While it has been shown \citep{Cash79,Nousek+89,Humphrey+09,Arnaud+11} that the $C$-statistic is preferable over $\chi^2$ for sources with only a few hundred counts or less, both the $C$-statistic and $\chi^2$ approaches converge when the number of counts in a spectrum is large. We therefore do not expect the ``test'' Pearson $\chi^2$ to be a reliable indicator of fit quality for low-count sources. However, the ``test'' $\chi^2$ of the best-fit models for those sources with $>$500 counts is consistently near unity.

We performed spectral fits for all sources with $>$50 net counts. For sources with $<$500 counts, the general shape of the source spectrum can be measured; for brighter sources with $\gtrsim$500 counts, further details of the X-ray spectrum (e.g., thermal emission features) begin to become apparent. The spectral fitting results for all 136 sources with $>$50 net counts are provided in Table~\ref{spectralfits}, and the 19 sources with $>$500 counts are discussed in more detail in the following subsection. We used \texttt{XSPEC} v12.8.1g to perform all spectral fitting, and all errors represent the 90\% uncertainties. All reported fluxes and luminosities are unabsorbed, unless otherwise noted.

Column (1) lists the source number in each galaxy. Column (2) provides the number of net counts observed in the 0.35-8 keV energy band. Sources with $>$500 net counts, discussed in detail below, are indicated with an asterisk. Column (3) lists the name of the best-fit model. Column (4) provides the amount of additional absorption beyond the Galactic column required by the spectrum. If no additional absorption was required, this field was left blank. Column (5) lists the power law photon index of the spectrum, if appropriate. This field is left blank for sources that did not require a power law component. Column (6) lists the temperature, in keV, of the thermal component of the spectrum, if appropriate. This field is left blank for sources that did not require a thermal component. Column (7) lists the degrees of freedom in each fit. Column (8) provides the $C$-statistic of the fit. Column (9) provides the Pearson $\chi^2$ value of the fit. Columns (10)-(11) provide the unabsorbed fluxes yielded by the best-fit model, in the 0.5-2 keV and 0.35-8 keV bands respectively. Column (12) provides the goodness of the fit as indicated by the \texttt{goodness} command; values near $\sim$40-60\% indicate the spectrum is well described by the model.

\subsection{Detailed Models of Bright Sources}
\begin{figure*}
\centering
\begin{tabular}{cc}
	\begin{overpic}[width=0.33\linewidth,angle=-90,clip=true]{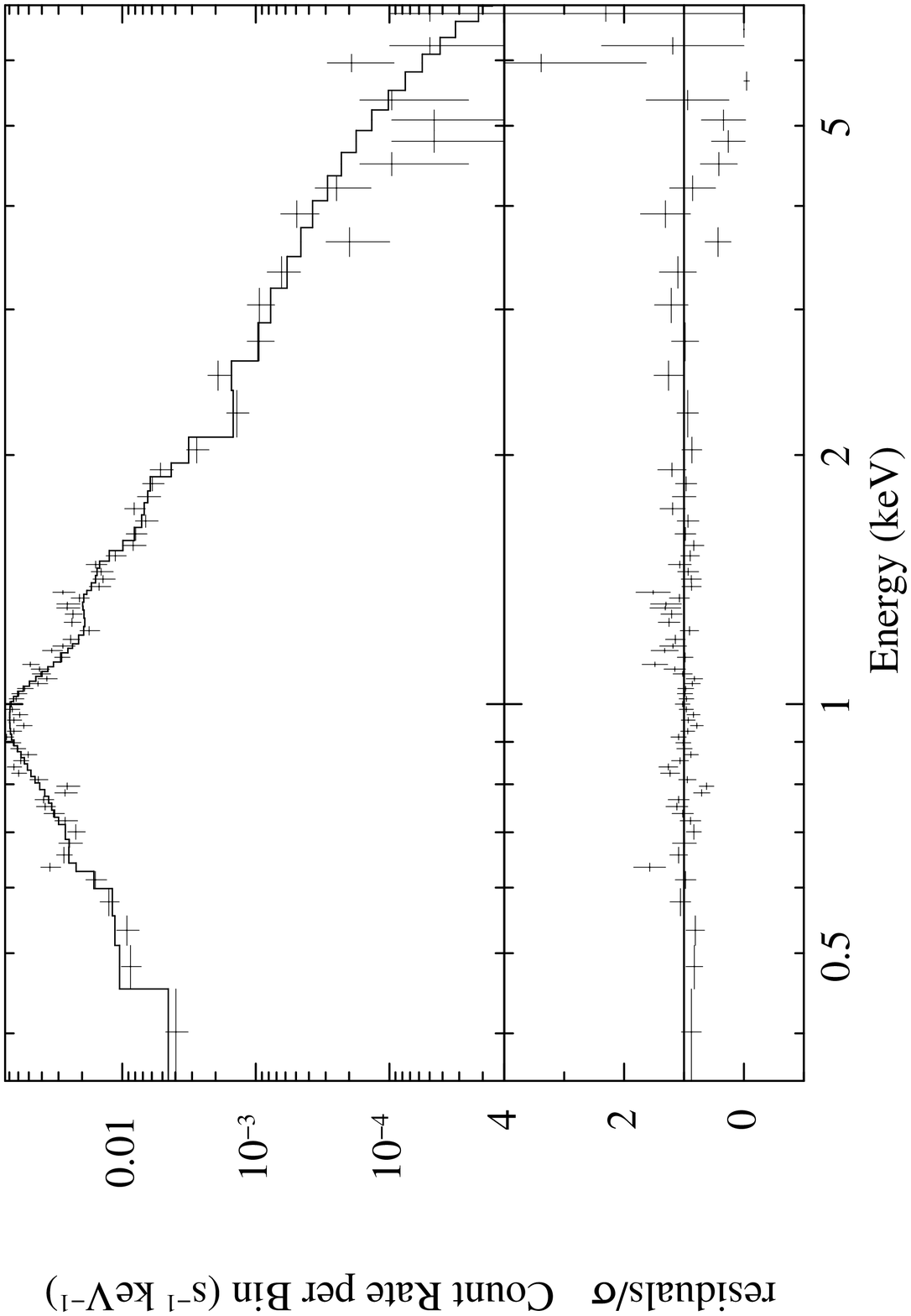} 
 		\put (79,60) {Source 23}
	\end{overpic}	&
	\begin{overpic}[width=0.3\linewidth,angle=-90,clip=true]{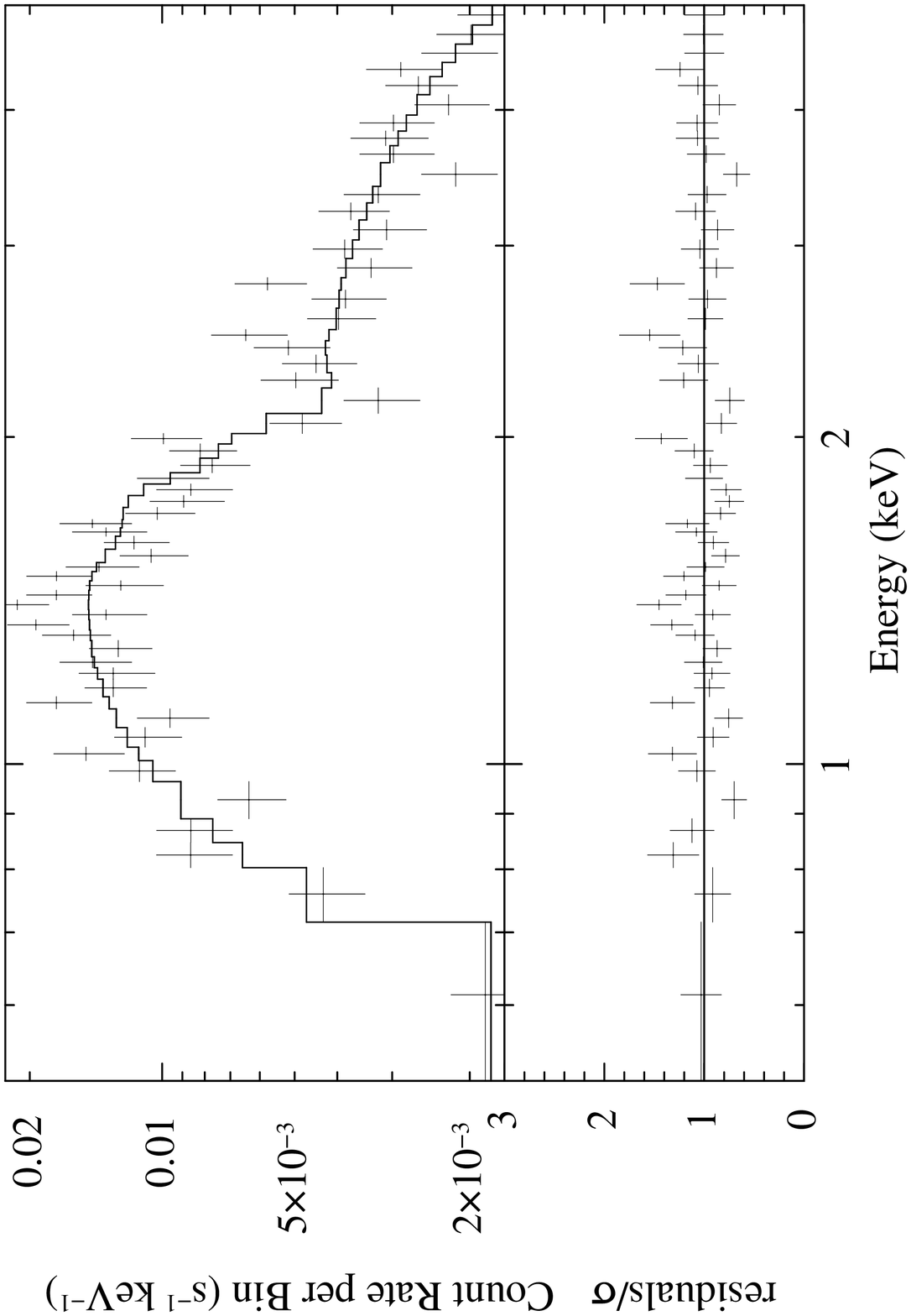} 
 		\put (79,60) {Source 62}
	\end{overpic}	\\
	\begin{overpic}[width=0.3\linewidth,angle=-90,clip=true]{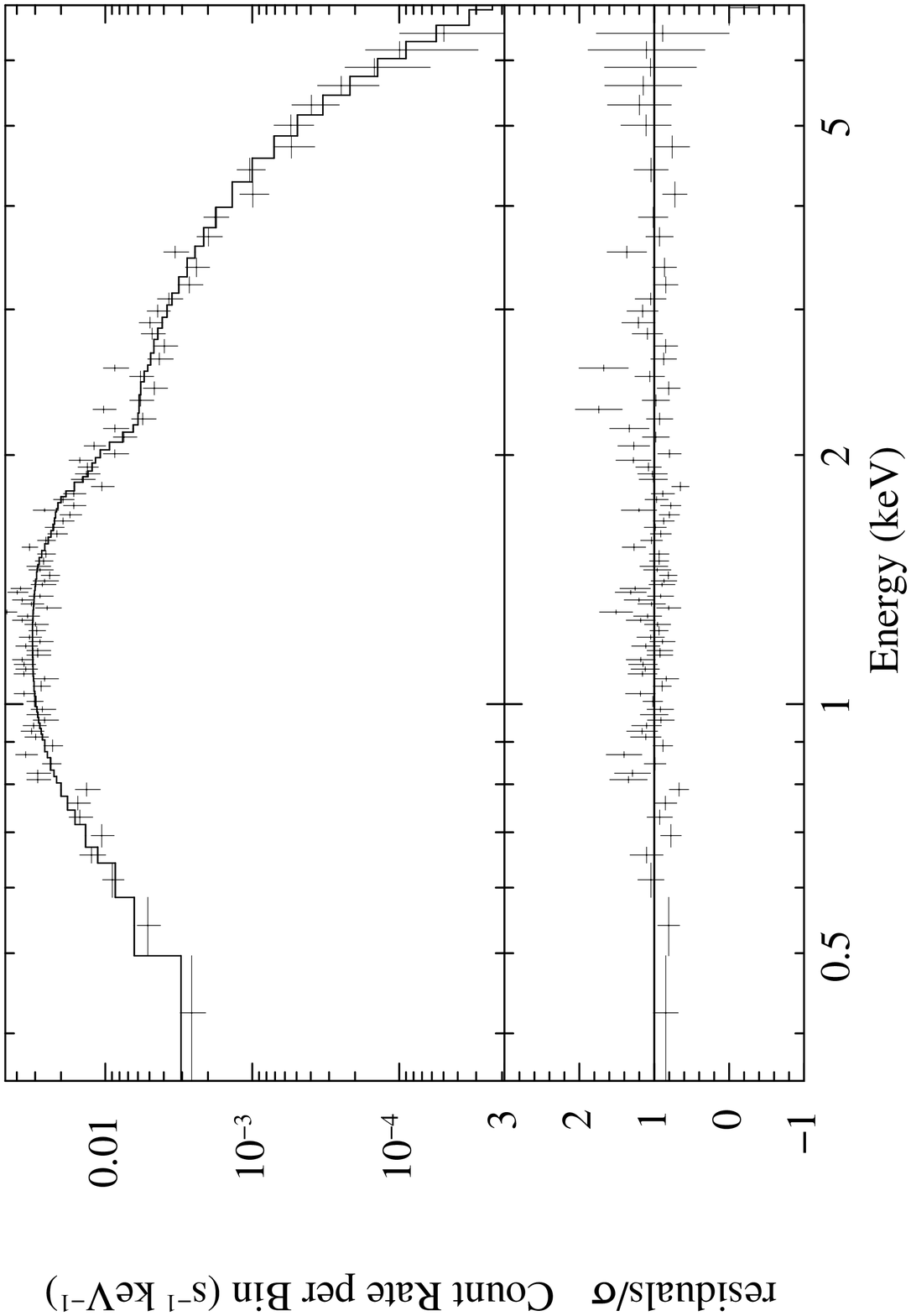} 
 		\put (79,60) {Source 63}
	\end{overpic}	&
	\begin{overpic}[width=0.32\linewidth,angle=-90,clip=true]{NGC55_src119_spectrum.eps} 
 		\put (79,60) {Source 119}
	\end{overpic}	\\
\multicolumn{2}{c}{
	\begin{overpic}[width=0.3\linewidth,angle=-90,clip=true]{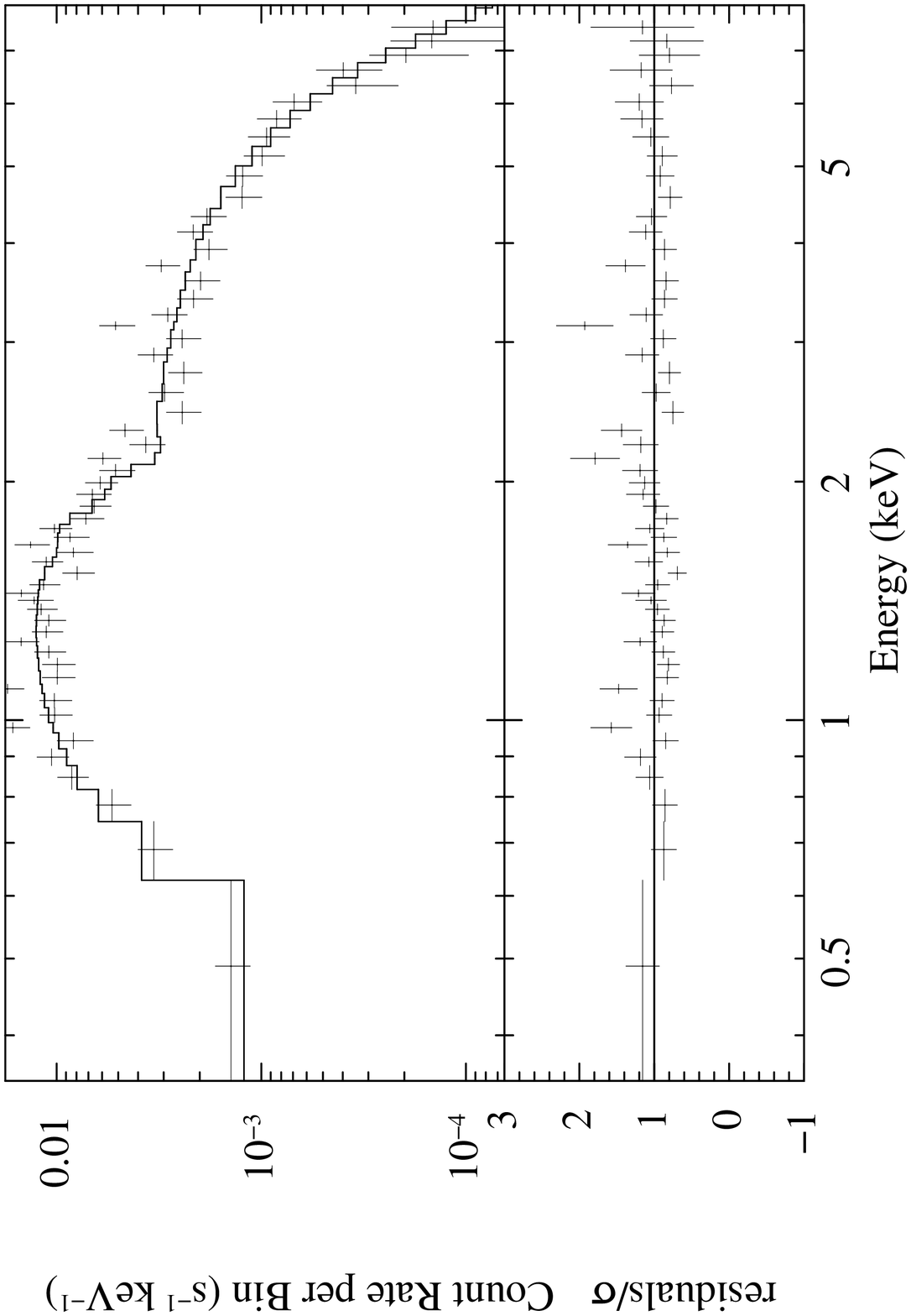} 
 		\put (79,60) {Source 122}
	\end{overpic}}		\\
\end{tabular}
\caption{The spectra (with best fit models superimposed) and residuals for the NGC~55 sources with more than 500 net counts. Spectra are binned for display purposes only.}
\label{bright_spec55}
\end{figure*}

Out of the combined 460 X-ray sources detected in NGC~55, NGC~2403, and NGC~4214, only 19 have more than five hundred net counts. In this subsection, we summarize the best fit spectral models found for each of these bright sources. A handful of these sources were sufficiently bright ($\gtrsim$10,000 counts) that photon pile-up presented a concern; for these cases, we use the \texttt{XSPEC} model \texttt{pileup} to account for photon pile-up \citep{Davis01}\footnote{See also \url{cxc.harvard.edu/ciao/download/doc/pileup\_abc.pdf}}. The spectra and corresponding best-fit models are shown in Figures~\ref{bright_spec55}-\ref{bright_spec4214}, where they have been binned for display purposes only. 

\subsubsection{NGC~55}

{\bf Source 23} has a total of 2,692 net counts. Although the spectrum was obviously dominated by thermal emission, a single-component \texttt{apec} model provided a poor fit. The addition of a second \texttt{apec} component and a power-law component were necessary to provide an acceptable fit. We find no evidence for additional absorption beyond the Galactic column. This source was observed using \XMM \citep[][their source 39]{Stobbart+06a}, and a Galactic RS CVn origin was suggested to explain the X-ray emission. Our low value of the absorption column and consistent thermal plasma temperatures support this conclusion. 

{\bf Source 62} has a total of 1,786 net counts detected in the archival \Chandra observations. A two-component model consisting of a power law and disk blackbody with additional absorption beyond the Galactic column was necessary to produce an acceptable fit. These results are comparable to those found by \cite[][their source 47]{Stobbart+06a} using \XMM. The inferred 0.35-8.0 keV X-ray luminosity of source 62 ($\sim$2.8$\times10^{38}$ \lum) is near the Eddington luminosity of a $\sim$2 \Msun compact object, consistent with an LMXB.

{\bf Source 63} contained 2,987 counts. Although the \XMM spectrum \cite[][their source 43]{Stobbart+06a} was best fit by an absorbed power law (\PL=$0.82\pm0.04$) and disk blackbody ($kT_{\rm in}=4.92^{+0.74}_{-0.78}$ keV) model, we did not require an additional disk blackbody to obtain an acceptable fit. Instead, our best fit model to the archival \Chandra data was a single-component power-law with an index of 0.9$\pm$0.1, comparable to the power-law component measured from the \XMM observations. 

{\bf Source 119} is the NGC~55 ULX designated XMMU J001528.9-391319 \citep{Stobbart+04}. It was detected in the archival \Chandra observations with 11,523 net counts. Due to the large number of counts detected for this source, we included pile-up in our spectral model. The estimated fraction of frames in the ACIS detector that have detected two or more pileup events is $\sim$5\% (see footnote above). The grade migration factor $\alpha$ was allowed to vary; we found a best-fit value of 0.28$\pm$0.04. Due to the large number of counts in the spectrum, our best fit model included three components: a power law, a thermal plasma, and a disk blackbody, all subject to absorption in excess of the Galactic column. The implied 0.35-8 keV luminosity for our best-fit model is 1.4$\times10^{39}$ \lum. Taken at face value, this luminosity implies a black hole mass of $\sim$12 \Msun, consistent with the \cite{Stobbart+04} result that this source is a dipping BH binary with a mass of $>$11 \Msun and whose luminosity varied from 8.9$\times10^{38}$ \lum to 1.6$\times10^{39}$ \lum in the 0.5-10 keV energy band. The thermal components account for $\sim$90\% of X-ray flux, somewhat different from the results of \citep{Stobbart+04}, who observed nearly $\sim100$\% of the X-ray flux originating in the power law component at the beginning of the \XMM observation and $<$27\% of the flux from the power law component towards the end of their observation. 

{\bf Source 122} has a total of 1,516 net counts in the archival \Chandra observations. We find the best fit model is an absorbed power law. The implied 0.35-8 keV luminosity of the source is $\sim2\times10^{38}$ \lum, consistent with a $\sim$1.4 \Msun NS accreting at the Eddington limit.

\subsubsection{NGC~2403}
{\bf Source 2} has a total of 904 net counts in the 0.35-8 keV energy band. A three component model consisting of a power law with two thermal emission components was necessary to produce an acceptable fit.  No absorption beyond the Galactic column is necessary to fit the spectrum. At the distance of NGC~2403, the best-fit model flux corresponds to a luminosity of 4.5$\times10^{37}$ \lum in the 0.35-8 keV band. 

{\bf Source 41} was designated source 20 in a previous analysis of Obs ID 2014 only \citep{Schlegel+03}. The spectrum was found to be equally well-fit with bremsstrahlung emission with $kT\sim3.5-4$ keV or a power law with \PL$\sim$1.9. A total of 10,033 net counts were detected in the 0.35-8 keV energy band from all available \Chandra observations. The spectrum was best fit by a power law plus a disk blackbody, and the results are similar to, but slightly softer than \citep{Schlegel+03}. An additional absorption component was required in our model. Due to the large number of counts, we included pile-up in our model; the pile-up fraction was $\sim$3\%. The 0.35-8.0 keV luminosity is 8.7$\times10^{38}$ \lum, making this a black hole candidate.

{\bf Source 42} is a ULX, originally discovered by the {\it Einstein} observatory \citep{Fabbiano+87}, with a BH mass $M\sim10-15$ \Msun \citep{Isobe+09}. Later observations with {\it ASCA} revealed a multi-color disk spectrum, with an innermost disk temperature of $kT_{\rm in}\sim1$ keV and an innermost disk radius of $R_{\rm in}\sim$130 km \citep{Kotoku+00}. Additional follow-up observations of this source with {\it Suzaku} and \Chandra Obs ID 4628 are consistent with this model \citep{Isobe+09}, although the \Chandra observation Obs ID 4630 is consistent with a power law model (\PL$\sim2.7$), bremsstrahlung (with $kT\sim$1.8 keV), or a multicolor disk model \citep[$kT_{\rm in}\sim$0.6 keV][]{Schlegel+03}. 

An acceptable fit to the 16,601 net counts in the 0.35-8 keV band for this source is reached using a power law plus a disk blackbody with a pileup fraction of $\sim$4\%. The best-fit model yields a corresponding 0.35-8 keV luminosity of ${\sim}2\times10^{39}$ \lum at the distance of NGC~2403. Taken at face value, this luminosity is consistent with a BH mass of $\sim$15 \Msun, consistent with the observations of \cite{Isobe+09}.

{\bf Source 51} has 658 net counts in the 0.35-8 keV energy band. This source was previously classified as a transient source within the nuclear star cluster of NGC~2403 \citep{Yukita+07}. Previous work has found the X-ray spectrum of this source to be consistent with a disk blackbody model \citep{Makishima+86,Makishima+00}.   We also find the best fitting model to the 0.35-8 keV spectrum is a disk blackbody, and with a relatively low 0.35-8.0 keV luminosity of only 4$\times$10$^{37}$ \lum.  We find no evidence for rapid variability, although the transient nature of the source is discussed in \cite{Yukita+07}. The relatively soft X-ray emission and high X-ray luminosity in earlier observations ($\sim$7$\times10^{38}$ \lum) are suggestive of a BH with $M\sim5$ \Msun. 

{\bf Source 62} required two thermal components and a power law to obtain a reasonable fit to the 5,433 detected counts. We note that this X-ray source is coincident with a foreground star, with USNO-B1.0 apparent magnitudes of $m_B=8.8$, $m_R$=8.1, and $m_I=7.9$. This source was also detected in the 2MASS catalog, with $J=7.5$, $H=7.2$, and $K=7.1$. With a $R-I$ color of $\sim0.2$, the star is likely a mid-to-early F type. Assuming an F-star absolute magnitude of $\sim$3.5 implies a distance to the source of $\sim$100 pc. We therefore estimate the 0.35-8 keV flux of Source 62 is $\sim7.5\times10^{29}$ \lum. Both the implied luminosity of Source 62 and the two-temperature thermal plasma fit to the spectrum are typical of other nearby F stars in the Milky Way \citep{Micela+96,Prosser+96,Dahm+07}.

{\bf Source 64} has a total of 772 net counts in the 0.35-8 keV energy band. The spectrum was best described with a two thermal emission components; the best fit model implies a 0.35-8.0 keV luminosity of 3$\times10^{37}$ \lum at the distance of NGC~2403. Due to the soft, thermal spectrum, we also compute the corresponding luminosity of Source 64 assuming it is a foreground star. Assuming a distance of 4 kpc (an appropriate distance for a Milky Way object in the direction of NGC~2403, which is $\sim$30$^{\circ}$ above the disk) yields a luminosity of $\sim$5$\times10^{31}$ \lum, typical of low mass stars.

{\bf Source 78} has 541 net counts, requires a power law plus a thermal emission component to provide an acceptable fit in the 0.35-8 keV energy band. This source was included in the USNO-B1.0 catalog with $m_R=11.7$ and $m_I=13.1$ and in the 2MASS catalog with $J=13.4$, $H=12.8$, and $K=12.7$. The source has been identified as a K4III giant at a heliocentric distance of $\sim$16 kpc \citep{Pickles+10}. At this distance, the corresponding X-ray luminosity is 1.5$\times10^{33}$ \lum in the 0.35-8 keV band, and the X-ray-to-optical flux ratio $log(f_X/f_V)$ of this source is -1.7.  A comparison to the \cite{Hunsch+98} catalog of X-ray emitting AFGKM giant stars shows that the implied X-ray luminosity of this source is ~2-3 orders of magnitude higher than for other K-type giant stars, and it is not consistent with X-ray emission from main-sequence K- and F-type stars \citep{Agueros+09}. It is therefore unlikely that this source is an isolated X-ray emitting star. The implied X-ray luminosity and X-ray-to-optical flux ratio are similar to Galactic bulge CVs \citep{vandenBerg+09}; the observed rapid X-ray variability (see Section~\ref{rapid_variability_section}) is further consistent with a CV origin \citep{Kuulkers+06}. The properties of this source are also consistent with a neutron star accreting from the wind of a late-type giant \citep[e.g., a symbiotic star, similar to GX 1+4;][]{Masetti+06}. However, our \HST imaging does not cover this source, making it impossible for us to investigate whether its optical properties support the CV-origin scenario.

{\bf Source 92} has a total of 5,375 net counts. An acceptable fit to the spectrum requires a power law plus a disk blackbody as well as additional absorption.  The fit yields a 0.35-8 keV luminosity at the distance of NGC~2403 of 4.3$\times10^{38}$ \lum. This source was designated source 1 in an earlier analysis \citep{Schlegel+03}.  The luminosity is near the Eddington limit for a $\sim$3\Msun compact object, putting it into a very interesting mass regime near the transition between neutron stars and black holes. Our results therefore support the interpretation of this source as a LMXB.

{\bf Source 135} has been identified as SN~2004dj, which has an estimated progenitor mass of 14-15 \Msun \citep{Maiz+04,Wang+05,Vinko+06}. This source, whose X-ray spectrum has evolved significantly with time, has been analyzed in detail by \cite{Chakraborti+12}. We perform a simpler analysis of the X-ray spectrum here for completeness. This source has a combined 1,258 net counts and requires 2 thermal components to provide an acceptable fit. At the distance of NGC~2403, the best-fit model predicts a 0.35-8 keV luminosity of $6.5\times10^{37}$ \lum. SN~2004dj is coincident with the compact cluster Sandage~96 \citep{Yamaoka04}, which is thought to host the progenitor star. For further discussion of the X-ray spectrum's evolution, we refer the reader to \cite{Chakraborti+12}.

{\bf Source 155} has 1,318 net counts in the 0.35-8 keV energy band. The spectrum is well-fit by a simple and hard absorbed power law. At the distance of NGC~2403, the best-fit model implies a 0.35-8 keV luminosity of $1.3\times10^{38}$ \lum. 

{\bf Source 162} has a spectrum that is well-fit by a simple absorbed power law. This source contains 957 net counts. At the distance of NGC~2403, the best-fit model implies a 0.35-8 keV luminosity of $7.3\times10^{37}$ \lum.

{\bf Source 170} has 599 counts and a spectrum that is well-fit by a simple absorbed power-law.  It's luminosity at the distance of NGC~2403 is about 5.5$\times10^{37}$.

{\bf Source 179} has 1,353 net counts. The 0.35-8 keV spectrum is also well-fit by an absorbed power law. At the distance of NGC~2403, the best-fit model implies a 0.35-8.0 keV luminosity of 1.9$\times10^{38}$ \lum. The spectrum of this source was previously found to be consistent with either a power law with a photon index of ${\sim}2$ (consistent with our best-fit model) or bremsstrahlung emission with a temperature of $kT{\sim}4$~keV \citep[][their source 28]{Schlegel+03}. 

\subsubsection{NGC~4214}

{\bf Source 16} was previously designated CXOU J121538.2+361921 \citep[][their source 11]{Dewi06} and has been identified as an X-ray binary with a period of 3.62 hours and an X-ray luminosity of a few 10$^{38}$ \lum. It has been proposed that the system consists of a slightly evolved helium star with $M\sim2-3$ \Msun and either a NS or low mass BH primary \citep{Dewi06,Ghosh+06}, although the favored interpretation is a NS-LMXB. This classification would make the system a direct progenitor of a double-NS binary. We detect 1,827 net counts in the 0.35-8 keV energy band for this source. The spectrum is well-fit by an absorbed power law, consistent with the analysis presented in \cite{Dewi06}. The best fit model implies a 0.35-8 keV luminosity of 2$\times10^{38}$ \lum at the distance of NGC~4214. Our measurements are therefore consistent with others of this source, and we support the interpretation as a helium star-NS LMXB.

\subsection{Radial Source Distributions}\label{radial}
Following the radial source distribution analysis in Paper~I, we assign an inclination-corrected, galactocentric distance to each X-ray source, assuming the galaxy center, inclination, and position angle for each host galaxy as given in Table~\ref{galbasic}. X-ray emitting foreground stars, found by visual inspection of both ground-based and \HST imaging (see Section~\ref{optical}) were removed. X-ray sources are then divided into radial bins based on their inclination-corrected distances from the center of their host galaxy. We use our 0.35-8 keV sensitivity maps (discussed in Section~\ref{sens_maps}) and the \lognlogs distribution of \cite{Cappelluti+09} to estimate the expected contamination by background AGN in each bin. The 2-8 keV AGN flux provided by \cite{Cappelluti+09} is converted to a 0.35-8 keV flux assuming the standard model (\PL=1.9 and \nH set to the Galactic value) with \texttt{WebPIMMS}\footnote{\url{https://heasarc.gsfc.nasa.gov/cgi-bin/Tools/w3pimms/w3pimms.pl}}. Since our observations are not sensitive to the full range of fluxes reported in the AGN \lognlogs distribution (in general, our flux sensitivity decreases with increasing de-projected distance from the galaxy center), we only use the detected X-ray sources with fluxes above the minimum flux of our least sensitive radial bin.

\begin{figure*}
\centering
\includegraphics[width=0.6\linewidth,clip=true]{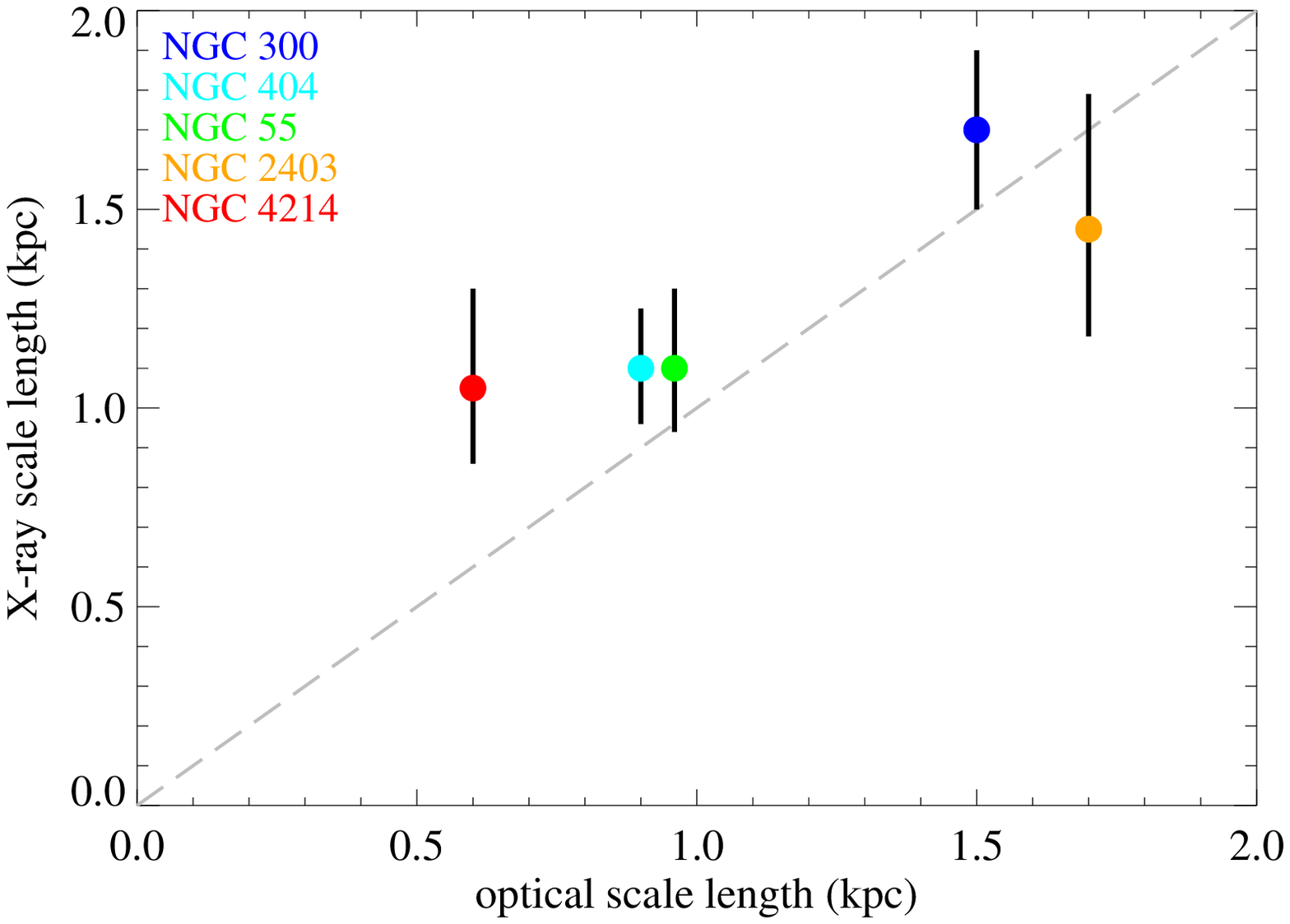} 
\caption{The X-ray scale lengths of each galaxy, derived from the radial source distribution for each galaxy, plotted against the optical scale lengths from the literature. The light gray dashed line shows a one-to-one correlation.}
\label{scale_length}
\end{figure*}

Each radial source distribution was then fit to an exponential disk profile (or, in the case of NGC~404, a de Vaucouleurs profile) plus a constant term to represent the background, and the disk scale length (or, for NGC~404, the effective radius) was measured. The X-ray scale lengths obtained for NGC~55, NGC~2403, and NGC~4214 are 1.10$^{+0.20}_{-0.16}$ kpc, 1.45$^{+0.34}_{-0.27}$ kpc, and 1.05$^{+0.25}_{-0.19}$ kpc, respectively. The X-ray effective radius found for NGC~404 is 1.10$^{+0.17}_{-0.15}$ kpc. The radial source distributions and the best fit profiles are shown in Figure~\ref{radial}. The radial source distribution and fit for NGC~300 are described in Paper~I. Figure~\ref{scale_length} compares our X-ray results to the optical scale lengths found in the literature: 0.96 kpc for NGC~55 \citep{Seth+05}, 1.7 kpc for NGC~2403 \citep{Barker+12}, 0.6 kpc for NGC~4214 \citep{Williams+11}, and an effective radius of 0.9 kpc for NGC~404 \citep{Baggett+98}. The best-fit line through the data yields a slope of 0.68$\pm$0.03, and a Spearman-Rank test yields a $\sim$5\% probability of the X-ray and optical scale lengths being correlated. NGC~4214, the only dwarf irregular galaxy in our sample, shows the largest discrepancy between the X-ray and optical scale lengths, indicating the X-ray sources are more widely dispersed from the center than the optical light. This may be partially due to the ages of the underlying stellar populations: although the inner regions of NGC~4214 have recently (within the last $\sim$100 Myr) experienced a vigorous burst of star formation \citep{Williams+11} resulting in a few HMXBs (which follow the light distribution), the majority of stars in NGC~4214 are old ($<$1\% of stars have ages $\lesssim$50 Myr). LMXBs formed from these older stars would be expected to follow the mass distribution of the galaxy, which may have a larger scale length compared to the brightest stars. If we remove the NGC~4214 scale length from our sample, the slope of the best-fit line through the remaining four galaxies becomes 0.82$\pm$0.06, and a Spearman-Rank test yields a 26\% probability of being correlated.

We additionally find the de-projected radius at which the observed X-ray source distribution reaches the background level. Beyond this distance, which we denote as $r_{\rm bkg}$, all observed X-ray sources are likely to be background AGN; within this distance, the observed X-ray sources will be a mix of background AGN and sources intrinsic to the host galaxy (e.g., XRBs and bright SNRs). Utilizing our sensitivity maps and observed radial source distributions, we estimate the fraction X-ray sources likely to be AGN as a function of de-projected distance from the galaxy (shown by the blue line in Figure~\ref{radial}). The expected AGN contribution as a function of radius is also summarized in Table~\ref{no_AGN}, where for each galaxy (column 1) we list the radii at which we expect AGN to make up $<$10\% (column 2), $\sim$25\% (column 3), $\sim$50\% (column 4), $\sim$75\% (column 5), and 100\% (e.g., $r_{\rm bkg}$; column 6) of X-ray sources.

\subsection{Time Variable Sources}\label{variability}
The three X-ray catalogs were systematically searched for sources showing both significant short-term (i.e., that occurs on timescales less than the observation exposure time) and long-term (i.e., occurring over multiple observations spanning months or years) variability in the 0.35-8 keV band. While rapid X-ray variability is routinely observed in both AGN and XRB systems \citep{Vaughan+03}, it is typically not observed in other sources of X-ray emission (i.e., SNRs). Unless otherwise noted, all fluxes or luminosities reported are unabsorbed. 

\subsubsection{Rapid Variability From \Chandra Observations}\label{rapid_variability_section}
We used a Kolmogorov-Smirnov (K-S) test to compare the cumulative photon arrival time distribution for each X-ray source to a uniform count rate model in each observation. The K-S test returns the probability (denoted by $\xi$) that both distributions were drawn from the same parent distribution. We consider sources with $\xi\leq10^{-3}$ rapidly variable; this value of $\xi$ corresponds to the $\sim$3$\sigma$ level. 

\begin{figure*}
\centering
 \includegraphics[width=0.85\linewidth,clip=true]{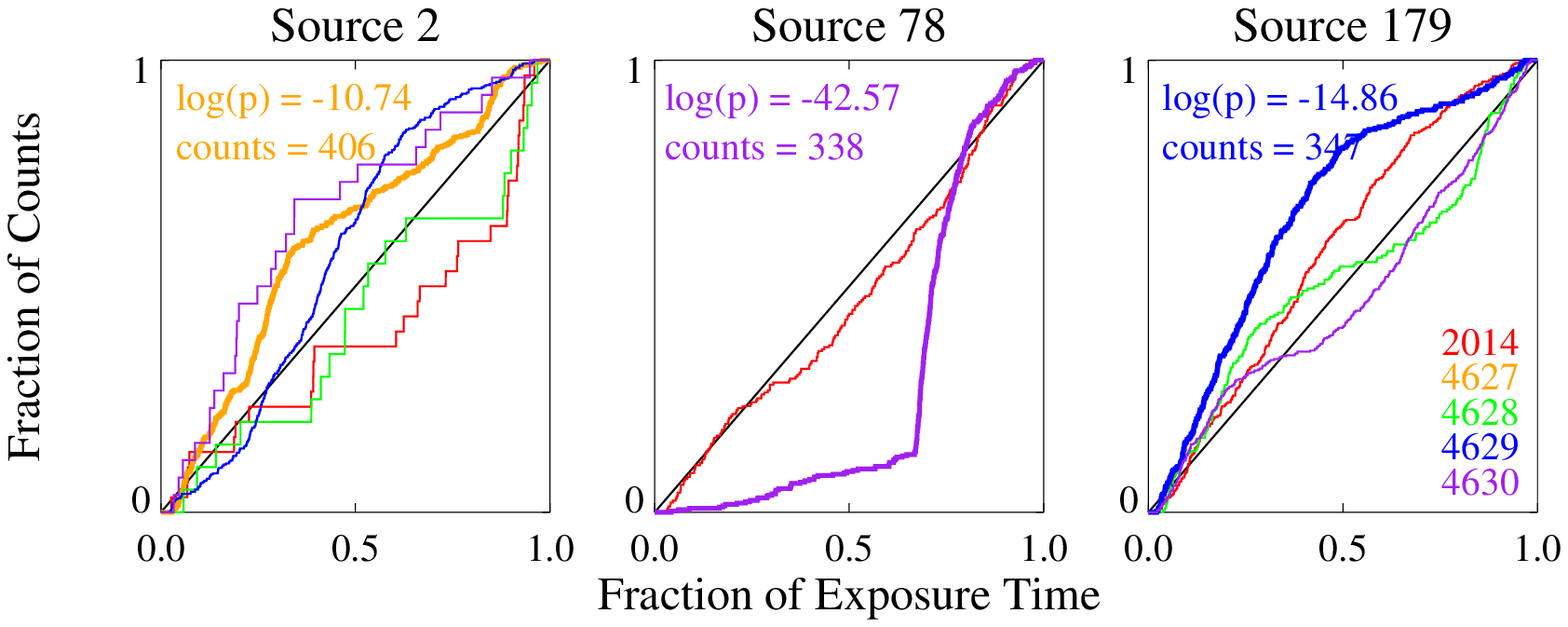} 
\caption{Cumulative photon arrival time distributions for X-ray sources in NGC~2403 showing evidence for rapid variability. The solid black line shows the expected distribution for a constant count rate.  Colored histograms and text indicate the cumulative distribution observed and statistics for different observations. The (color encoded) \Chandra Observation ID numbers are given in the rightmost panel. Each panel lists the (logarithmic) KS probability for no variability and the net 0.35-8 keV counts observed for each source.}
\label{short_2403}
\end{figure*}

\begin{figure*}
\centering
\includegraphics[width=1\linewidth,clip=true]{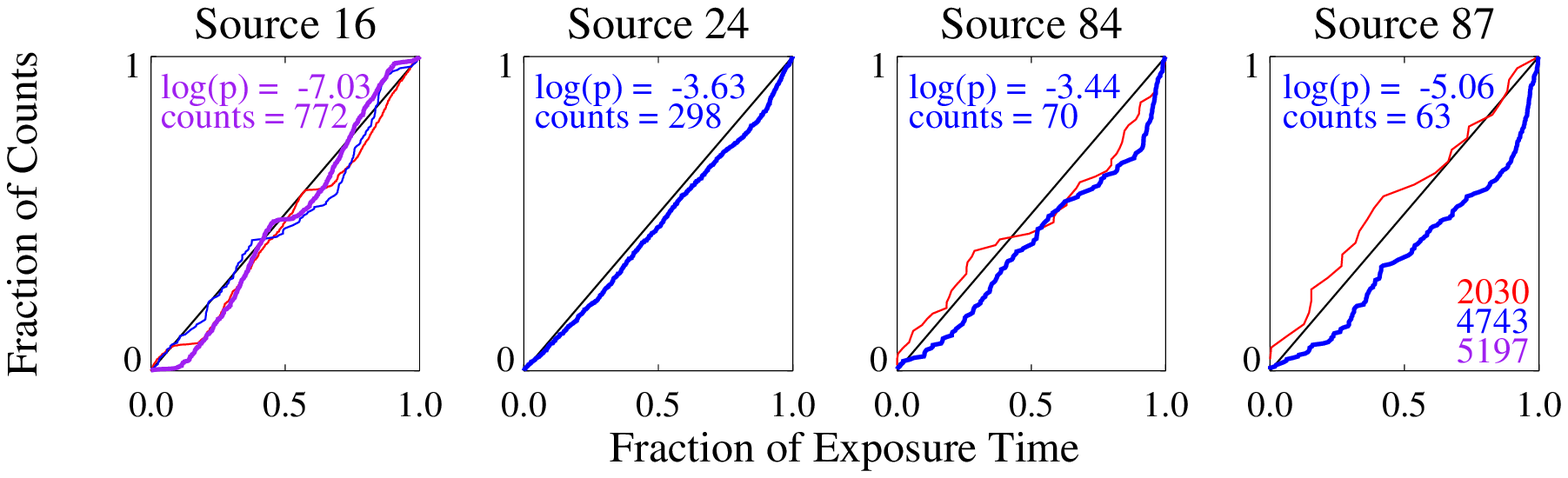} 
\caption{Cumulative photon arrival time distributions for X-ray sources in NGC~4214 showing evidence for rapid variability. The solid black line shows the expected distribution for a constant count rate. Colored histograms and text indicate the cumulative distribution observed and statistics for different observations. The (color encoded) \Chandra Observation ID numbers are given in the rightmost panel. Each panel lists the (logarithmic) KS probability for no variability and the net 0.35-8 keV counts observed for each source in the observation with the lowest KS probability.}
\label{short_4214}
\end{figure*}

In NGC~55, we find no sources with K-S probabilities indicative of short-term variability. We find three and four such rapidly varying X-ray sources in NGC~2403 and NGC~4214, respectively. In Figure~\ref{short_2403} and \ref{short_4214}, we show the cumulative arrival times of photons detected in each exposure for these short-term variable sources. The $\xi$ values of these sources are reported in Table~\ref{table_rapid_variable}.

\subsubsection{Long-Term Variability From \Chandra Observations}\label{rapid_variability_section}
To investigate long-term variability of the X-ray sources in each galaxy, we utilize the variability threshold $\eta$ \citep[as in ][]{Tullmann+11}, defined as $\eta=(flux_{\rm max}-flux_{\rm min})/\Delta flux$, where $flux_{\rm max}$ and $flux_{\rm min}$ are the maximum and minimum fluxes from detections of the source, respectively. The flux error $\Delta flux$ is calculated using the Gehrels approximation \citep[appropriate for low count data; ][]{Gehrels86}. We consider any sources with $\eta\ge5$ variable. There are 87 sources in NGC~55, 72 sources in NGC~2403, and 48 sources in NGC~4214 that meet this criterion.

Table~\ref{table_variable} provides the long-term \Chandra variability $\eta$ values for sources with $\eta\ge5$. We additionally list the corresponding maximum to minimum flux ratios for these sources. Column (1) lists the source number. Columns (2)-(3) provide $\eta$ (indicative of long-term variability), and the ratio of maximum to minimum observed flux, respectively, for sources in NGC~55. Columns (4)-(5) provide the same information for sources in NGC~2403, and columns (6)-(7) provide the same information for sources in NGC~4214. Only sources showing evidence for variability are included in the table, and these values only refer to variability within or between multiple \Chandra observations. 

\subsubsection{Long-Term Variability Using \XMM}
In addition to our \Chandra survey, multiple archival observations by \XMM are available that contain our catalog sources within the field of view. We utilize \XMM fluxes provided by FLIX\footnote{See: \url{www.ledas.ac.uk/flix/flix.html}}, an upper limit server for \XMM data provided by the \XMM Survey Science Center. FLIX allows the user to input source coordinates and provides estimates of the flux at the source position, or a flux upper limit where no source was detected, in a variety of energy bands. All fluxes are then converted to the 0.35-8 keV energy band assuming the same spectral model as for our sensitivity maps (a power law with \PL = 1.9 absorbed by the Galactic column). All \XMM upper limits correspond to the 3$\sigma$ level.

Tables~\ref{NGC55_longterm}-\ref{NGC4214_longterm} list {\it all} available archival X-ray fluxes from both \Chandra and \XMM (including upper-limits for exposures during which the source was not detected). For many sources, this included upper limits. The tables follow approximately the same form, with the first column providing the source number. The next several column headings provide the dates on which each \Chandra or \XMM observation was performed, listed in chronological order; these columns provide the 0.35-8 keV flux (or 3$\sigma$ upper limit on the 0.35-8 keV flux) observed in each exposure. The following two columns provide the ratio of the maximum to minimum fluxes, calculated two ways. The first ratio uses only source detections. For some sources, there is a flux upper limit in an observation that is below the faintest source detection; in these cases, we calculate a lower limit on the flux ratio from the lowest measured upper limit.

\begin{figure*}
\centering
\includegraphics[width=0.65\linewidth,clip=true]{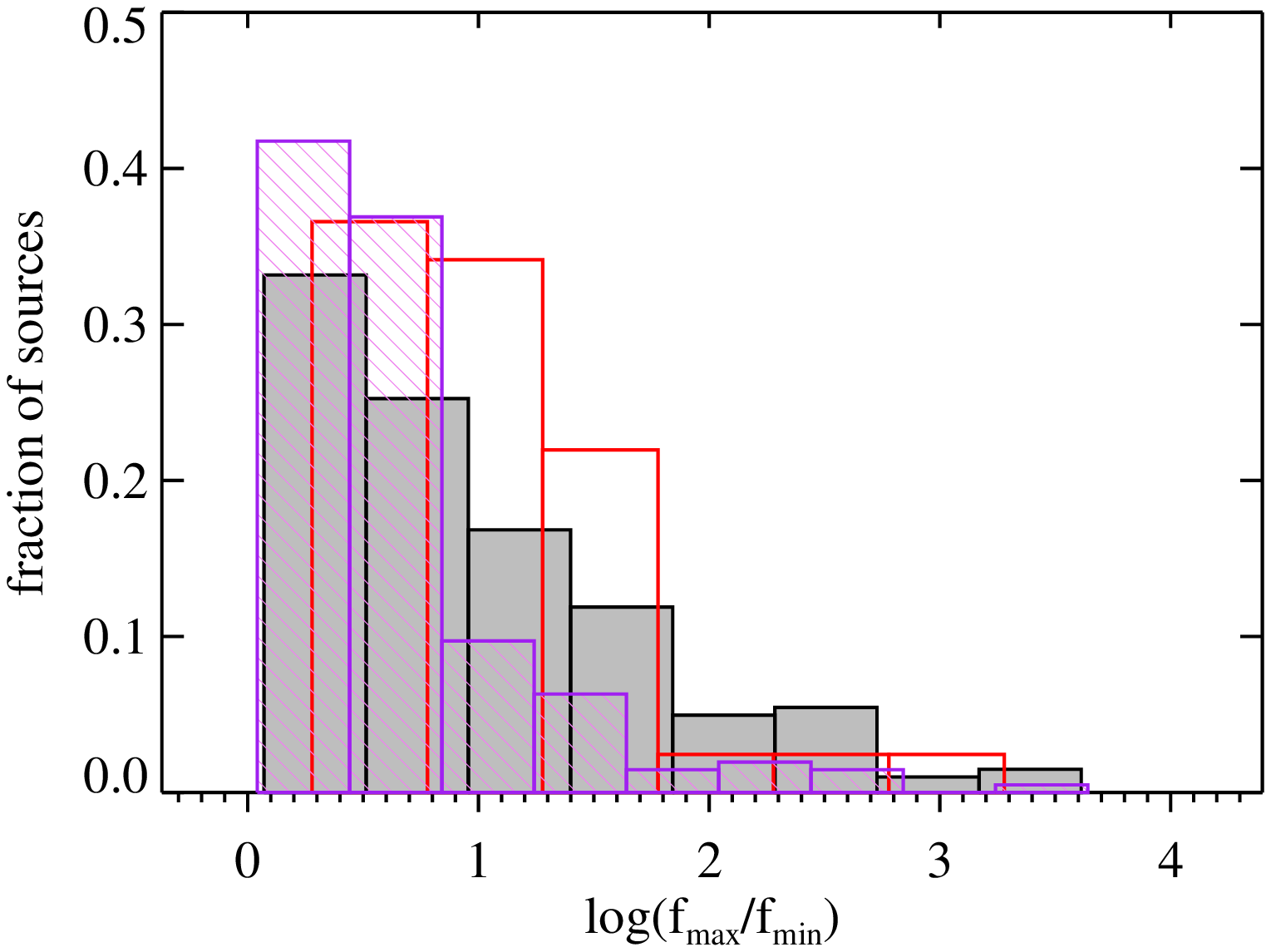} 
\caption{Histogram showing the maximum to minimum flux ratios from source detections (black) and upper-limits from source non-detections (red) using all available archival observations. The purple histogram shows the flux ratios using only the archival \Chandra observations. The median flux ratio derived from source detections is $\sim$5.8; the median flux ratio when utilizing upper-limits from source non-detections is 8.9. The \Chandra-only flux ratio distribution has a median value of $\sim$3.5.}
\label{flux_ratio}
\end{figure*}

Figure~\ref{flux_ratio} shows a histogram of these two flux ratios: the ratios derived from source detections only is shown in black, and the ratios found by using a flux upper limit is shown in red. The median flux ratio from detections is $\sim$5.8, while for ratio lower limits it is $\sim$8.9.

X-ray bursting or transient behavior (whereby the flux changes by many orders of magnitude) is observed in both LMXBs and HMXBs \citep{Bozzo+08,Fragos+08,Paul+11,Revn+11}. Some of the sources in our catalogs may exhibit such behavior, especially when a source was detected in at least one exposure but not another (despite the source's position being within the field of view). While previous work on the Antennae has shown the XLFs to be unaffected by such variability \citep{Zezas+07}, these observations only included sources with X-ray luminosities above a few $10^{37}$ \lum and may not be universal for lower-luminosity sources like those detected in our survey (with luminosities down to $\sim10^{36}$ \lum). 

The heterogeneity of observation exposure times and instruments makes it difficult to identify such transient or bursting sources on an individual basis. When we examine our catalog as a whole, combined with the archival \XMM measurements, we find that $\sim$39\% of sources with multiple observations have a maximum-to-minimum flux ratio of at least a factor of ten. The typical flux uncertainty for faint sources ($<10^{-14}$ \flux) in our catalog is $\sim$30\%, and the uncertainty decreases to $\sim5-8$\% for sources with fluxes $>10^{-13}$ \flux. For sources with a maximum flux of $<10^{-14}$ \flux, an order of magnitude decrease in brightness corresponds to a $\sim3\sigma$ change. For a flux ratio of $\sim$10 to be significant at the 5$\sigma$ level, the maximum flux of the source must exceed $\sim2\times10^{-14}$ \flux. Therefore, only those sources with a maximum flux $>2\times10^{-14}$ \flux and a flux ratio $>$10 are variable at the 5$\sigma$ level.

When considering the flux ratio lower limits for sources below the detection limit in an observation, we find $\sim$44\% of sources show an order of magnitude change or more. However, when only the flux ratios derived from the archival \Chandra observations are used, we find only $\sim$16\% of sources exhibit a change in flux greater than ten. This suggests that many of these sources with a factor of $\sim$10 change in flux is likely due to uncertainties in the flux or detector differences. Only $\sim$11\% of sources in our catalog have flux ratios of more than a hundred when all available data is considered, 7\% of sources have flux ratios from lower limits of one hundred, and 4\% of sources show this level of variability from the \Chandra-only observations. For these sources, the long-term variability is almost certainly real. 

\begin{figure*}
\centering
\includegraphics[width=0.65\linewidth,clip=true]{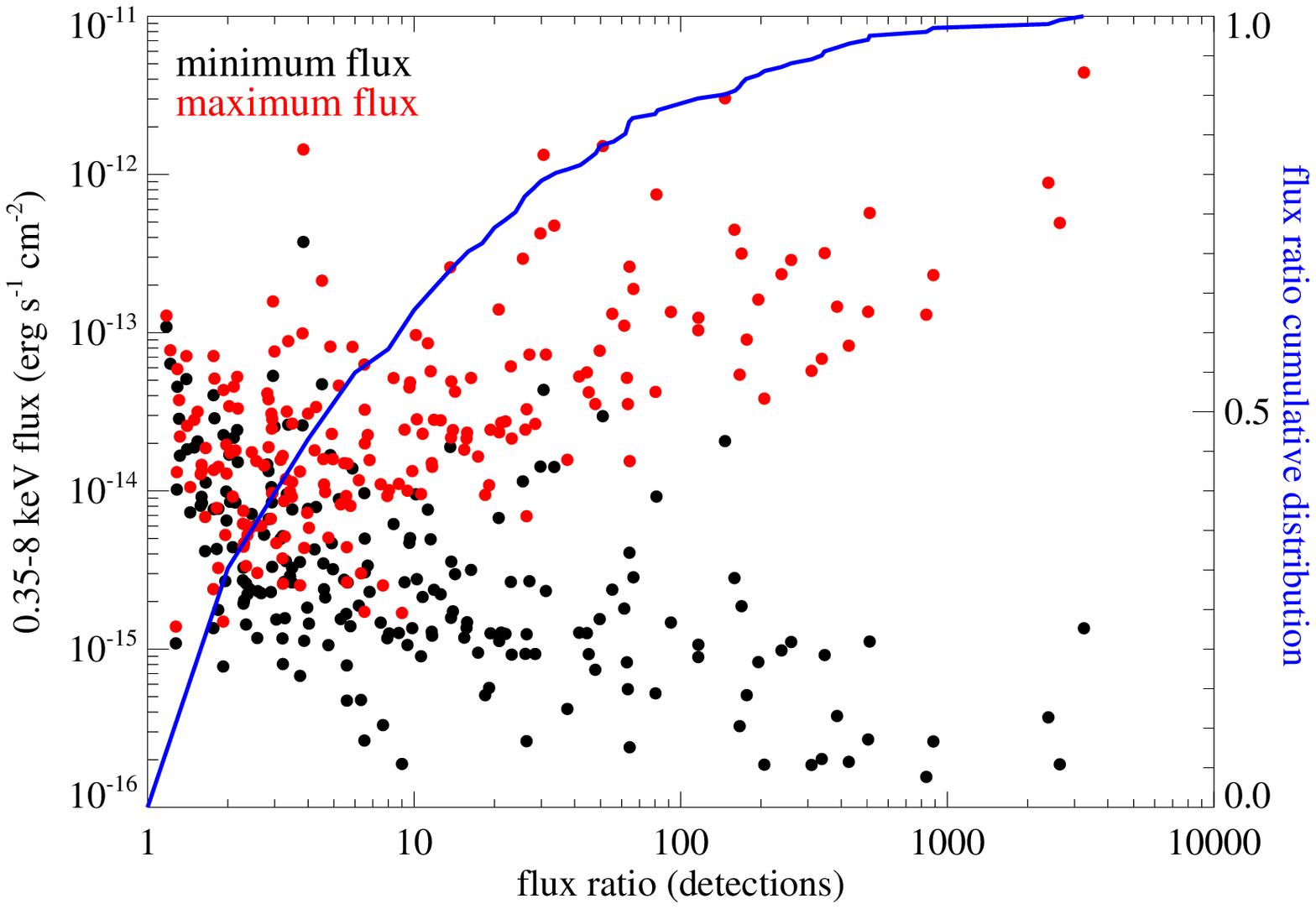} 
\caption{The minimum flux (black) and maximum flux (red) as a function of detected flux ratio. The blue line shows the cumulative distribution of the flux ratio determined from source detections.}
\label{ratio_vs_flux}
\end{figure*}

Figure~\ref{ratio_vs_flux} shows the maximum and minimum detected fluxes of source in our catalog as a function of their flux ratio (determined by detections). The flux ratio compared to the minimum flux of each source shows a clear decreasing trend, while the opposite is observed for the same comparison against the maximum flux. These trends are not surprising: in order for a source to have a large flux ratio, it {\it must} have fluctuated from a relatively low flux to a substantially larger one. For sources with a flux ratio below an order of magnitude there is significant overlap between the minimum and maximum observed fluxes around $\sim$10$^{-14}$ \flux, revealing the flux range where most of our detections lie.

\section{Multiwavelength Observations and Source Classification}\label{optical}
\subsection{Optical Counterparts from \HST}
To aid in source classification, we searched for optical counterpart candidates for each X-ray source that was contained within one or more of our \HST fields. For many of our X-ray sources, we found multiple optical counterpart candidates within the \Chandra X-ray error circle (typically, with radii on the order of $\sim$0\as5). We kept only those optical sources that met the quality cuts described in \cite{Williams+09}: i.e., sources with a sufficiently high signal-to-noise ratio, not flagged as unusable, meeting predetermined crowding and sharpness thresholds, etc. The \HST exposures reach a typical limiting magnitude of 26 in $F814W$ and 27 in $F606W$, making late B-type main sequence stars (and in some cases, early A-type main sequence stars) detectable in our observations. Figures~\ref{optical_55}-\ref{optical_4214} show ``postage stamp'' finding charts for each X-ray source covered by an \HST pointing. The 90\% \Chandra positional uncertainty, centered on the source position, is indicated in the figures. 

To determine the likely source classification, we utilized color magnitude diagrams (CMDs) and X-ray color magnitude diagrams (XCMDs) as in Paper~I. We provide a brief summary of our approach here, and the reader is referred to Paper~I for further details. We use the optical colors from our \HST observations to separate optical counterpart candidates into main sequence (`MS'; our exposures are sensitive to A-type stars and earlier) or red giant branch (`RGB') subgroups. Many authors have shown it is possible to use a combination of X-ray and optical colors to discriminate between XRBs and background AGN \citep[e.g., ][]{Horn+01,Shty+05}. 

\cite{McGowan+08} found that known pulsars and HMXBs located in the Small Magellanic Cloud occupied a region of the XCMD defined by a combination of optical colors and X-ray-to-optical flux ratio. In order to directly compare with HMXBs in the SMC, we use their definition of the logarithmic X-ray-to-optical flux ratio, $log(f_X/f_V)$, where $f_X$ is the 2-10 keV flux and $f_V$ is the optical flux. We compute this flux ratio for each source within the error circle, using the observed flux in either $F555W$ or $F606W$ for the $V$-band, whichever is available. Although the $F606W$ filter has a larger bandpass than $F555W$, we estimate that for the majority of stars accessible in our \HST exposures the effect on $log(f_X/f_V)$ will be small, $\sim$0.1. To estimate the 2-10 keV flux, we convert the measured 2-8 keV to the 2-10 keV energy band using WebPIMMs\footnote{See \url{http://heasarc.gsfc.nasa.gov/cgi-bin/Tools/w3pimms/w3pimms.pl}.} and the best-fit spectral model (for sufficiently bright sources) or the assumed power law with Galactic absorption used in our sensitivity map computation.

The region of the XCMD bounded by $B-V\lesssim0$ and $log(f_X/f_V)\lesssim1$ contains likely pulsars and HMXBs, while LMXBs and AGN preferentially fall outside this region. We therefore separate the candidate optical counterparts into likely `AGN' or 'HMXB' subgroups. If {\it any} of the optical counterpart candidates is consistent with the HMXB subgroup, we classify the X-ray source as an HMXB candidate; otherwise, the source is placed into a different category (e.g., an AGN).

\subsection{Comparisons with Multiwavelength Catalogs}
We have also attempted to place constraints on the nature of the X-ray sources contained in the CLVS catalogs by examining archival multi-wavelength data. We searched for counterparts in the following point source catalogs: the {\it GALEX} GR6 data release (NUV, FUV), USNO-B1.0 ($B$, $R$), 2MASS All Sky ($J$, $H$, $K$), and NVSS (1.4 GHz). X-ray source positions were also correlated with SIMBAD objects. Tables~\ref{multiwavelength55}-\ref{multiwavelength4214} summarize the multiwavelength observations available for each of the X-ray sources with at least one multi-wavelength detection. All three tables follow approximately the same format, with column (1) providing the source number. Column (2) provides the number of potential optical \HST sources falling within the \Chandra error circle. Column (3) provides the identifying information from the SIMBAD catalog. The next several columns provide the magnitudes of multiwavelength counterparts from a variety of public source catalogs: USNO-B1.0, SDSS, 2MASS, and {\it GALEX}.

UV emission can be a good tracer of young stars and background AGN. It is possible that some LMXBs will additionally exhibit detectable levels of UV emission, although their optical colors are expected to be significantly redder than those of an AGN. The optical USNO-B1.0 catalog provides photometric data for objects in five optical bandpasses and provides a $\sim$85\% accuracy for distinguishing foreground Milky Way stars from non-stellar objects \citep{Monet+03}, making it possible to separate these sources from clusters and nearby AGN. The infrared data are also sensitive to Milky Way stars, red giants, associations of older stars, and background galaxies. Radio emission is indicative of radio-loud background AGN; however, a large fraction of AGN are known to be radio-quiet \citep{Kellermann+89}, so the lack of radio emission is not used to disqualify an AGN origin for a given X-ray source. Although many SNRs, both Galactic and in nearby galaxies have been detected at radio frequencies, most have too low a radio surface brightness to be detectable at distances of a few Mpc \citep[e.g.,][]{Long+10,Gordon+99}.

\subsection{Source Classification}
We combine all of the X-ray and multi-wavelength data available for each source to determine a likely source classification. While the detection of optical counterpart candidates for those X-ray sources with overlapping \HST fields allows some sources to be assigned a classification based on their multiwavelength properties, many X-ray sources are faint or do not have optical coverage, making their physical nature ambiguous. We therefore attempt to additionally combine statistical quantities (i.e., using the radial source distribution, location in spiral arms, inter-arm regions, or background regions) with individual source properties (variability, X-ray spectral shape, and optical colors) to determine the most likely source classification. For a fair comparison with the NGC~300 X-ray source catalog (see Paper~I for details), we use the \texttt{xclass} IDL routine\footnote{See \url{http://www.astro.washington.edu/users/bbinder/xclass/}} to carry out our final source classifications. Since the primary goal of the CLVS is to study the HMXB populations of these galaxies, the \texttt{xclass} classifier was designed to quickly find sources with properties consistent with HMXBs (e.g., blue optical counterpart, associated with spiral arms, X-ray spectrum or hardness most consistent with a power law \PL$\sim$0.5-2.5, etc.). Some X-ray sources that are not classified as HMXBs (most of which are then classified as AGN) will undoubtedly possess individually interesting properties; thus, some users may decide to perform a follow-up analysis on a different output category.

The final source classifications for NGC~55, NGC~2403, and NGC~4214 are displayed in Figure~\ref{final_classes} and are summarized in Table~\ref{classification}. Column (1) lists the source number. Column (2) provides the preliminary source classification based on the hardness ratios alone. Column (3) lists the galactocentric distance of the source (in kpc). Column (4) lists whether or not the source appears associated with the galaxy (i.e., coincident with a spiral arm). Column (5) lists whether evidence for X-ray variability was observed and, if so, what type (`rapid' and/or `long'). Column (6) lists whether the source had its X-ray spectrum fit and, if so, what general shape the spectrum showed (`po', `diskbb', etc.). Column (7) provides the source type as determined by the X-ray to optical flux ratio. Column (8) summarizes what publicly available multiwavelength information is available for the source. Column (9) lists any known counterparts to the source (e.g., foreground star, SNR, etc.) from SIMBAD. Column (10) lists our final classification of the source based on all available data.

\section{Summary}
We have constructed comprehensive X-ray point source catalogs of NGC~55, NGC~2403, and NGC~4214 as part of the \Chandra Local Volume Survey. When combined with the catalogs of NGC~300 \citep{Binder+12} and NGC~404 \citep{Binder+13}, the \Chandra Local Volume Survey contains 629 X-ray sources down to a limiting unabsorbed luminosity of $\sim5\times10^{35}$ \lum in the 0.35-8 keV band. We have presented X-ray hardness ratios, spectral analysis, radial source distributions, and an analysis of the temporal variability for the X-ray sources detected in NGC~55, NGC~2403, and NGC~4214.  To constrain the nature of each X-ray source, cross-correlations with multi-wavelength data were generated, and we searched overlapping \HST fields for optical counterparts to our X-ray detections. Based on all of this information, each X-ray source was classified as a candidate X-ray binary, background AGN, supernova remnant, or foreground star.  The catalogs constructed for the \Chandra Local Volume Survey will be used in a follow-up paper analyze the properties of the source populations in more depth, discuss the implications of our data for scenarios of HMXB evolution and to make connections between those populations and to the morphology, star formation history, and metallicity of the host galaxy.

\acknowledgements
We would like to thank the anonymous referee for the helpful comments and recommendations that improved this manuscript. B. B. and B. F. W. acknowledge support from \Chandra grant AR2-13005X. TJG and PPP acknowledge acknowledge support under NASA contract NAS8-03060 with the Chandra X-ray Center. This research has made use of the NASA/IPAC Extragalactic Database (NED) which is operated by the Jet Propulsion Laboratory, California Institute of Technology, under contract with the National Aeronautics and Space Administration. This research has made use of the VizieR catalogue access tool, CDS, Strasbourg, France. The original description of the VizieR service was published in A\&AS 143, 23

\bibliography{apjmnemonic,data_v4}
\bibliographystyle{apj}


\begin{table*}[ht]
\centering
\caption{Summary of CLVS Galaxy Properties}
\begin{tabular}{ccccccc}
\hline \hline
& Property								& NGC~55		& NGC~300	& NGC~404	& NGC~2403	& NGC~4214	\\
\hline
(1)	& Distance (Mpc)						& 2.1$^a$			& 2.0$^b$		& 3.05$^b$	& 3.3$^c$		& 2.9$^c$		\\
(2)	& Morphology$^d$						& SB(s)m			& SA(s)d		& SA0		& SAB(s)cd	& IAB(s)m		\\
(3)	& R.A. (J2000)$^e$						& 00:14:53.6		& 00:54:53.5	& 01:09:27.0	& 07:36:51.4 	& 12:15:39.2	\\
(4)	& Decl. (J2000)$^e$						& -39:11:48		& -37:41:04	& +35:43:05	& +65:36:09	& +36:19:37	\\
(5)	& $m-M_0^e$							& 26.42			& 26.45		& 27.74		& 27.73		& 27.59		\\
(6)	& $M_B^d$							& -17.76			& -17.68		& -16.63		& -18.61		& -17.39		\\
(7)	& Inclination$^f$ ($^{\circ}$)				& 83				& 45			& 19			& 56			& 38			\\
(8)	& Position Angle$^e$ ($^{\circ}$)			& 108			& 111		& 0			& 128		& 0			\\
(9)	& Major linear diameter$^f$ (kpc) 			& 12.05			& 12.95		& 3.25		& 19.43		& 7.05		\\ 
(10)	& Stellar mass (log($M_*/$\Msun))			 & 10.48$^d$ 		& 9.63$^9$ 	& 8.83$^i$ 	& 10.27$^j$	& 9.18$^f$	\\
(11)	& \nH$^k$ (10$^{20}$ cm$^{-2}$)			 & 1.37			& 4.19		& 5.13		& 4.36		& 1.99		\\
(12)	& 12+log(O/H)							& 8.05$^l$		& 8.41$^m$	& 8.6	$^n$		& 8.48$^o$	& 8.36$^p$	\\
(13)	& total unabs. log($L_X$) (\lum, 0.35-8 keV)	& 39.34$^q$		& 38.49$^q$	& 37.85$^r$ 	& 40.23$^s$	& 38.87$^t$	\\
\hline \hline
\label{galbasic}
\end{tabular}
\tablecomments{$^a$\cite{Tikhonov+05}, $^b$\cite{Dalcanton+09}, $^c$\cite{Freedman+88}, $^d$\cite{RC3}, $^e$NED, $^f$\cite{Kara+04}, $^g$\cite{Munoz+07}, $^h$\cite{Puche+90}, $^i$\cite{Thilker+10}, $^j$\cite{deBlok+08}, $^k$Column density through the ISM of the Milky Way \citep{Kalberla+05}, $^l$\cite{Lee+06}, $^m$\cite{Bresolin+09}, $^n$\cite{Bresolin13}, $^o$\cite{Berg+13}, $^p$\cite{Engelbracht+08}, $^q$\cite{Zang+97}, $^r$\cite{Komossa+99}, $^s$\cite{Fraternali+02b}, $^t$\cite{Ott+05}}
\end{table*}

\begin{table*}[ht]
\centering
\caption{Summary of \Chandra Observations}
\begin{tabular}{cccccccc}
\hline \hline
Galaxy		& Observation 		& Date		& Effective Exposure		& Fraction of Total	& R. A.$^a$	& Decl.$^a$	& Detector \\
			& ID				&			& Time (ks)			& Exposure Time	& (J2000)		& (J2000)		&		\\
(1)			& (2)				& (3)			& (4)					& (5)				& (6)			& (7)			& (8)		\\
\hline
\multirow{3}{*}{NGC~55}		& 2255			& 11 Sep 2001 		& 47		& 78.6\%		& 00:15:08.5		& -39:13:13.3		& ACIS-I \\
						& 4744			& 29 June 2004	& 9.5		& 99.7\%		& 00:14:54.0		& -39:11:49.0 		& ACIS-I \\
						& Total			&				& 56.5	& 82.5\%		&				&				&	\\ 
\smallskip \\
\multirow{6}{*}{NGC~2403}	& 2014			& 17 Apr 2001		& 35		& 98.8\%		& 07:36:51.9		& +65:36:00.6		& ACIS-S \\
						& 4627			& 9 Aug. 2004		& 31		& 75.6\%		& 07:37:17.1		& +65:35:58.3		& ACIS-S \\
						& 4628			& 23 Aug. 2004		& 42		& 89.7\%		& 07:37:17.1		& +65:35:58.3		& ACIS-S \\
						& 4629			& 3 Oct. 2004		& 40		& 90.9\%		& 07:37:17.1		& +65:35:58.3		& ACIS-S \\
						& 4630			& 22 Dec. 2004		& 42		& 86.5\%		& 07:37:17.1		& +65:35:58.3		& ACIS-S \\
						& Total			&				& 190	& 88.4\%			&				& \\ 
\smallskip \\
\multirow{4}{*}{NGC~4214}	& 2030			& 16 Oct 2001		& 25		& 96.2\%		& 12:15:38.7		& +36:19:41.9		& ACIS-S \\
			& 4743						& 3 Apr. 2004		& 26		& 96.3\%		& 12:15:38.9		& +36:19:40.0		& ACIS-S \\
			& 5197						& 30 July 2004		& 28		& 98.6\%		& 12:15:38.9		& +36:19:40.0		& ACIS-S \\
			& Total						&				& 79		& 97.5\%		&				&				& \\
\hline \hline
\label{obs}
\end{tabular}
\tablecomments{$^a$Observation coordinates are those of the \Chandra ACIS aim-point.}
\end{table*}

\begin{table*}[ht]
\centering
\renewcommand{\tabcolsep}{0.08cm}
\caption{Alignment of \Chandra and \HST Images to 2MASS}
\begin{tabular}{ccccccccc}
\hline \hline
\multirow{2}{*}{Galaxy}	& \multirow{2}{*}{Description}	& \HST Obs. 	& \# of X-ray 		& R.A. 		& Decl.    		& \# Sources Used	& $r.m.s.$,	& $r.m.s.$,	 \\
		&			& ID		 	 & Sources in Field	& (J2000) 		& (J2000)		& For Alignment	& R. A.$^a$	& Decl.$^a$	 \\
 (1)		& (2)			& (3)  	    	& (4)				 	& (5) 		& (6)			& (7)				& (8)			& (9) \\
\hline
\multirow{7}{*}{NGC~55}		& \Chandra	& merged	& 154	& 00:15:09.96	& -39:13:12.9	& 5				& 0\as0297			& 0\as0942				\\
						& WIDE-1		& 11307	& 2		& 00:14:16.54	& -39:09:45.0	& 4				& 0\as0783 (0\as0725)	& 0\as2265 (0\as2060)		\\
						& WIDE-3		& 11307	& 1		& 00:14:30.03	& -39:10.18.8	& 4				& 0\as0533 (0\as0443)	& 0\as1174 (0\as0701)		\\
						& WIDE-4		& 11307	& 3		& 00:14:38.26	& -39:10:41.9	& 4				& 0\as0401 (0\as0270)	& 0\as2437 (0\as0877)		\\
						& WIDE-5		& 11307	& 4		& 00:15:11.55	& -39:12:58.3	& 4				& 0\as3832 (0\as3820)	& 0\as1197 (0\as0740)		\\
						& DISK		& 9765	& 9		& 00:15:31.06	& -39:14:13.4	& 5				& 0\as0323 (0\as0126)	& 0\as1262 (0\as0840)		\\
						& FIELD		& 9765	& 22		& 00:14:53.73	& -39:11:48.1	& 6				& 0\as0454 (0\as0343)	& 0\as1301 (0\as0898)		\\
 \\
\multirow{6}{*}{NGC~2403}	& \Chandra	& merged	& 190 	& 07:37:12.85 & +65:36:21.0 	& 4				& 0\as0048			& 0\as0296				\\
						& HALO-1		& 10523	& 1		& 07:37:54.67	& +65:31:29.2	& 4				& 0\as0127 (0\as0118)	& 0\as0636 (0\as0563)		\\
						& HALO-6		& 10523	& 8		& 07:37:31.38	& +65:40:29.5	& 6				& 0\as0569 (0\as0567)	& 0\as0412 (0\as0287)		\\
						& HALO-7		& 10523	& 2		& 07:38:01.5	& +65:43:49.1	& 4				& 0\as0098 (0\as0086)	& 0\as0473 (0\as0369)		\\
						& X-1		& 10579	& 18		& 07:36:42.36	& +65:35:36.3	& 4				& 0\as0069 (0\as0050)	& 0\as0419 (0\as0296)		\\
						& PR			& 10182	& 37		& 07:36:58.6	& +65:36:10.9	& 3				& 0\as0702 (0\as0700)	& 0\as0487 (0\as0387)		\\
\\
\multirow{3}{*}{NGC~4214}	& \Chandra	& merged	& 116 	& 12:15:39.51	& 36:17:22.6	& 4				& 0\as0101			& 0\as0541				\\
						& N4214		& 11986	& 4		& 12:15:32.63	& 36:21:40.7	& 3				& 0\as0147 (0\as0108)	& 0\as0541 (0\as0012)		\\
						& DEEP		& 10915	& 4		& 12:15:23.22	& 36:21:51.9	& 4				& 0\as0134 (0\as0088)	& 0\as0566 (0\as0166)		\\
\hline \hline
\end{tabular}
\tablecomments{$^a$Alignment errors listed show the uncertainties in each individual field and the \Chandra image added in quadrature. The error in parentheses is the uncertainty reported by \texttt{ccmap} for the individual fields.}
\label{align}
\end{table*}

\begin{table*}[ht]
\centering
\caption{Number of X-ray Sources at each \AEx Iteration}
\begin{tabular}{ccccc}
\hline \hline
Galaxy		& Initial	& First pass	& Second pass		& Final 	\\
(1)			& (2)		& (3)			& (4)				& (5)		\\
\hline
NGC~55		& 906	& 370		& 220			& 154	\\
NGC~300		& 799	& 224		& 126			& 95		\\
NGC~404		& 762	& 129		& 104			& 74		\\
NGC~2403	& 1308	& 648		& 298			& 190	\\
NGC~4214	& 1177	& 474		& 211			& 116	\\
\hline
Total			& 4952	& 1845		& 959			& 629	\\
\hline \hline
\label{ae_passes}
\end{tabular}
\end{table*}

\begin{table*}[ht]
\centering
\caption{Expanded X-ray Source Lists for All Galaxies in the \Chandra Local Volume Survey}
\begin{tabular}{ccccccccc}
\hline \hline
Source	& Source ID		& R.A. 	& Decl.	& Positional	& $\theta^b$	& net counts	& log(\pns)	& $f_{\rm 0.35-8 keV}^a$	\\
No.		& (CXOLV J+...)	& (J2000)	& (J2000)	& Error (\asn)	& (\am)		& (0.35-8 keV)	& (0.35-8 keV)	& (10$^{-15}$ \flux)		\\
(1)		& (2)				& (3)		& (4)		& (5)			& (6)			& (7)			& (8)			& (9)					\\
\hline
\multicolumn{9}{c}{NGC~300} \\
\hline
1 & 005411.98-373951.8 & 13.5499370 & -37.664408 & 0.70 &   8.2 &    22.8$^{+   6.6}_{-   5.5}$ &  -9.1 &     4.0 \\
2 & 005412.31-373359.6 & 13.5513190 & -37.566583 & 0.93 &   9.3 &    18.2$^{+   6.6}_{-   5.5}$ &  -4.8 &     3.4 \\
3 & 005413.98-373710.5 & 13.5582680 & -37.619588 & 0.36 &   7.9 &    70.8$^{+   9.8}_{-   8.8}$ & $<$ -10 &    12.4 \\
4 & 005415.57-373316.4 & 13.5649070 & -37.554563 & 0.59 &   9.2 &    45.1$^{+   8.5}_{-   7.4}$ & $<$ -10 &    11.8 \\
5 & 005419.92-373744.5 & 13.5830010 & -37.629032 & 0.61 &   6.6 &    12.1$^{+   5.0}_{-   3.8}$ &  -6.0 &     2.0 \\
6 & 005421.10-374241.2 & 13.5879360 & -37.711458 & 0.51 &   7.5 &    31.1$^{+   7.1}_{-   6.0}$ & $<$ -10 &     6.4 \\
7 & 005422.16-374024.6 & 13.5923720 & -37.673501 & 0.67 &   6.3 &     9.4$^{+   4.7}_{-   3.6}$ &  -3.8 &     1.5 \\
8 & 005422.52-374312.1 & 13.5938400 & -37.720045 & 0.59 &   7.5 &    22.5$^{+   6.4}_{-   5.3}$ & $<$ -10 &     3.9 \\
9 & 005422.52-373850.6 & 13.5938410 & -37.647406 & 0.45 &   6.0 &    16.1$^{+   5.5}_{-   4.3}$ &  -9.0 &     2.7 \\
10 & 005425.06-374358.2 & 13.6044550 & -37.732836 & 0.53 &   7.6 &    29.4$^{+   7.1}_{-   6.0}$ & $<$ -10 &     5.4 \\
\hline
\multicolumn{9}{c}{NGC~404} \\
\hline
1 & 010838.25+354027.3 & 17.1593960 &  35.674264 & 0.23 &  10.0 &   655.3$^{+  27.2}_{-  26.1}$ & $<$ -10 &    95.3 \\
2 & 010841.45+354056.8 & 17.1727100 &  35.682458 & 0.47 &   9.3 &   116.5$^{+  13.1}_{-  12.0}$ & $<$ -10 &    13.2 \\
3 & 010842.38+354300.8 & 17.1765910 &  35.716903 & 0.55 &   8.9 &    67.4$^{+  10.6}_{-   9.5}$ & $<$ -10 &     7.8 \\
4 & 010847.72+353948.3 & 17.1988680 &  35.663419 & 0.69 &   8.4 &    33.2$^{+   8.3}_{-   7.2}$ &  -9.5 &     3.7 \\
5 & 010848.00+354437.4 & 17.2000240 &  35.743725 & 0.37 &   7.9 &    34.2$^{+   7.1}_{-   6.0}$ & $<$ -10 &    12.3 \\
6 & 010848.45+354433.4 & 17.2018810 &  35.742616 & 0.37 &   7.8 &    37.5$^{+   7.4}_{-   6.3}$ & $<$ -10 &    10.1 \\
7 & 010848.48+353906.9 & 17.2020350 &  35.651937 & 0.22 &   8.5 &   454.3$^{+  22.8}_{-  21.8}$ & $<$ -10 &    51.6 \\
8 & 010850.31+353843.8 & 17.2096470 &  35.645503 & 0.60 &   8.3 &    46.0$^{+   9.1}_{-   8.0}$ & $<$ -10 &     5.1 \\
9 & 010853.61+354501.9 & 17.2234150 &  35.750537 & 0.58 &   6.9 &    26.2$^{+   6.8}_{-   5.7}$ & $<$ -10 &     4.2 \\
10 & 010854.69+354239.7 & 17.2279010 &  35.711041 & 0.41 &   6.4 &    40.0$^{+   7.9}_{-   6.8}$ & $<$ -10 &     4.2 \\
\hline
\multicolumn{9}{c}{NGC~55} \\
\hline
1 & 001503.50-391533.8 &  3.7645870 & -39.259401 & 0.19 &   2.7 &     9.8$^{+   4.3}_{-   3.1}$ & $<$ -10 &     1.6 \\
2 & 001502.09-391646.1 &  3.7587260 & -39.279490 & 0.25 &   3.9 &    11.6$^{+   4.6}_{-   3.4}$ & $<$ -10 &     2.1 \\
3 & 001458.59-392211.3 &  3.7441560 & -39.369828 & 0.45 &   9.2 &    81.1$^{+  10.7}_{-   9.6}$ & $<$ -10 &    15.0 \\
4 & 001459.11-391901.6 &  3.7463080 & -39.317128 & 0.47 &   6.4 &    18.7$^{+   5.7}_{-   4.6}$ & $<$ -10 &     2.8 \\
5 & 001459.34-391723.1 &  3.7472760 & -39.289762 & 0.23 &   4.8 &    30.0$^{+   6.6}_{-   5.5}$ & $<$ -10 &     4.3 \\
6 & 001502.18-391425.2 &  3.7591020 & -39.240335 & 0.17 &   1.9 &     7.9$^{+   4.0}_{-   2.8}$ & $<$ -10 &     2.9 \\
7 & 001451.70-392417.8 &  3.7154570 & -39.404951 & 0.62 &  11.6 &   103.3$^{+  12.2}_{-  11.1}$ & $<$ -10 &    20.7 \\
8 & 001448.96-392229.9 &  3.7040110 & -39.374988 & 0.38 &  10.1 &   168.6$^{+  14.5}_{-  13.5}$ & $<$ -10 &    33.0 \\
9 & 001457.84-391555.7 &  3.7410040 & -39.265498 & 0.20 &   3.7 &    17.4$^{+   5.3}_{-   4.2}$ & $<$ -10 &     2.6 \\
10 & 001500.62-391414.2 &  3.7525970 & -39.237285 & 0.18 &   2.2 &     3.0$^{+   2.9}_{-   1.6}$ &  -5.2 &     1.1 \\
\hline
\multicolumn{9}{c}{NGC~2403} \\
\hline
1 & 073706.49+652736.5 &  114.2770800 &  65.460154 & 0.72 &   8.2 &    30.0$^{+   8.3}_{-   7.2}$ &  -6.7 &     3.8 \\
2 & 073707.39+653455.8 &  114.2808300 &  65.582175 & 0.02 &   1.9 &   904.7$^{+  31.1}_{-  30.1}$ & $<$ -10 &    31.4 \\
3 & 073705.08+653146.7 &  114.2711900 &  65.529650 & 0.21 &   4.5 &    54.9$^{+   8.9}_{-   7.8}$ & $<$ -10 &     3.5 \\
4 & 073701.86+652751.6 &  114.2577900 &  65.464334 & 0.82 &   8.0 &    22.2$^{+   7.3}_{-   6.2}$ &  -5.2 &     3.1 \\
5 & 073707.11+653515.8 &  114.2796600 &  65.587735 & 0.16 &   1.6 &     8.7$^{+   4.3}_{-   3.1}$ &  -5.7 &     0.5 \\
6 & 073703.99+653241.6 &  114.2666600 &  65.544909 & 0.37 &   3.6 &    10.8$^{+   4.6}_{-   3.4}$ &  -6.9 &     1.4 \\
7 & 073706.42+653451.8 &  114.2767500 &  65.581057 & 0.08 &   1.9 &    41.5$^{+   7.7}_{-   6.6}$ & $<$ -10 &     1.4 \\
8 & 073701.68+653236.3 &  114.2570200 &  65.543435 & 0.27 &   3.8 &    20.3$^{+   6.2}_{-   5.1}$ &  -8.0 &     0.9 \\
9 & 073656.80+653029.4 &  114.2366800 &  65.508181 & 0.29 &   5.8 &    82.2$^{+  11.2}_{-  10.1}$ & $<$ -10 &     5.6 \\
10 & 073659.27+653149.7 &  114.2469800 &  65.530482 & 0.29 &   4.5 &    36.4$^{+   7.9}_{-   6.8}$ & $<$ -10 &     2.3 \\
\hline
\multicolumn{9}{c}{NGC~4214} \\
\hline
1 & 121537.80+361613.2 &  183.9075100 &  36.270343 & 0.41 &   3.7 &     7.0$^{+   4.0}_{-   2.8}$ &  -4.7 &     2.0 \\
2 & 121531.56+360947.4 &  183.8815200 &  36.163172 & 0.92 &   9.2 &    28.3$^{+   7.6}_{-   6.4}$ &  -8.1 &     5.5 \\
3 & 121533.96+361428.5 &  183.8915300 &  36.241258 & 0.36 &   3.1 &     5.8$^{+   3.6}_{-   2.4}$ &  -7.1 &     1.8 \\
4 & 121530.29+361203.6 &  183.8762200 &  36.201004 & 0.39 &   5.6 &    32.7$^{+   6.9}_{-   5.8}$ & $<$ -10 &    10.6 \\
5 & 121533.02+361531.2 &  183.8876100 &  36.258681 & 0.37 &   4.2 &    19.8$^{+   6.0}_{-   4.9}$ &  -9.3 &     2.5 \\
6 & 121524.05+361252.1 &  183.8502100 &  36.214481 & 0.69 &   5.5 &     8.7$^{+   4.3}_{-   3.1}$ &  -5.7 &     3.4 \\
7 & 121535.34+361752.0 &  183.8972800 &  36.297805 & 0.19 &   2.1 &    11.1$^{+   4.6}_{-   3.4}$ &  -9.1 &     1.0 \\
8 & 121529.43+361545.4 &  183.8726600 &  36.262624 & 0.24 &   3.0 &    16.1$^{+   5.2}_{-   4.1}$ & $<$ -10 &     2.3 \\
9 & 121528.30+361644.1 &  183.8679200 &  36.278920 & 0.26 &   3.3 &    33.1$^{+   7.1}_{-   6.0}$ & $<$ -10 &     3.1 \\
10 & 121536.06+361847.3 &  183.9002900 &  36.313160 & 0.18 &   2.0 &    10.3$^{+   4.4}_{-   3.3}$ &  -9.3 &     0.9 \\
\hline \hline
\end{tabular}
\tablecomments{Table~\ref{srclist} is published in its entirety in the electronic edition of the journal. Only the first ten entries are shown for each galaxy for guidance regarding its form and content. \newline
$^a$Unabsorbed X-ray flux, assuming the power law with $\Gamma=1.9$ absorbed by the appropriate Galactic column given in Table~\ref{galbasic}; $^b\theta$ is the off-axis angle from the \Chandra ACIS aim point. Sources marked with a $^*$ are ``marginally detected'' sources and are not included in subsequent analysis (see the text for further details).}
\label{srclist}
\end{table*}

\begin{table*}[ht]
\renewcommand{\tabcolsep}{2.5pt}
\centering
\caption{Limiting Fluxes and Luminosities of Merged Observations}
\begin{tabular}{ccccccccccccc}
\hline \hline
Galaxy	& Energy band	& \multicolumn{2}{c}{70\%}			&& \multicolumn{2}{c}{90\%}		&& \multicolumn{2}{c}{95\%}		&&  \multicolumn{2}{c}{minimum$^a$}	 \\
		& (keV)		& $f_X^b$	& $L_X^b$		&& $f_X$		& $L_X$		&& $f_X$		& $L_X$		&& $f_X$		& $L_X$	\\ \\ \cline{3-4} \cline{6-7} \cline{9-10} \cline{12-13}
(1)		& (2)			& (3)			& (4)					&& (5)		& (6)				&& (7)			& (8)			&& (9)		& (10)	\\
\hline
\multirow{3}{*}{NGC~55}		& 0.35-8	& 3.1		& 1.6		& & 4.7	& 2.5		& & 5.4	& 2.8  	& & 0.7	& 0.4 \\
						& 0.5-2	& 0.8		& 0.4		& & 0.6	& 0.3		& & 2.2	& 1.2  	& & 0.2 	& 0.1 \\
						& 2-8	& 1.2		& 1.0		& & 2.8	& 1.5		& & 3.5	& 1.9		& & 0.4 	& 0.2 \\
\hline
\multirow{3}{*}{NGC~2403}	& 0.35-8	& 1.3		& 1.7		& & 2.5	& 3.2		& & 3.9	& 5.0 	& & 0.4 	& 0.5 \\
						& 0.5-2	& 0.5		& 0.7		& & 0.7	& 0.9 	& & 1.3 	& 1.7 	& & 0.1 	& 0.1 \\
						& 2-8	& 0.8		& 1.1		& & 1.4	& 1.9		& & 2.1	& 2.8 	& & 0.3 	& 0.4 \\
\hline 
\multirow{3}{*}{NGC~4214}	& 0.35-8	& 1.7		& 1.7		& & 3.4	& 3.4		& & 3.8	& 3.8 	& & 0.5 	& 0.5 \\
						& 0.5-2	& 0.3		& 0.3		& & 1.1	& 1.1		& & 1.2	& 1.2		& & 0.1 	& 0.1 \\
						& 2-8	& 1.0		& 1.0		& & 2.0	& 2.0		& & 2.7	& 2.7		& & 0.3 	& 0.3 \\
\hline \hline
\label{sens}
\end{tabular}
\tablecomments{\newline
$^a$The flux and luminosity of the faintest source meeting our \pns threshold.\newline
$^b$The units of flux are $10^{-15}$ \flux and the units of luminosity are $10^{36}$ \lum.}
\end{table*}

\begin{table*}[ht]
\centering
\caption{Hardness Ratio Source Classification Scheme}
\begin{tabular}{ccc}
\hline \hline
Classification & Definition & \# Sources$^a$ \\
(1) & (2) & (3) \\
\hline
X-ray binary or AGN (`XRB/AGN')	& $-0.4 < HR2 < 0.4$, $-0.4 < HR1 < 0.4$	& 298 \\
Absorbed source (`ABS')			& $HR1 > 0.4$						& 69 \\
Supernova remnant (`SNR')		& $HR2 < 0.1$, $HR1 < -0.4$			& 53 \\
Indeterminate hard source (`HARD')	& $HR2 > 0.4$, $-0.4 < HR1 < 0.4$		& 19 \\
Indeterminate soft source (`SOFT')	& $HR2 < -0.4$, $-0.4 < HR1 < 0.4$		& 18 \\
Indeterminate source (`INDET')		& $HR2 > 0.1$, $HR1 < -0.4$			& 3 \\
\hline \hline
\label{color_class}
\end{tabular}
\tablecomments{$^a$The total number of sources in NGC~55, NGC~2403, and NGC~4214 with hardness ratios placing them in each category.}
\end{table*}

\begin{table*}[ht]
\centering
\caption{Hardness Ratios of X-ray Sources}
\begin{tabular}{cccccccccccccc}
\hline \hline
\multicolumn{4}{c}{NGC~55}			& & \multicolumn{4}{c}{NGC~2403}	& & \multicolumn{4}{c}{NGC~4214}	\\  \cline{1-4} \cline{6-9} \cline{11-14}
No. 	& HR1 	& HR2 	& ID			& & No. 	&HR1	& HR2	& ID	& &  No. 	& HR1	& HR2	& ID		 \\
(1)	& (2)		& (3)		& (4)			& & (5)	& (6)		& (7)		&(8)	& & (9)	& (10)	& (11)	& (12)	\\
\hline
1 &  0.22 $\pm$  0.37 & -0.38 $\pm$  0.29 & XRB 	&& 1 & 0.21 $\pm$  0.07 &  0.41 $\pm$  0.12 & HARD	& & 1 & -0.59 $\pm$  0.20 &  0.16 $\pm$  0.17 & SNR \\
2 &  0.45 $\pm$  0.22 & -0.49 $\pm$  0.22 & ABS 	&& 2 & -0.13 $\pm$  0.03 & -0.16 $\pm$  0.02 & XRB	& & 2 & 0.31 $\pm$  0.12 & -0.17 $\pm$  0.14 & XRB \\ 
3 &  0.30 $\pm$  0.08 & -0.22 $\pm$  0.09 & XRB 	&& 3 & 0.32 $\pm$  0.10 & -0.34 $\pm$  0.09 & XRB		& & 3 & 0.21 $\pm$  0.25 &  0.18 $\pm$  0.39 & XRB \\ 
4 &  0.38 $\pm$  0.11 &  0.13 $\pm$  0.20 & XRB 	&& 4 & 0.36 $\pm$  0.12 & -0.23 $\pm$  0.14 & XRB		& & 4 & 0.06 $\pm$  0.14 & -0.08 $\pm$  0.14 & XRB \\ 
5 &  0.47 $\pm$  0.13 & -0.35 $\pm$  0.15 & ABS 	&& 5 & 0.16 $\pm$  0.24 &  0.20 $\pm$  0.37 & XRB		& & 5 & -0.44 $\pm$  0.17 & -0.19 $\pm$  0.10 & SNR \\ 
6 &  0.54 $\pm$  0.28 & -0.45 $\pm$  0.38 & ABS 	&& 6 & -0.43 $\pm$  0.20 & -0.01 $\pm$  0.16 & SNR	& & 6 & 0.24 $\pm$  0.19 &  0.07 $\pm$  0.26 & XRB \\ 
7 &  0.25 $\pm$  0.07 & -0.23 $\pm$  0.08 & XRB 	&& 7 & 0.50 $\pm$  0.09 & -0.09 $\pm$  0.16 & ABS		& & 7 & 0.31 $\pm$  0.32 & -0.30 $\pm$  0.35 & XRB \\ 
8 &  0.41 $\pm$  0.05 & -0.22 $\pm$  0.07 & ABS 	&& 8 & 0.33 $\pm$  0.13 & -0.05 $\pm$  0.18 & XRB		& & 8 & -0.17 $\pm$  0.21 & -0.23 $\pm$  0.15 & XRB \\ 
9 & -0.12 $\pm$  0.23 & -0.30 $\pm$  0.15 & XRB 	&& 9 & 0.13 $\pm$  0.08 & -0.16 $\pm$  0.08 & XRB 	& & 9 & 0.44 $\pm$  0.12 & -0.15 $\pm$  0.18 & ABS \\ 
10 & -0.22 $\pm$  0.48 & -0.00 $\pm$  0.34 & XRB && 10 & 0.19 $\pm$  0.12 & -0.13 $\pm$  0.12 & XRB		& & 10 & -0.07 $\pm$  0.15 &  0.67 $\pm$  0.22 & HARD \\ 
\hline \hline
\end{tabular}
\label{HRtable}
\tablecomments{Table~\ref{HRtable} is published in its entirety in the electronic edition of the journal. Only the first ten entries are shown for each galaxy for guidance regarding its form and content.}
\end{table*}

\begin{table*}[ht]
\centering
\caption{Percentage (\%) of Preliminary Classifications Based on Hardness Ratio Analysis}
\begin{tabular}{cccccc}
\hline \hline
Galaxy		& XRB			& HARD			& SOFT			& SNR			& ABS			\\
(1)			& (2)				& (3)				& (4)				& (5)				& (6)				\\
\hline
NGC~55		& 58.4$\pm$4.7	& 5.8$\pm$0.5		& 6.4$\pm$0.5		& 9.1$\pm$0.7		& 20.1$\pm$1.6	\\
NGC~300		& 62.1$\pm$6.4	& 8.4$\pm$0.9		& 4.2$\pm$0.4		& 8.4$\pm$0.9		& 16.8$\pm$1.7	\\
NGC~404		& 60.8$\pm$7.1	& 5.4$\pm$0.6		& 9.5$\pm$1.1		& 4.1$\pm$0.5		& 20.2$\pm$2.4	\\
NGC~2403	& 65.8$\pm$4.8	& 3.2$\pm$0.2		& 3.2$\pm$0.2		& 14.7$\pm$1.1	& 12.6$\pm$0.9	\\
NGC~4214	& 67.2$\pm$6.2	& 3.4$\pm$0.3		& 3.4$\pm$0.3		& 11.2$\pm$1.0	& 14.7$\pm$1.4	\\
\hline
Total			& 62.9$\pm$13.3	& 5.2$\pm$1.1		& 5.3$\pm$1.1		& 9.5$\pm$2.0		& 16.9$\pm$3.6	\\
\hline \hline
\label{HRclass_percent}
\end{tabular}
\end{table*} 

\clearpage

\begin{scriptsize}
\begin{center}
\renewcommand{\tabcolsep}{0.05cm}
\begin{longtable}{cccccccccccc}
\caption{Best-Fit Spectral Models for X-ray Sources with $\geq$50 Counts} \\

\hline \hline
Source	& Net Counts	& Best-fit$^a$	& \nH$^b$			& \PL		& $kT$		& $dof^c$		& $C^d$	&$\chi^{2,e}$	& $f_X^f$ (0.5-2 keV)	& $f_X^f$ (0.35-8 keV)	& ``goodness$^g$''	\\ \cline{10-11}
No.		& (0.35-8 keV)	& Model		& (10$^{21}$ cm$^{-2}$)	&			& (keV)		&			&		&			& \multicolumn{2}{c}{(10$^{-15}$ \flux)}	 		& (\%)			\\
(1)		& (2)			& (3)			& (4)					& (5)			& (6)			& (7)			& (8)		& (9)			& (10)				& (11)				& (12)			\\
\hline

\endfirsthead

\multicolumn{12}{c}{{\bfseries \tablename\ \thetable{} -- continued from previous page}} \\
\hline \hline
Source	& Net Counts	& Best-fit$^a$	& \nH$^b$			& \PL		& $kT$		& $dof^c$		& $C^d$	&$\chi^{2,e}$	& $f_X^f$ (0.5-2 keV)	& $f_X^f$ (0.35-8 keV)	& ``goodness$^g$''	\\ \cline{10-11}
No.		& (0.35-8 keV)	& Model		& (10$^{21}$ cm$^{-2}$)	&			& (keV)		&			&		&			& \multicolumn{2}{c}{(10$^{-15}$ \flux)}	 		& (\%)			\\
(1)		& (2)			& (3)			& (4)					& (5)			& (6)			& (7)			& (8)		& (9)			& (10)				& (11)				& (12)			\\
\hline
\endhead

\hline \multicolumn{12}{c}{{Continued on next page}} \\ 
\endfoot

\hline \hline
\multicolumn{12}{l}{Table~\ref{spectralfits} is published in its entirety in the electronic edition of the journal. Only the first ten entries are shown for each galaxy.} \\
\multicolumn{12}{l}{$^a$The best-fit model: {\it po} refers to a power law; {\it apec} is the thermal emission model; {\it diskbb} is a disk blackbody; {\it brems} refers} \\
\multicolumn{12}{l}{to bremsstrahlung emission. All models include a Galactic absorbing column.} \\
\multicolumn{12}{l}{$^b$Intrinsic source absorption, if beyond the Galactic column was required.} \\
\multicolumn{12}{l}{$^c$Degrees of freedom.} \\
\multicolumn{12}{l}{$^d$Cash statistic.} \\
\multicolumn{12}{l}{$^e$Pearson $\chi^2$.} \\
\multicolumn{12}{l}{$^f$ Unabsorbed X-ray fluxes.} \\
\multicolumn{12}{l}{$^g$ Results of the \texttt{XSPEC} ``goodness'' command, run using 5000 realizations.}\\
\multicolumn{12}{l}{Sources marked with an asterisk contain more than 500 net counts and are discussed further in the text.} \\
\multicolumn{12}{l}{$^\dagger$Spectral model includes pile-up component.} \\
\endlastfoot

\hline
\\[1pt]
\multicolumn{12}{c}{NGC~55} \\
\\[1pt]
\hline
3	& 81		& po			& ...				& 2.0$\pm$0.5			& ...					& 522	& 254	& 883	& 7.51	& 14.53	& 47	\\
7	& 103	& po			& ...				& 2.1$\pm$0.5			& ...					& 316	& 177	& 280	& 11.37	& 21.48	& 64	\\
8	& 169	& po			& ...				& 1.3$\pm$0.3			& ...					& 521	& 355	& 739	& 13.03	& 49.16	& 56	\\
11	& 79		& po			& $<$2.2			& 2.3$^{+0.8}_{-0.6}$	& ...					& 316	& 176	& 309	& 8.07	& 13.47	& 45	\\
15	& 54		& diskbb		& ...				& ...					& 0.9$^{+0.4}_{-0.3}$	& 521	& 180	& 727	& 4.15	& 7.40	& 55	\\
23$^*$ & 2692	& po+2apec	& ...				& 2.7$\pm$0.2			& 0.3$\pm$0.1; 1.0$\pm$0.1 & 449	& 358	& 477	& 261.0	& 307.4	& 62	\\
37	& 123	& po+apec	& ...				& 2.9$\pm$0.1			& 1.0$\pm$0.1			& 314	& 327	& 331	& 13.05	& 16.60	& 65	\\
44	& 205	& po			& ...				& 1.3$\pm$0.2			& ...					& 521	& 356	& 509	& 112.54	& 386.60	& 45	\\
52	& 65		& apec		& ...				& ...					& 3.4$\pm$1.3			& 521	& 247	& 460	& 3.75	& 7.94	& 47	\\
54	& 439	& po+apec	& 2.8$^{+0.4}_{-0.2}$ & 1.4$\pm$0.3		& 0.3$\pm$0.1			& 450	& 410	& 442	& 52.48	& 138.98	& 40	\\
\hline
\\[1pt]
\multicolumn{12}{c}{NGC~2403} \\
\\[1pt]
\hline
2$^*$ & 904	& po+2apec	& ...				&1.4$\pm$0.2			& $<$0.3; 0.67$^{+0.1}_{-0.9}$ & 520 & 515 	& 541	& 13.88	& 34.76	& 37	\\
3	& 55		& po			& ...				& 1.7$\pm$0.5			& ...					& 315	& 170	& 340	& 1.48	& 3.70	& 43	\\
9	& 82		& diskbb		& ...				& ...					& 0.5$^{+0.2}_{-0.1}$	& 373	& 201	& 301	& 2.16	& 4.53	& 65	\\
11	& 275	& po+diskbb	& ...				& 1.1$^{+0.6}_{-0.7}$	& 0.2$\pm$0.1			& 314	& 305	& 303	& 15.63	& 35.55	& 61	\\
12	& 90		& diskbb		& 27.3$^{+1.8}_{-1.1}$ & ...				& 0.6$\pm$0.3			& 509	& 298	& 473	& 31.80	& 41.71	& 61	\\
19	& 115	& po			& ...				& 2.6$^{+0.5}_{-0.4}$	& ...					& 442	& 238	& 417	& 5.11	& 7.33	& 52	\\
20	& 364	& po			& ...				& 2.1$\pm$0.2			& ...					& 316	& 274	& 341	& 5.51	& 10.33	& 65	\\
23	& 57		& diskbb		& ...				& ...					& 0.6$^{+0.5}_{-0.2}$	& 316	& 217	& 466	& 2.94	& 3.83	& 60	\\
26	& 51		& po			& ...				& 1.2$\pm$0.6			& ...					& 267	& 141	& 297	& 5.17	& 21.26	& 56	\\
29	& 161	& apec		& 35.5$^{+12.8}_{-9.5}$ & ...				& 4.0$\pm$1.9			& 352	& 322	& 468	& 45.86	& 90.33	& 71	\\

\hline
\\[1pt]
\multicolumn{12}{c}{NGC~4214} \\
\\[1pt]
\hline
16$^*$ & 1827	& po	 		& 1.5$\pm$0.4 		& 1.8$\pm$0.1			& ...					& 317	& 312	& 286	& 92.29	& 213.38	& 54	\\
17	& 347	& apec+apec	& ...				& ...					& 3.9$\pm$0.9; 0.6$^{+0.2}_{-0.4}$ & 519 & 362 & 471	& 23.99	& 47.92	& 58	\\
22	& 198	& po			& 1.3$^{+1.6}_{-0.6}$ & 2.1$^{+0.7}_{-0.5}$	& ...					& 318	& 319	& 305	& 27.38	& 51.48	& 43	\\
24	& 78		& po			& ...				& 3.1$^{+0.8}_{-0.6}$	& ...					& 453	& 421	& 419	& 12.19	& 14.94	& 59	\\
27	& 58		& po			& ...				& 1.0$^{+0.3}_{-0.2}$	& ...					& 385	& 252	& 832	& 2.39	& 11.94	& 60	\\
28	& 60		& brems		& $<$27.9			& ...					& $<$1.3				& 476	& 202	& 437	& 13.34	& 14.03	& 75	\\
30	& 52		& po			& ...				& 2.5$\pm$0.8			& ...					& 454	& 219	& 396	& 3.94	& 5.57	& 66	\\
32	& 62		& apec		& ...				& 					& 6.3$\pm$1.6			& 318	& 176	& 335	& 2.31	& 6.35	& 66	\\
33	& 94		& po			& ...				& 2.5$\pm$0.5			& ...					& 522	& 296	& 469	& 11.81	& 17.84	& 65	\\
35	& 72		& po			& ...				& 1.9$\pm$0.4			& ...					& 521	& 213	& 324	& 7.80	& 16.48	& 57	\\
\label{spectralfits}
\end{longtable}
\end{center}
\end{scriptsize}

\begin{table*}[ht]
\centering
\caption{AGN Contribution as a Function of De-Projected Radius}
\begin{tabular}{cccccc}
\hline \hline
Galaxy		& $<$10\%	& 25\%	& 50\%	& 75\%	& 100$^a$\%	\\
(1)			& (2)			& (3)		& (4)		& (5)		& (6)		\\
\hline	
NGC~55		& 2.0 kpc		& 3.0 kpc	& 3.8 kpc	& 4.3 kpc	& 4.7 kpc	\\
NGC~2403	& 0.8 kpc		& 2.1 kpc	& 3.2 kpc	& 3.9 kpc	& 4.5 kpc	\\
NGC~4214	& 0.9 kpc		& 1.9 kpc	& 2.7 kpc	& 3.4 kpc	& 4.0 kpc	\\
NGC~404		& 0.5 kpc		& 0.7 kpc	& 1.8 kpc	& 2.6 kpc	& 3.3 kpc	\\ 
\hline \hline
\end{tabular}\label{no_AGN}
\tablecomments{$^a$The radius at which 100\% of X-ray sources are expected to be AGN is referred to in the text as $r_{\rm bkg}$.}
\end{table*}

\begin{table*}[ht]
\centering
\caption{Rapidly Variable X-ray Sources From \Chandra Observations}
\begin{tabular}{ccc}
\hline \hline
Galaxy	& Source	& $\xi^a$	\\
(1)    	 	& (2)		& (3)		\\
\hline
NGC~2403	& 2		& 10$^{-10.74}$	\\
NGC~2403	& 78		& 10$^{-42.57}$	\\
NGC~2403	& 179	& 10$^{-14.86}$	\\
NGC~4214	& 16		& 10$^{-7.03}$		\\
NGC~4214	& 24		&10$^{-3.63}$		\\
NGC~4214	& 84		& 10$^{-3.44}$		\\
NGC~4214	& 87		& 10$^{-5.06}$		\\
\hline\hline
\end{tabular}\label{table_rapid_variable}
\tablecomments{$^a$The K-S probability of the source being constant within either \Chandra observation.}
\end{table*}

\begin{table*}[ht]
\centering
\caption{Long-Term Variability of X-ray Sources From \Chandra Observations}
\begin{tabular}{ccccccccccc}
\hline \hline
\multicolumn{3}{c}{NGC~55}					&& \multicolumn{3}{c}{NGC~2403}					&& \multicolumn{3}{c}{NGC~4214}	\\  \cline{1-3} \cline{5-7} \cline{9-11}
No.	& $\eta^a$	& flux$_{max}$/flux$_{min}^b$	&& No.	& $\eta^a$	& flux$_{max}$/flux$_{min}^b$	&& No.	& $\eta^a$	& flux$_{max}$/flux$_{min}^b$	\\
(1)     & (2)		& (3)						&& (4)	& (5)			& (6)						&& (7)	& (8)			& (9)				\\
\hline
2	& 6.4		& 1.7		&& 3	& 119  	& 27.1	&& 2	&  9.0 	& 3.9	  	\\
3	& 5.1		& 1.9 	&& 4	& 6.5 	& 2.4 	&& 3	& 19.5 	& 2.7		\\
6	& 5.8		& 1.6	 	&& 7	& 5.2		& 5.2		&& 10 & 9.9	& 3.3		\\
10	& 11.8	& 2.3		&& 12 & 20.1	& 1.6		&& 12 & 12.4   	& 7.3 	\\
12	& 22.5	& 36.5	&& 13 & 8.6	& 2.0		&& 14 & 13.0	& 1.5		\\
13	& 38.5	& 3.4 	&& 15 & 10.6	& 5.3		&& 17 & 57.6	& 4.5		\\
14	& 11.1	& 2.7	 	&& 16 & 6.6 	& 1.6		&& 18 & 16.5 	& 1.4		\\
16	& 11.7	& 2.3	 	&& 18 & 7.6	& 3.0		&& 19 & 7.4 	& 2.3		\\
18	& 39.2	& 3.4 	&& 20 & 13.9 	& 3.1		&& 20 & 49.7	& 6.1	 	\\
20	& 2255	& 142	&& 21 & 20.8 	& 9.2		&& 21 & 47.8 	& 18.5	\\
\hline\hline
\end{tabular}\label{table_variable}
\tablecomments{Table~\ref{table_variable} is published in its entirety in the electronic edition of the journal. Only the first ten entries are shown for each galaxy for guidance regarding its form and content. $^a$$\eta$ is the variability index defined by \cite{Tullmann+11} and is sensitive to long-term variability. Only sources with $\eta\ge5$, indicative of a variability between the individual \Chandra exposures, are shown.}
\end{table*}

\begin{table*}[ht]
\centering
\caption{Long-Term Variability Data for NGC~55 X-ray Sources}
\begin{tabular}{cccccccc}
\hline \hline
\multirow{2}{*}{Source}	& \multicolumn{4}{c}{0.35-8 keV flux ($10^{-15}$ \flux)}						&& \multicolumn{2}{c}{$f_{\rm max}/f_{\rm min}$} \\ \cline{2-5} \cline{7-8}
		& 11 Sept. 2001$^a$	& 14 Nov. 2001$^b$	& 15 Nov. 2001$^c$	& 29 June 2004$^d$	&& detections	& upper limits		\\
(1)		& (2)					& (3)				& (4)				& (5)					&& (6)		& (7)			\\
\hline
3		& 2.45				& 24.33			& $<$9.40			& 1.26		&& 19.3	& ...			\\
6		& 3.31				& 9.71			& $<$6.37			& 5.15		&& 2.9	& ...			\\
7		& 3.17				& 51.74			& 29.14			& $<$3.01		&& 16.3	& $>$17.2		\\
8		& 0.83				& 48.11			& 51.81			& $<$3.01		&& 62.4	& ...			\\
11		& 3.05				& 19.94			& 8.25			& $<$3.01		&& 6.5	& $>$6.6		\\
12		& 0.24				& 14.36			& 15.41			& 8.74		&& 64.2	& ...			\\
13		& 10.20				& $<$6.84			& 13.13			& $<$3.01		&& 1.3	& $>$4.4		\\
14		& 2.26				& $<$7.09			& $<$6.28			& 6.02		&& 2.7	& ...			\\
15		& 2.74				& 14.98			& 5.72			& 3.98		&& 5.5	& ...			\\
16		& 3.27				& $<$7.56			& 5.78			& 7.48		&& 2.3	& ...			\\
\hline \hline
\end{tabular}
\tablecomments{Table~\ref{NGC55_longterm} is published in its entirety in the electronic edition of the journal. Only the first ten entries are shown for guidance regarding its form and content. \newline
$^a$MJD 52176; \Chandra Obs. ID 2255. $^b$MJD 52227; \XMM Obs. ID 0028740201. $^c$MJD 52228; \XMM Obs. ID 0028740101. $^d$MJD 53185; \Chandra Obs. ID 4744.}
\label{NGC55_longterm}
\end{table*}

\begin{table*}[ht]
\centering
\caption{Long-Term Variability Data for NGC~2403 X-ray Sources}
\begin{tabular}{cccccccccccc}
\hline \hline
\multirow{3}{*}{Source}	& \multicolumn{8}{c}{0.35-8 keV flux ($10^{-15}$ \flux)}														&&\multicolumn{2}{c}{$f_{\rm max}/f_{\rm min}$}  \\ \cline{2-9}   \cline{11-12}
		& 17 Apr.		& 30 Apr. 		& 11 Sept. 	& 9 Aug. 		& 23 Aug. 		& 12 Sept. 	& 3 Oct. 		& 22 Dec. 		&&  \multirow{2}{*}{detections}	& \multirow{2}{*}{upper limits}	\\	
		& 2001$^a$	& 2003$^b$	& 2003$^c$	& 2004$^d$	& 2004$^e$	& 2004$^f$	& 2004$^g$	&2004$^h$	&& 						&					\\
(1)		& (2)			& (3)			& (4)			& (5)			& (6)			& (7)			& (8)			& (9)			&& (10)		& (11)		\\
\hline
3	& 5.27		& $<$35.09	& $<$39.04	& 72.68		& 2.69		& 4.93		& 63.31		& 3.38		&& 27.0	& ...			\\
4	& 5.26		& $<$34.49	& $<$40.10	& ...			& 4.61		& $<$5.34		& 2.22		& ...			&& 2.4	& ...			\\
7	& 1.09		& 54.21		& 51.21		& ...			& 0.64		& 24.60		& 1.71		& 0.33		&& 164.3	& ...			\\
12	& 62.95		& $<$42.22	& 88.37		& ...			& ...			& 26.23		& 38.26		& ...			&& 3.4	& ...			\\
13	& 6.50		& $<$22.39	& $<$26.73	& ...			& 12.85		& $<$4.84		& $<$10.66	& ...			&& 2.0	& $>$2.7		\\
15	& 1.55		& $<$52.63	& $<$62.05	& ...			& 6.30		& $<$12.71	& $<$8.22		& ...			&& 4.1	& ...			\\
16	& 8.37		& $<$28.00	& $<$26.33	& ...			& 13.39		& 8.95		& $<$13.39	& ...			&& 1.6	& ...			\\
18	& 4.67		& $<$34.24	& $<$29.51	& ...			& 4.25		& $<$5.29		& $<$1.54		& ...			&& 1.1	& $>$3.0		\\
20	& 6.17		& $<$27.15	& 51.66		& ...			& 12.91		& 19.98		& 19.22		& ...			&& 8.4	& ...			\\
21	& 8.83		& $<$29.23	& $<$49.17	& 8.82		& 24.37		& $<$4.87		& 2.65		& 13.30		&& 9.2	& ...			\\
\hline \hline
\end{tabular}
\tablecomments{Table~\ref{NGC2403_longterm} is published in its entirety in the electronic edition of the journal. Only the first ten entries are shown for guidance regarding its form and content. \newline
$^a$MJD 52016; \Chandra Obs. ID 2014. $^b$MJD 52759; \XMM Obs. ID 0150651101. $^c$MJD 52893; \XMM Obs. ID 0150651201. $^d$MJD 53226; \Chandra Obs. ID 4627. $^e$MJD 53240; \Chandra Obs. ID 4628. $^f$MJD 53260; \XMM Obs. ID 0164560901. $^g$MJD 53281; \Chandra Obs. ID 4629. $^h$MJD 53361; \Chandra Obs. ID 4630.
}
\label{NGC2403_longterm}
\end{table*}

\begin{table*}[ht]
\centering
\caption{Long-Term Variability Data for NGC~4214 X-ray Sources}
\begin{tabular}{cccccccc}
\hline \hline
Source	& \multicolumn{4}{c}{0.35-8 keV flux ($10^{-15}$ \flux)}						&& \multicolumn{2}{c}{$f_{\rm max}/f_{\rm min}$} \\ \cline{2-5}  \cline{7-8}
		& 16 Oct. 2001$^a$	& 22 Nov. 2001$^b$	& 3 Apr. 2004$^c$	& 30 Jul. 2004$^d$	&& detections	& upper limits		\\
(1)		& (2)				& (3)				& (4)				& (5)				&& (6)		& (7)				\\
\hline
2	& 0.83		& 161.80		& 2.31		& ...			&& 194.9		& ...			\\
3	& 14.49		& $<$36.74	& ...			& 5.30		&& 2.7		& ...			\\
10	& 5.13		& $<$21.73	& 2.76		& 1.57		&& 3.3		& ...			\\
12	& 3.83		& 42.23		& 1.65		& 0.53		&& 79.7		& ...			\\
14	& 20.55		& $<$60.51	& ...			& 31.65		&& 1.5		& ...			\\
17	& 213.01		& $<$91.70	& 47.22		& 196.15		&& 4.5		& ...			\\
18	& 71.12		& $<$46.42	& 50.87		& 60.43		&& 1.4		& ...			\\
19	& 4.44		& $<$50.11	& 1.94		& ...			&& 2.3		& ...			\\
20	& 14.04		& 188.95		& 2.84		& 17.44		&& 66.5		& ...			\\
21	& 0.51		& $<$30.64	& 9.44		& ...			&& 18.9		& ...			\\
\hline \hline
\end{tabular}
\tablecomments{Table~\ref{NGC4214_longterm} is published in its entirety in the electronic edition of the journal. Only the first ten entries are shown for guidance regarding its form and content. \newline
$^a$MJD 52198; \Chandra Obs. ID 2030. $^b$MJD 52235; \XMM Obs. ID 0035940201. $^c$MJD 53098; \Chandra Obs. ID 4743. $^d$MJD 53216; \Chandra Obs. ID 5197.}
\label{NGC4214_longterm}
\end{table*}

\begin{table*}[ht]
\centering
\caption{Public Multiwavelength Data for NGC~55 X-ray Sources}
\begin{tabular}{ccccccccccccccccc}
\hline \hline	
Source	& \# \HST		& & SIMBAD	& & \multicolumn{5}{c}{USNO-B1.0}				& & \multicolumn{3}{c}{2MASS}	& & \multicolumn{2}{c}{{\it GALEX}}	\\ 
									\cline{6-10} 							\cline{12-14}					\cline{16-17}										
No.		& matches	& & ID		& & $B1$	& $R1$	& $B2$	& $R2$	& $I$		& & $J$	& $H$	& $K$		& & FUV	& NUV				\\
(1)		& (2)			& & (3)		& & (4)	 & (5)	& (6)		& (7)		& (8)		& & (9)	& (10)	& (11)		& & (12)	& (13)				\\
\hline
1		&			& &			& &		&		&		&		&		& &		&		&		& & 19.40	&		\\
2		&			& &			& &		& 19.88	& 17.83	&		&		& &		&		&		& & 22.24	& 21.28	\\
3		&			& &			& &		&		&		&		&		& &		&		&		& & 24.18	& 23.41	\\
5		&			& &			& &		&		&		&		&		& &		&		&		& & 		& 22.66	\\
7		&			& & ROSAT	& &		& 19.93	& 20.45	& 23.47	&		& &		&		&		& & 23.29	& 20.25	\\
8		&			& &			& &		& 20.17	& 20.22	& 20.55	&		& &		&		&		& &		& 22.31	\\
10		&			& &			& &		&		&		&		&		& &		&		&		& & 16.47	&		\\
11		&			& &			& &		&		&		&		&		& &		&		&		& & 21.67	&		\\
12		&			& &			& &		&		&		&		&		& &		&		&		& & 23.44	& 22.82	\\
13		&			& &			& &		&		&		&		&		& &		&		&		& & 21.88	& 22.66	\\
\hline \hline
\end{tabular}
\tablecomments{Table~\ref{multiwavelength55} is published in its entirety in the electronic edition of the journal. Only the first ten entries are shown for guidance regarding its form and content.\newline
$^a$SIMBAD counterparts are listed as: SNR (supernova remnant), radio (radio source), H~II (H~II region), galaxy, Cl* (star cluster), QSO, SN (supernova designation), star, ULX, ROSAT (historically known {\it ROSAT} source), EINSTEIN (historically known {\it Einstein} source), GC (globular cluster), cephied, em. gal (emission-line galaxy), gal. group (galaxy group), RS CVN (cataclysmic variable). }
\label{multiwavelength55}
\end{table*}

\begin{table*}[ht]
\centering
\caption{Public Multiwavelength Data for NGC~2403 X-ray Sources}
\begin{tabular}{ccccccccccccccccc}
\hline \hline	
Source	& \# \HST		& & SIMBAD	& & \multicolumn{5}{c}{USNO-B1.0}				& & \multicolumn{3}{c}{2MASS}	& & \multicolumn{2}{c}{{\it GALEX}}	\\ 
									\cline{6-10} 							\cline{12-14}					\cline{16-17}				
No.		& matches	& & ID		& & $B1$	& $R1$	& $B2$	& $R2$	& $I$		& & $J$	& $H$	& $K$		& & FUV	& NUV				\\
(1)		& (2)			& & (3)		& & (4)	 & (5)	& (6)		& (7)		& (8)		& & (9)	& (10)	& (11)		& & (12)	& (13)				\\
\hline
1		&			& &			& &		&		&		&		&		& & 		&		&		& & 21.03	& 21.02	\\
2		& 2			& &			& &		&		&		&		&		& & 		&		&		& &		&		\\
3		&			& &			& &		&		&		&		&		& & 		&		&		& & 17.47	& 17.32	\\
5		& 3			& &			& &		&		&		&		&		& & 		&		&		& &		&		\\
7		& 11			& &			& &		&		&		&		&		& & 		&		&		& & 16.01	& 15.80	\\
9		&			& &			& &		&		&		&		&		& & 		&		&		& & 20.92	& 19.01	\\
11		&			& &			& &		&		&		&		&		& & 		&		&		& & 23.46	& 22.37	\\
12		&			& &			& & 20.85	& 20.06	&		&		&		& &		&		&		& &		&		\\
13		&			& &			& &		&		&		&		&		& & 		&		&		& &		&		\\
14		&			& &			& & 20.99	&		& 21.77	&		&		& &		&		&		& & 23.48	& 22.83	\\
\hline \hline
\end{tabular}
\tablecomments{Table~\ref{multiwavelength2403} is published in its entirety in the electronic edition of the journal. Only the first ten entries are shown for guidance regarding its form and content.\newline
$^a$SIMBAD counterparts are listed as: SNR (supernova remnant), radio (radio source), H~II (H~II region), galaxy, Cl* (star cluster), QSO, SN (supernova designation), star, ULX, ROSAT (historically known {\it ROSAT} source), EINSTEIN (historically known {\it Einstein} source), GC (globular cluster), cephied, em. gal (emission-line galaxy), gal. group (galaxy group), RS CVN (cataclysmic variable).}
\label{multiwavelength2403}
\end{table*}

\begin{table*}[ht]
\centering
\caption{Public Multiwavelength Data for NGC~4214 Sources}
\setlength{\tabcolsep}{1.5pt}
\begin{tabular}{ccccccccccccccccccccccc}
\hline \hline	
Source	& \HST		& & SIMBAD	& & \multicolumn{5}{c}{USNO-B1.0}				& & \multicolumn{5}{c}{SDSS}		& & \multicolumn{3}{c}{2MASS}	& & \multicolumn{2}{c}{{\it GALEX}}		\\ 
									\cline{6-10} 								\cline{12-16}					\cline{18-20}					\cline{22-23}				
No.		& matches	& & ID		& & $B1$	& $R1$	& $B2$	& $R2$	& $I$		& & $u$	& $g$	& $r$	& $i$		& $z$	& & $J$	& $H$	& $K$		& & FUV	& NUV		\\
(1)		& (2)			& & (3)		& & (4)	 & (5)	& (6)		& (7)		& (8)		& & (9)	& (10)	& (11)	& (12)	& (13)	& & (14)	& (15)	& (16)		& & (17)	& (18)		\\
\hline
1		&			& &			& & 18.25	& 19.79	&		& 19.46	&		& & 22.25	& 22.28	& 21.79	& 22.47	& 22.59	& &  		&		&		& &  		&		\\
2		&			& &			& &		&		&		&		&		& & 22.97	& 23.68	& 23.64	& 22.37	& 20.15	& & 		&		&		& &   		&		\\
3		&			& &			& &		&		&		&		&		& & 22.44	& 22.72	& 23.05	& 21.62	& 21.70	& & 		&		&		& &		& 22.64	\\
4		&			& &			& &		&		&		&		&		& & 		&		&		&		&		& &		&		&		& & 		& 24.34	\\
5		&			& &			& & 		& 19.47	& 20.59	& 19.44	& 17.21	& & 23.11	& 20.81	& 19.31	& 17.86	& 17.08	& & 15.83	& 15.12	& 14.60	& & 20.15	&		\\
6		&			& &			& &		&		&		&		&		& & 23.14	& 22.86	& 22.18	& 21.64	& 21.32	& &  		&		&		& & 		& 23.53	\\
8		&			& &			& &		&		&		&		&		& & 		&		&		&		&		& &		&		&		& & 		& 21.32	\\
10		&			& &			& & 		&		&		& 19.91	& 18.63	& &  		&		&		&		&		& &  		&		&		& &  		&		\\
13		&			& &			& & 		& 20.25	& 20.25	& 20.76	&		& & 20.57	& 20.54	& 20.54	& 20.15	& 20.23	& &  		&		&		& & 22.65	& 21.79	\\
14		&			& &			& &		&		&		&		&		& & 21.82	& 21.41	& 21.12	& 20.62	& 20.25	& &  		&		&		& & 		& 21.99	\\ 
\hline \hline
\end{tabular}
\tablecomments{Table~\ref{multiwavelength4214} is published in its entirety in the electronic edition of the journal. Only the first ten entries are shown guidance regarding its form and content.\newline
$^a$SIMBAD counterparts are listed as: SNR (supernova remnant), radio (radio source), H~II (H~II region), galaxy, Cl* (star cluster), QSO, SN (supernova designation), star, ULX, ROSAT (historically known {\it ROSAT} source), EINSTEIN (historically known {\it Einstein} source), GC (globular cluster), cephied, em. gal (emission-line galaxy), gal. group (galaxy group), RS CVN (cataclysmic variable).}
\label{multiwavelength4214}
\end{table*}

\begin{table*}[ht]
\centering
\caption{Final X-ray Source Classification}
\begin{tabular}{cccccccccc}
\hline \hline
Source 	& Class		& $d$ 	& Assoc. with	& Variability	& X-ray		& $f_x/f_o$	& Multi-					& Known			& Final  		\\
No.		& from HRs	& (kpc)	& galaxy?		&			& Spectrum	& Class		& wavelength				& Counterpart$^a$	& Classification \\
 (1)		& (2)			& (3)	    	& (4)		 	& (5) 		& (6)			& (7)			& (8)						& (9)				& (10)		 \\
\hline
\multicolumn{10}{c}{NGC~55} \\
\hline
1		& XRB		& 2.3		&			& 			&			&			& FUV					&			& AGN 		\\
2		& ABS		& 3.0		&			& long		&			&			& optical, FUV, NUV			&			& AGN		\\
3		& XRB		& 6.3		&			& long		& po			&			& FUV, NUV				&			& AGN 		\\
4		& XRB		& 4.4		&			& 			&			&			& 						&			& AGN		\\
5		& ABS		& 3.2		&			& 			&			&			& NUV					&			& AGN		\\
6		& ABS		& 0.6		& \checkmark	& long		&			&			& 						&			& HMXB 		\\
7		& SOFT		& 7.6		&			& 			& po			&			& optical, FUV, NUV			& ROSAT		& AGN 		\\
8		& ABS		& 6.4		&			& 			& po			&			& optical, NUV				&			& AGN		\\
9		& XRB		& 2.3		&			& 			&			&			& 						&			& AGN		\\
10		& XRB		& 0.2		& \checkmark	& long		&			&			& FUV					&			& LMXB?		\\
\hline
\multicolumn{10}{c}{NGC~2403} \\
\hline
1		& XRB		& 0.5		&			&			&			& 			& FUV, NUV				&			& AGN		\\
2		& XRB		& 1.4		& \checkmark	& rapid		& po+2apec	& AGN		& 						&			& LMXB		\\
3		& SOFT		& 3.1		& \checkmark	& long		& po			& 			& FUV, NUV				&			& LMXB?		\\
4		& ABS		& 1.7		&			& long		&			&  			&						&			& AGN		\\
5		& XRB		& 1.0		& \checkmark	&			&			& AGN		&						&			& LMXB		\\
6		& SNR		& 2.8		& \checkmark	&			&			&  			&						&			& HMXB		\\
7		& ABS		& 1.3		& \checkmark	& long		&			& AGN		& FUV, NUV				&			& LMXB		\\
8		& ABS		& 3.5		& \checkmark	&			&			&  			&						&			& HMXB		\\
9		& XRB		& 4.9		&			&			& diskbb		&  			& FUV, NUV				&			& AGN		\\
10		& XRB		& 0.8		& \checkmark	&			&			&  			&						&			& HMXB		\\
\hline
\multicolumn{10}{c}{NGC~4214} \\
\hline
1		& SNR		& 0.8		&  			& 			& 			& 		& optical				& 				& star?		\\
2		& XRB		& 1.5		& 			& long 		& 			&  		& optical				& 				& AGN		\\
3		& XRB		& 1.8		& 			& long		&  			& 		& optical				& 				& AGN		\\
4		& XRB		& 0.0		& 			& 			&  			& 		& NUV				& 				& AGN		\\
5		& SNR		& 0.6		& 			& 			&  			& 		& optical, FUV			& 				& star?		\\
6		& XRB		& 0.6		& 			& 			&  			& 		& optical, NUV			& 				& AGN		\\
7		& XRB		& 1.2		&  \checkmark	& 			&  			& 		& 					&				& HMXB		\\
8		& XRB		& 0.8		& 			& 			&  			& 		& NUV				& 				& AGN		\\
9		& ABS		& 0.6		& 			& 			&  			& 		& 					&				& AGN		\\
10		& HARD		& 1.5		&  \checkmark	& long		&  			& 		& optical				& 				& LMXB		\\
\hline \hline
\end{tabular}
\tablecomments{Table~\ref{classification} is published in its entirety in the electronic edition of the journal. Only the first ten entries are shown for each galaxy for guidance regarding its form and content. \newline
$^a$SIMBAD counterparts are listed as: SNR (supernova remnant), radio (radio source), H~II (H~II region), galaxy, Cl* (star cluster), QSO, SN (supernova designation), star, ULX, ROSAT (historically known {\it ROSAT} source), EINSTEIN (historically known {\it Einstein} source), GC (globular cluster), cephied, em. gal (emission-line galaxy), gal. group (galaxy group), RS CVN (cataclysmic variable).}
\label{classification}
\end{table*}

\clearpage


\begin{figure*}
\centering
\begin{tabular}{cc}
	\begin{overpic}[width=0.3\linewidth,angle=-90,clip=true]{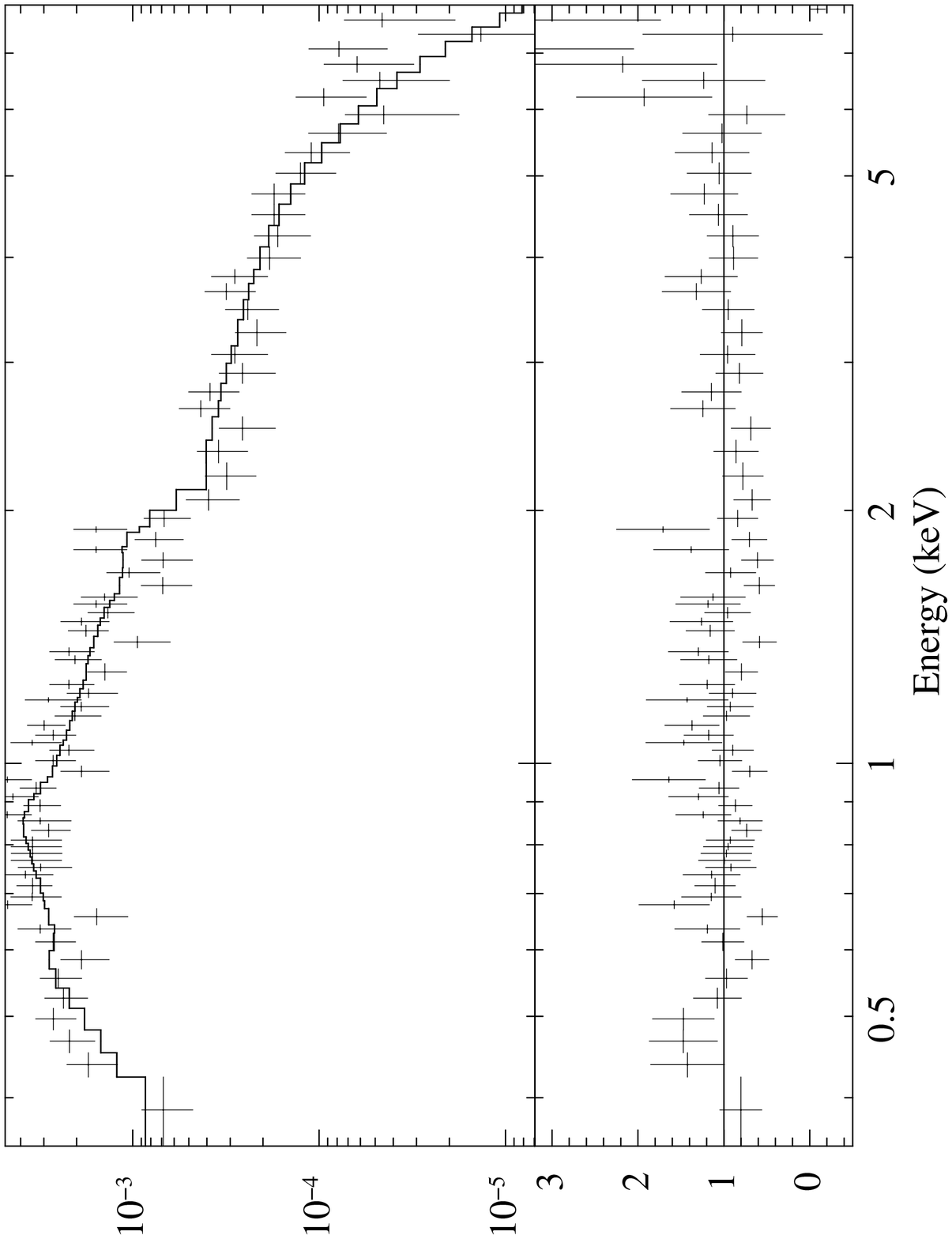} 
 		\put (79,62) {Source 2}
	\end{overpic}	&
	\begin{overpic}[width=0.3\linewidth,angle=-90,clip=true]{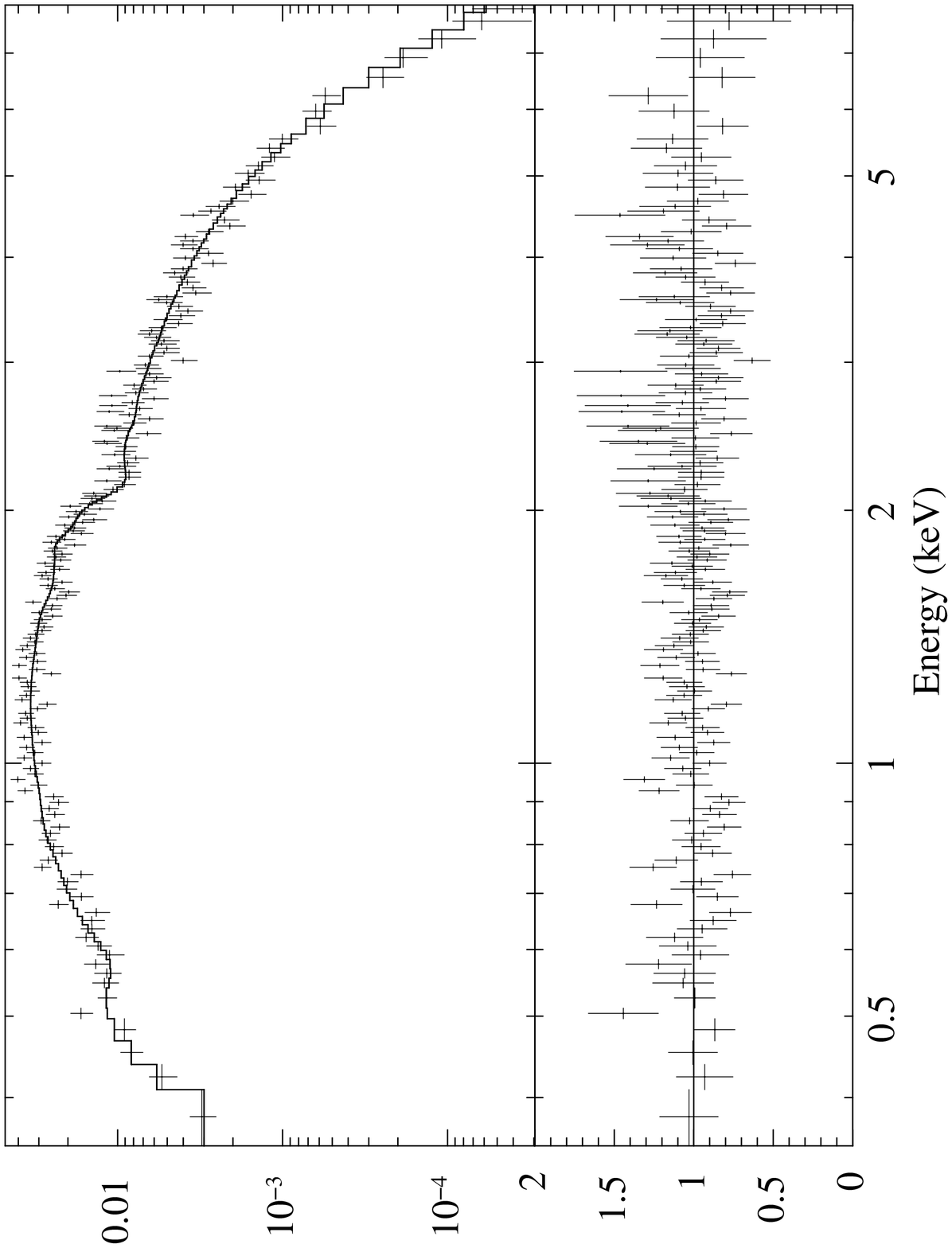} 
 		\put (79,62) {Source 41}
	\end{overpic}	\\

	\begin{overpic}[width=0.3\linewidth,angle=-90,clip=true]{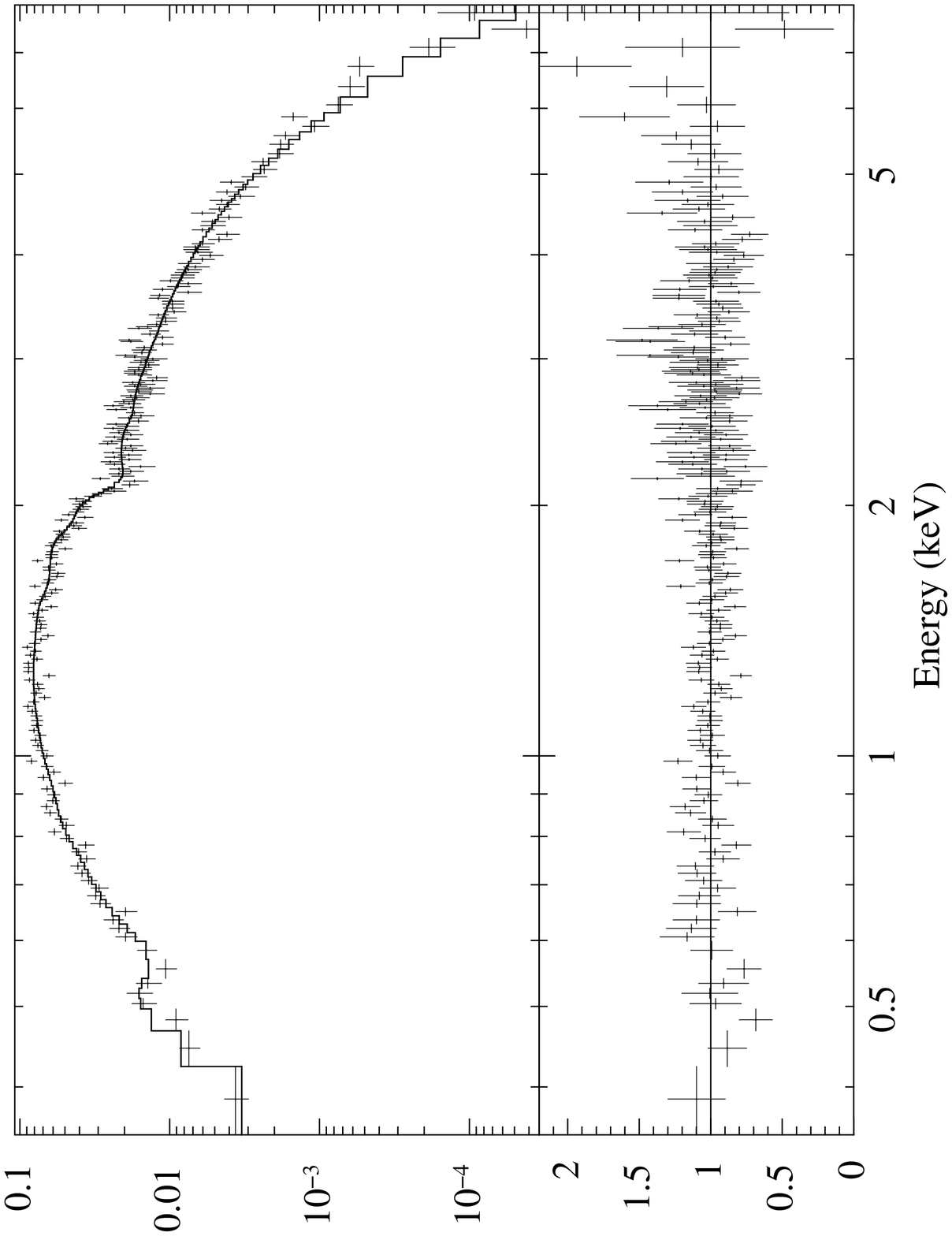} 
 		\put (79,62) {Source 42}
	\end{overpic}	&
	\begin{overpic}[width=0.3\linewidth,angle=-90,clip=true]{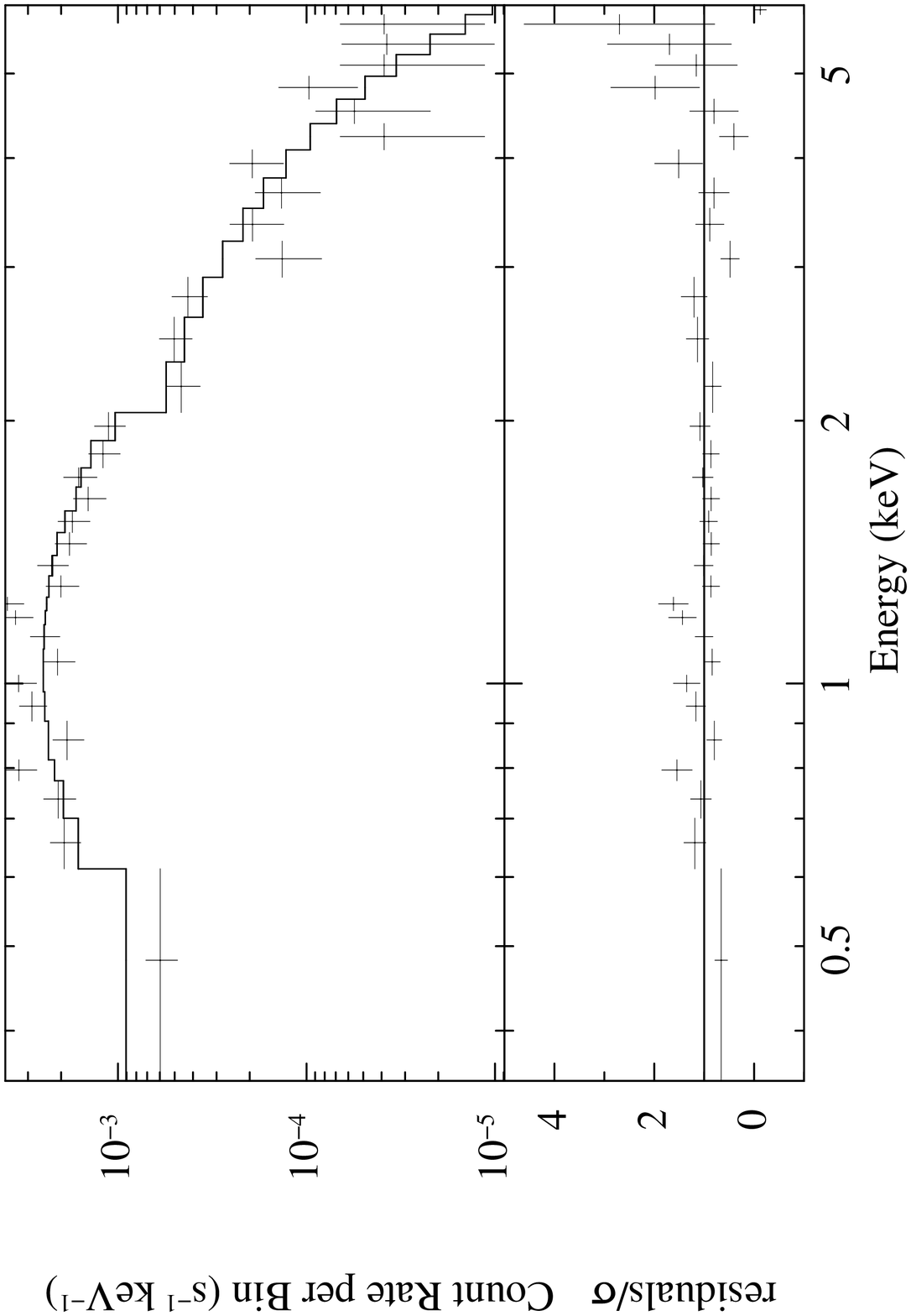} 
 		\put (79,62) {Source 51}
	\end{overpic}	\\
	
	\begin{overpic}[width=0.3\linewidth,angle=-90,clip=true]{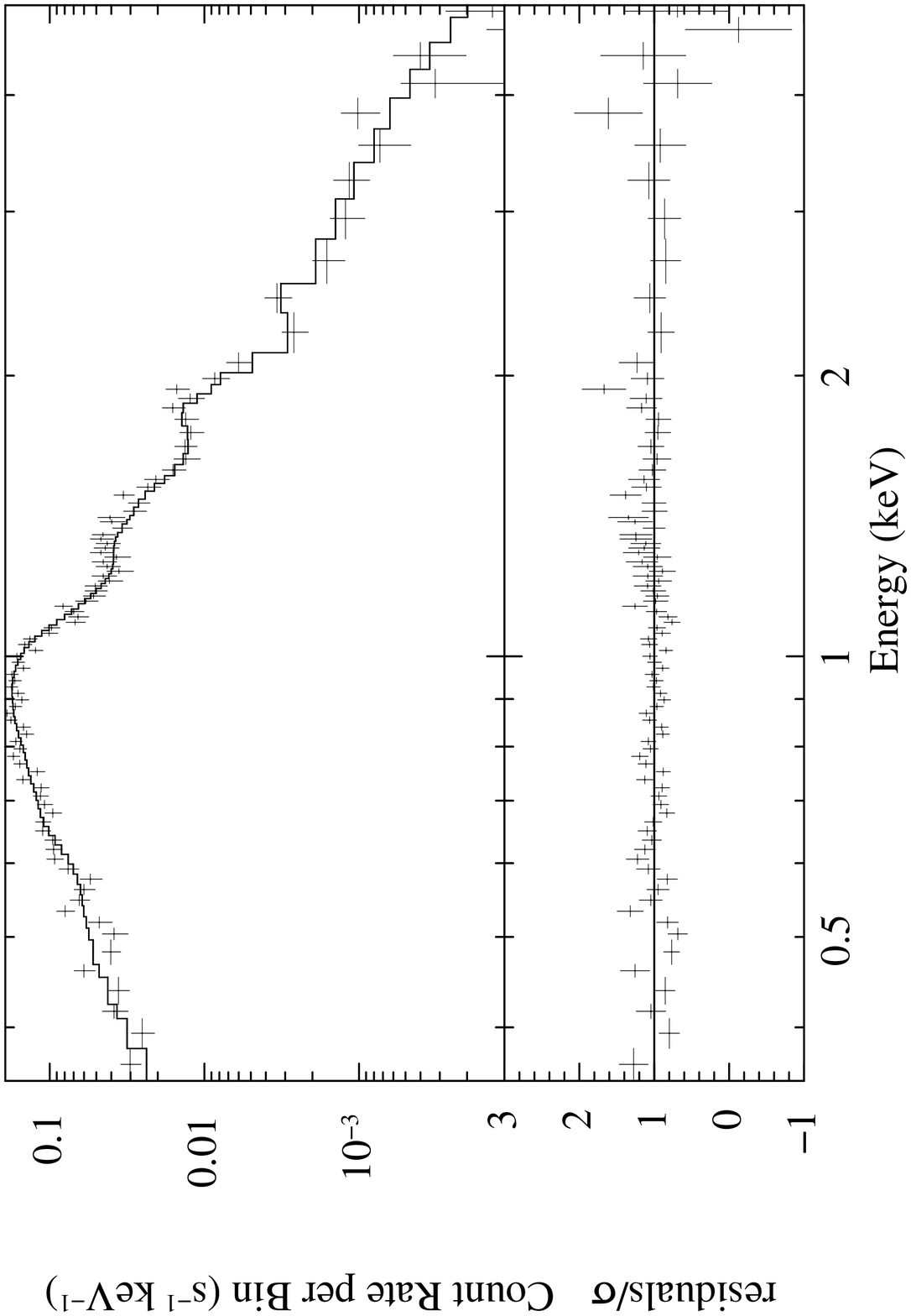} 
 		\put (79,60) {Source 62}
	\end{overpic}	&
	\begin{overpic}[width=0.3\linewidth,angle=-90,clip=true]{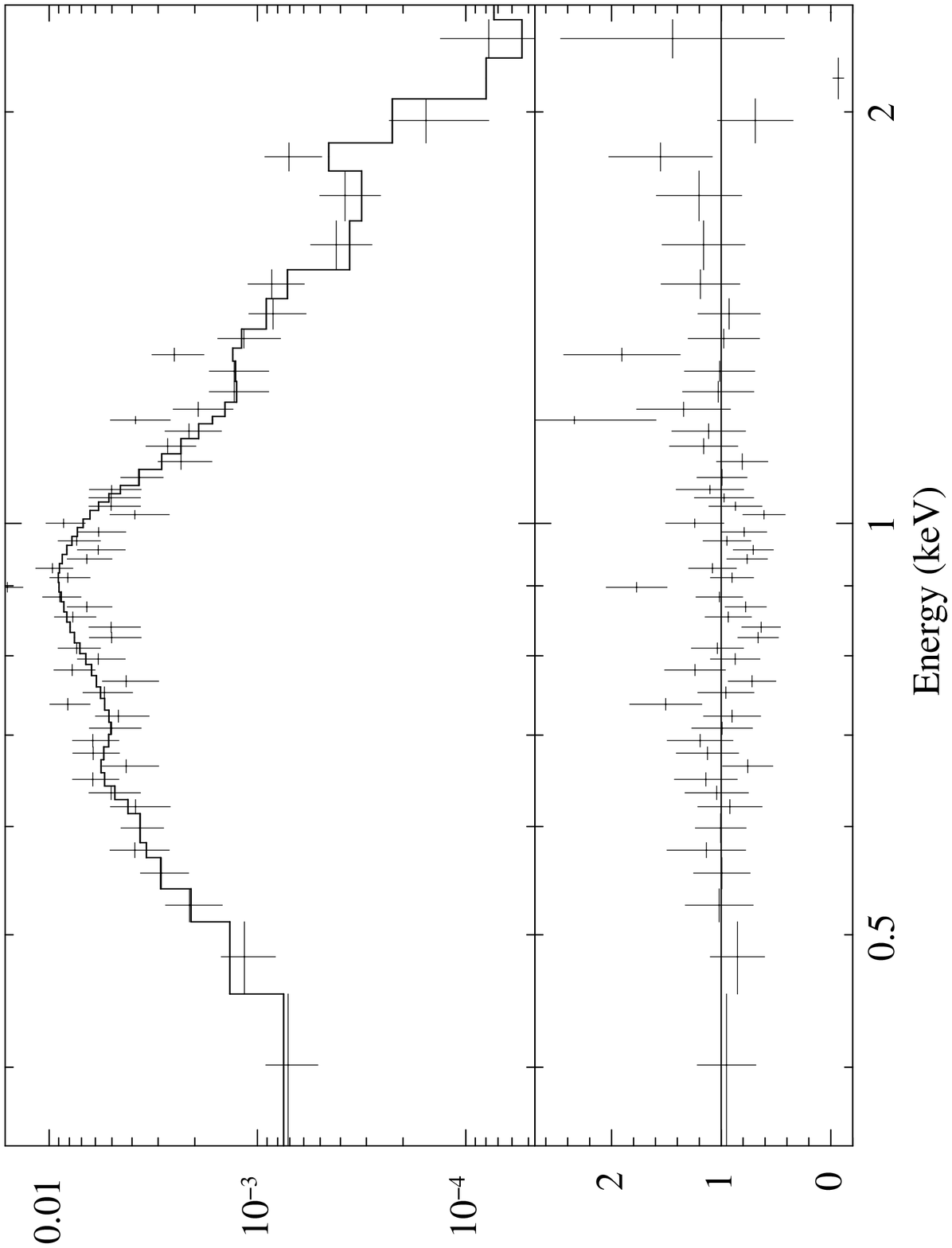} 
 		\put (79,62) {Source 64}
	\end{overpic}	\\
	
	\begin{overpic}[width=0.3\linewidth,angle=-90,clip=true]{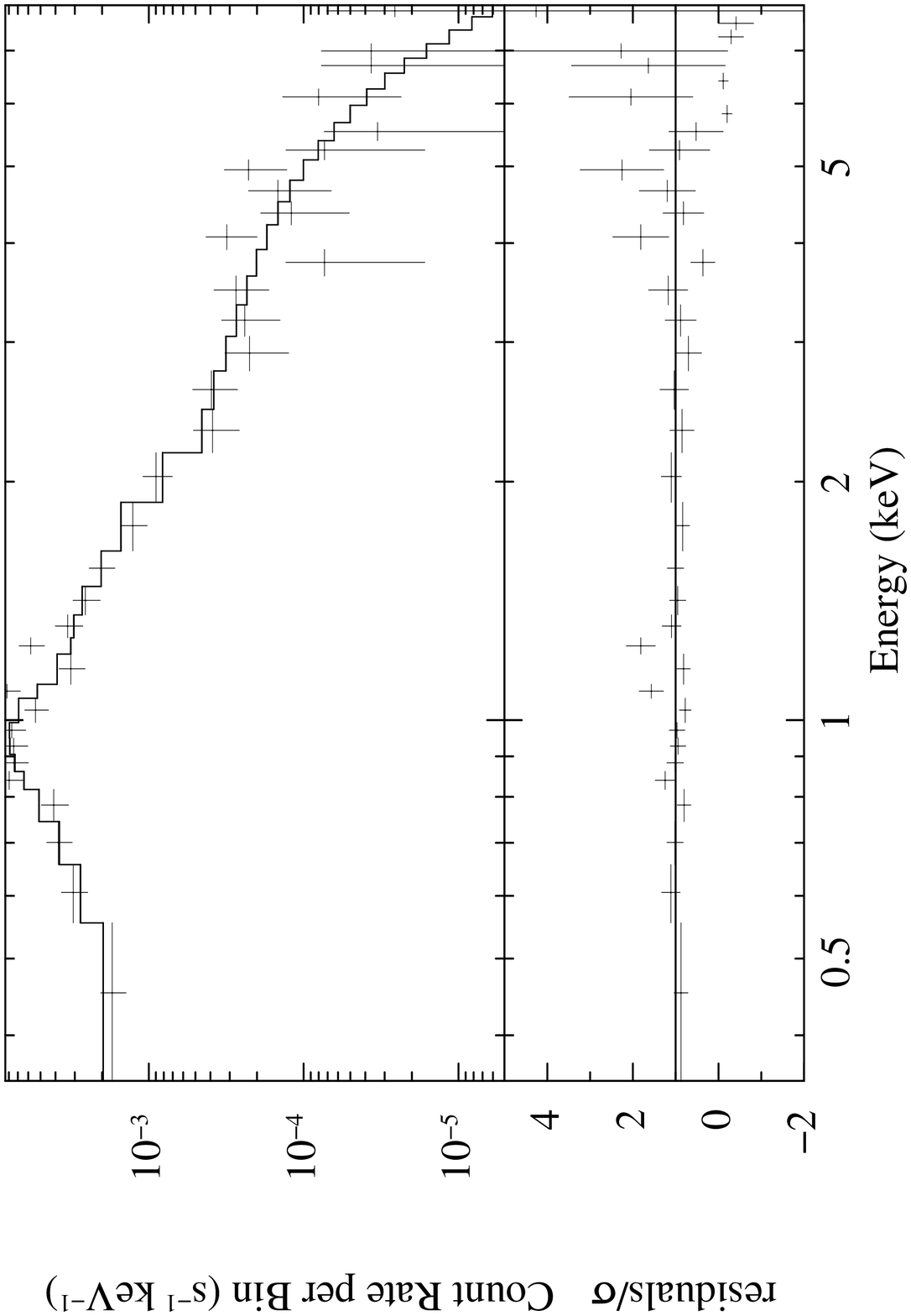} 
 		\put (79,60) {Source 78}
	\end{overpic}	&
	\begin{overpic}[width=0.3\linewidth,angle=-90,clip=true]{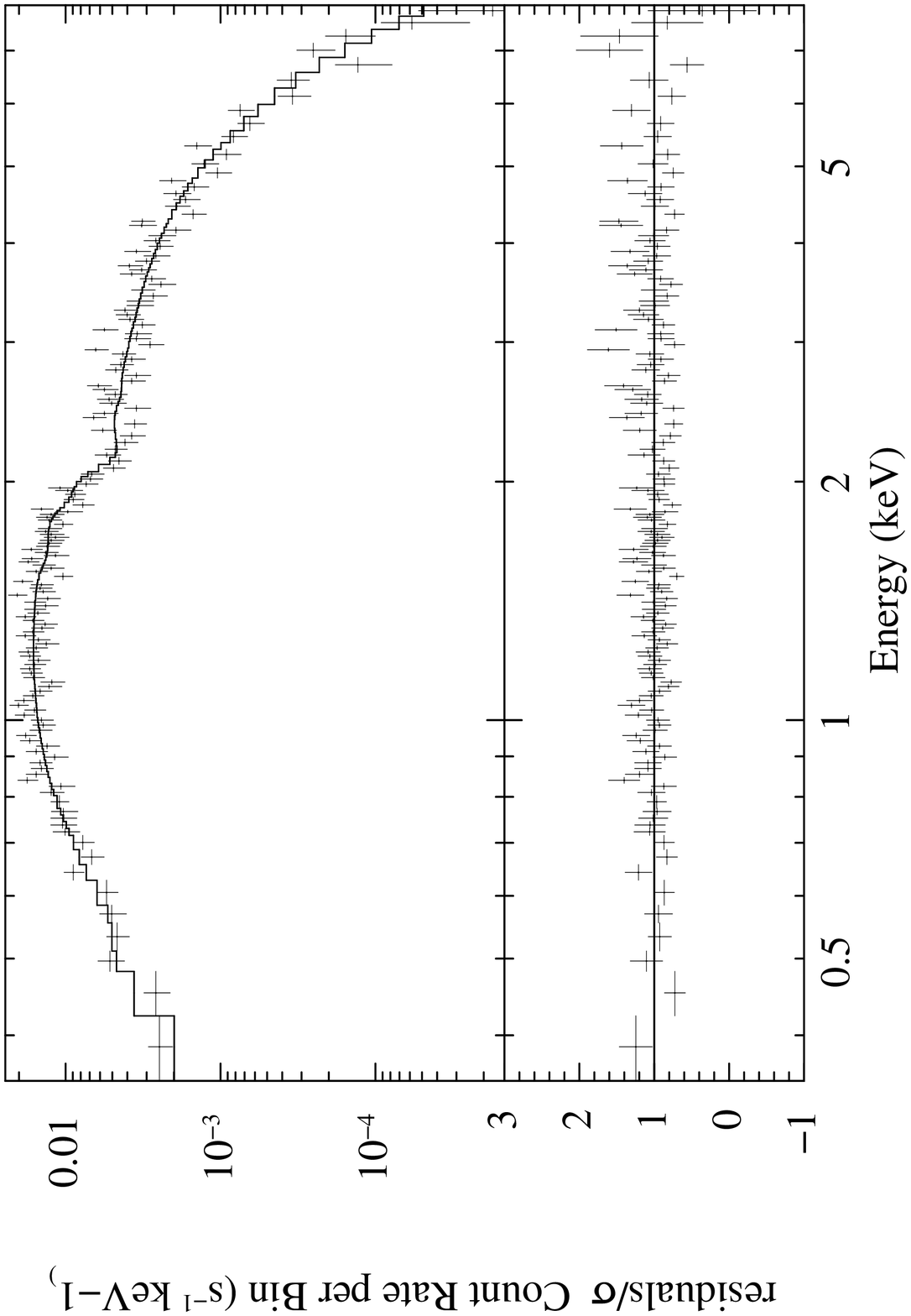} 
 		\put (79,60) {Source 92}
	\end{overpic}	\\
\end{tabular}
\caption{The spectra (with best fit models superimposed) and residuals for the NGC~2403 sources with more than 500 net counts. Spectra are binned for display purposes only.}
\label{bright_spec2403}
\end{figure*}

\begin{figure*}
\setcounter{figure}{3}
\centering
\begin{tabular}{cc}
	\begin{overpic}[width=0.3\linewidth,angle=-90,clip=true]{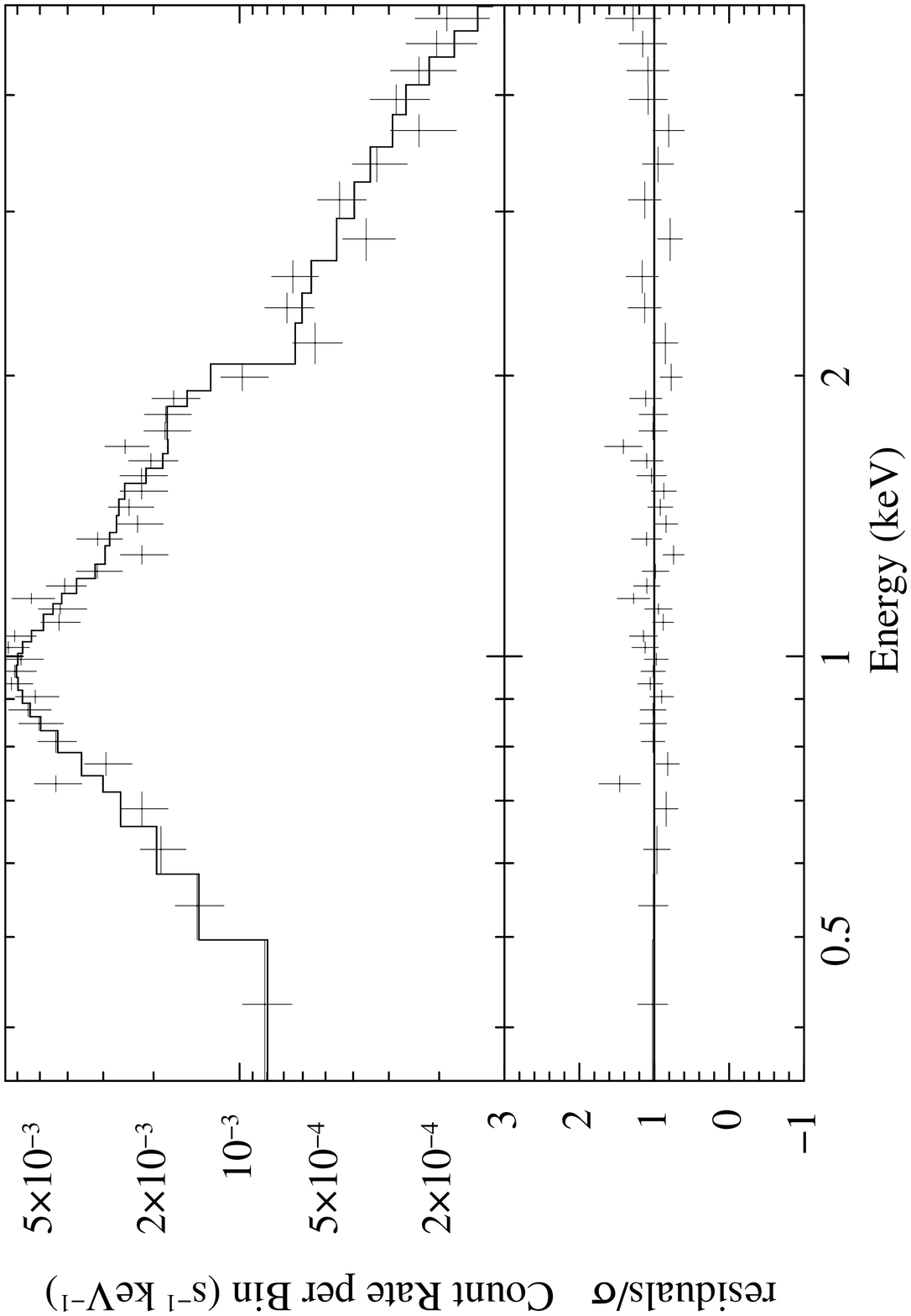} 
 		\put (79,60) {Source 135}
	\end{overpic}	&
	\begin{overpic}[width=0.3\linewidth,angle=-90,clip=true]{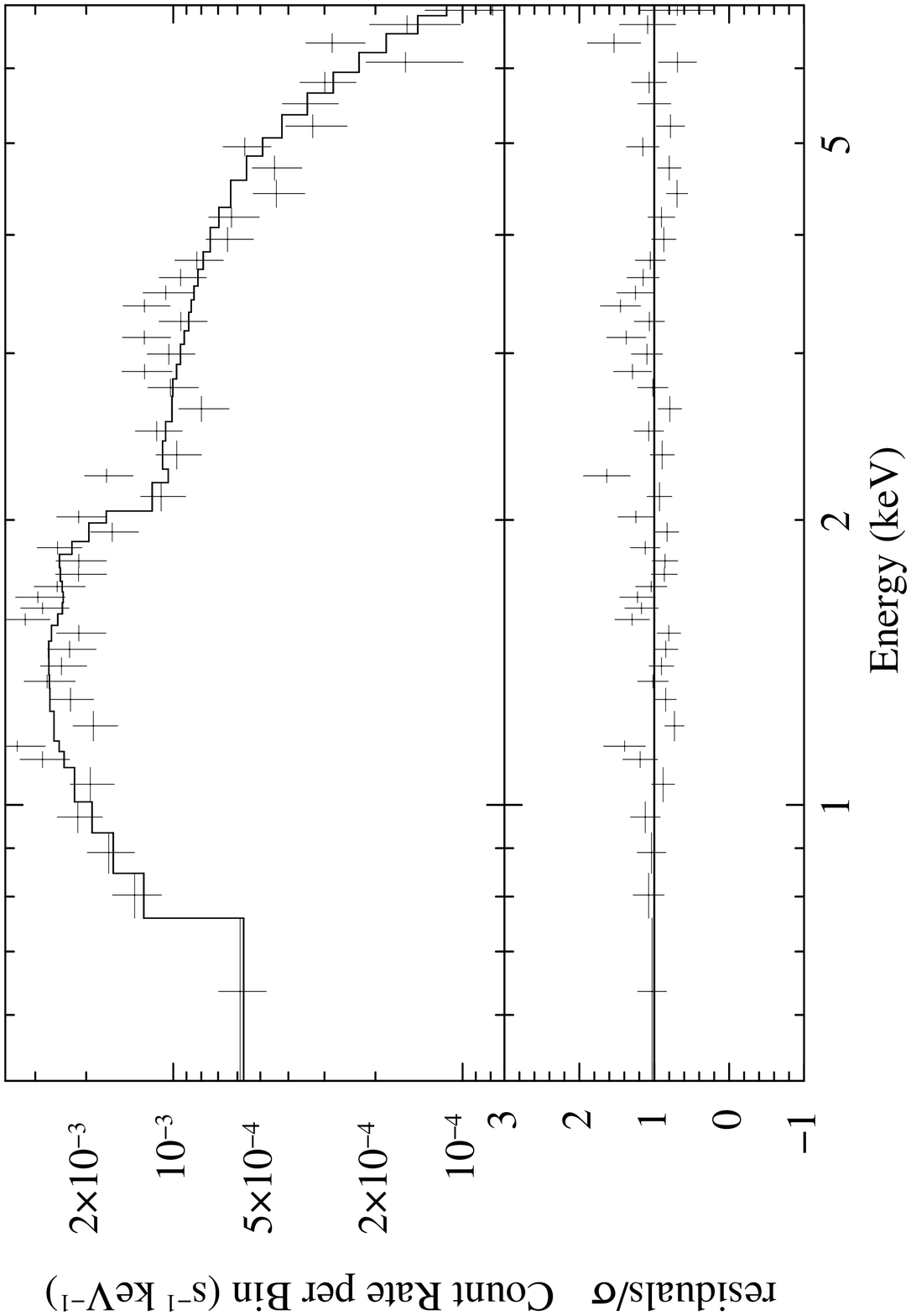} 
 		\put (79,60) {Source 155}
	\end{overpic}	\\
	\begin{overpic}[width=0.3\linewidth,angle=-90,clip=true]{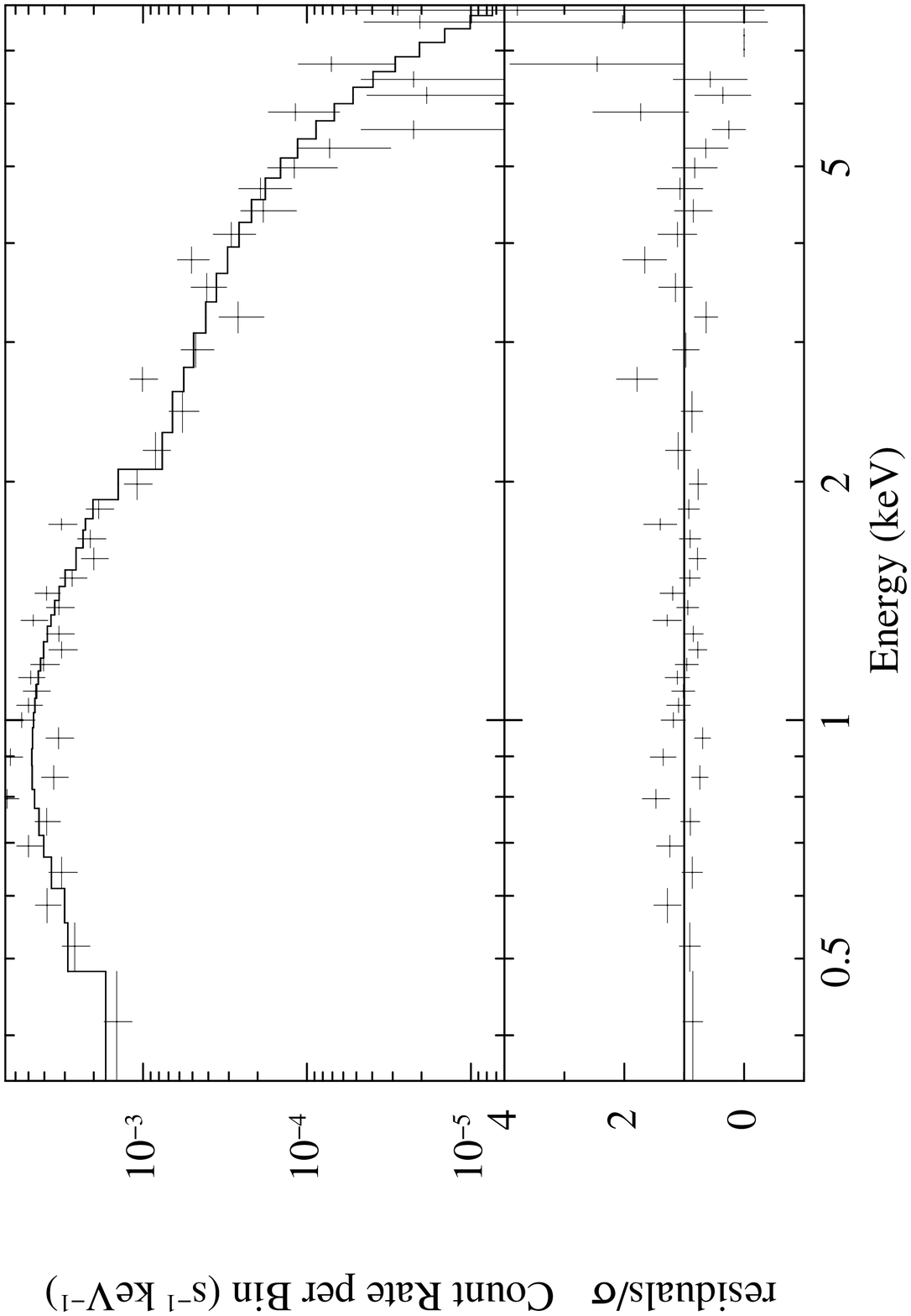} 
 		\put (79,60) {Source 162}
	\end{overpic}	&
	\begin{overpic}[width=0.3\linewidth,angle=-90,clip=true]{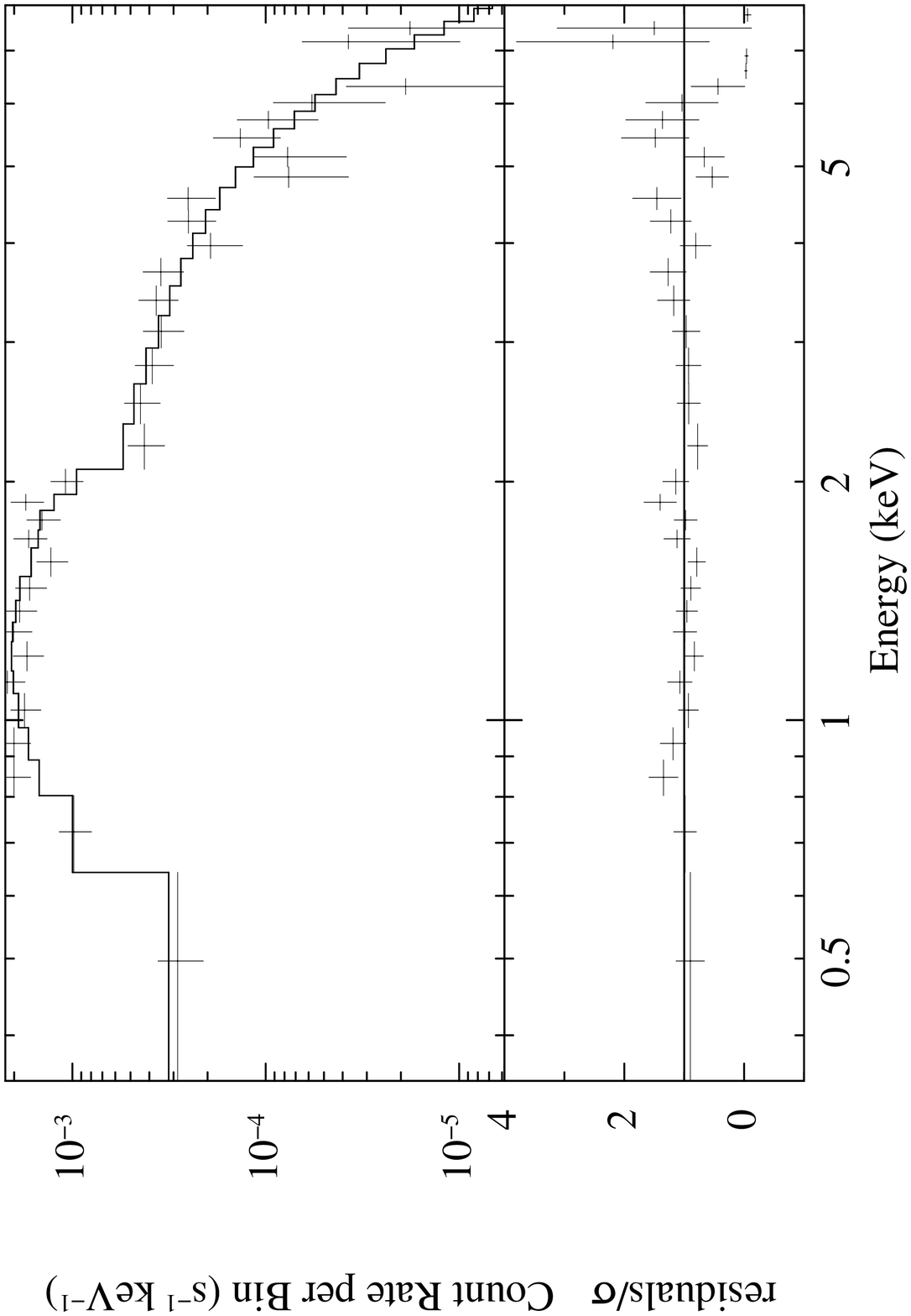} 
 		\put (79,60) {Source 170}
	\end{overpic}	\\	
\multicolumn{2}{c}{
	\begin{overpic}[width=0.3\linewidth,angle=-90,clip=true]{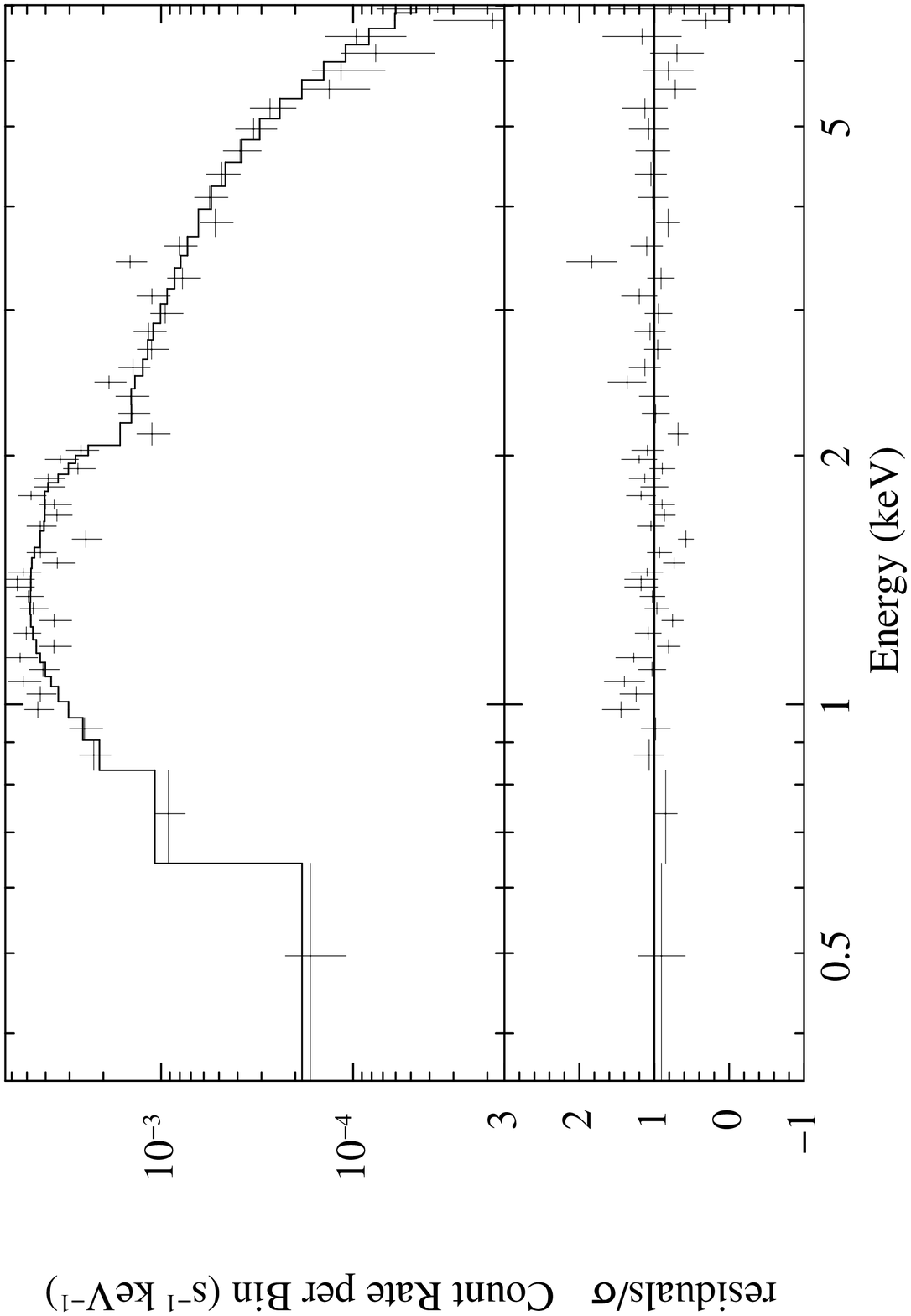} 
 		\put (79,60) {Source 179}
	\end{overpic}}	\\
\end{tabular}
\caption{{\it (Continued)} The spectra (with best fit models superimposed) and residuals for the NGC~2403 sources with more than 500 net counts. Spectra are binned for display purposes only.}
\label{bright_spec2403}
\end{figure*}

\begin{figure*}
\centering
\begin{tabular}{c}
\begin{overpic}[width=0.3\linewidth,angle=-90,clip=true]{NGC4214_src16_spectrum.eps}
 		\put (79,60) {Source 16}
	\end{overpic}
\end{tabular}
\caption{The spectrum (with best fit models superimposed) and residuals for source 16, the only NGC~4214 source with more than 500 net counts. Spectra are binned for display purposes only.}
\label{bright_spec4214}
\end{figure*}

\begin{figure*}
\centering
\begin{tabular}{cc}
\includegraphics[width=0.5\linewidth,clip=true]{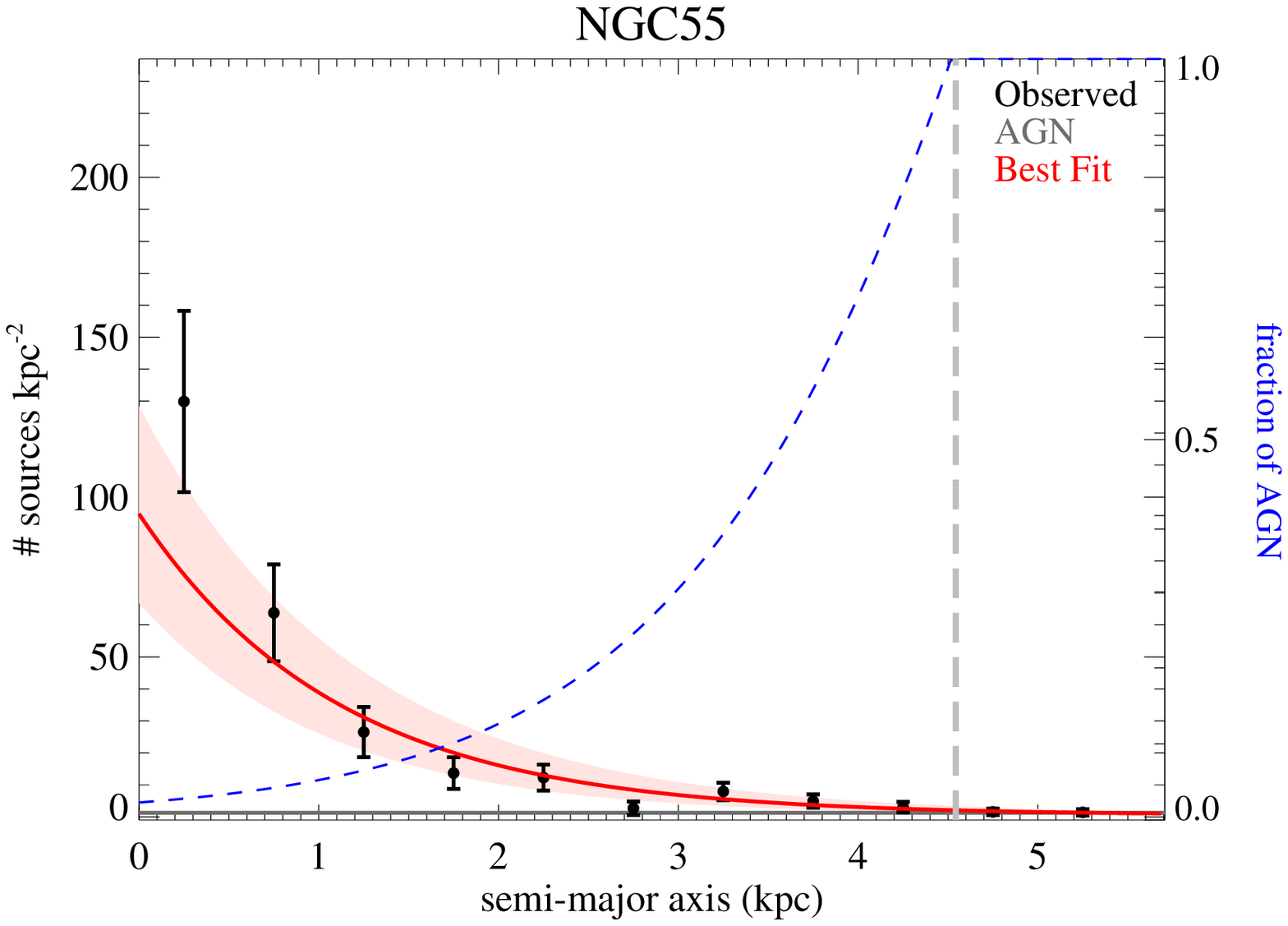} & 
\includegraphics[width=0.5\linewidth,clip=true]{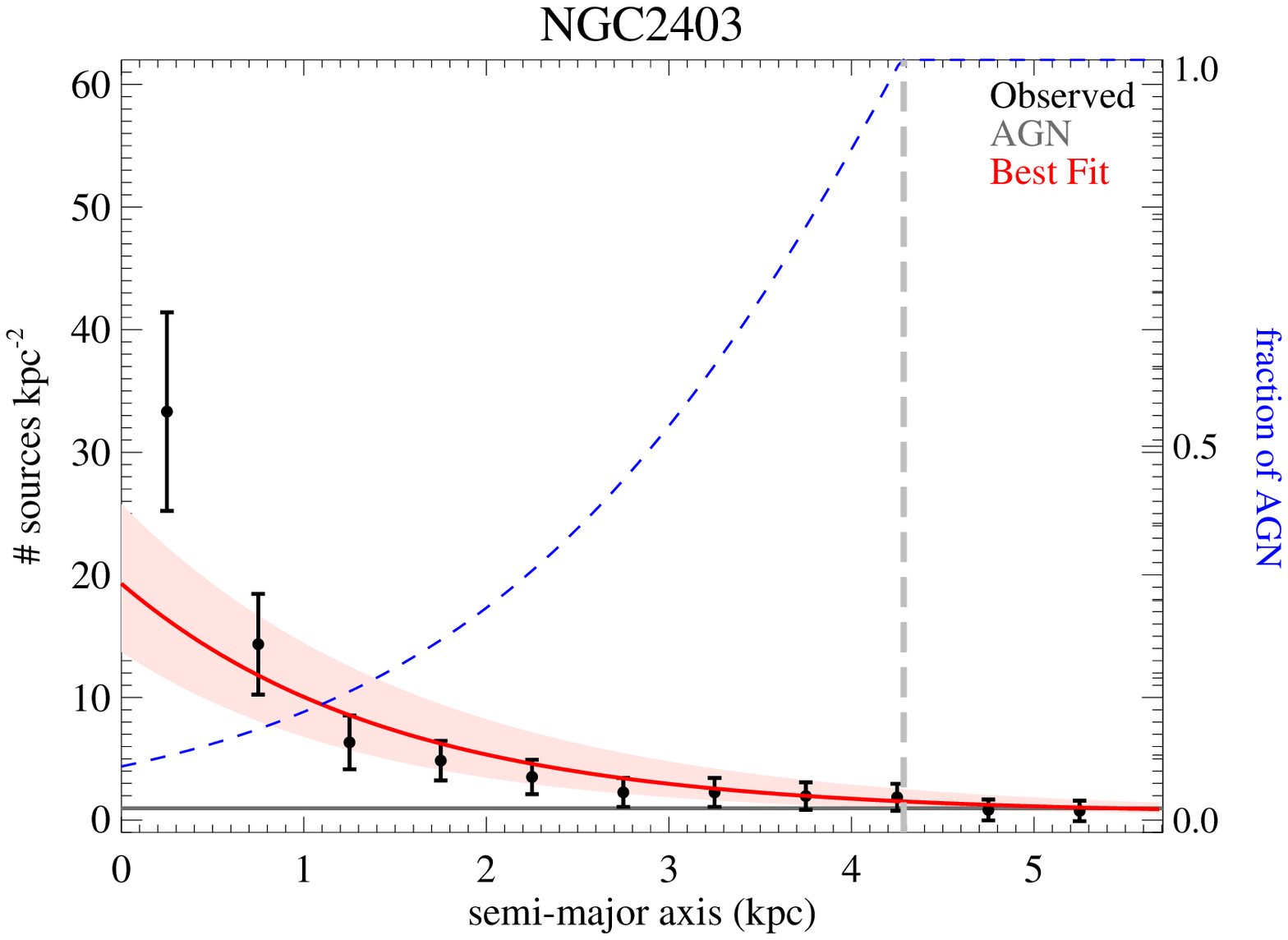} \\
\includegraphics[width=0.5\linewidth,clip=true]{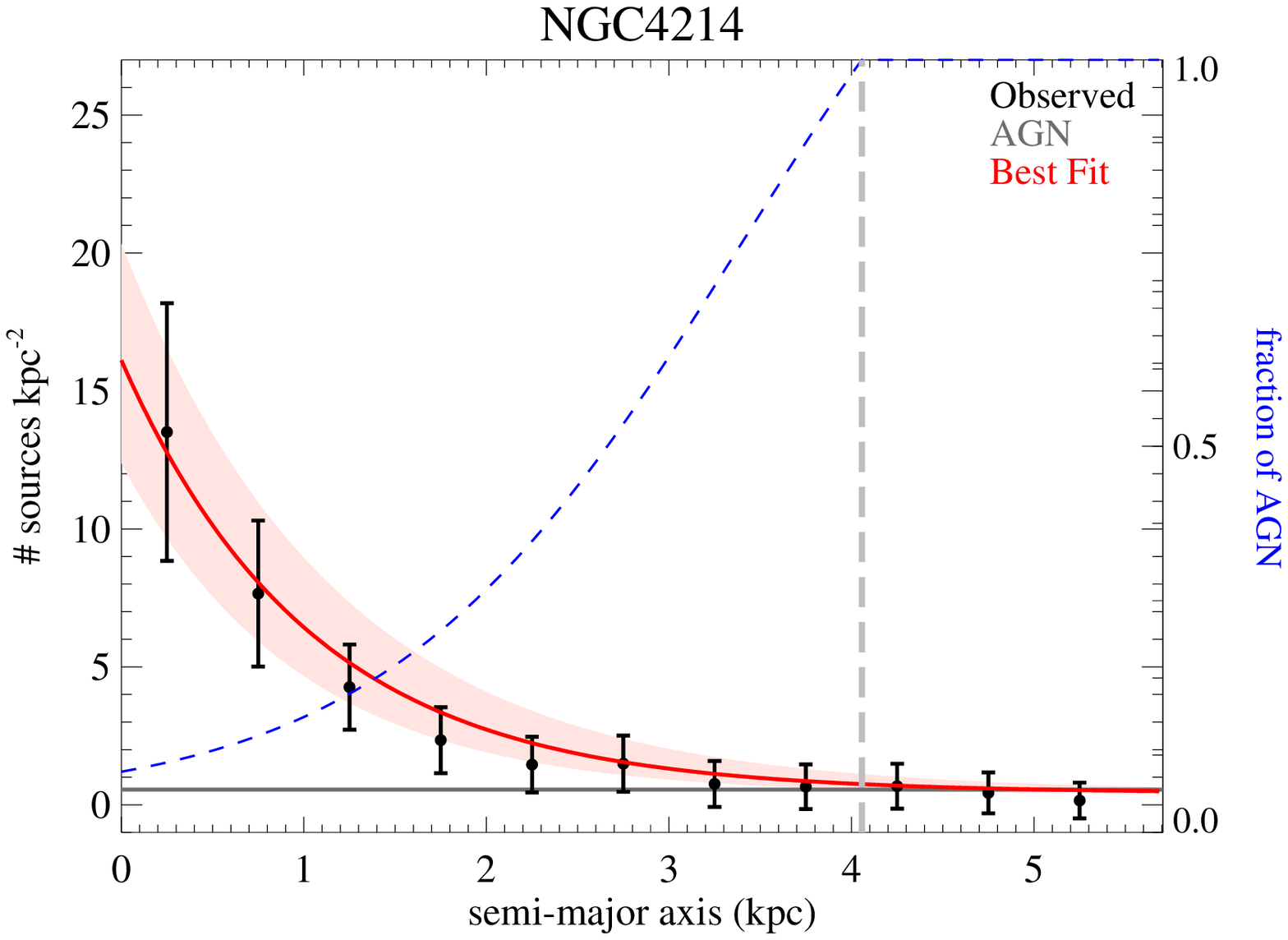} &
\includegraphics[width=0.5\linewidth,clip=true]{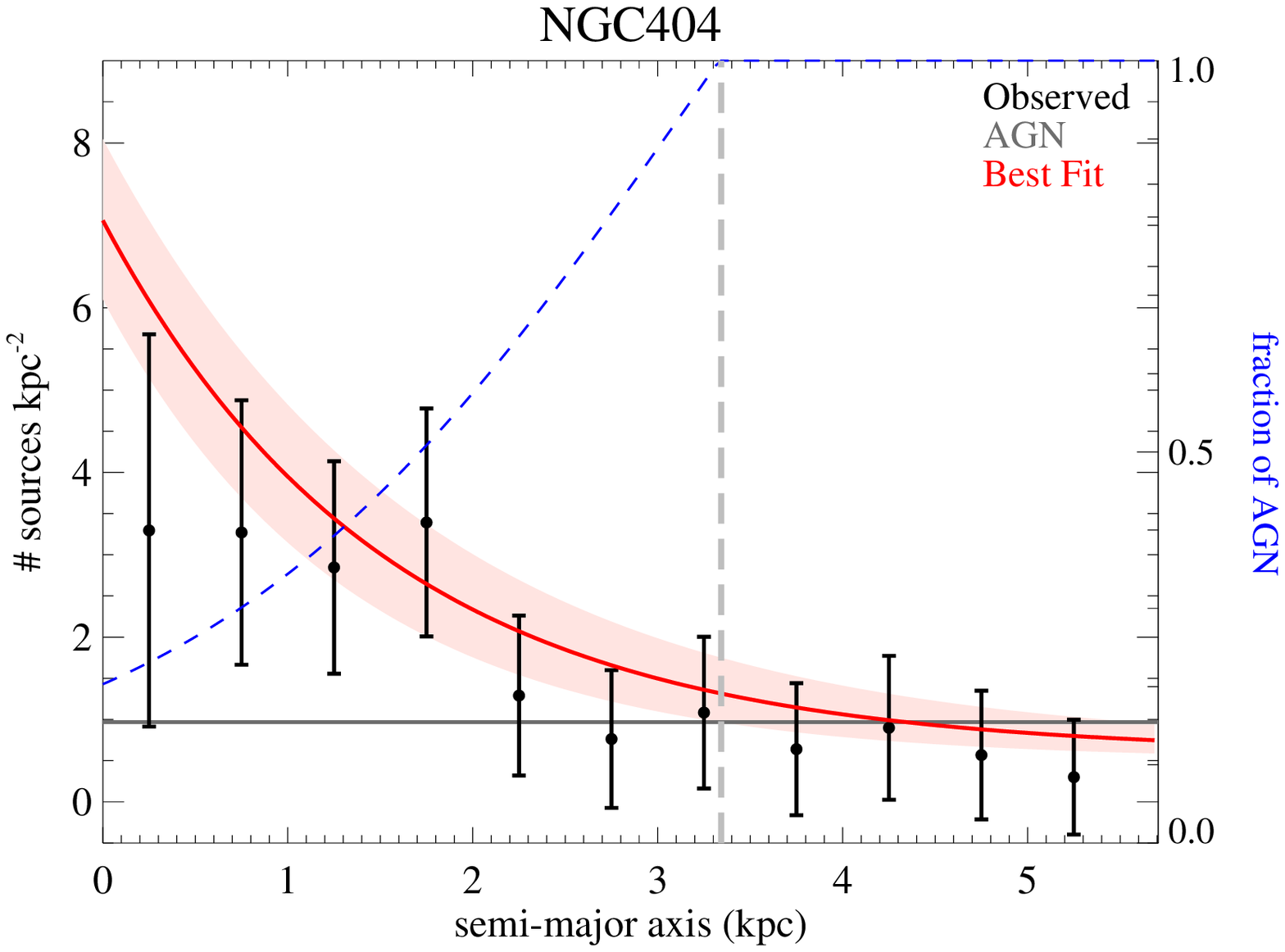} \\
\end{tabular}
\caption{Radial source distributions. The total distributions (e.g., for all X-ray sources) are shown in black. The red line shows the best-fit exponential disk model (or a de Vaucouleurs model for NGC~404) to the data, and the 90\% confidence region is shown in pink. The horizontal gray line indicates the expected number of AGN at each (de-projected) distance, and the vertical, dashed gray line shows the galactocentric distance at which the source density reaches the background level. The dashed blue line shows the predicted fraction of X-ray sources that are AGN as a function of distance. {\it Top left}: NGC~55; {\it top right}: NGC~2043; {\it bottom left}: NGC~4214; {\it bottom right: NGC~404.}}
\label{radial}
\end{figure*}

\clearpage

\begin{figure*}
\centering
\begin{tabular}{cccc}
Source 27 & Source 29 & Source 30 & Source 32 \\
\includegraphics[width=0.23\linewidth,clip=true,trim=0.5cm 3cm 0.5cm 2.5cm]{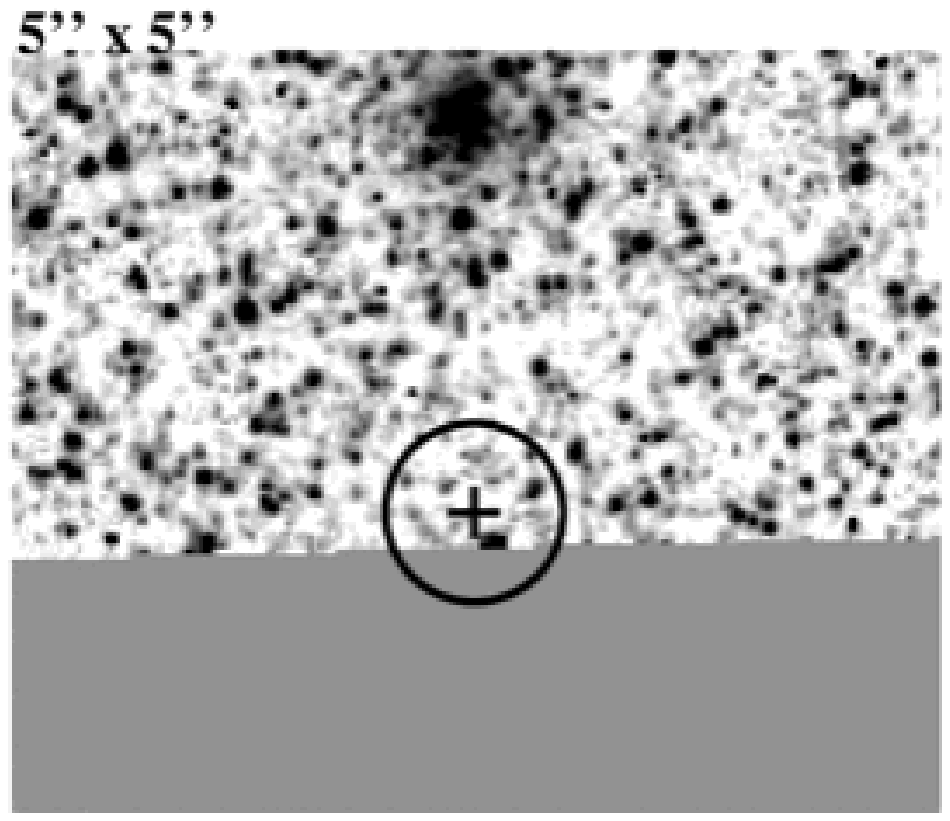} & 
\includegraphics[width=0.23\linewidth,clip=true,trim=0.5cm 3cm 0.5cm 2.5cm]{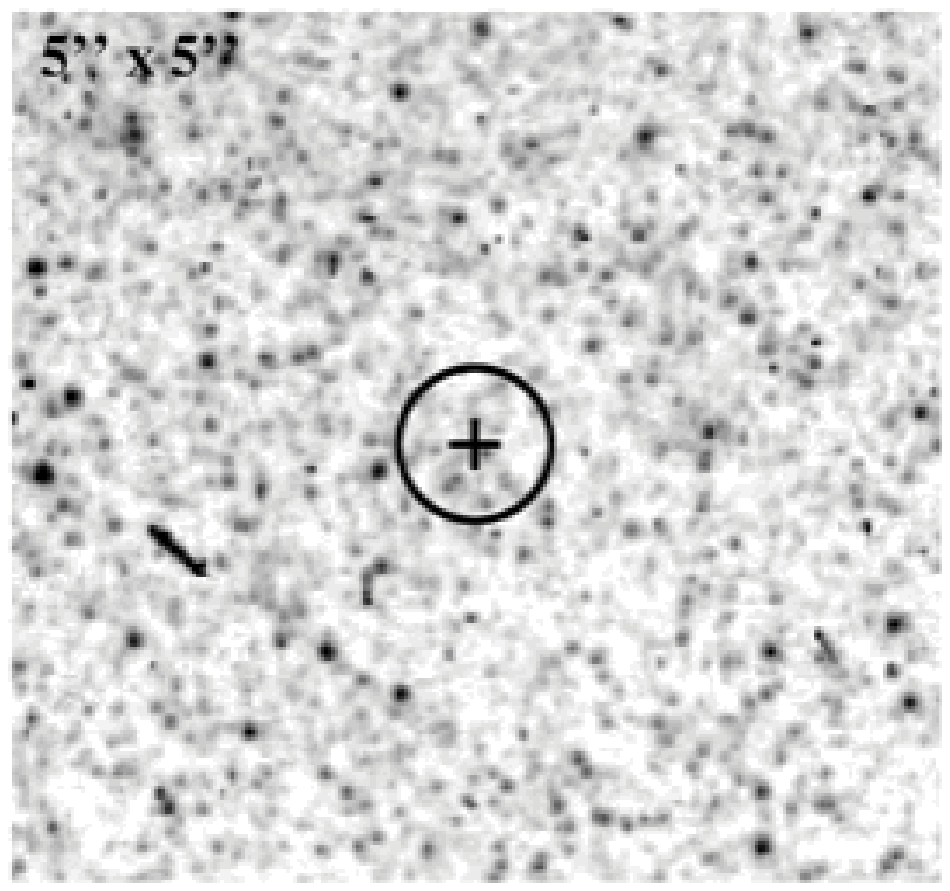} & 
\includegraphics[width=0.23\linewidth,clip=true,trim=0.5cm 3cm 0.5cm 2.5cm]{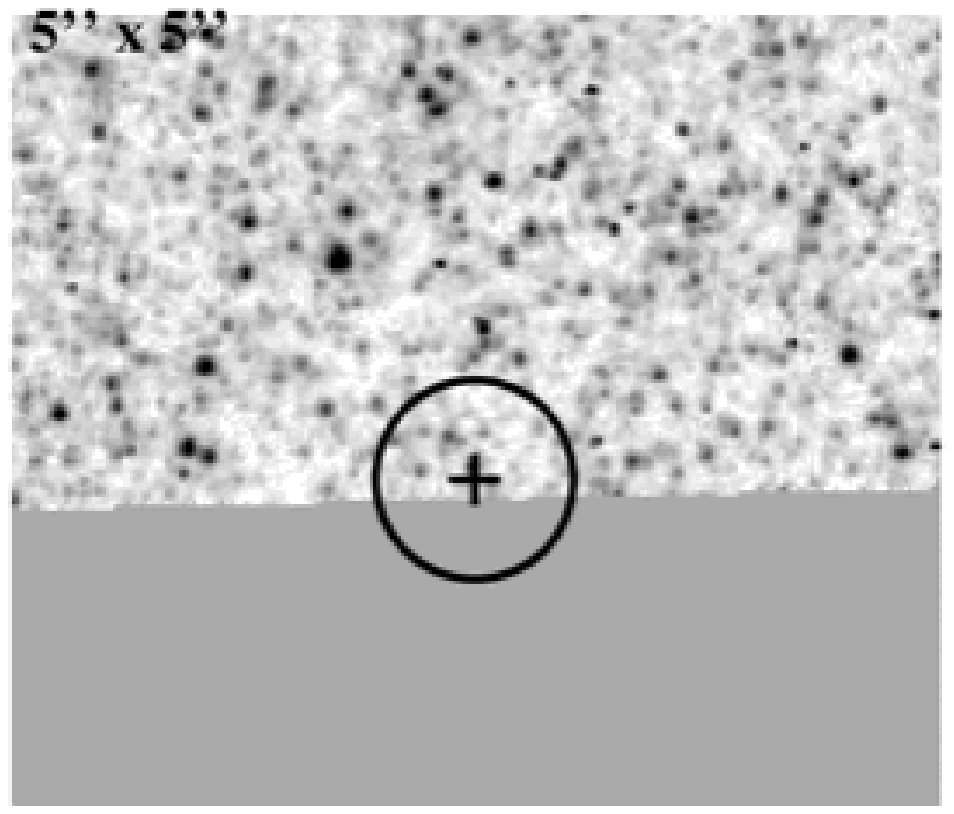} & 
\includegraphics[width=0.23\linewidth,clip=true,trim=0.5cm 3cm 0.5cm 2.5cm]{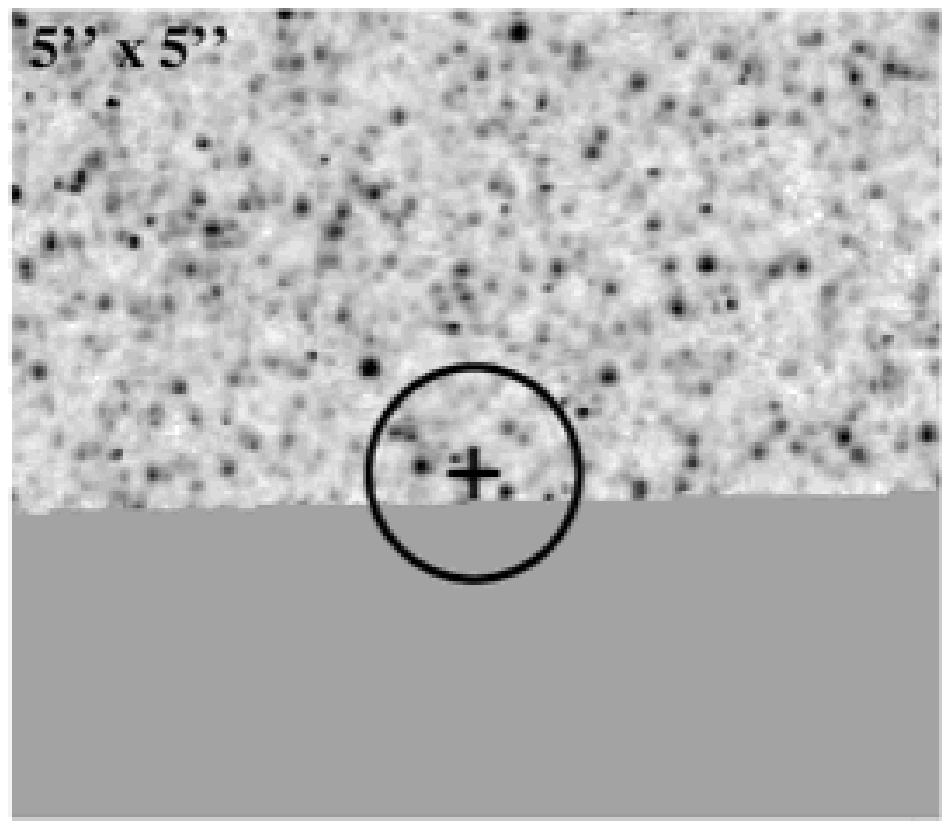} \\ 

Source 35 & Source 36 & Source 38 & Source 41 \\ 
\includegraphics[width=0.23\linewidth,clip=true,trim=0.5cm 3cm 0.5cm 0.5cm]{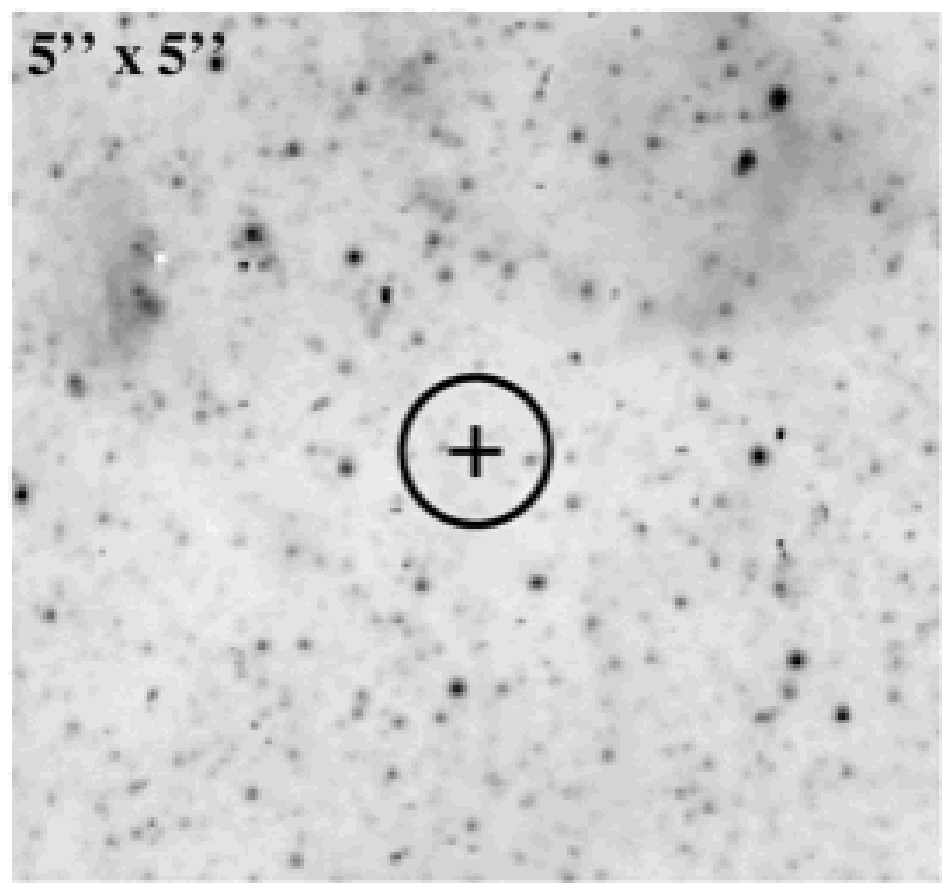} & 
\includegraphics[width=0.23\linewidth,clip=true,trim=0.5cm 3cm 0.5cm 0.5cm]{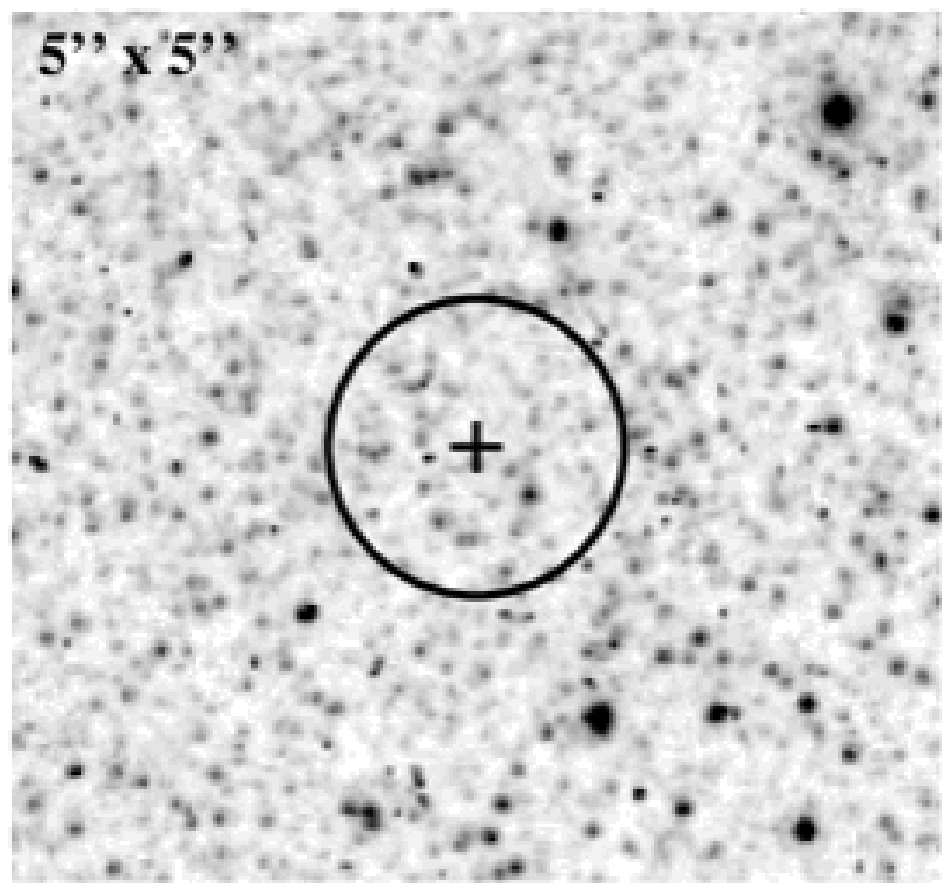} & 
\includegraphics[width=0.23\linewidth,clip=true,trim=0.5cm 3cm 0.5cm 0.5cm]{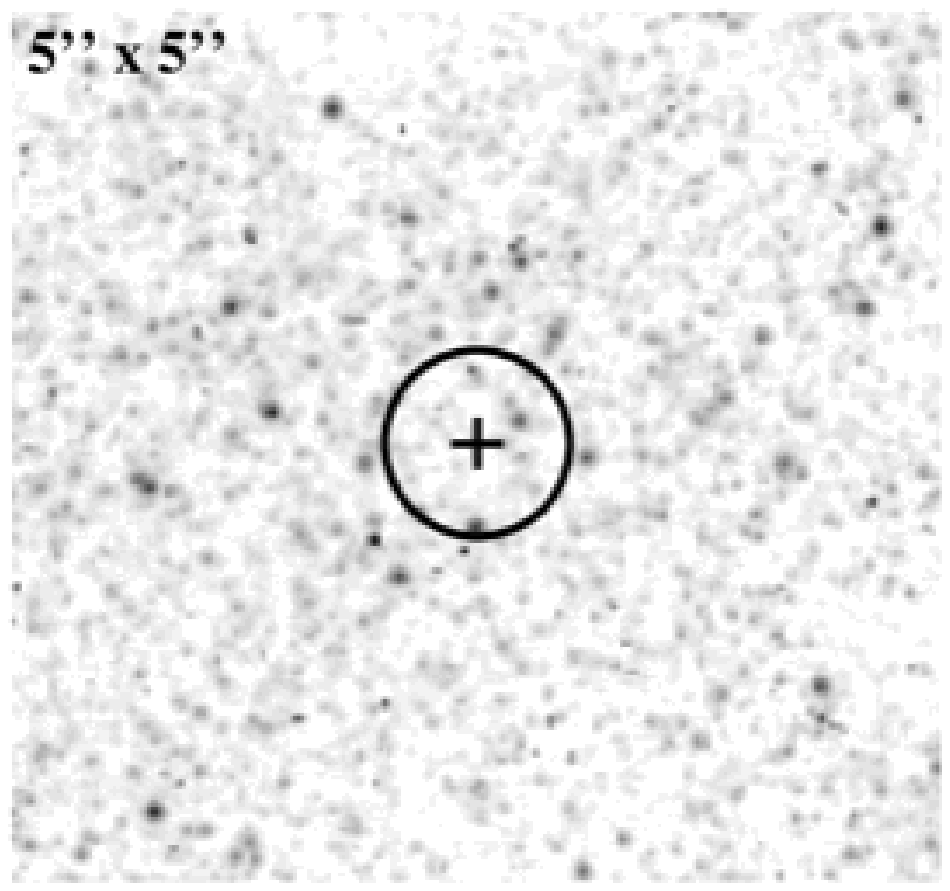} & 
\includegraphics[width=0.23\linewidth,clip=true,trim=0.5cm 3cm 0.5cm 0.5cm]{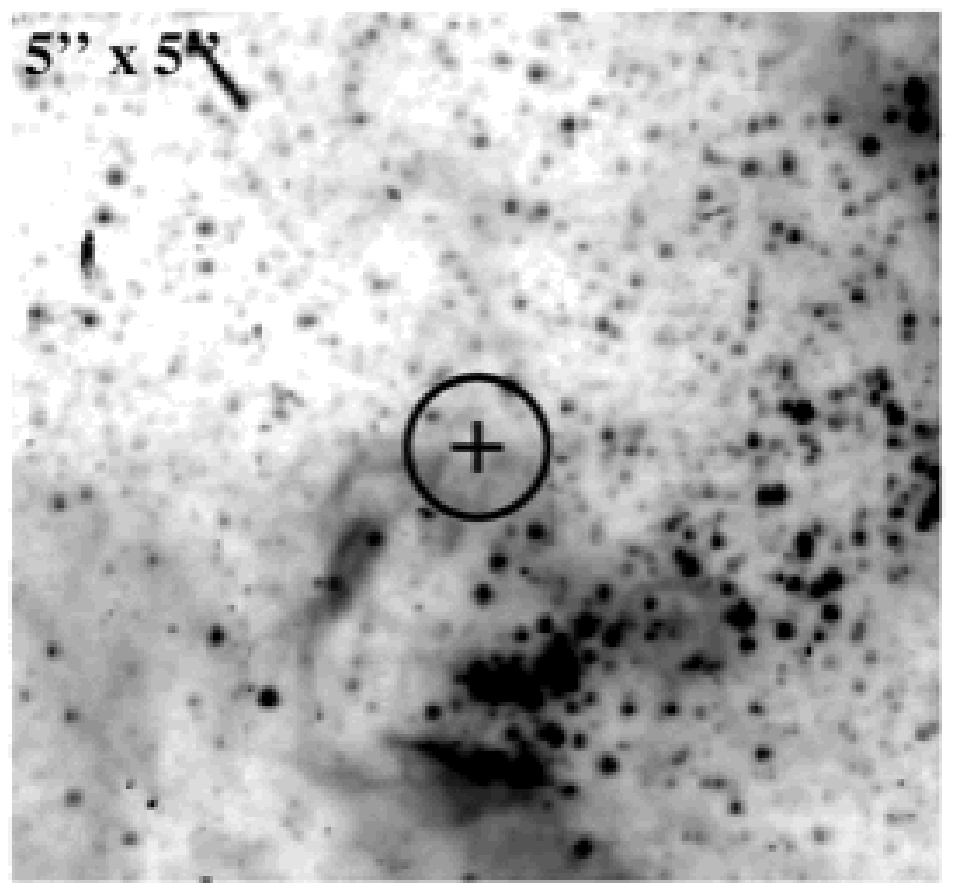} \\ 

Source 42 & Source 43 & Source 45 & Source 46 \\ 
\includegraphics[width=0.23\linewidth,clip=true,trim=0.5cm 3cm 0.5cm 2.5cm]{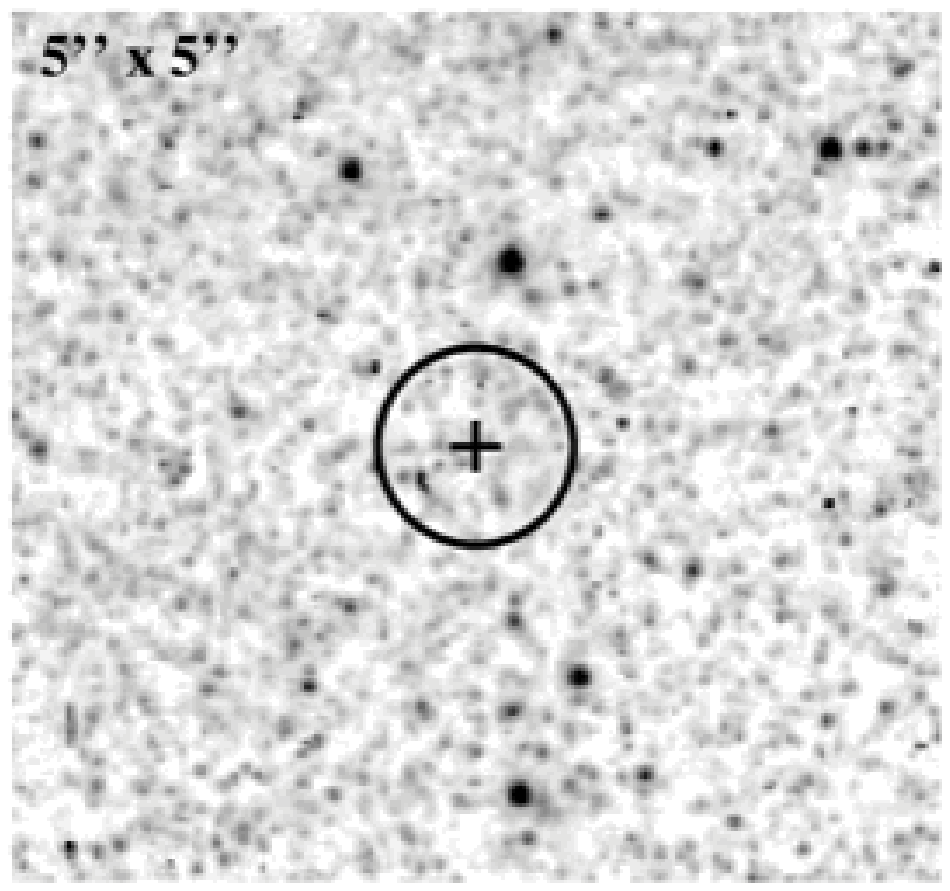} & 
\includegraphics[width=0.23\linewidth,clip=true,trim=0.5cm 3cm 0.5cm 2.5cm]{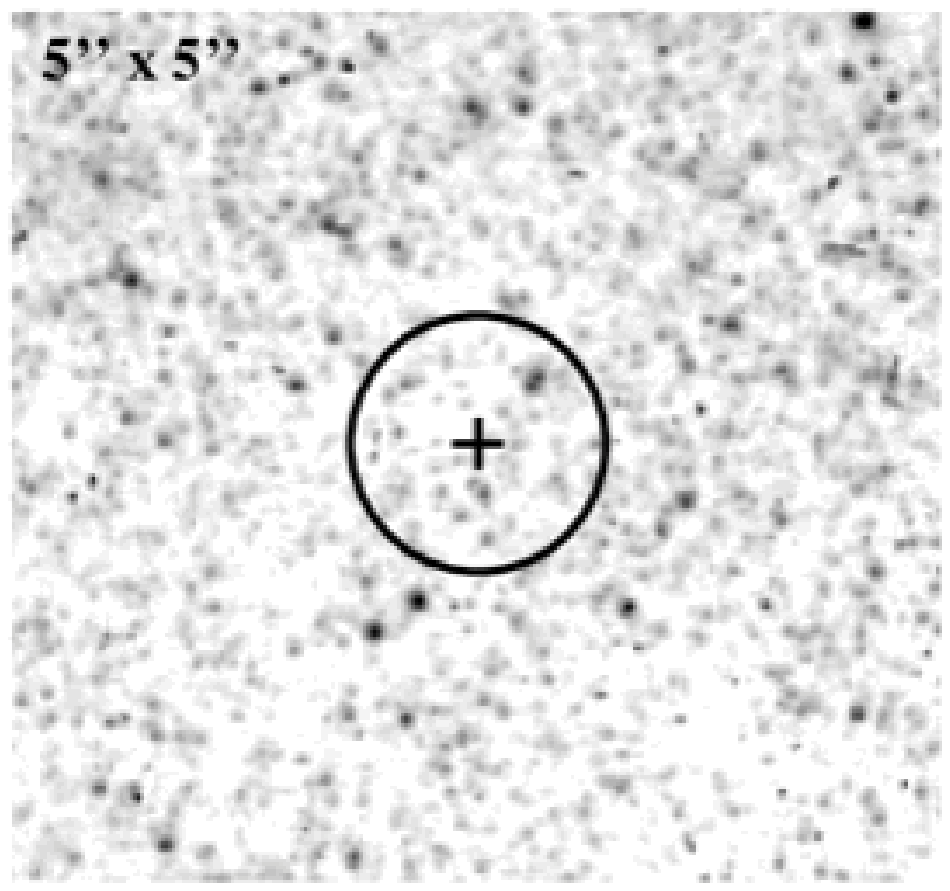} & 
\includegraphics[width=0.23\linewidth,clip=true,trim=0.5cm 3cm 0.5cm 2.5cm]{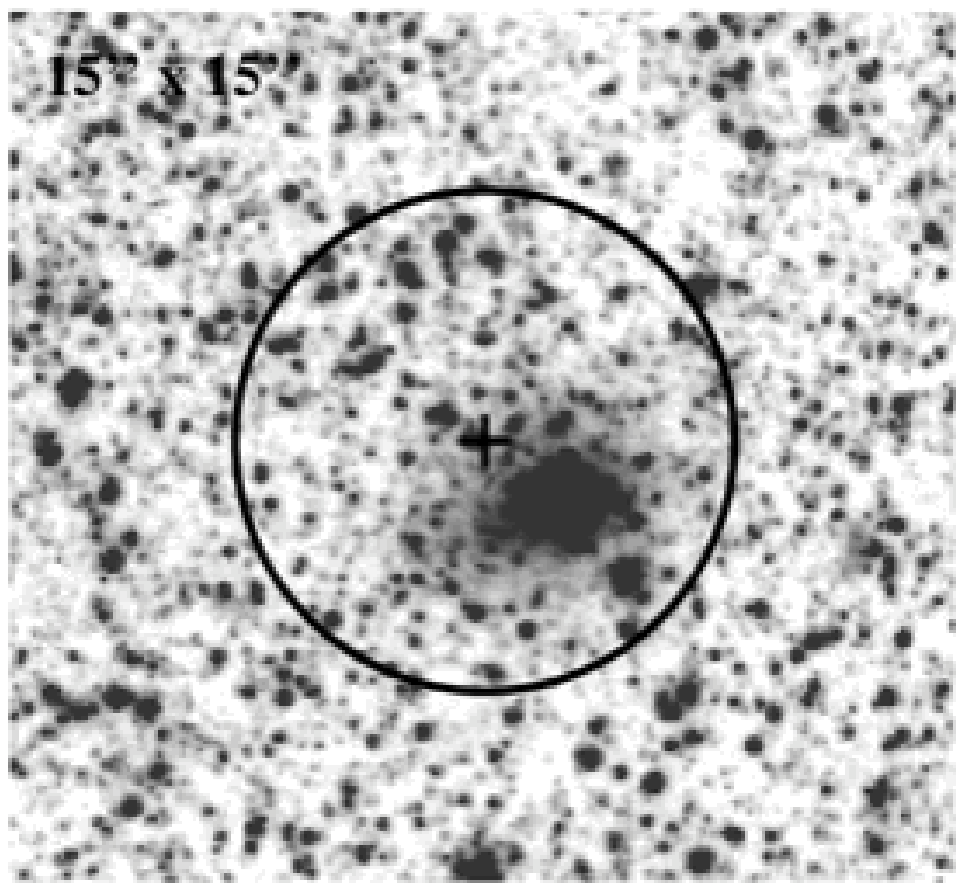} & 
\includegraphics[width=0.23\linewidth,clip=true,trim=0.5cm 3cm 0.5cm 2.5cm]{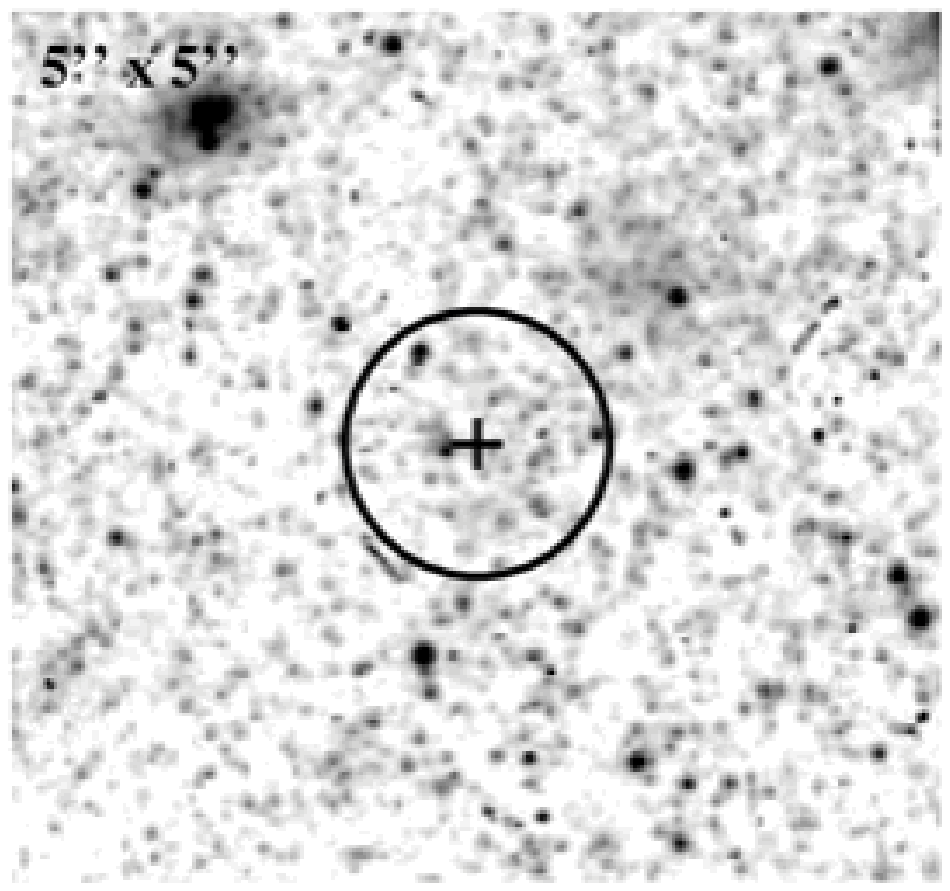} \\ 

Source 47 & Source 48 & Source 49 & Source 50 \\
\includegraphics[width=0.23\linewidth,clip=true,trim=0.5cm 3cm 0.5cm 2.5cm]{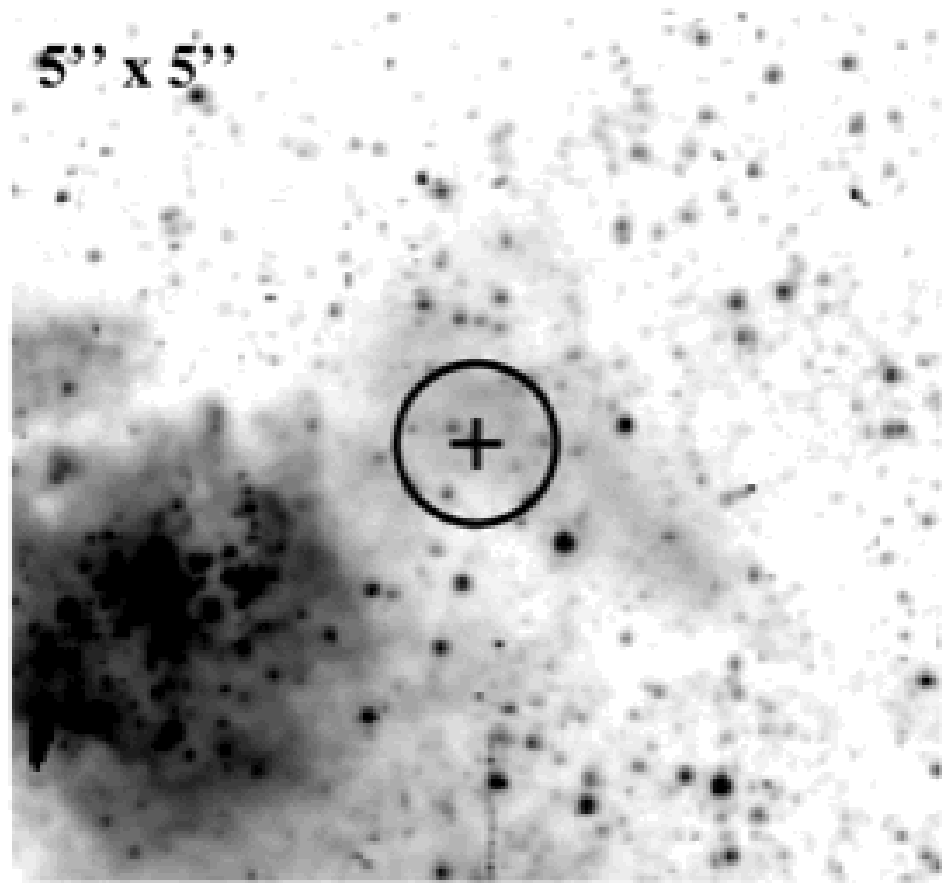} & 
\includegraphics[width=0.23\linewidth,clip=true,trim=0.5cm 3cm 0.5cm 2.5cm]{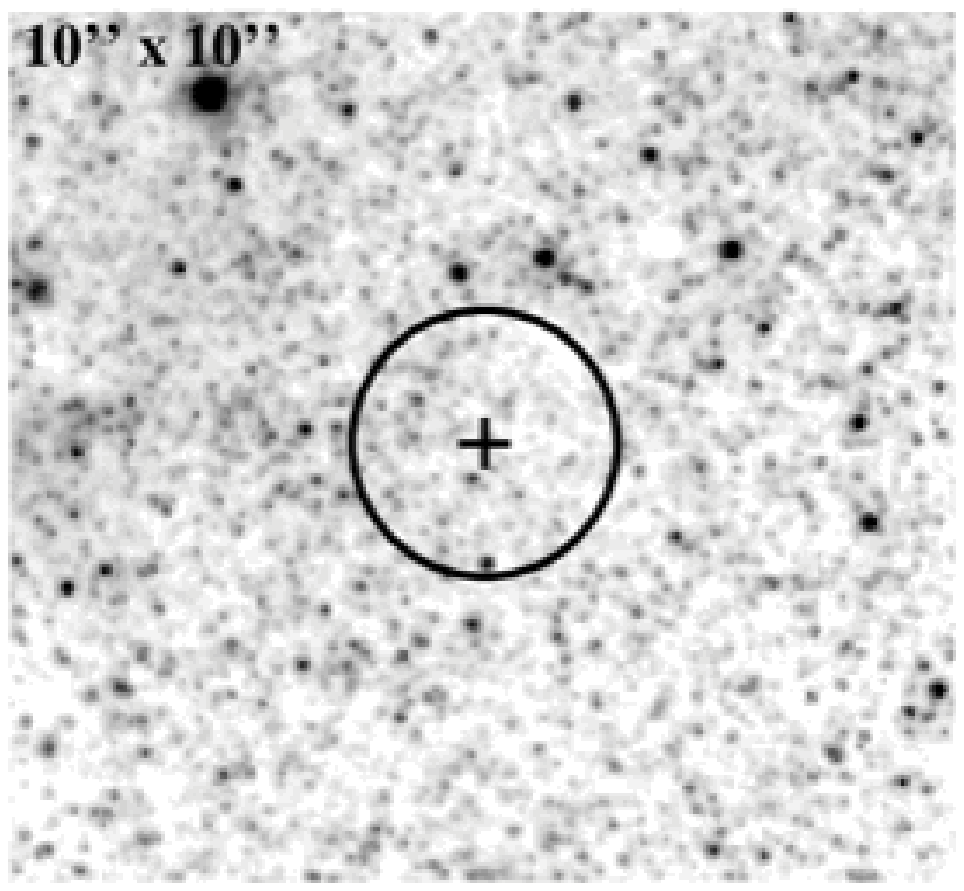} & 
\includegraphics[width=0.23\linewidth,clip=true,trim=0.5cm 3cm 0.5cm 2.5cm]{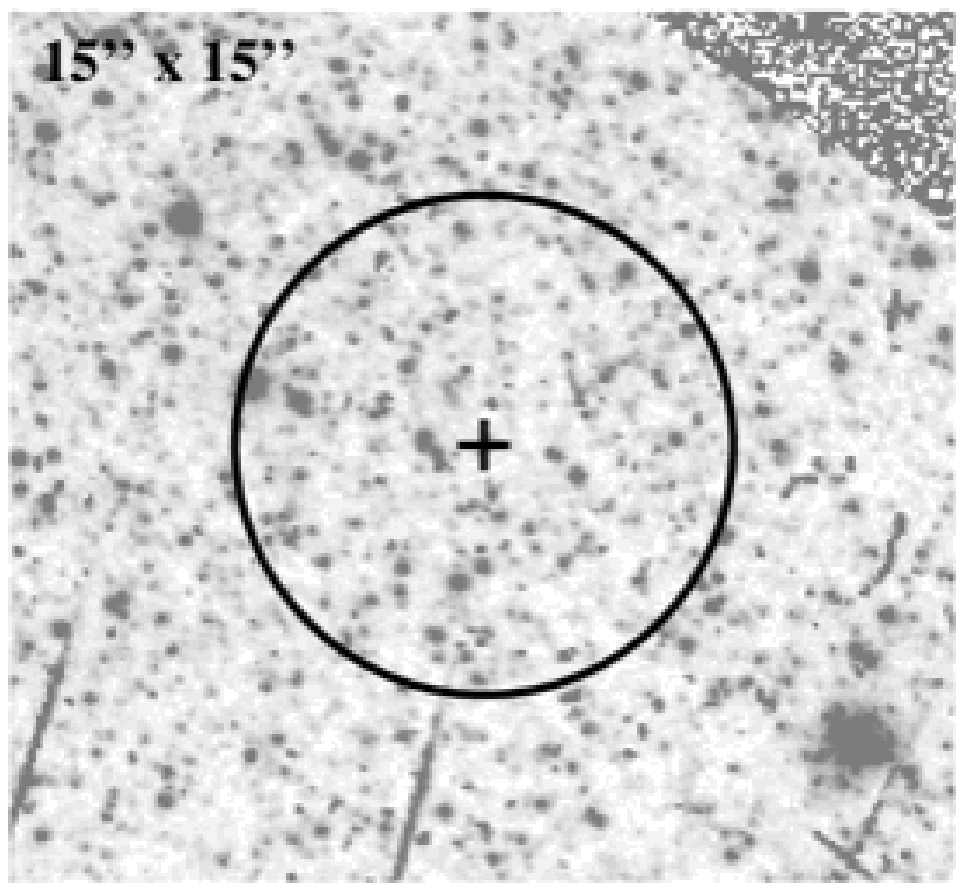} & 
\includegraphics[width=0.23\linewidth,clip=true,trim=0.5cm 3cm 0.5cm 2.5cm]{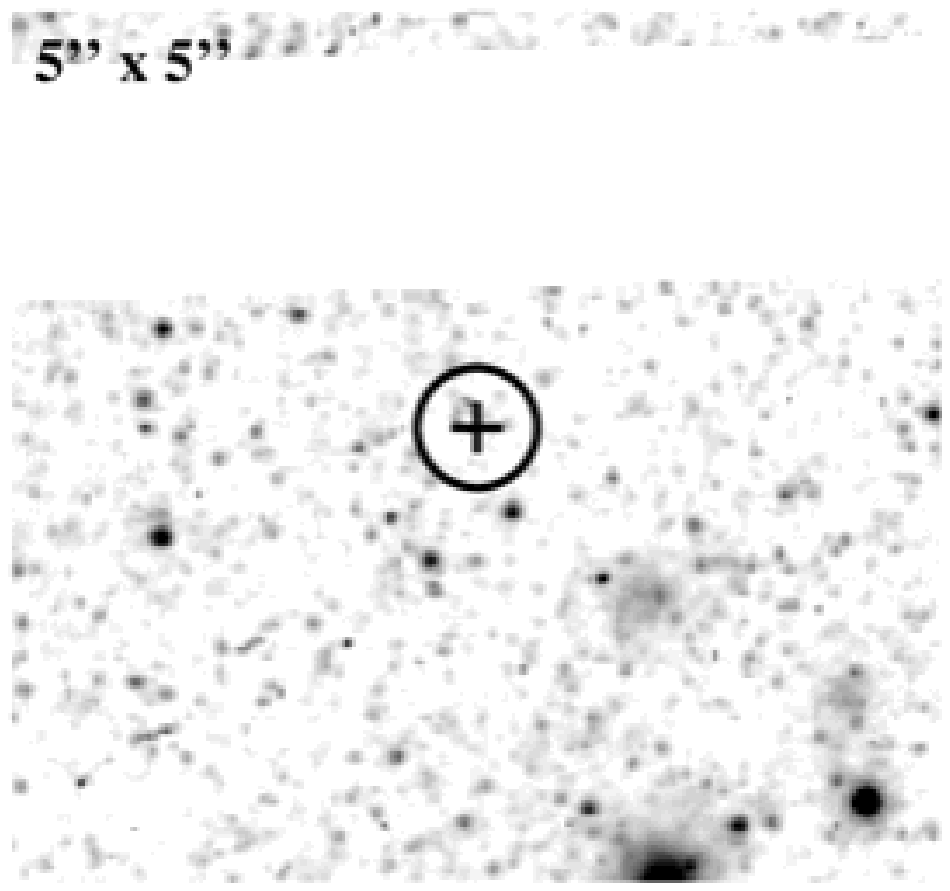} \\ 

Source 51 & Source 52 & Source 53 & Source 56 \\ 
\includegraphics[width=0.23\linewidth,clip=true,trim=0.5cm 3cm 0.5cm 2.5cm]{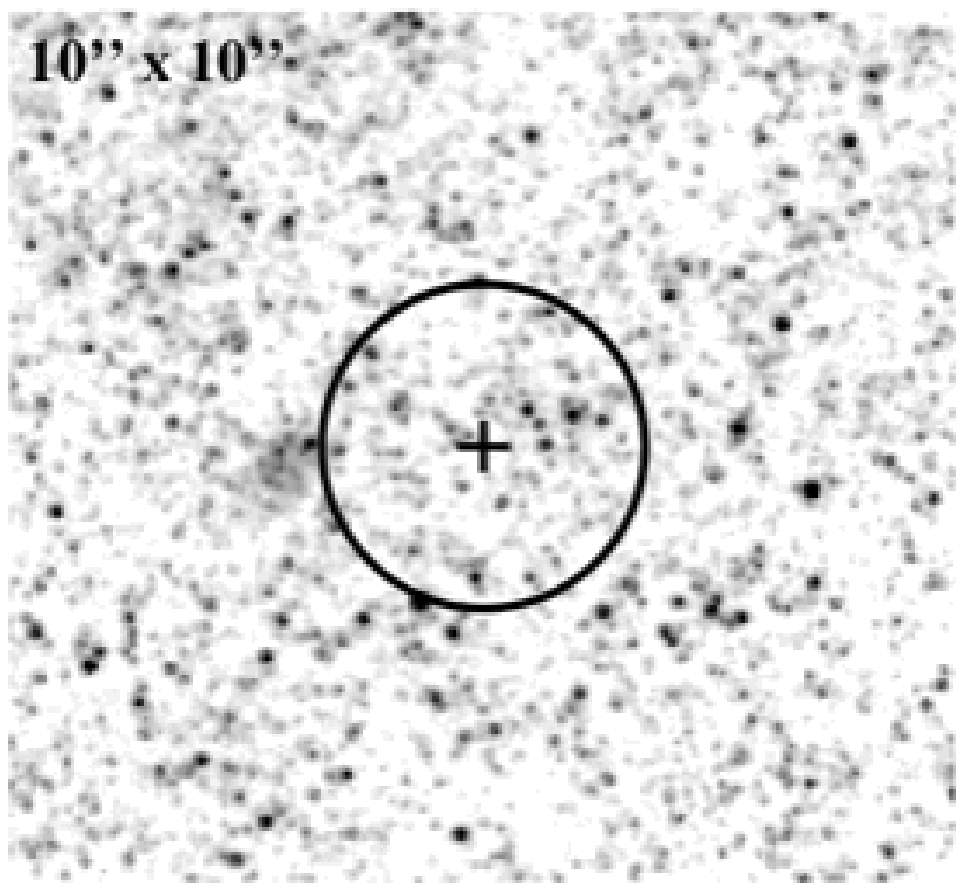} & 
\includegraphics[width=0.23\linewidth,clip=true,trim=0.5cm 3cm 0.5cm 2.5cm]{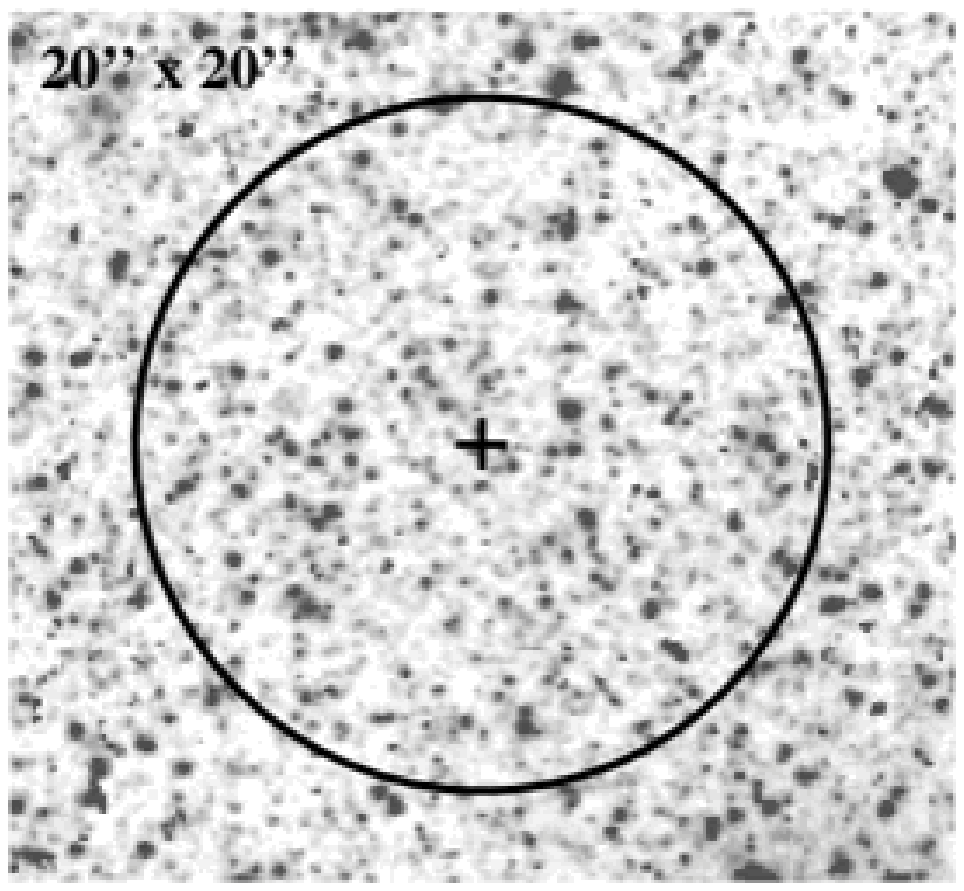} & 
\includegraphics[width=0.23\linewidth,clip=true,trim=0.5cm 3cm 0.5cm 2.5cm]{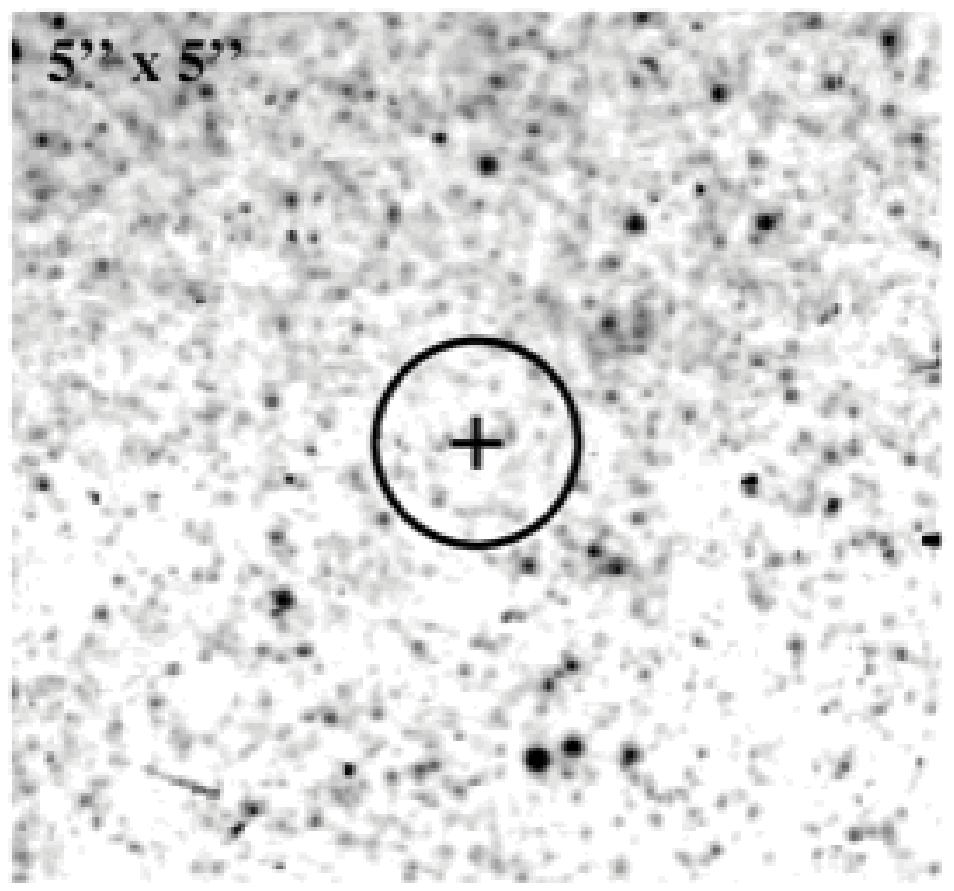} & 
\includegraphics[width=0.23\linewidth,clip=true,trim=0.5cm 3cm 0.5cm 2.5cm]{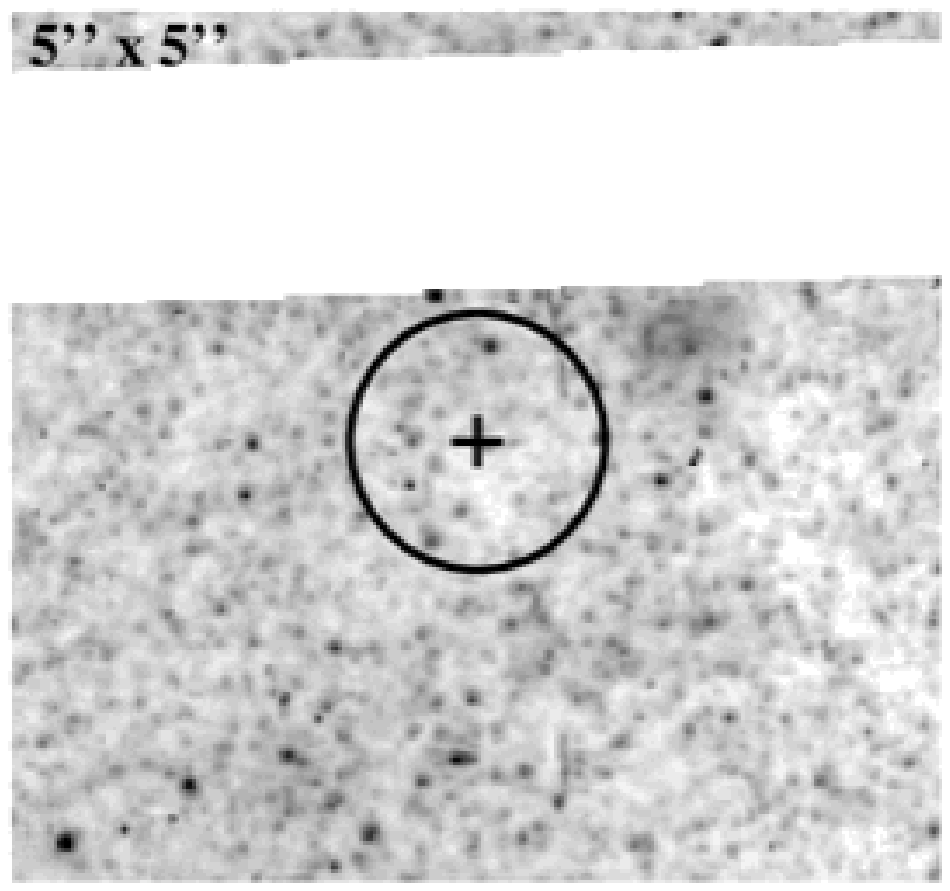} \\ 
\end{tabular}
\caption{Optical \HST images for X-ray sources detected in NGC~55. The box size (5\asn$\times$ 5\asn, 10\asn$\times$ 10\asn, or 15\asn$\times$ 15\asn) is given in the top-left corner of each image.  The circle shows the \Chandra 90\% error circle, centered on the source position.} 
\label{optical_55}
\end{figure*}

\setcounter{figure}{11}
\begin{figure*}
\centering
\begin{tabular}{cccc}
Source 57 & Source 58 & Source 59 & Source 61 \\ 
\includegraphics[width=0.23\linewidth,clip=true,trim=0.5cm 3cm 0.5cm 2.5cm]{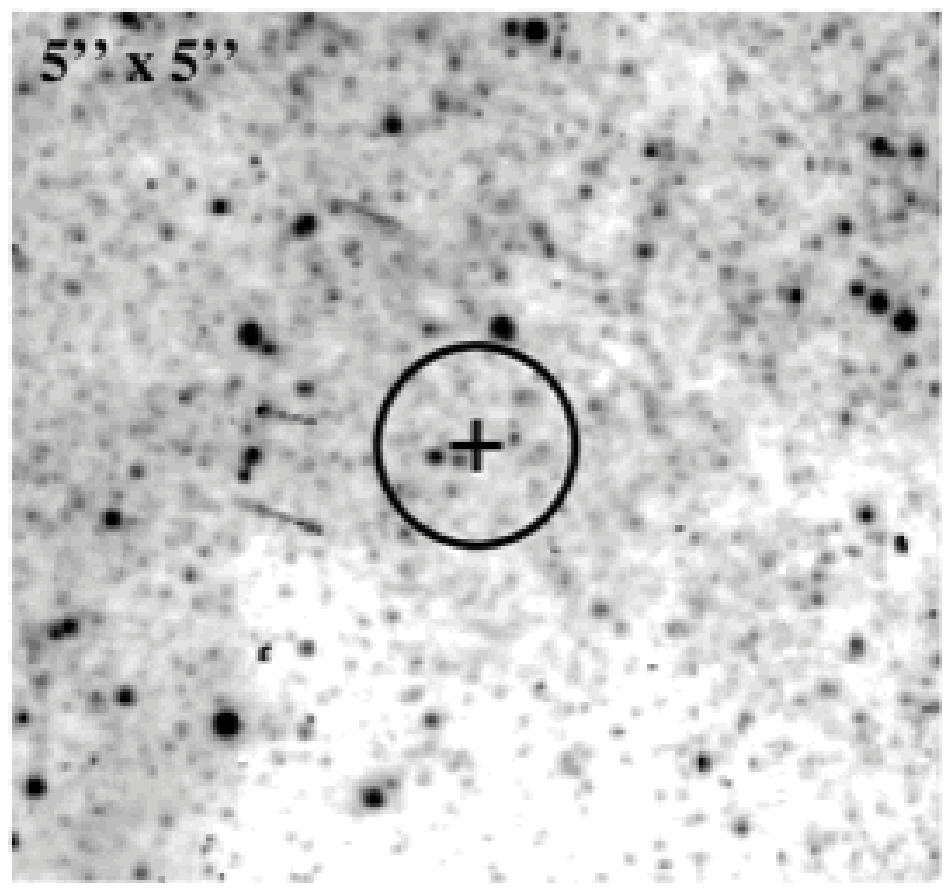} & 
\includegraphics[width=0.23\linewidth,clip=true,trim=0.5cm 3cm 0.5cm 2.5cm]{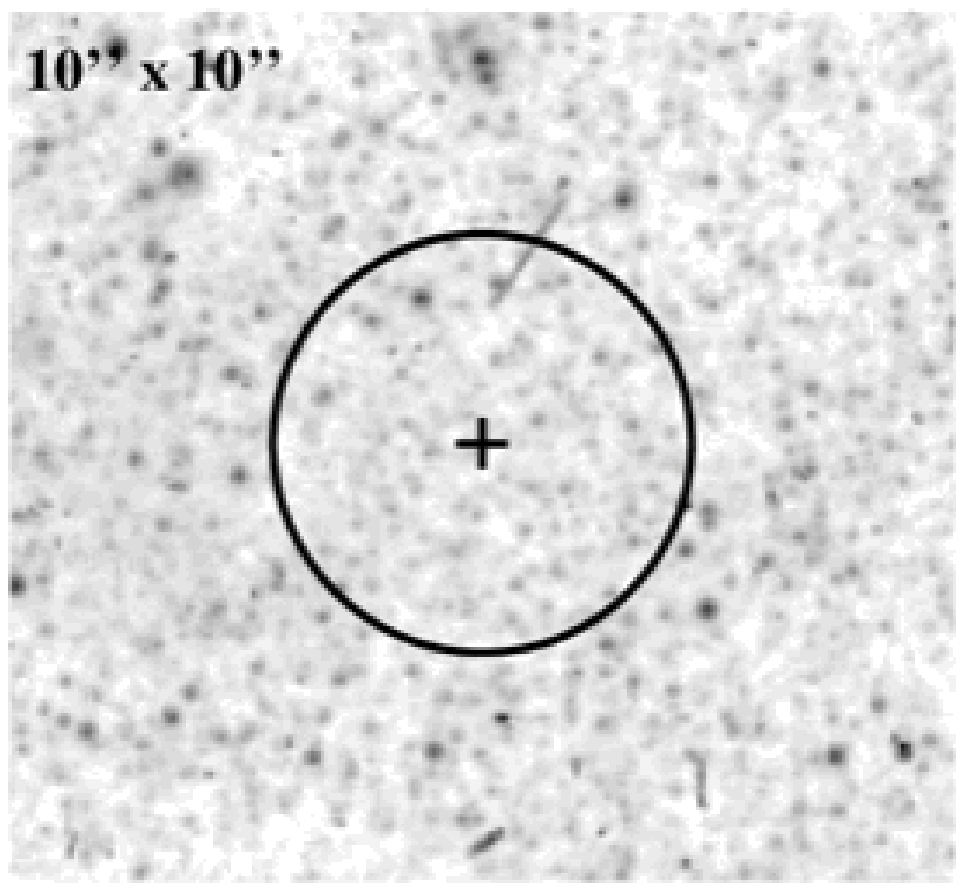} & 
\includegraphics[width=0.23\linewidth,clip=true,trim=0.5cm 3cm 0.5cm 2.5cm]{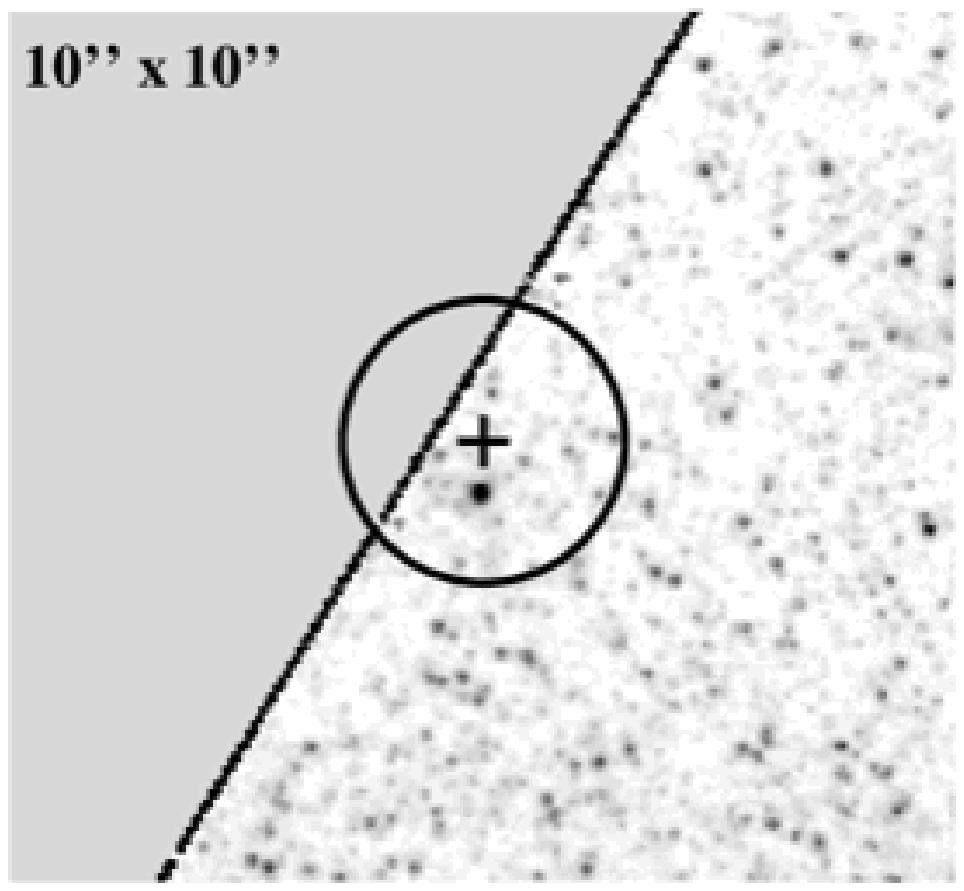} & 
\includegraphics[width=0.23\linewidth,clip=true,trim=0.5cm 3cm 0.5cm 2.5cm]{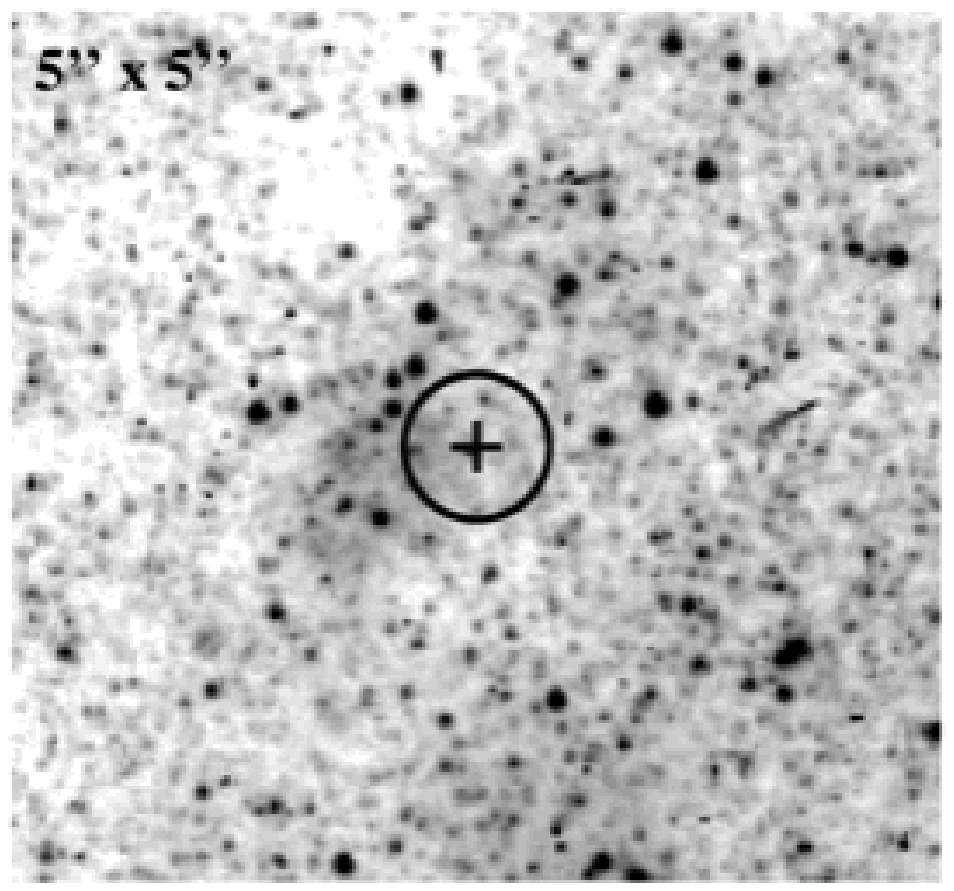} \\ 

Source 62 & Source 63 & Source 65 & Source 71 \\ 
\includegraphics[width=0.23\linewidth,clip=true,trim=0.5cm 3cm 0.5cm 2.5cm]{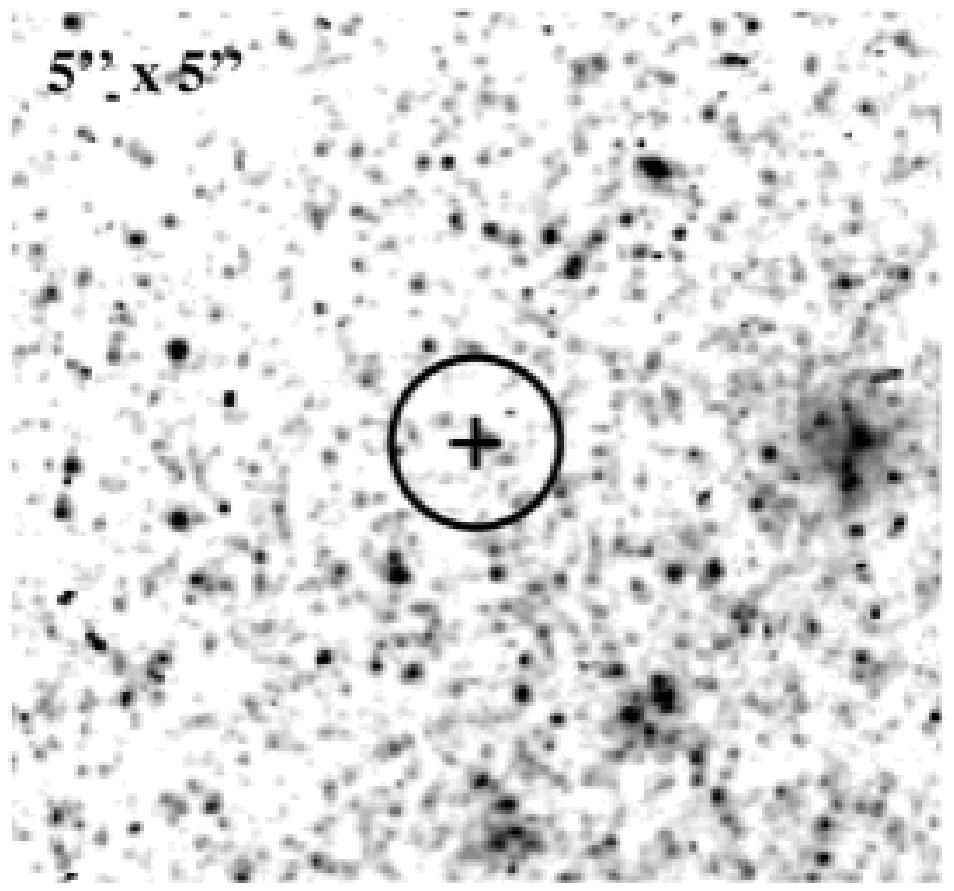} & 
\includegraphics[width=0.23\linewidth,clip=true,trim=0.5cm 3cm 0.5cm 2.5cm]{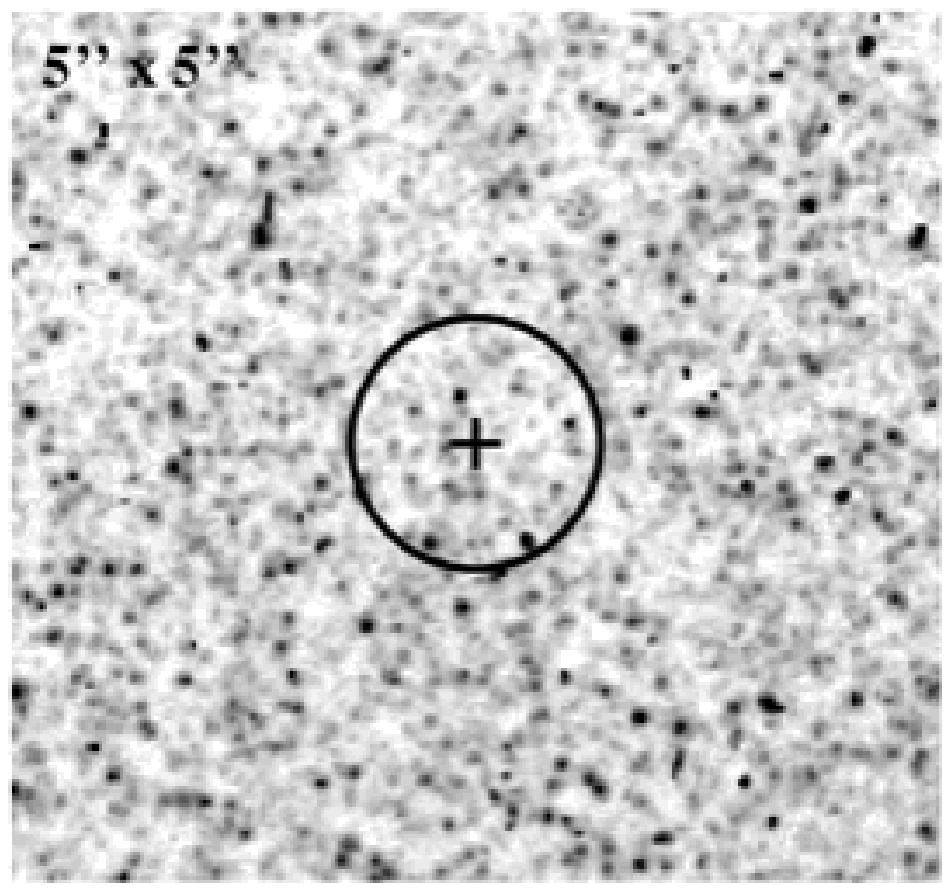} & 
\includegraphics[width=0.23\linewidth,clip=true,trim=0.5cm 3cm 0.5cm 2.5cm]{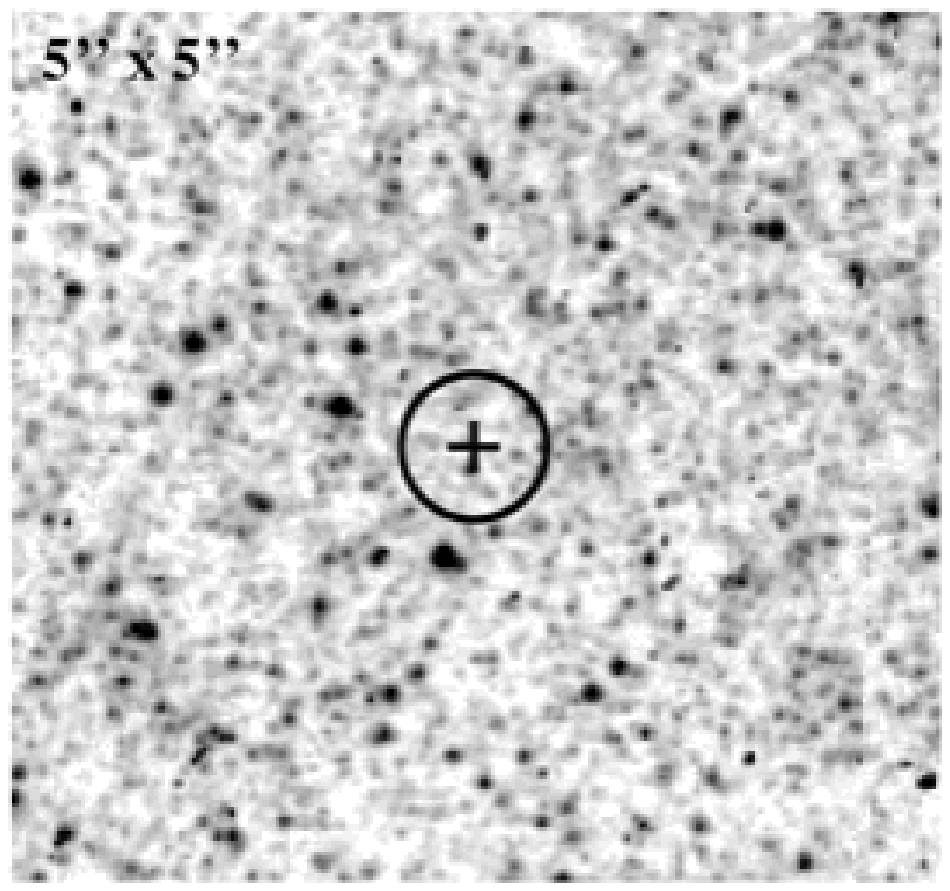} & 
\includegraphics[width=0.23\linewidth,clip=true,trim=0.5cm 3cm 0.5cm 2.5cm]{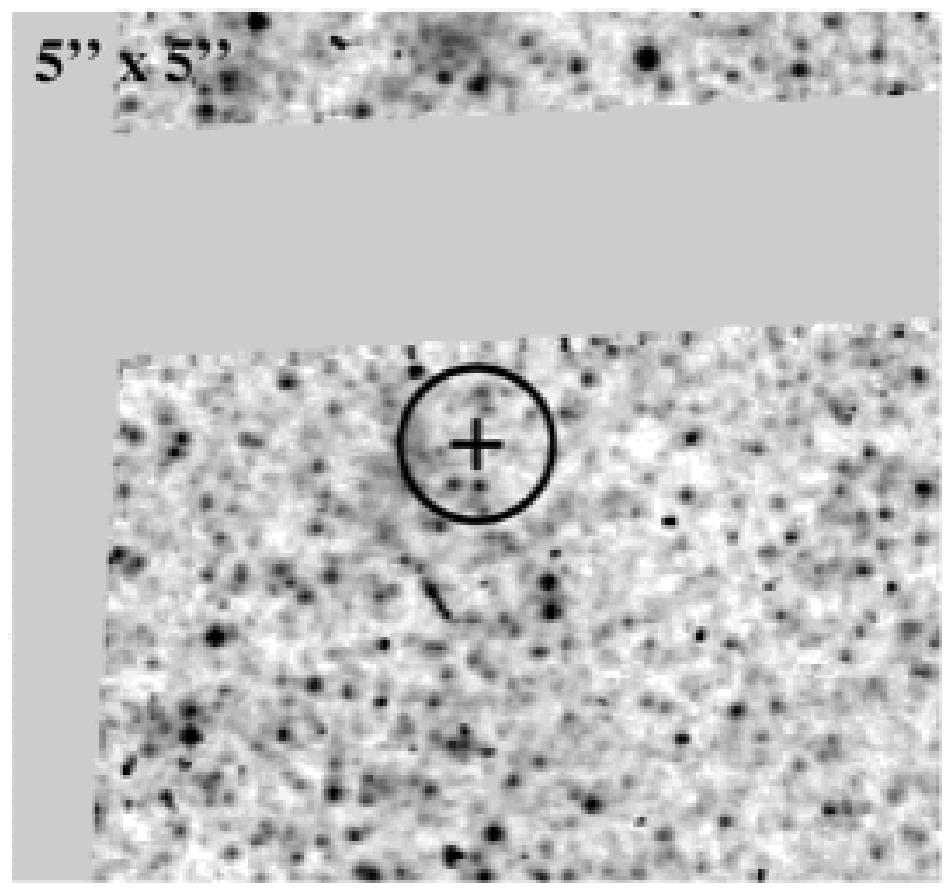} \\ 

Source 100 & Source 114 & Source 116 & Source 117 \\ 
\includegraphics[width=0.23\linewidth,clip=true,trim=0.5cm 3cm 0.5cm 2.5cm]{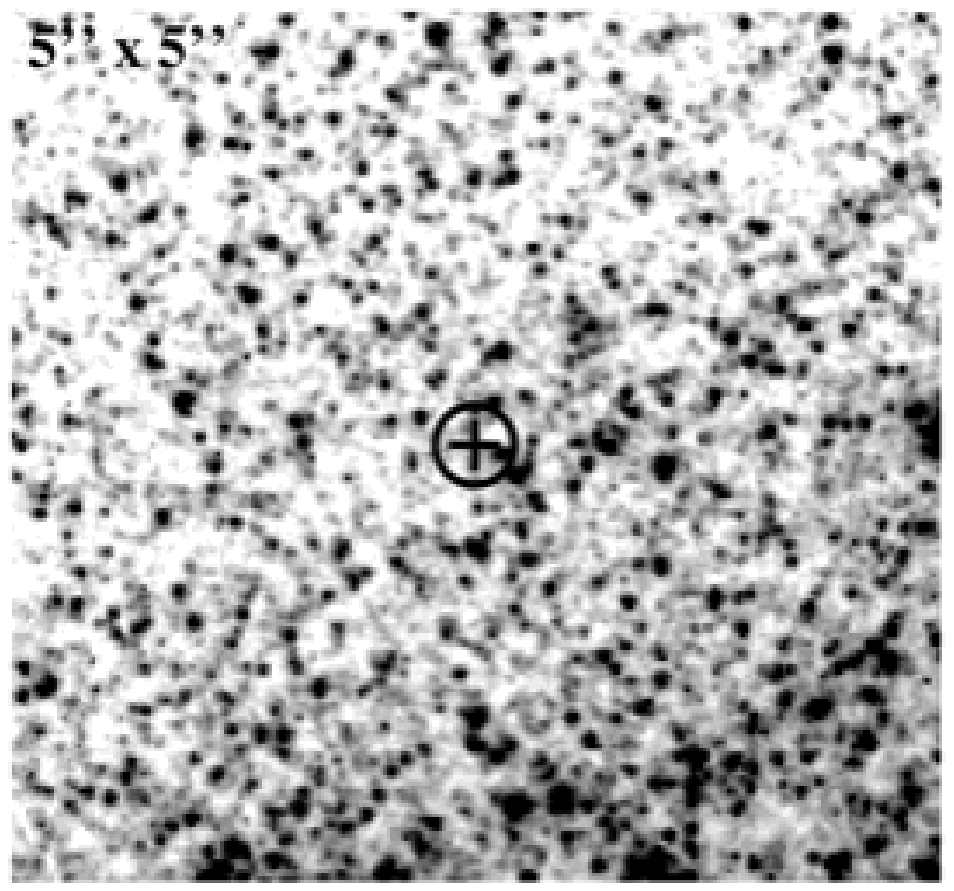} & 
\includegraphics[width=0.23\linewidth,clip=true,trim=0.5cm 3cm 0.5cm 2.5cm]{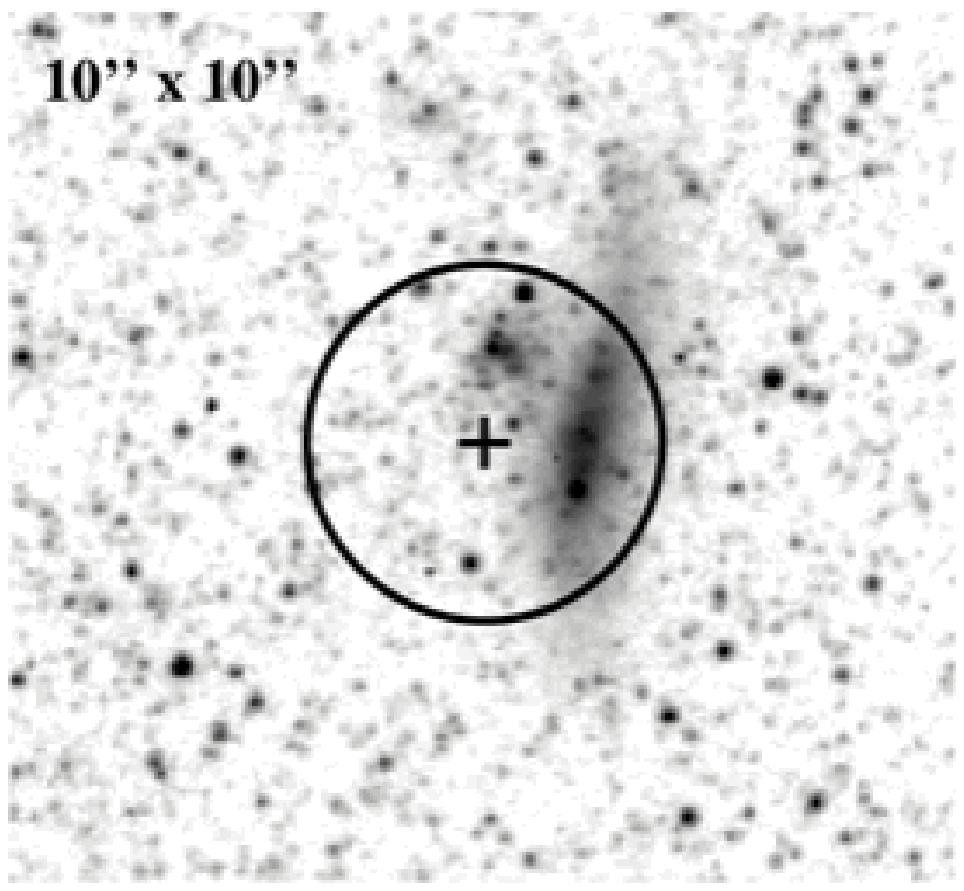} & 
\includegraphics[width=0.23\linewidth,clip=true,trim=0.5cm 3cm 0.5cm 2.5cm]{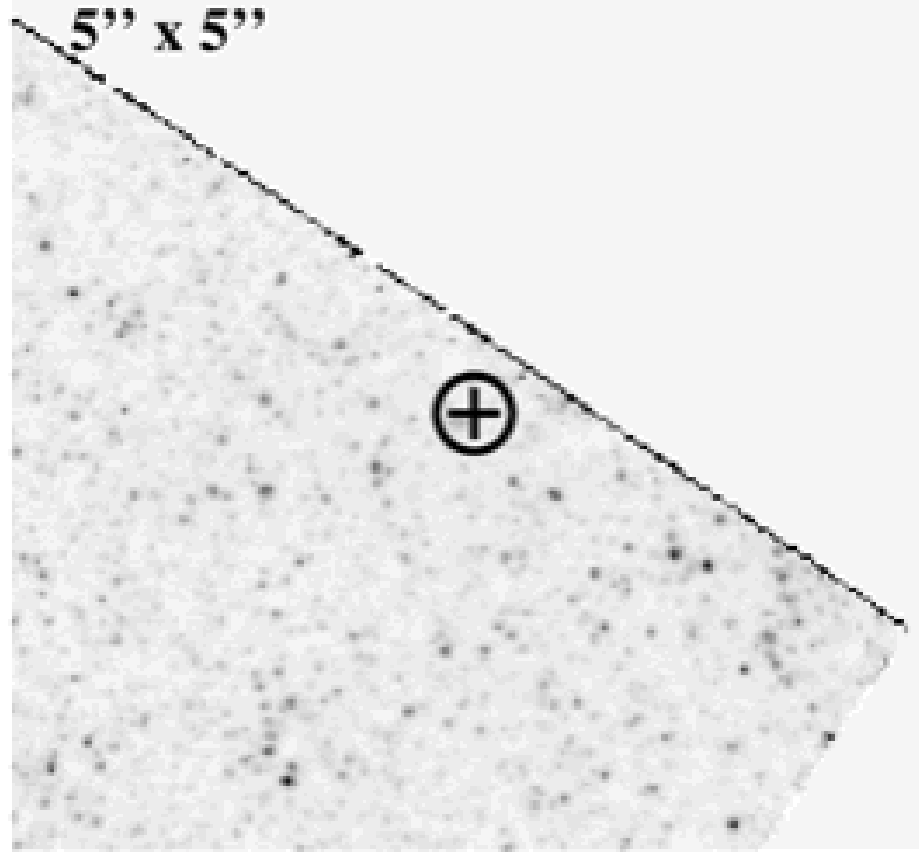} & 
\includegraphics[width=0.23\linewidth,clip=true,trim=0.5cm 3cm 0.5cm 2.5cm]{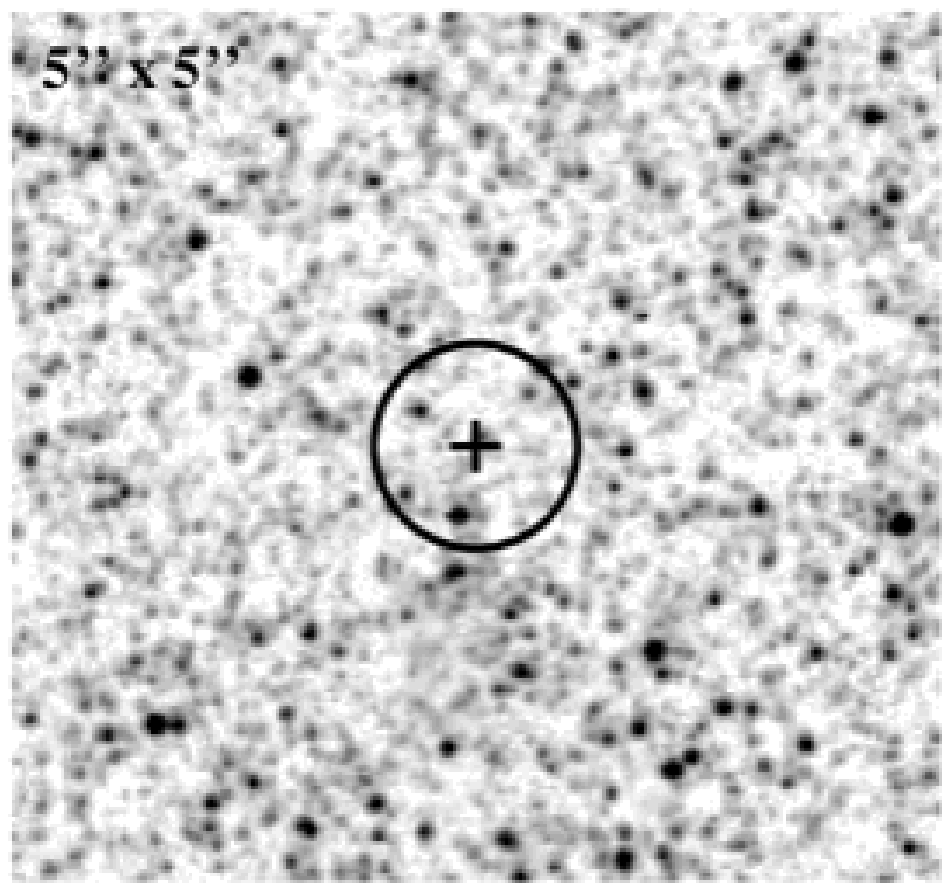} \\ 

Source 119 & Source 120 & Source 122 & Source 123 \\ 
\includegraphics[width=0.23\linewidth,clip=true,trim=0.5cm 3cm 0.5cm 2.5cm]{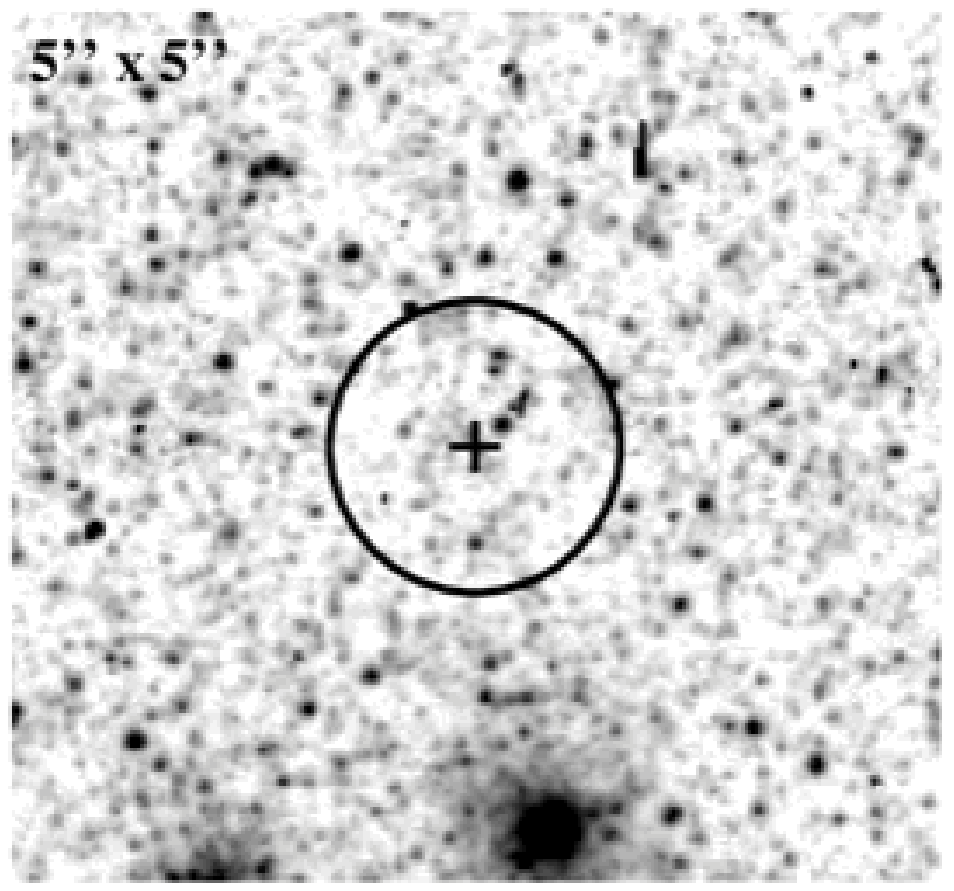} & 
\includegraphics[width=0.23\linewidth,clip=true,trim=0.5cm 3cm 0.5cm 2.5cm]{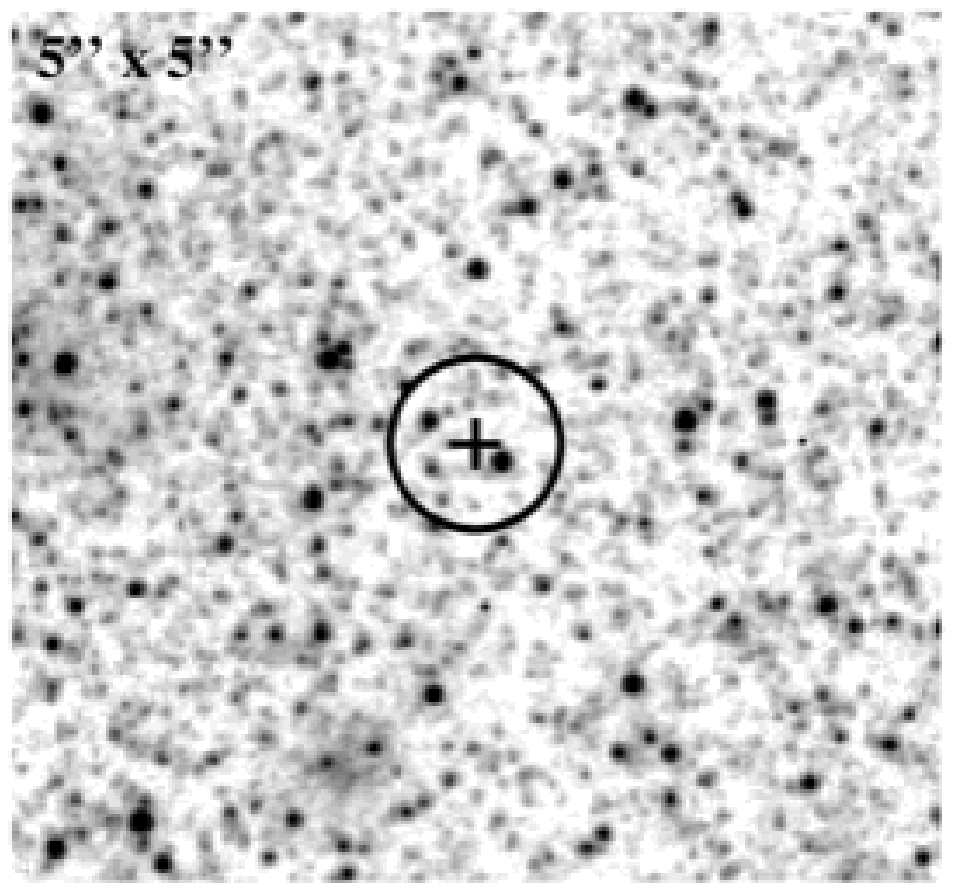} & 
\includegraphics[width=0.23\linewidth,clip=true,trim=0.5cm 3cm 0.5cm 2.5cm]{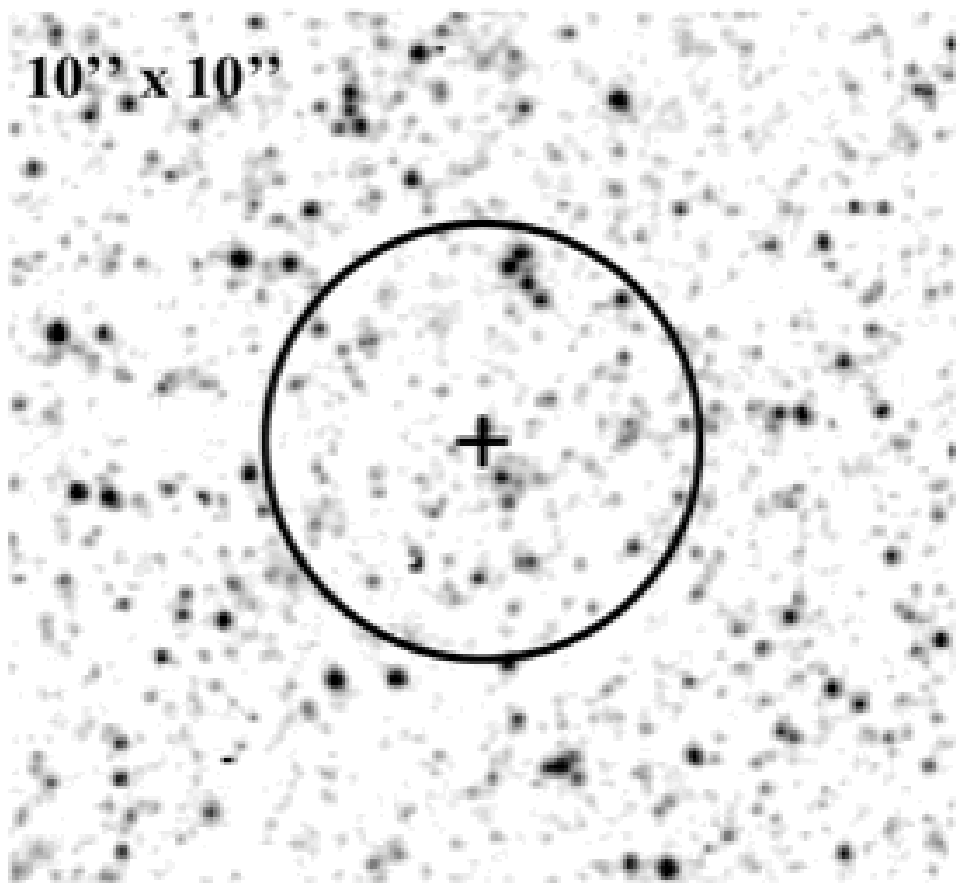} & 
\includegraphics[width=0.23\linewidth,clip=true,trim=0.5cm 3cm 0.5cm 2.5cm]{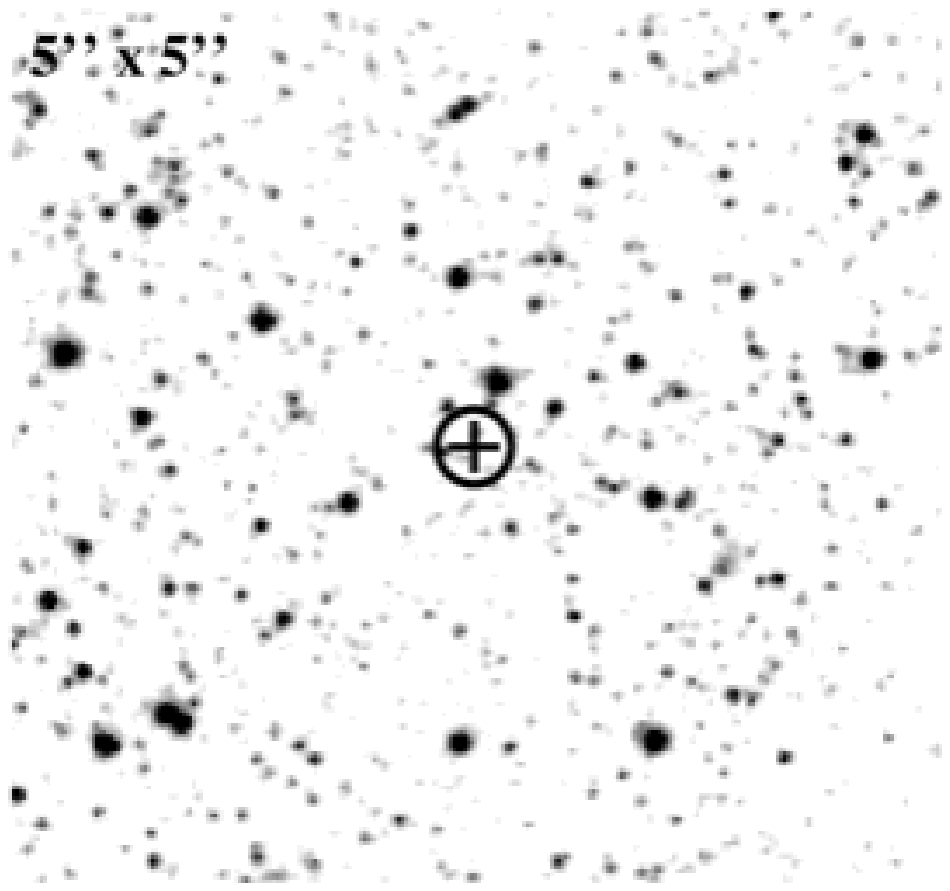} \\ 

Source 124 & Source 130 & Source 133 & Source 134 \\ 
\includegraphics[width=0.23\linewidth,clip=true,trim=0.5cm 3cm 0.5cm 2.5cm]{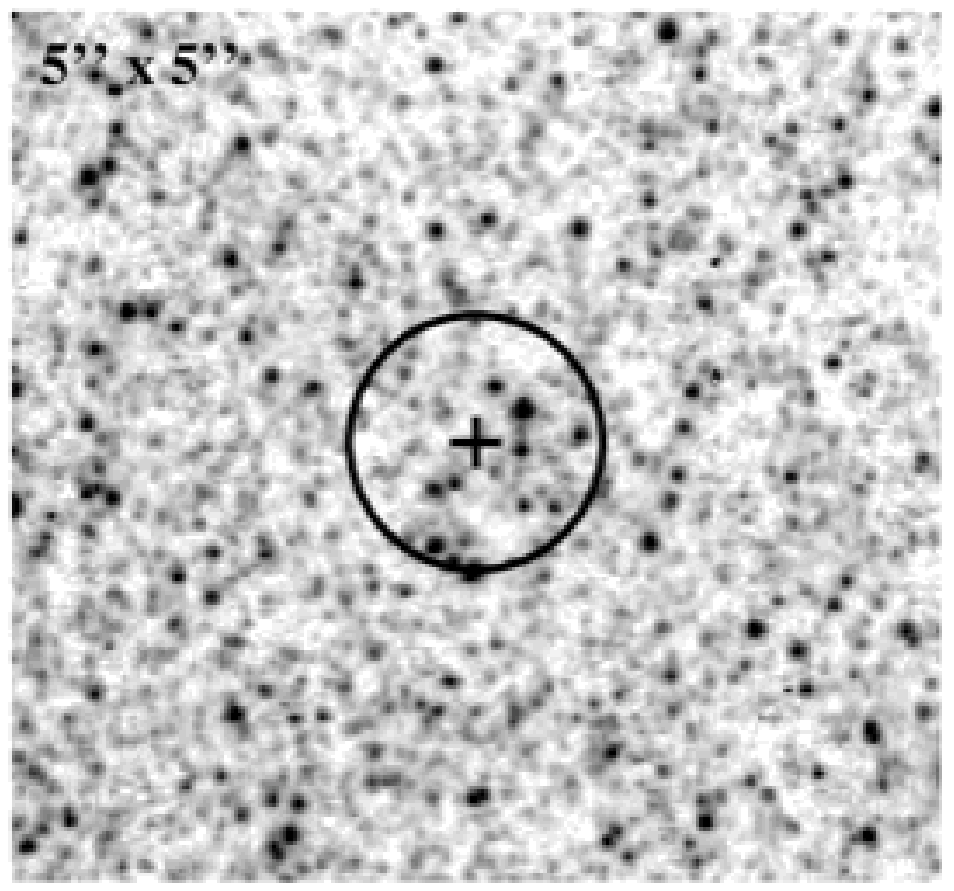} & 
\includegraphics[width=0.23\linewidth,clip=true,trim=0.5cm 3cm 0.5cm 2.5cm]{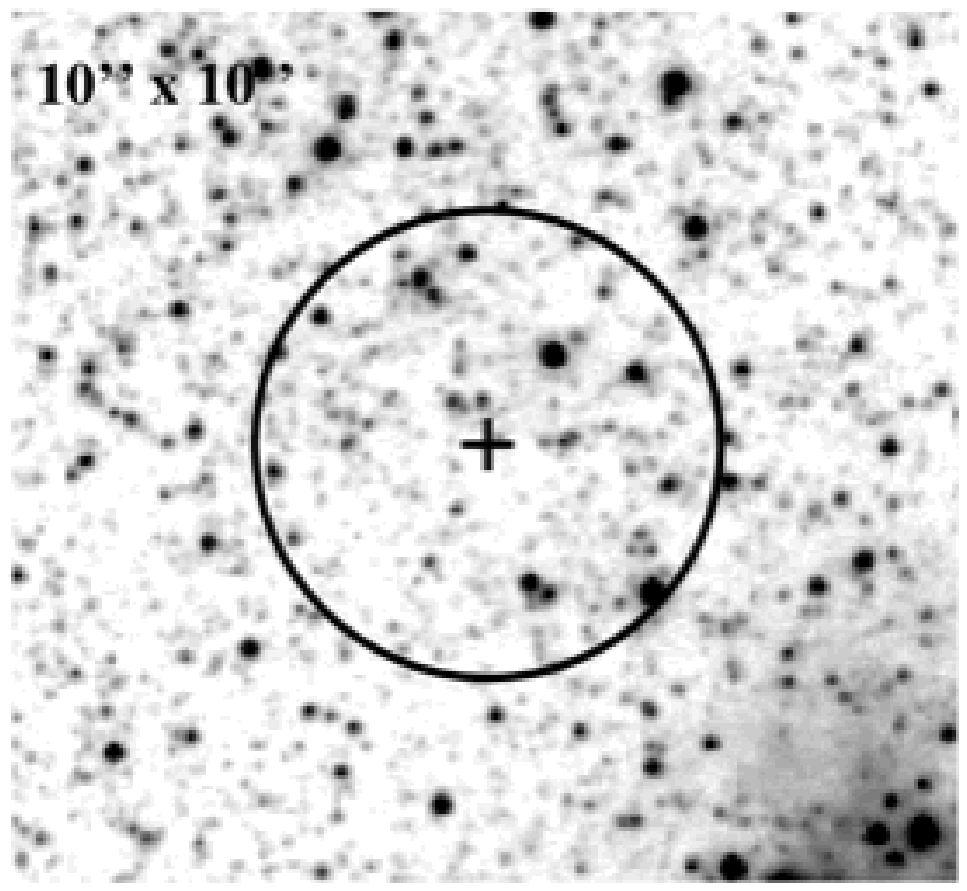} & 
\includegraphics[width=0.23\linewidth,clip=true,trim=0.5cm 3cm 0.5cm 2.5cm]{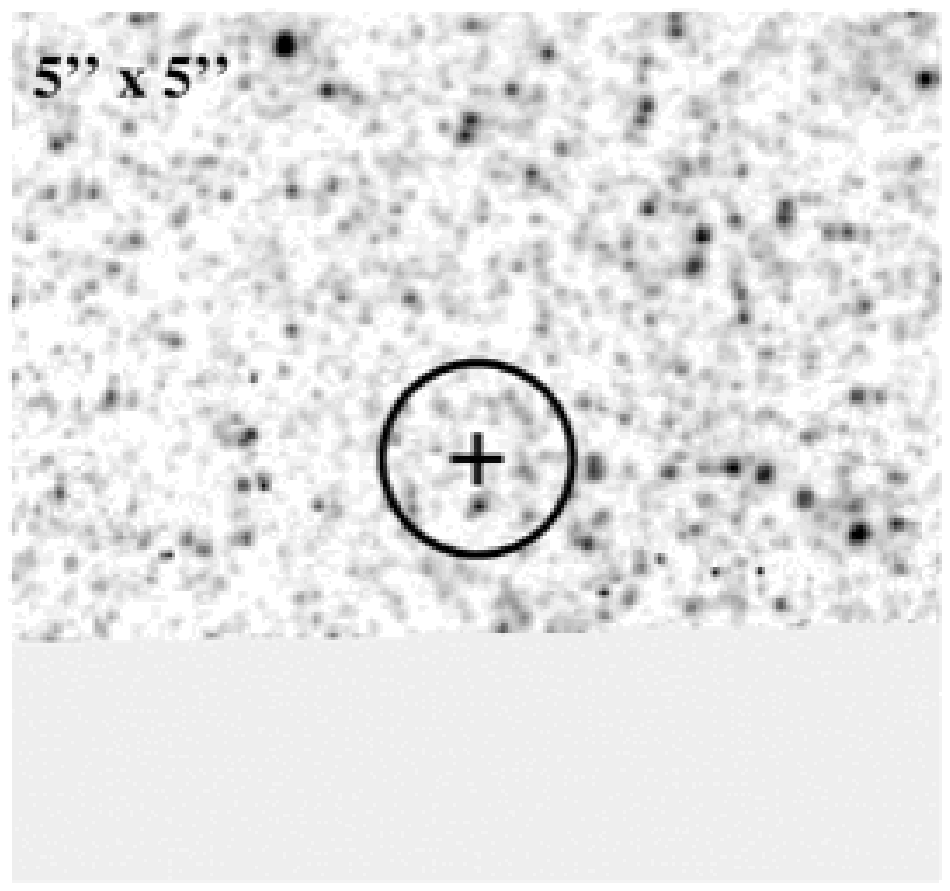} & 
\includegraphics[width=0.23\linewidth,clip=true,trim=0.5cm 3cm 0.5cm 2.5cm]{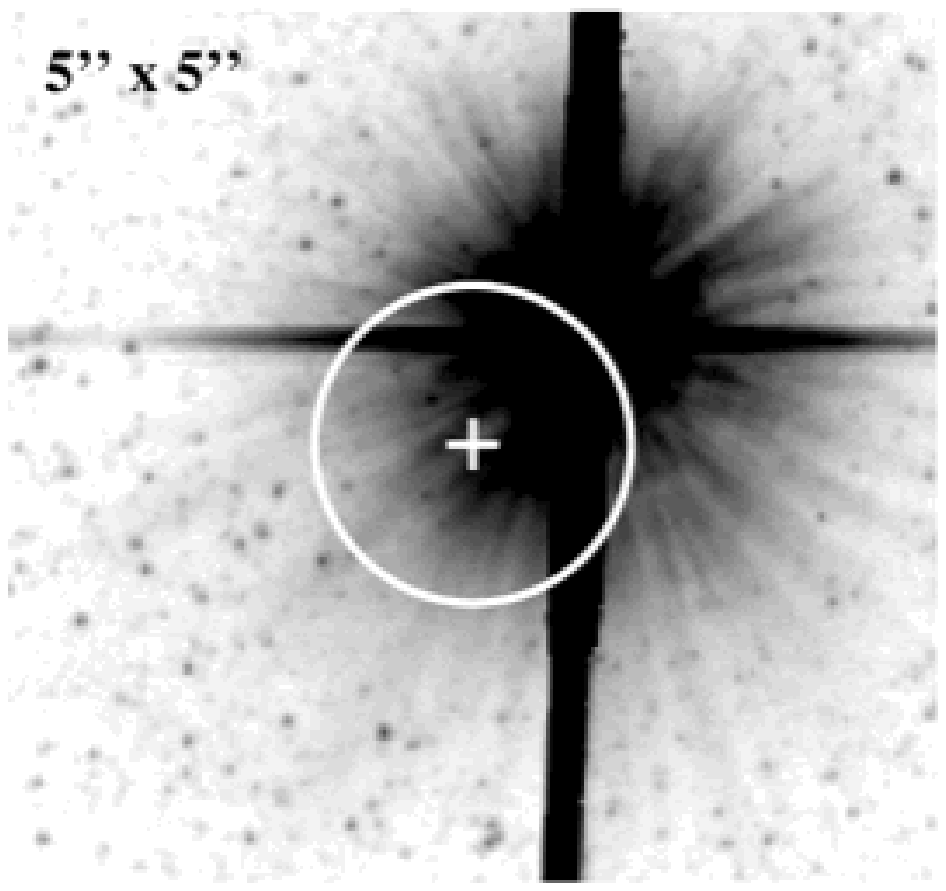} \\ 
\end{tabular}
\caption{{\it (Continued)} Optical \HST images for X-ray sources detected in NGC~55. The box size (5\asn$\times$ 5\asn, 10\asn$\times$ 10\asn, or 15\asn$\times$ 15\asn) is given in the top-left corner of each image. The circle shows the \Chandra 90\% error circle, centered on the source position.}
\label{optical_55}
\end{figure*}

\setcounter{figure}{11}
\begin{figure*}
\centering
\begin{tabular}{cccc}
Source 138 &   &   &   \\ 
\includegraphics[width=0.23\linewidth,clip=true,trim=0.5cm 3cm 0.5cm 2.5cm]{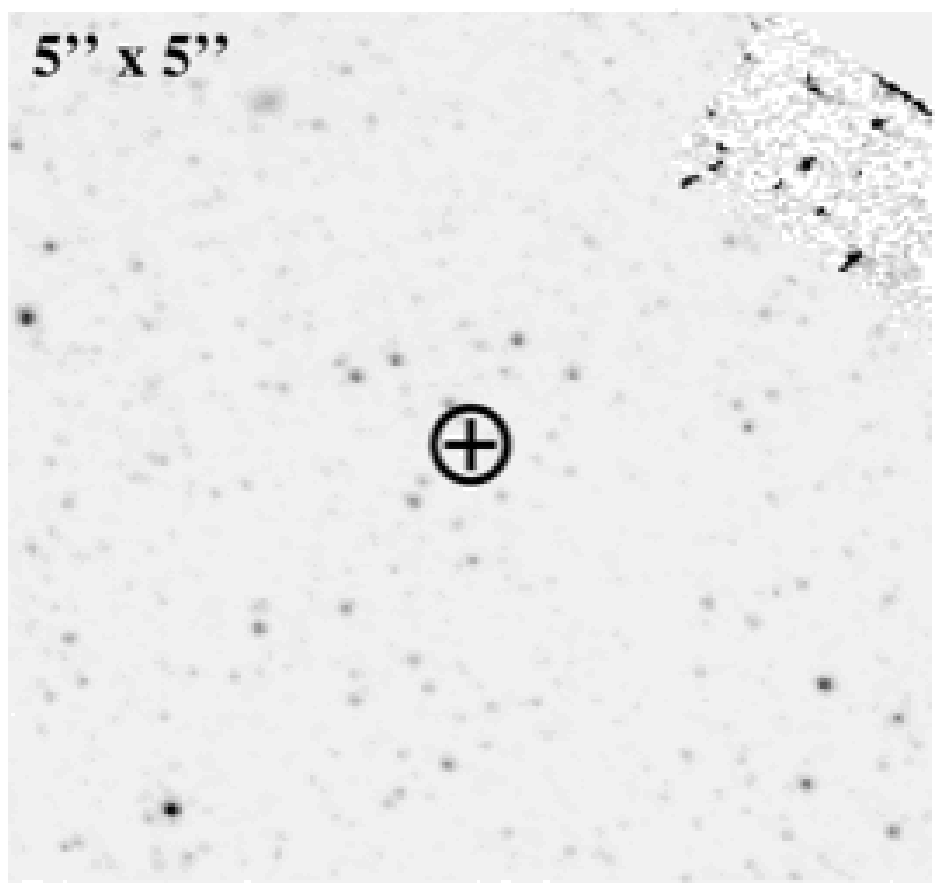} & & & \\ 
\end{tabular}
\caption{{\it (Continued)} Optical \HST images for X-ray sources detected in NGC~55. The box size (5\asn$\times$ 5\asn, 10\asn$\times$ 10\asn, or 15\asn$\times$ 15\asn) is given in the top-left corner of each image. The circle shows the \Chandra 90\% error circle, centered on the source position.} 
\label{optical_55}
\end{figure*}

\clearpage

\begin{figure*}
\centering
\begin{tabular}{cccc}
Source 2 & Source 5 & Source 7 & Source 16 \\
\includegraphics[width=0.23\linewidth,clip=true,trim=0.5cm 3cm 0.5cm 2.5cm]{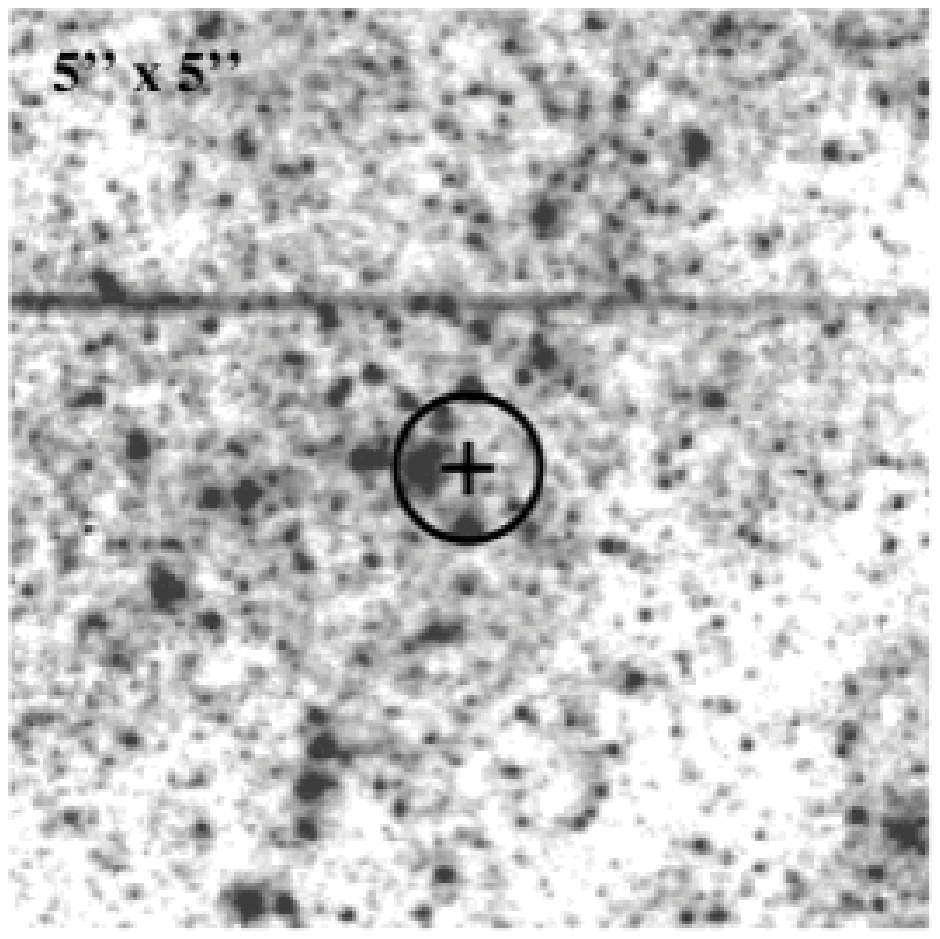} & 
\includegraphics[width=0.23\linewidth,clip=true,trim=0.5cm 3cm 0.5cm 2.5cm]{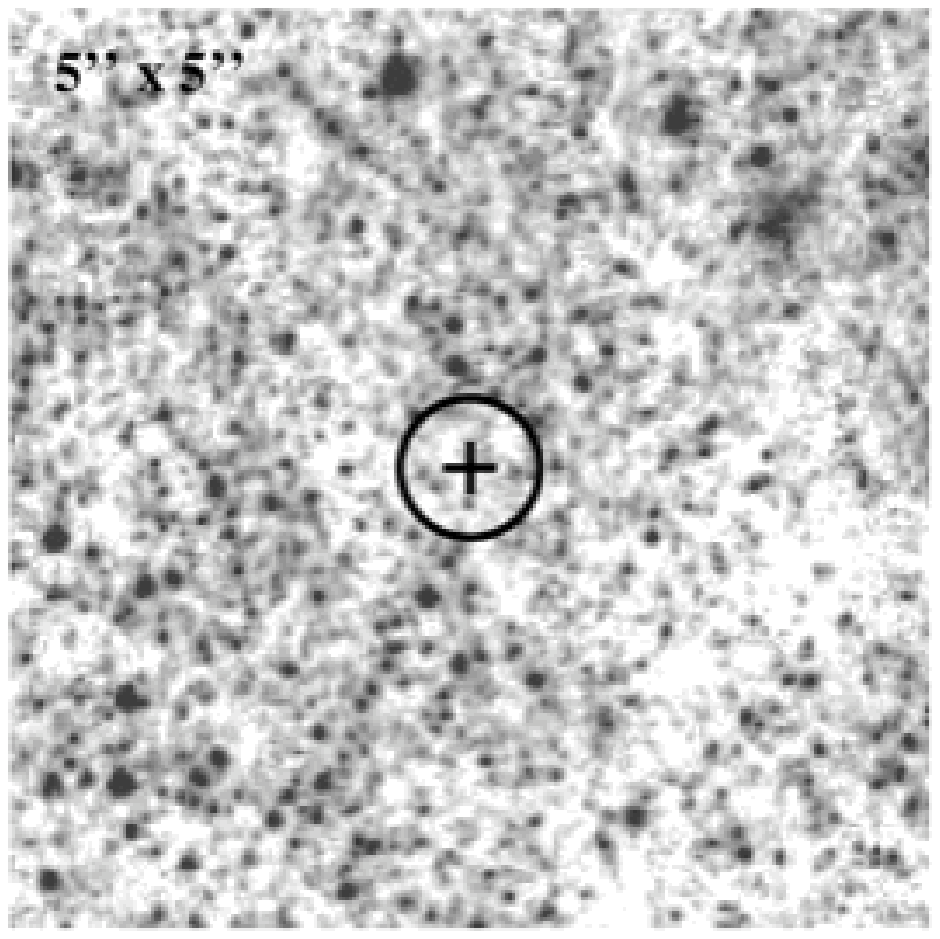} & 
\includegraphics[width=0.23\linewidth,clip=true,trim=0.5cm 3cm 0.5cm 2.5cm]{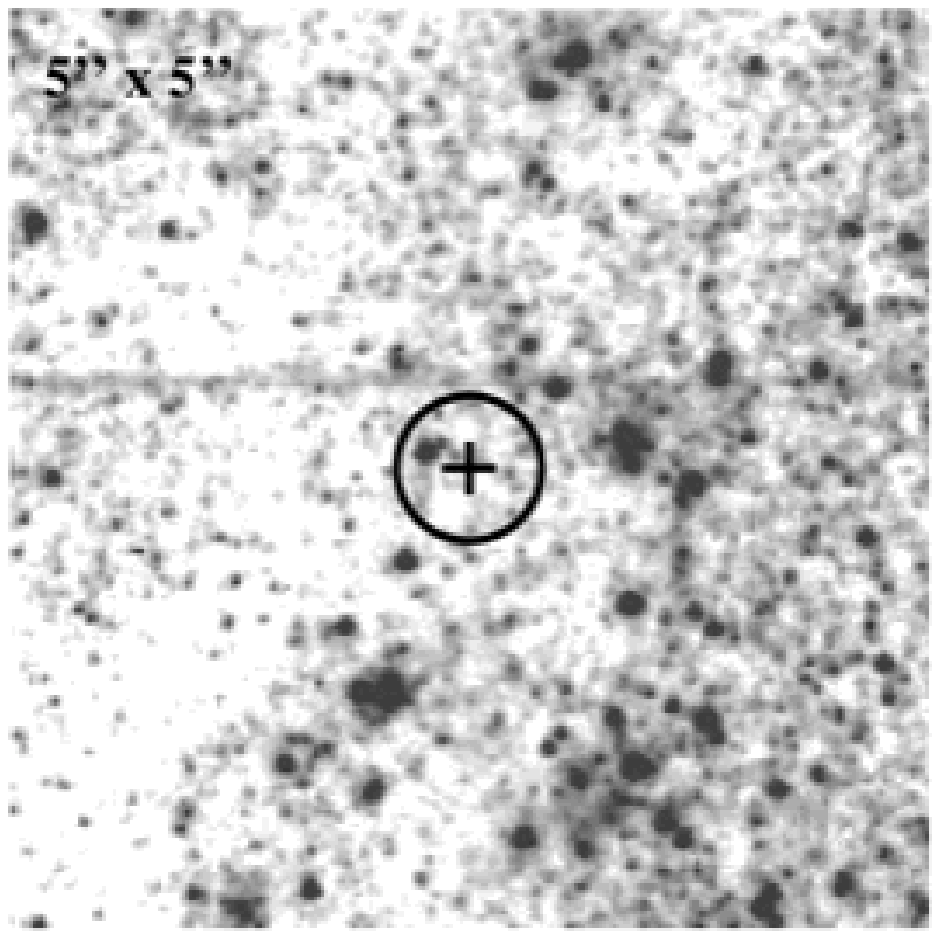} & 
\includegraphics[width=0.23\linewidth,clip=true,trim=0.5cm 3cm 0.5cm 2.5cm]{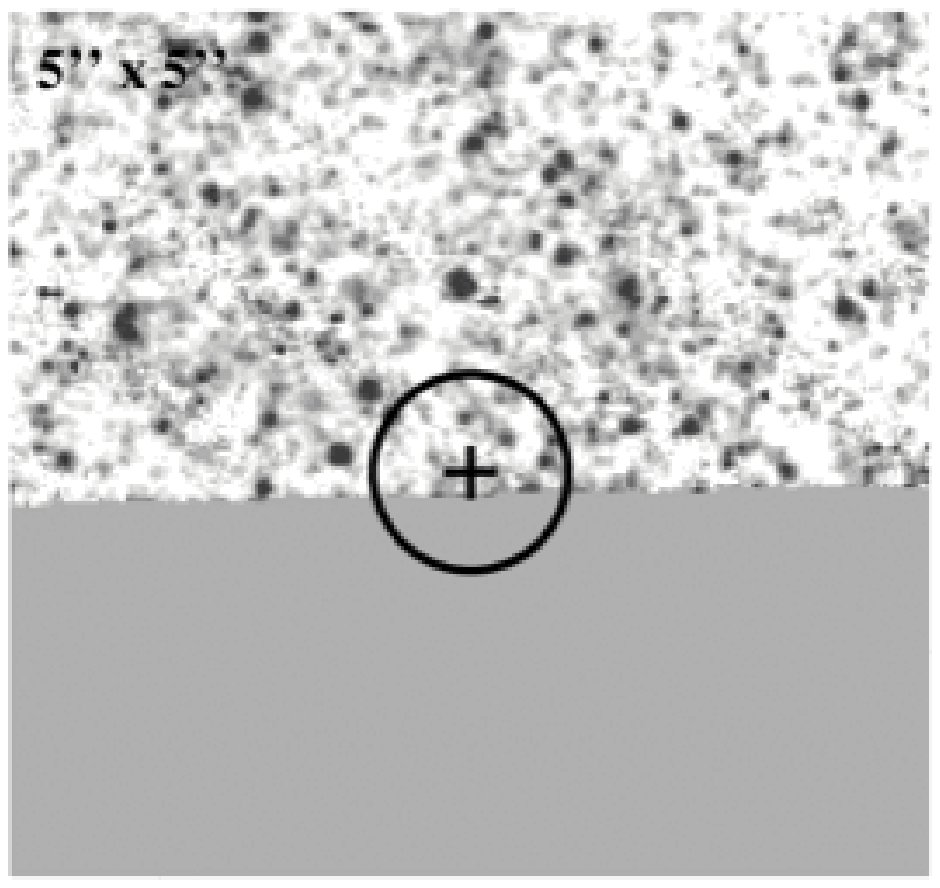} \\ 

Source 18 & Source 20 & Source 22 & Source 38 \\ 
\includegraphics[width=0.23\linewidth,clip=true,trim=0.5cm 3cm 0.5cm 2.5cm]{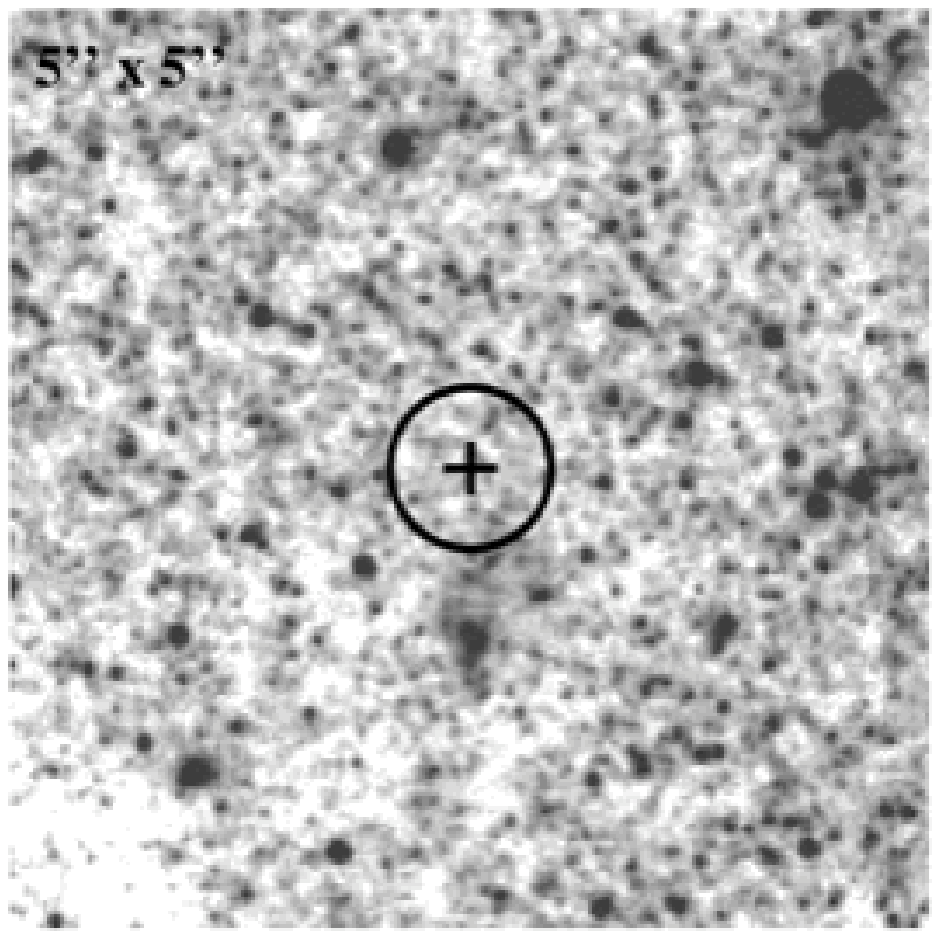} & 
\includegraphics[width=0.23\linewidth,clip=true,trim=0.5cm 3cm 0.5cm 2.5cm]{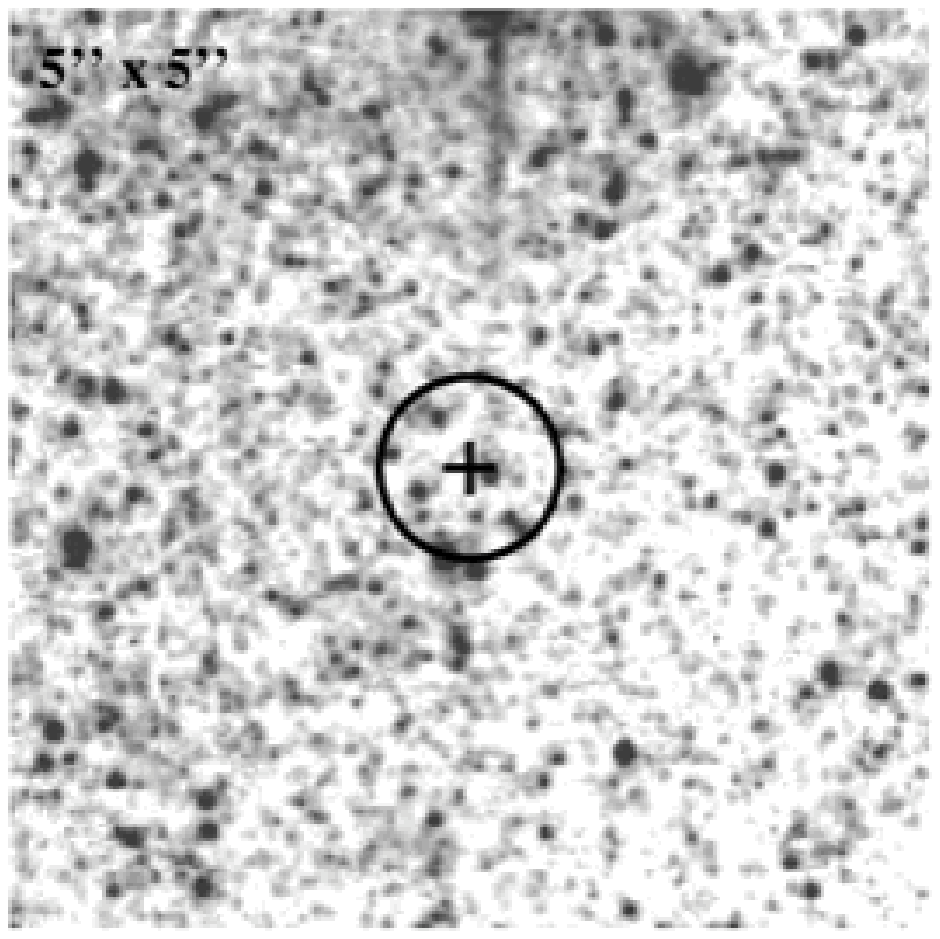} & 
\includegraphics[width=0.23\linewidth,clip=true,trim=0.5cm 3cm 0.5cm 2.5cm]{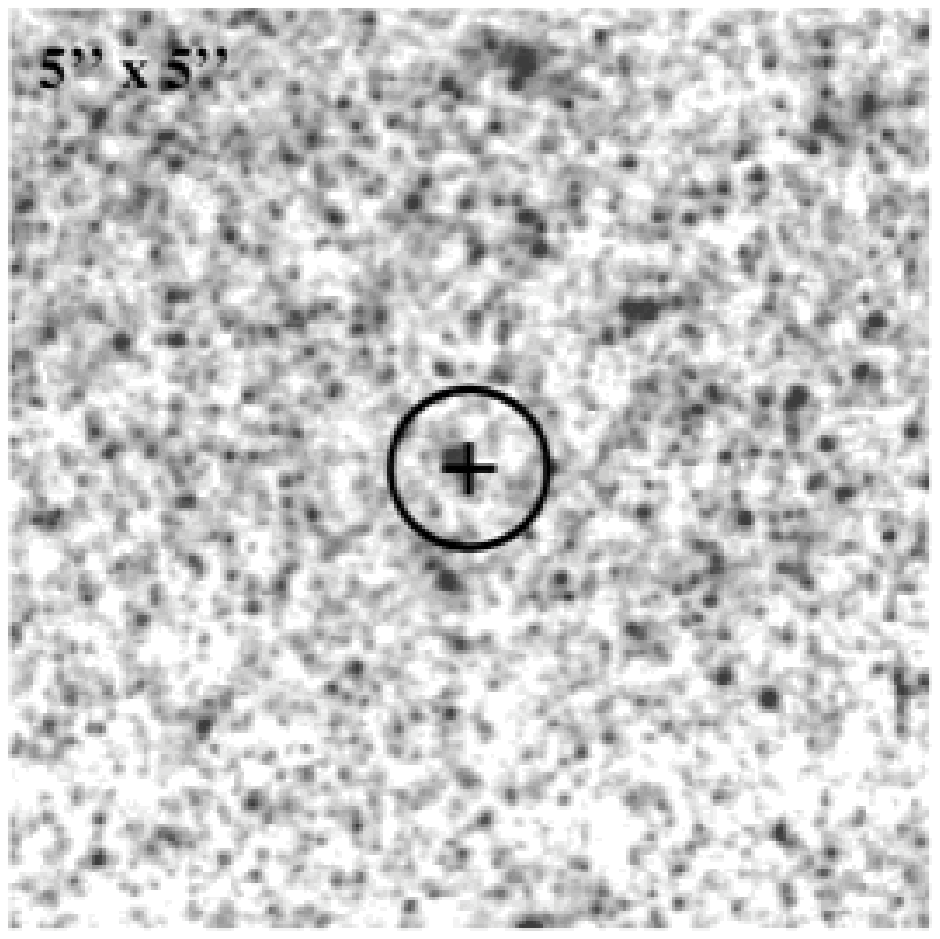} & 
\includegraphics[width=0.23\linewidth,clip=true,trim=0.5cm 3cm 0.5cm 2.5cm]{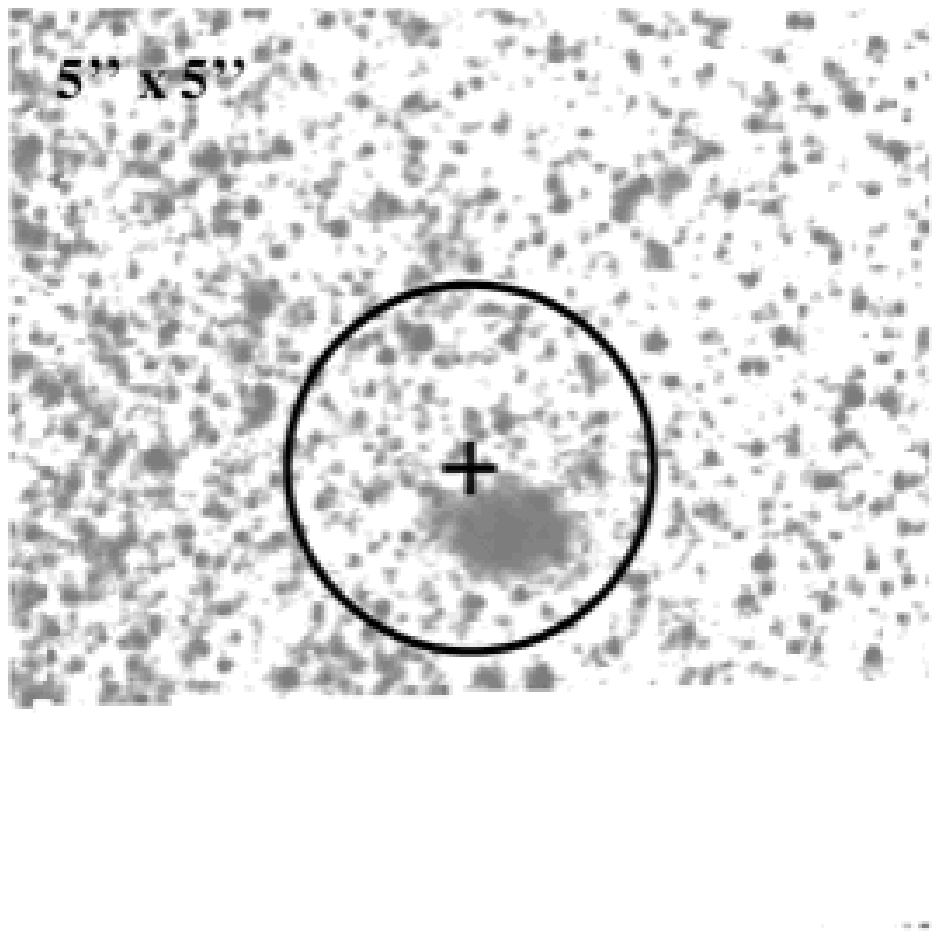} \\ 

Source 39 & Source 40 & Source 41 & Source 42 \\ 
\includegraphics[width=0.23\linewidth,clip=true,trim=0.5cm 3cm 0.5cm 2.5cm]{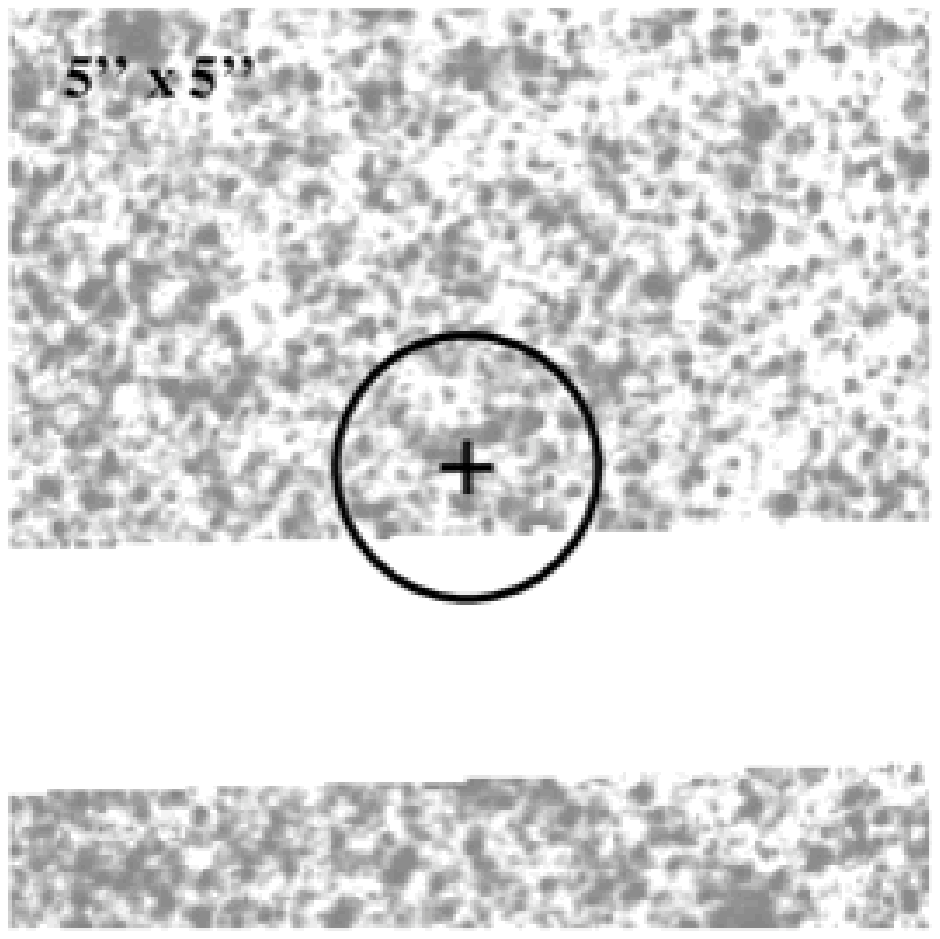} & 
\includegraphics[width=0.23\linewidth,clip=true,trim=0.5cm 3cm 0.5cm 2.5cm]{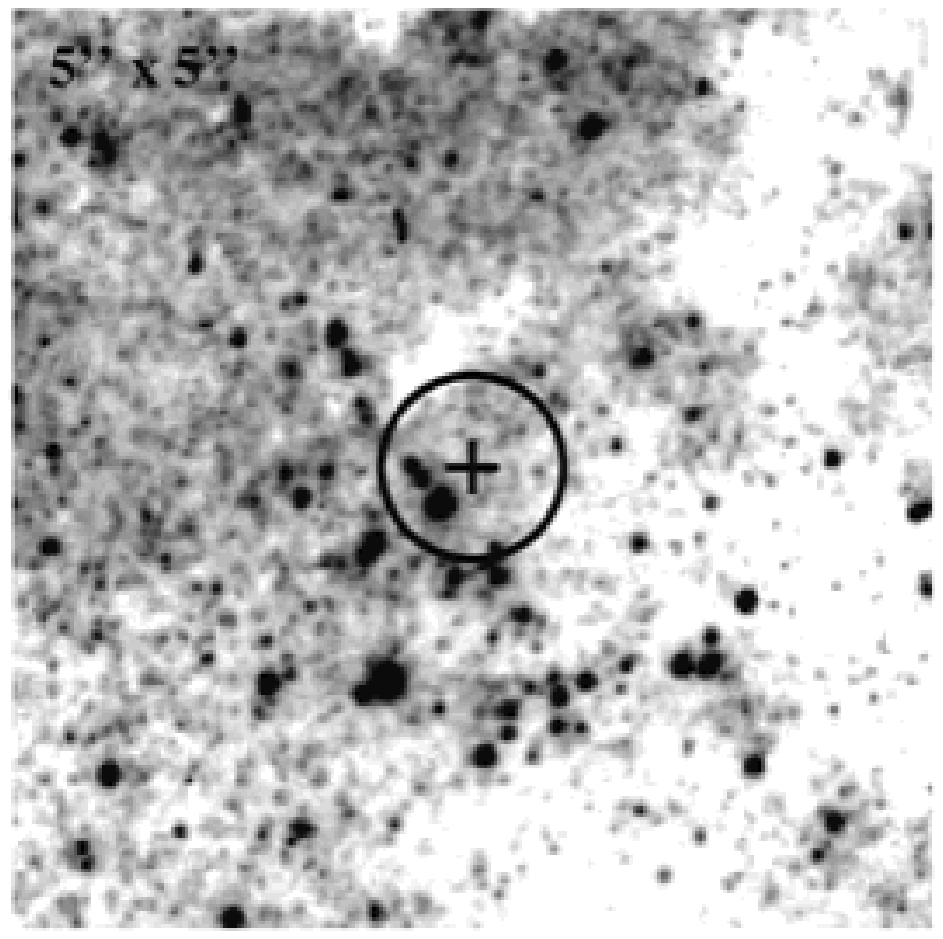} & 
\includegraphics[width=0.23\linewidth,clip=true,trim=0.5cm 3cm 0.5cm 2.5cm]{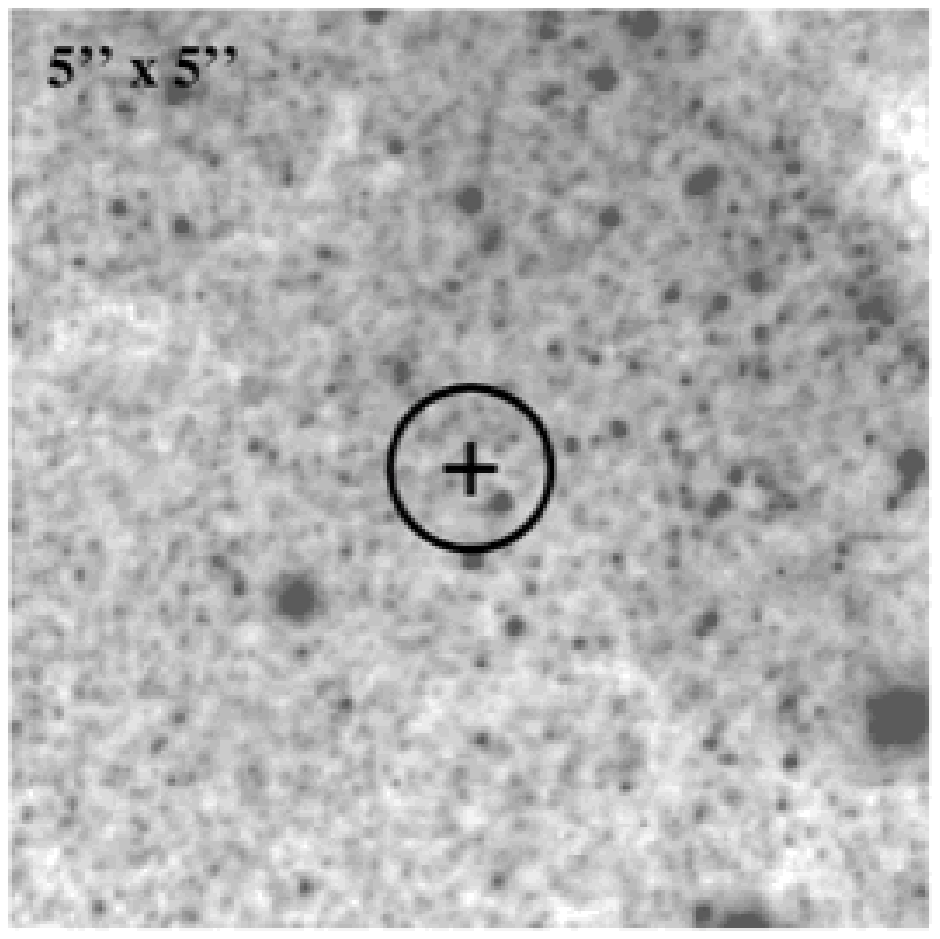} & 
\includegraphics[width=0.23\linewidth,clip=true,trim=0.5cm 3cm 0.5cm 2.5cm]{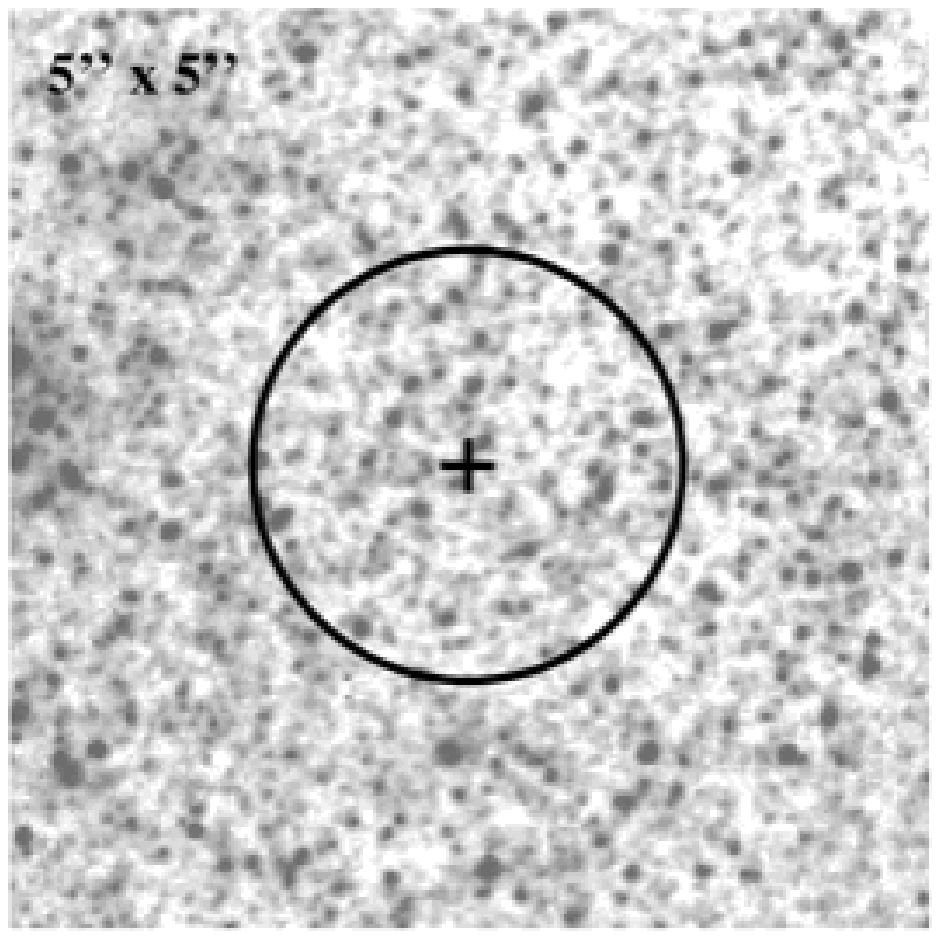} \\ 

Source 44 & Source 45 & Source 46 & Source 47 \\ 
\includegraphics[width=0.23\linewidth,clip=true,trim=0.5cm 3cm 0.5cm 2.5cm]{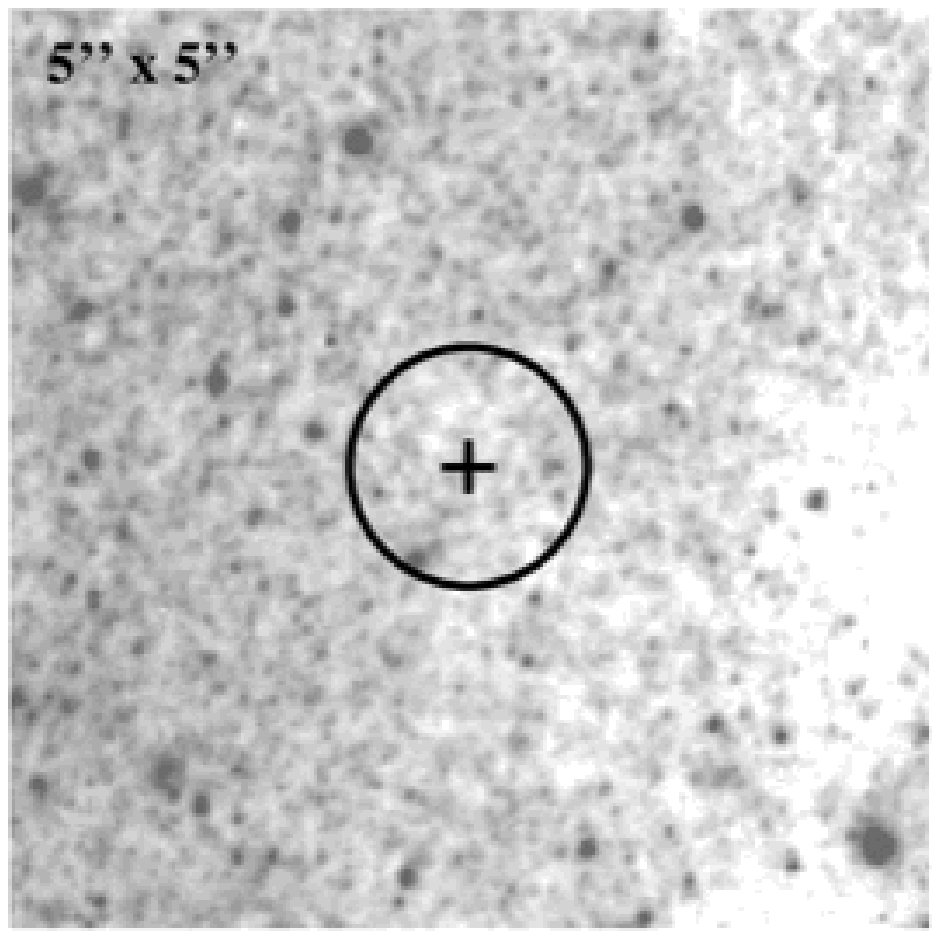} & 
\includegraphics[width=0.23\linewidth,clip=true,trim=0.5cm 3cm 0.5cm 2.5cm]{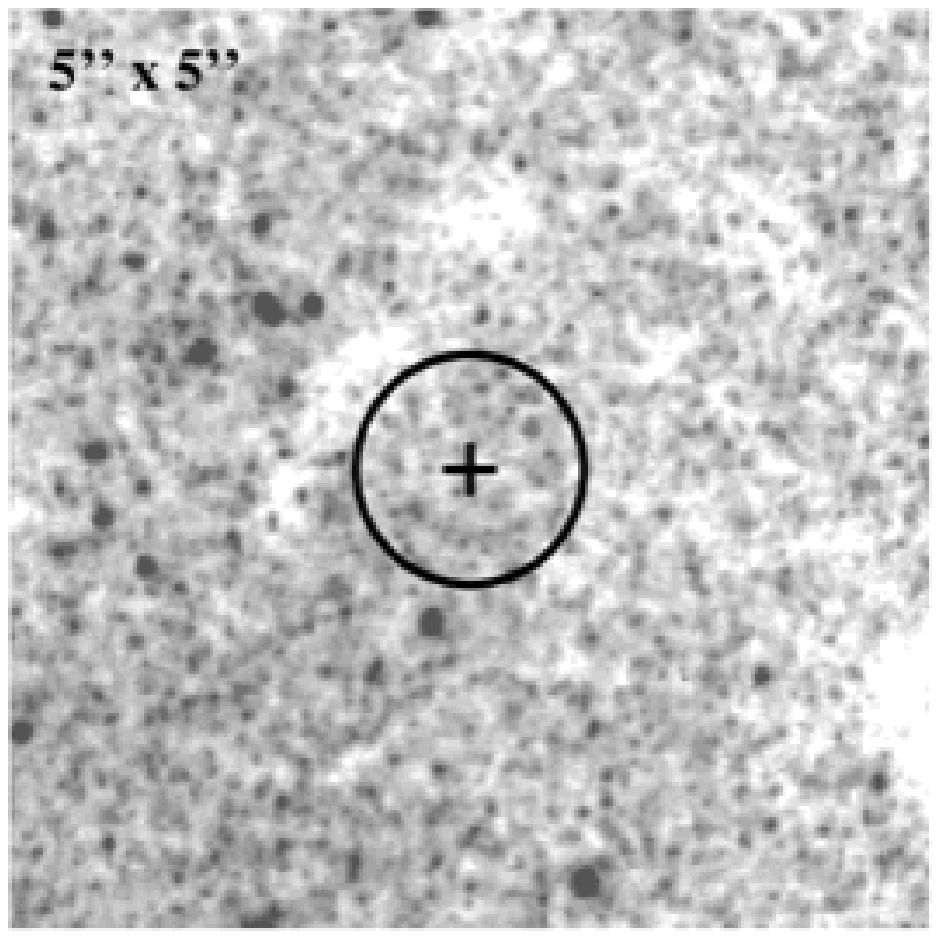} & 
\includegraphics[width=0.23\linewidth,clip=true,trim=0.5cm 3cm 0.5cm 2.5cm]{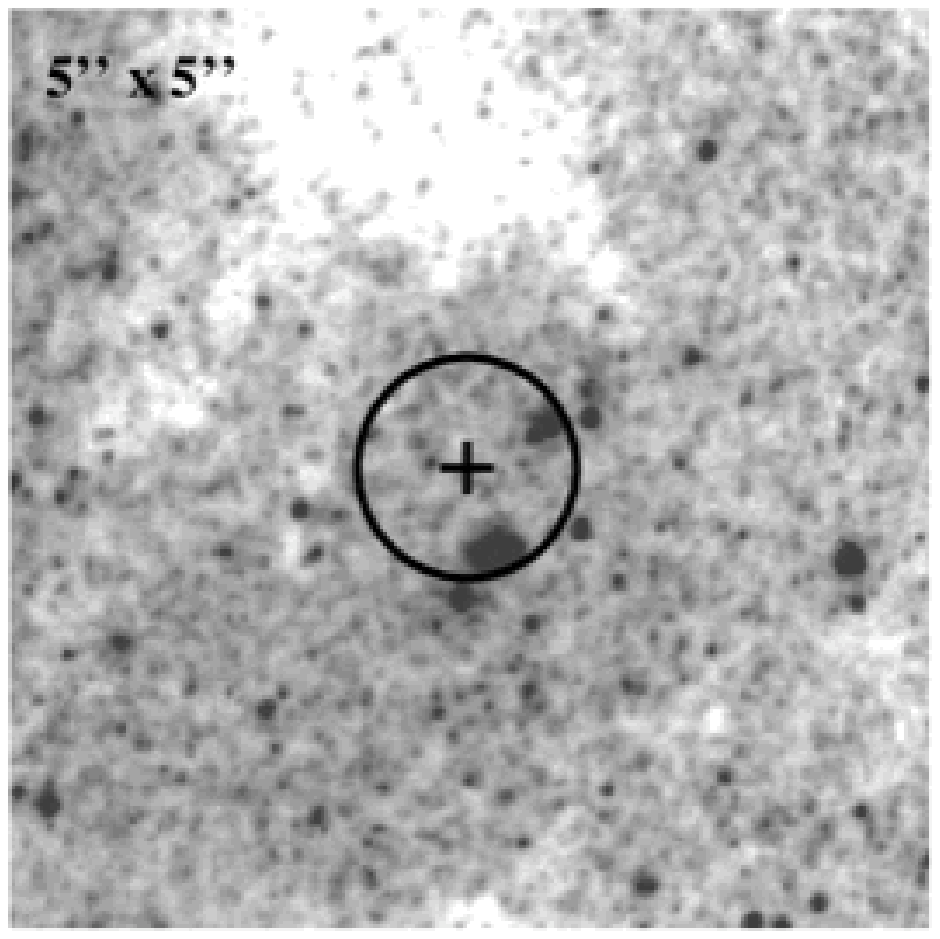} & 
\includegraphics[width=0.23\linewidth,clip=true,trim=0.5cm 3cm 0.5cm 2.5cm]{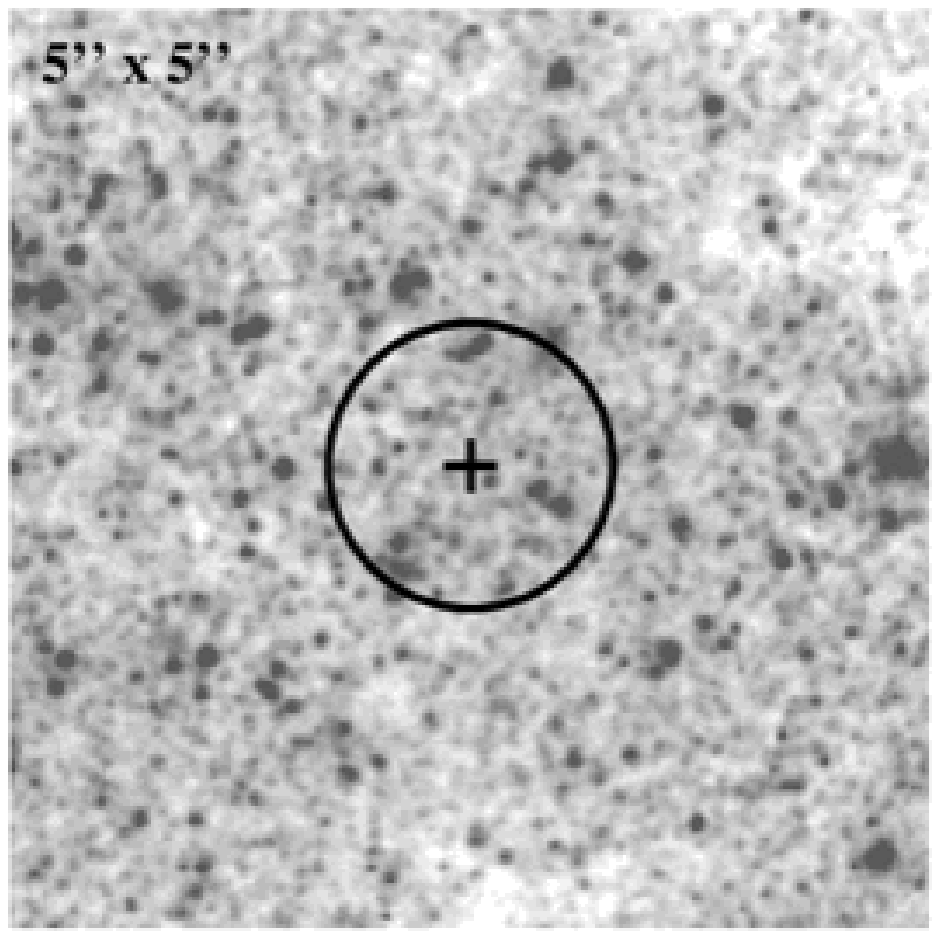} \\ 

Source 48 & Source 49 & Source 50 & Source 51 \\ 
\includegraphics[width=0.23\linewidth,clip=true,trim=0.5cm 3cm 0.5cm 2.5cm]{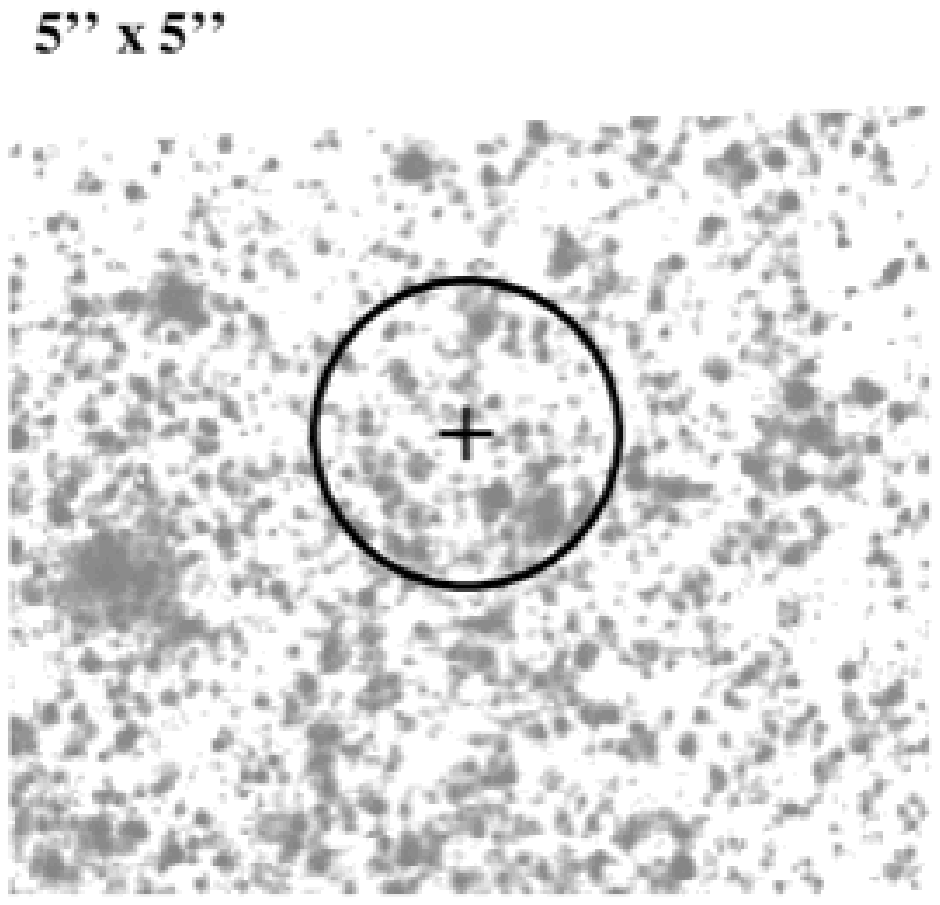} & 
\includegraphics[width=0.23\linewidth,clip=true,trim=0.5cm 3cm 0.5cm 2.5cm]{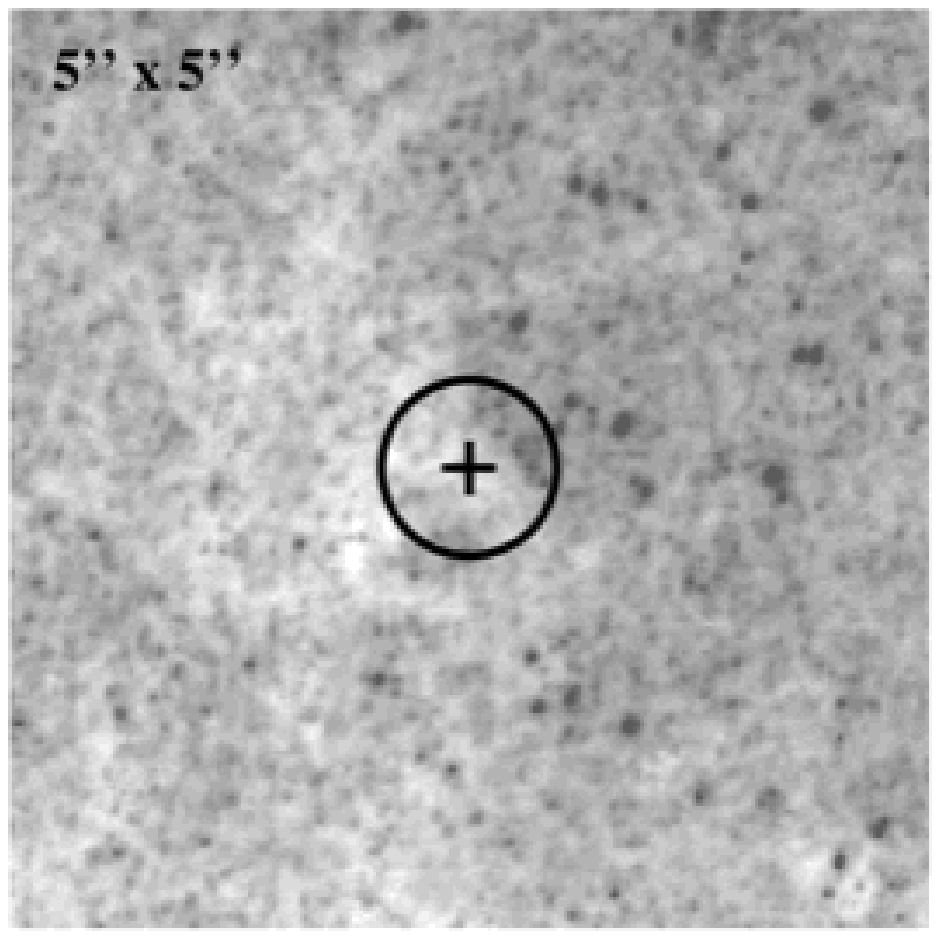} & 
\includegraphics[width=0.23\linewidth,clip=true,trim=0.5cm 3cm 0.5cm 2.5cm]{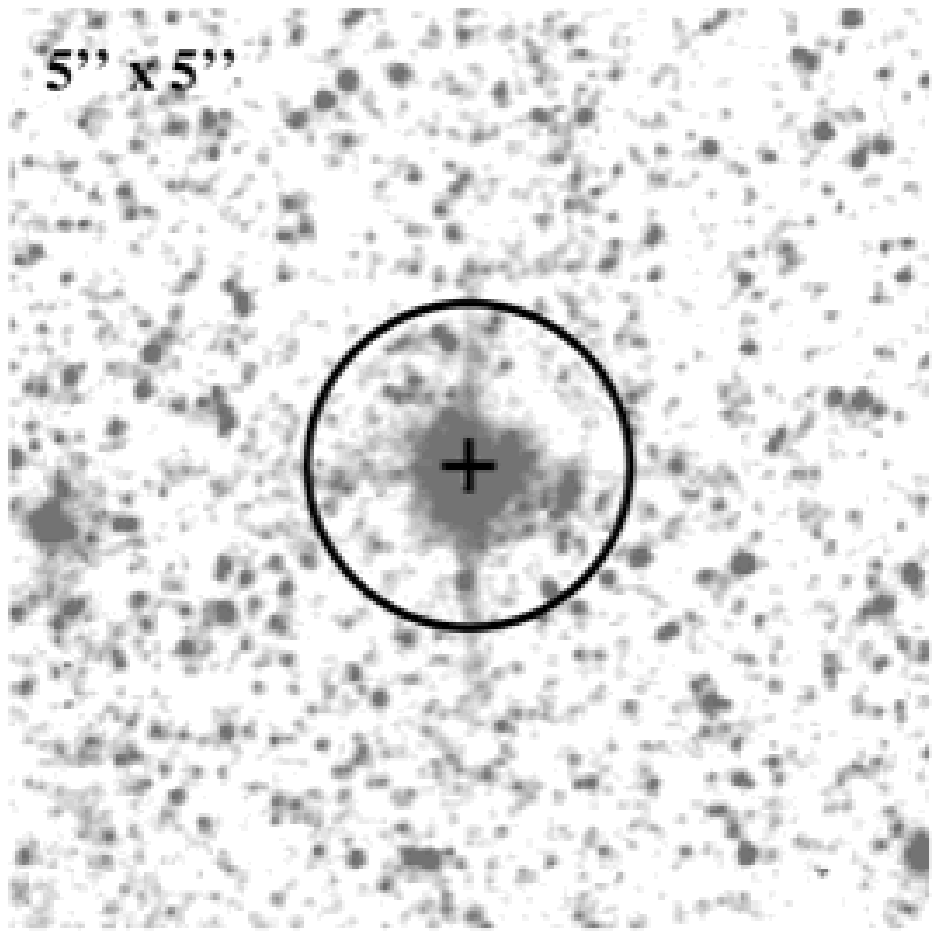} & 
\includegraphics[width=0.23\linewidth,clip=true,trim=0.5cm 3cm 0.5cm 2.5cm]{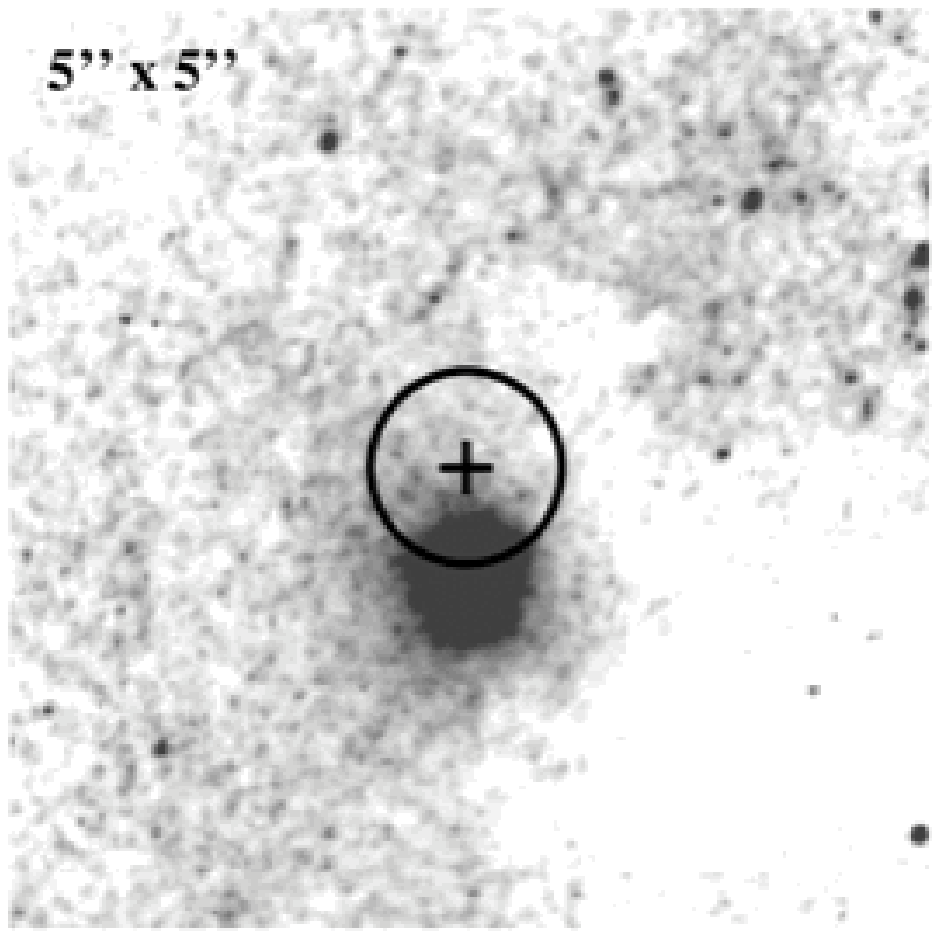} \\ 
\end{tabular}
\caption{Optical \HST images for X-ray sources detected in NGC~2403. The box size (5\asn$\times$ 5\asn, 10\asn$\times$ 10\asn, or 15\asn$\times$ 15\asn) is given in the top-left corner of each image. The circle shows the \Chandra 90\% error circle, centered on the source position.} 
\label{optical_2403}
\end{figure*}

\setcounter{figure}{12}
\begin{figure*}
\centering
\begin{tabular}{cccc}
Source 53 & Source 54 & Source 55 & Source 56 \\ 
\includegraphics[width=0.23\linewidth,clip=true,trim=0.5cm 3cm 0.5cm 2.5cm]{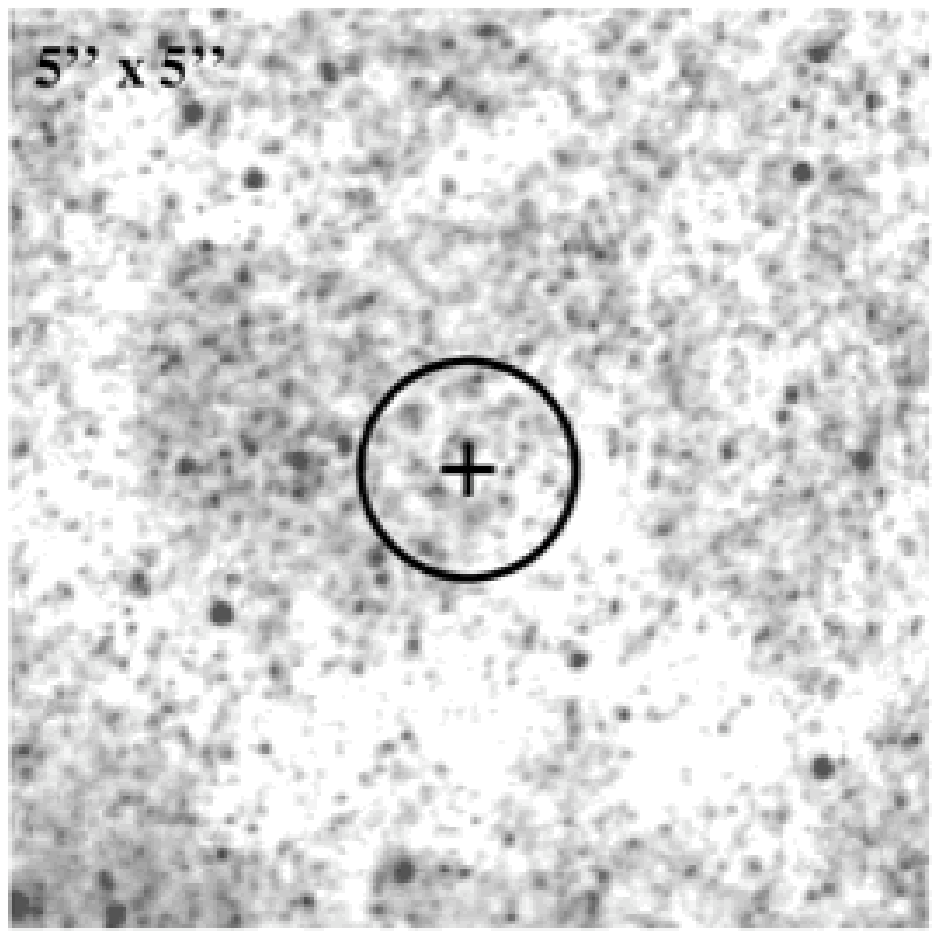} & 
\includegraphics[width=0.23\linewidth,clip=true,trim=0.5cm 3cm 0.5cm 2.5cm]{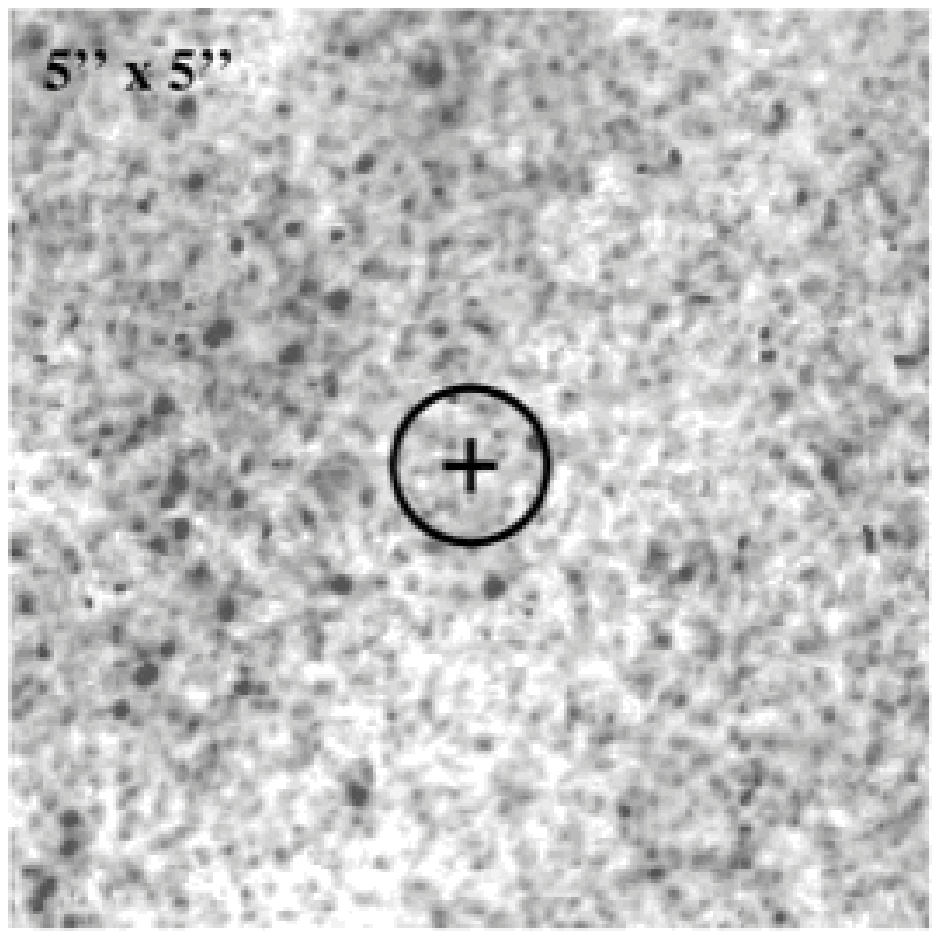} & 
\includegraphics[width=0.23\linewidth,clip=true,trim=0.5cm 3cm 0.5cm 2.5cm]{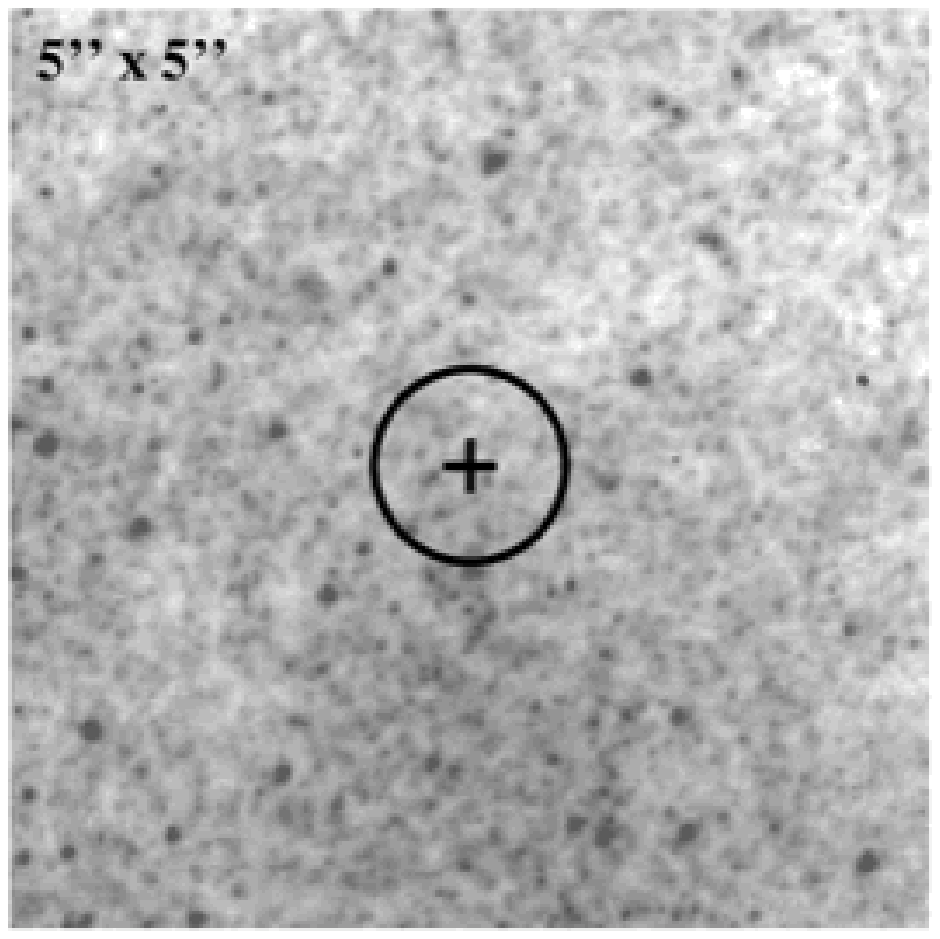} & 
\includegraphics[width=0.23\linewidth,clip=true,trim=0.5cm 3cm 0.5cm 2.5cm]{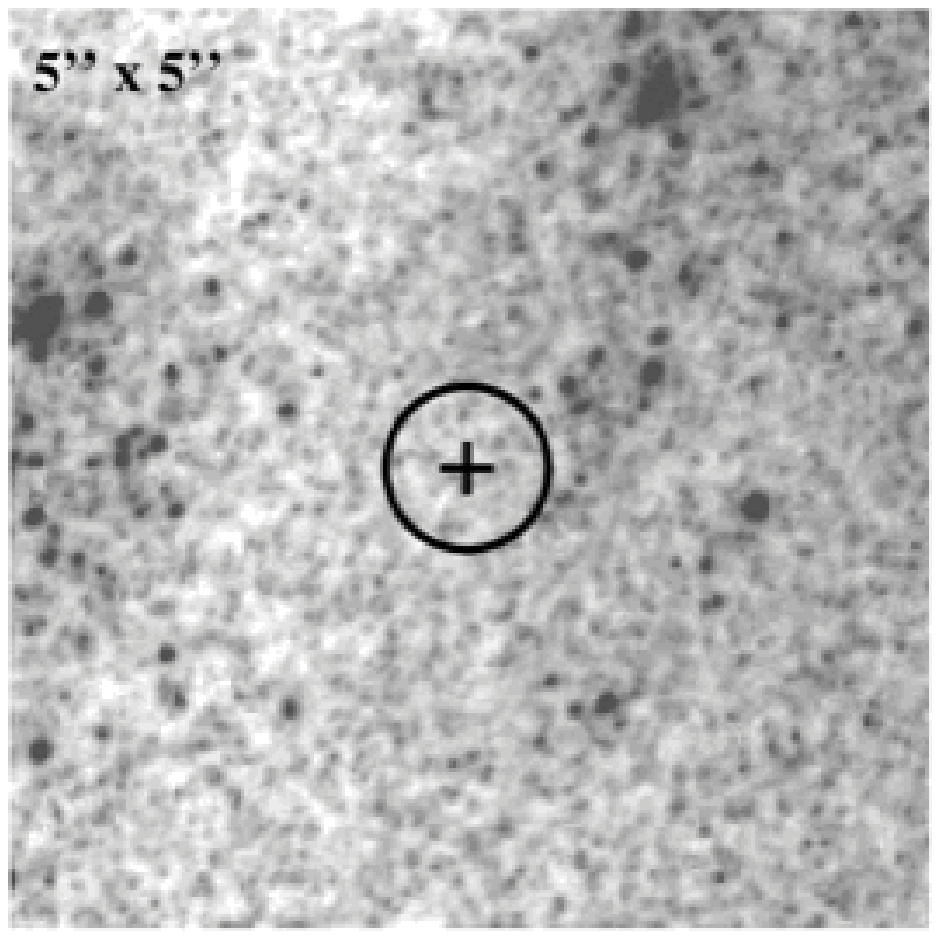} \\ 

Source 57 & Source 59 & Source 60 & Source 61 \\ 
\includegraphics[width=0.23\linewidth,clip=true,trim=0.5cm 3cm 0.5cm 2.5cm]{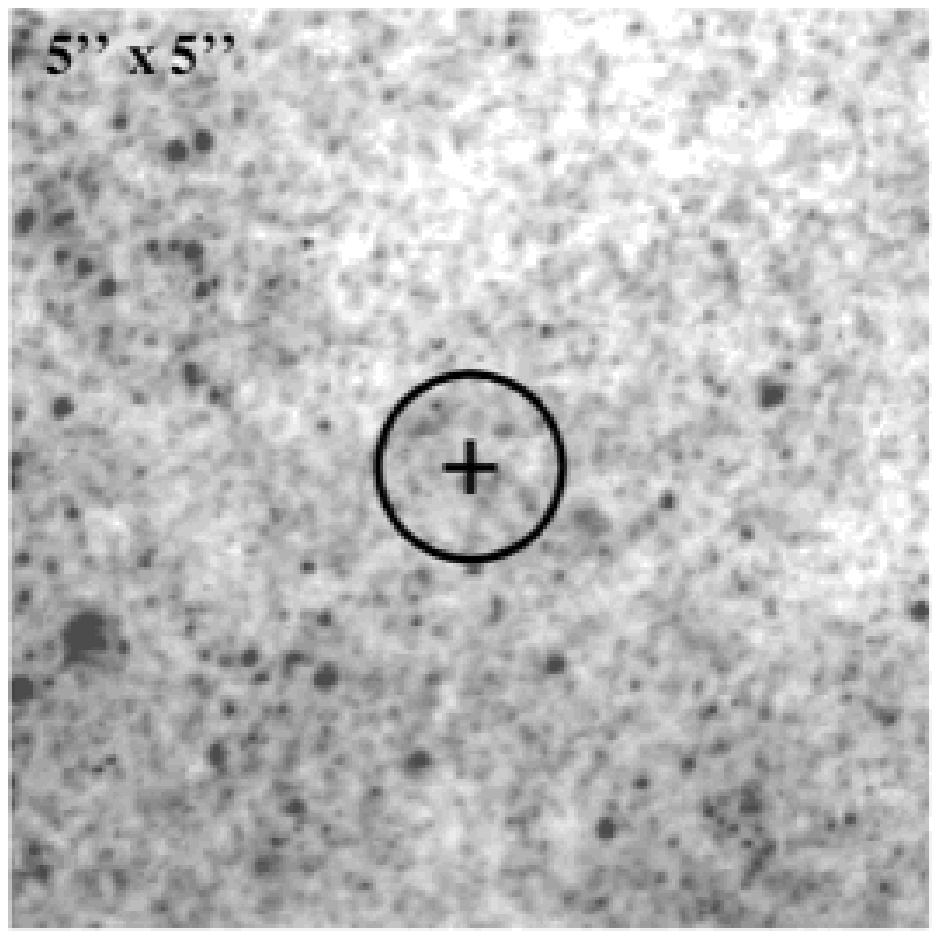} & 
\includegraphics[width=0.23\linewidth,clip=true,trim=0.5cm 3cm 0.5cm 2.5cm]{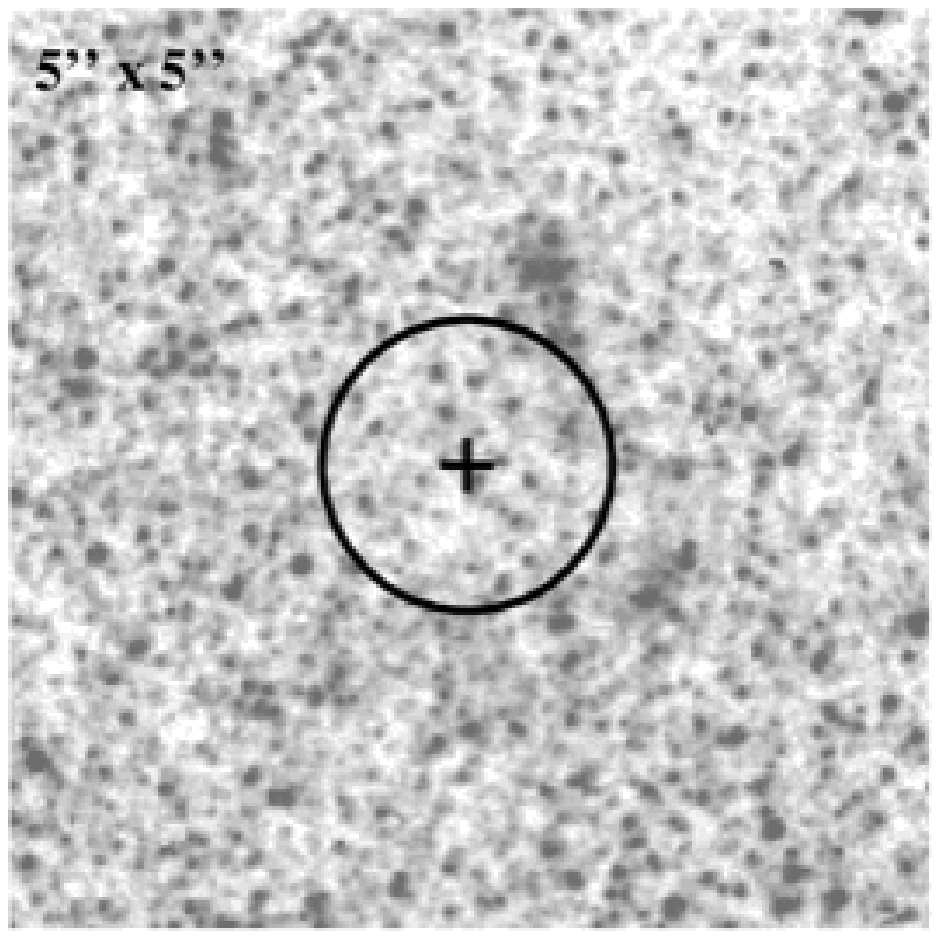} & 
\includegraphics[width=0.23\linewidth,clip=true,trim=0.5cm 3cm 0.5cm 2.5cm]{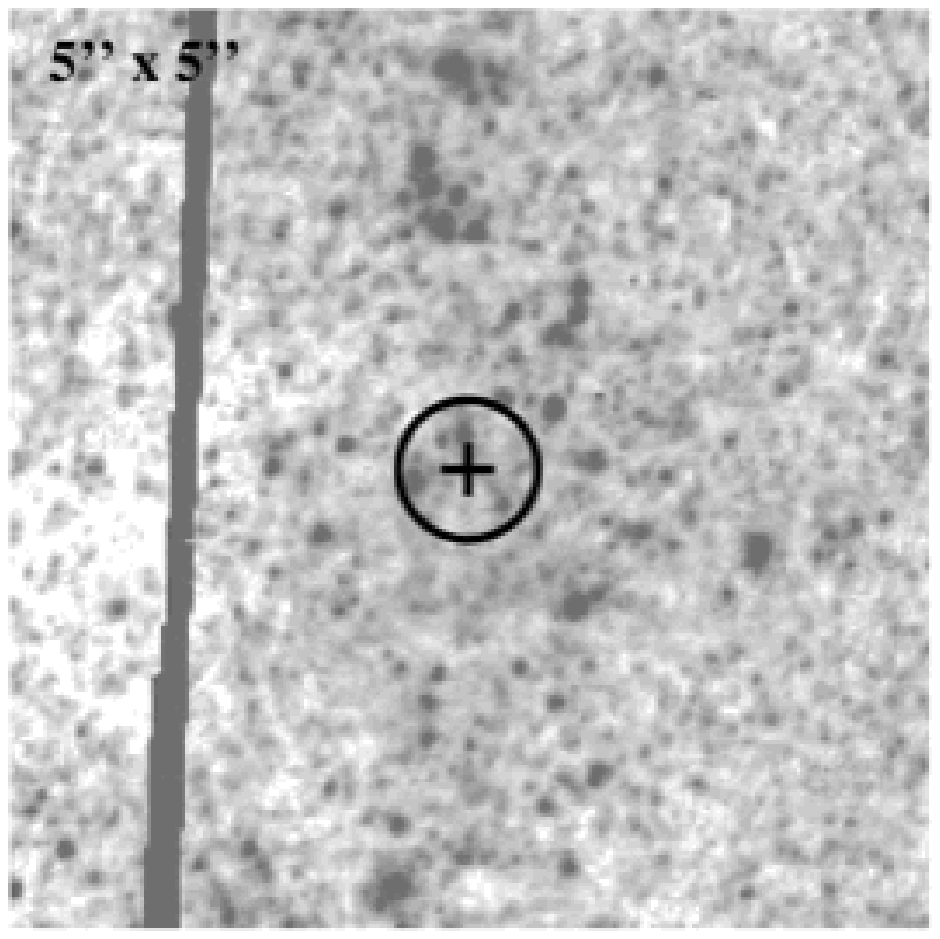} & 
\includegraphics[width=0.23\linewidth,clip=true,trim=0.5cm 3cm 0.5cm 2.5cm]{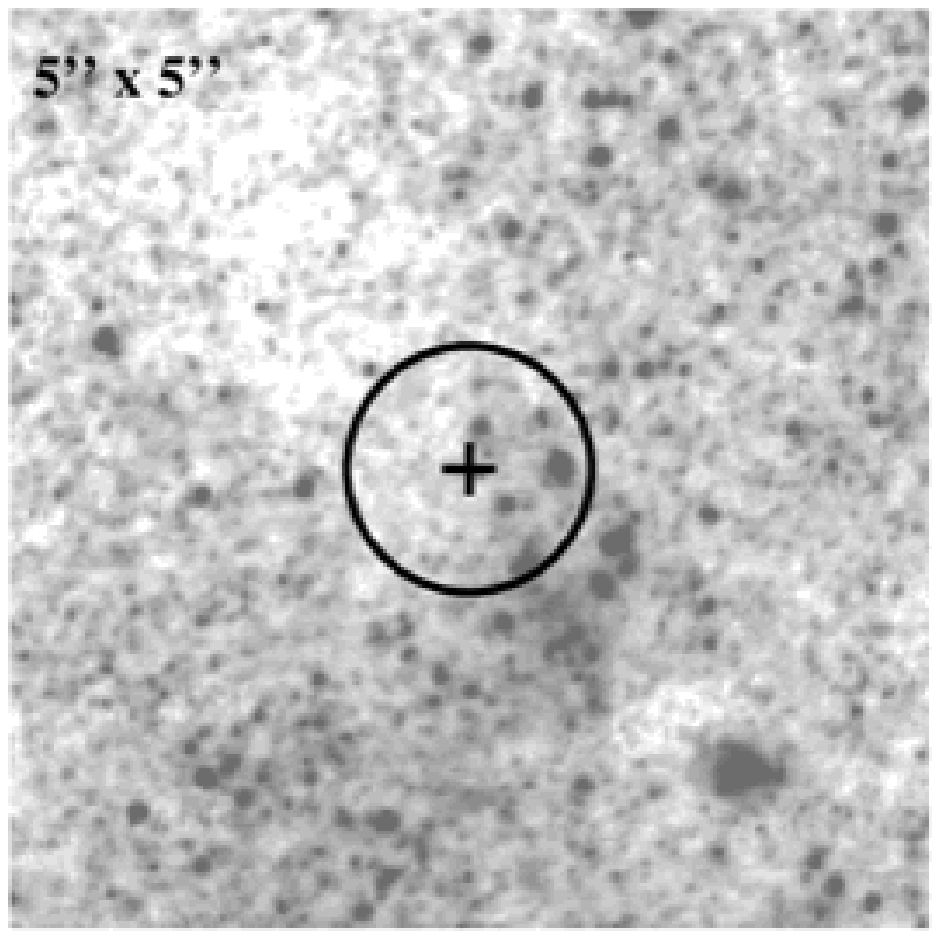} \\ 

Soure 63 & Source 64 & Source 67 & Source 69 \\ 
\includegraphics[width=0.23\linewidth,clip=true,trim=0.5cm 3cm 0.5cm 2.5cm]{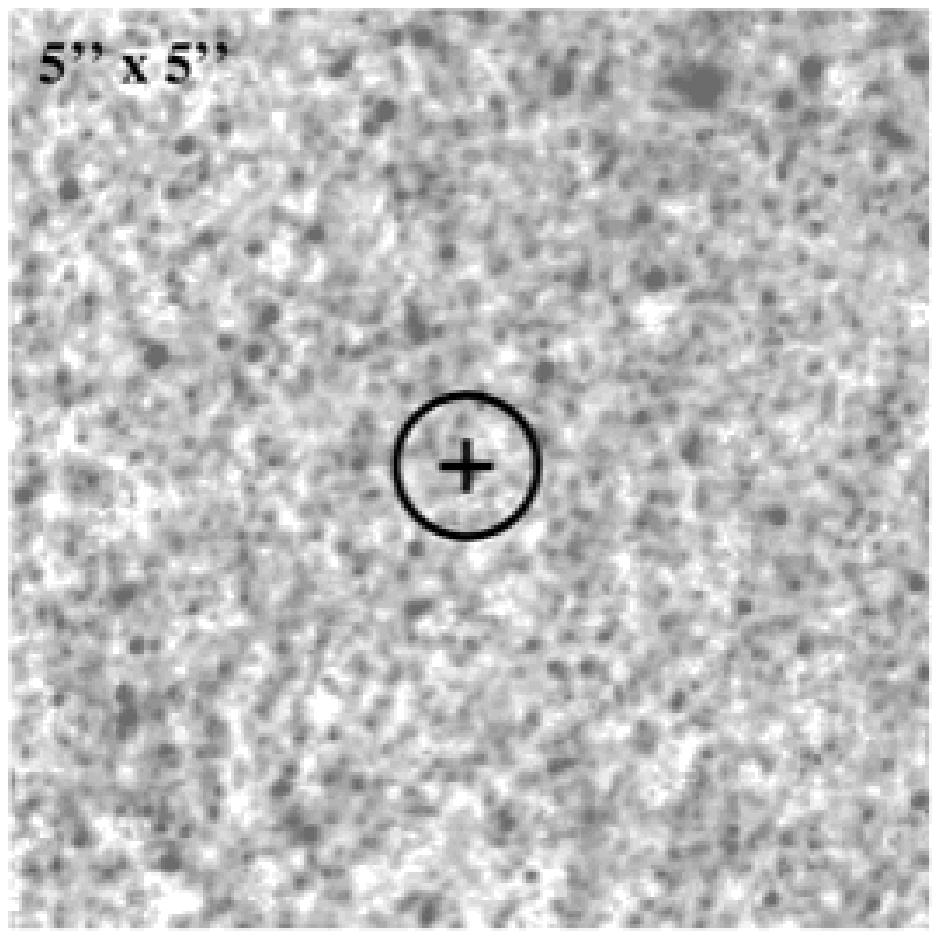} & 
\includegraphics[width=0.23\linewidth,clip=true,trim=0.5cm 3cm 0.5cm 2.5cm]{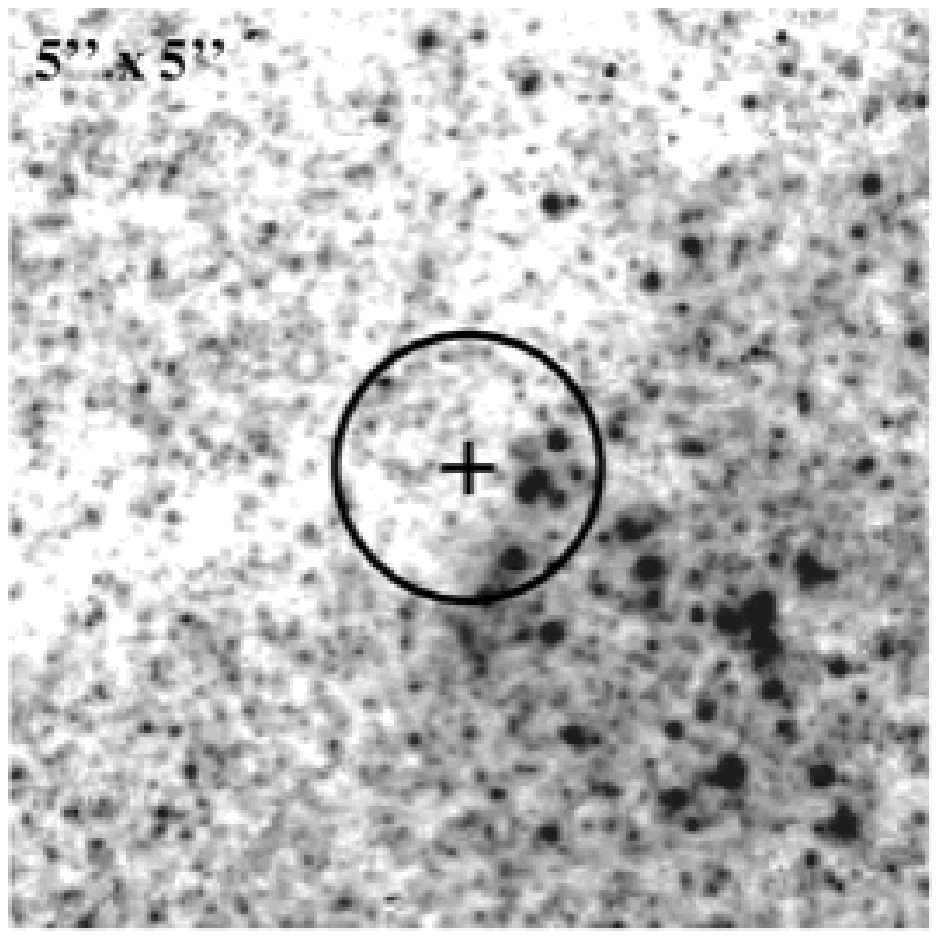} & 
\includegraphics[width=0.23\linewidth,clip=true,trim=0.5cm 3cm 0.5cm 2.5cm]{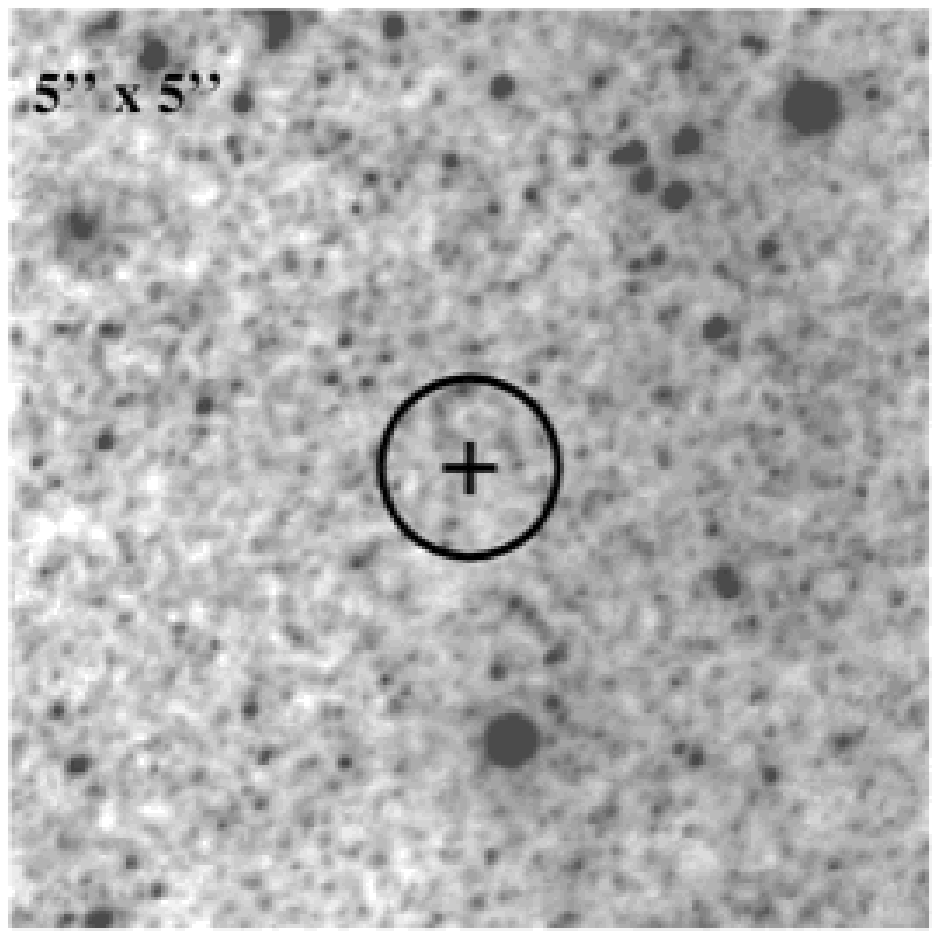} & 
\includegraphics[width=0.23\linewidth,clip=true,trim=0.5cm 3cm 0.5cm 2.5cm]{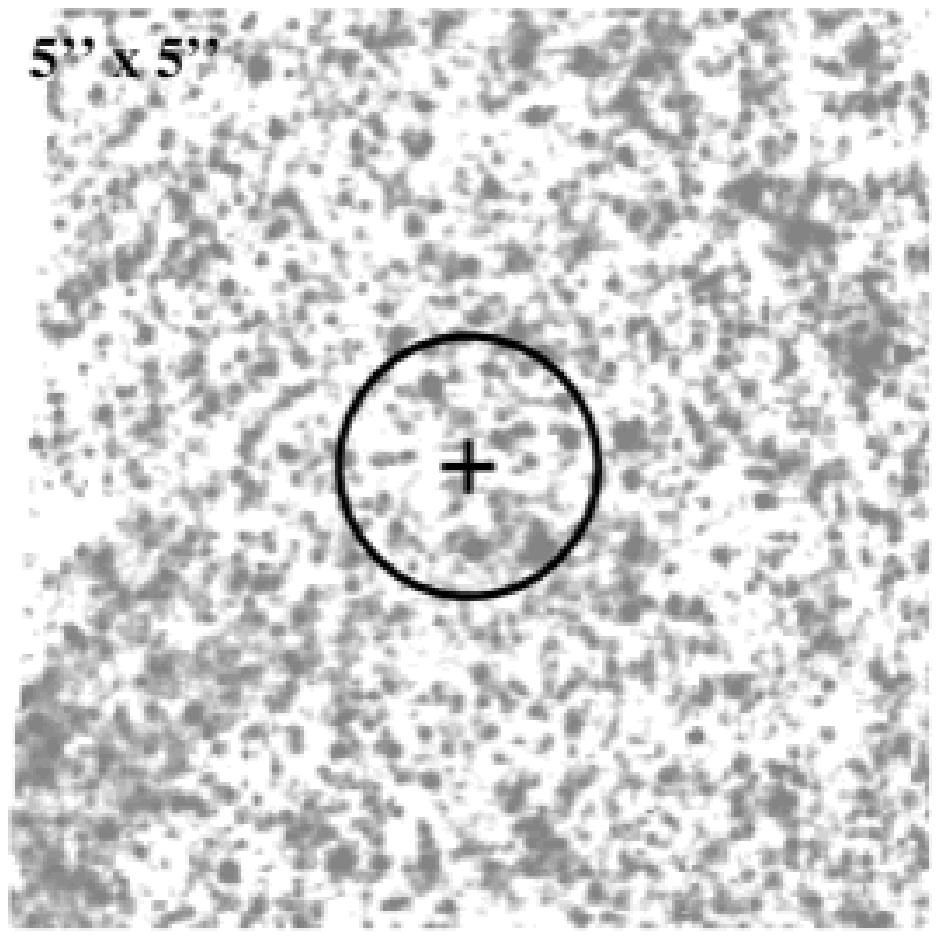} \\ 

Source 70 & Source 75 & Source 81 & Source 84 \\ 
\includegraphics[width=0.23\linewidth,clip=true,trim=0.5cm 3cm 0.5cm 2.5cm]{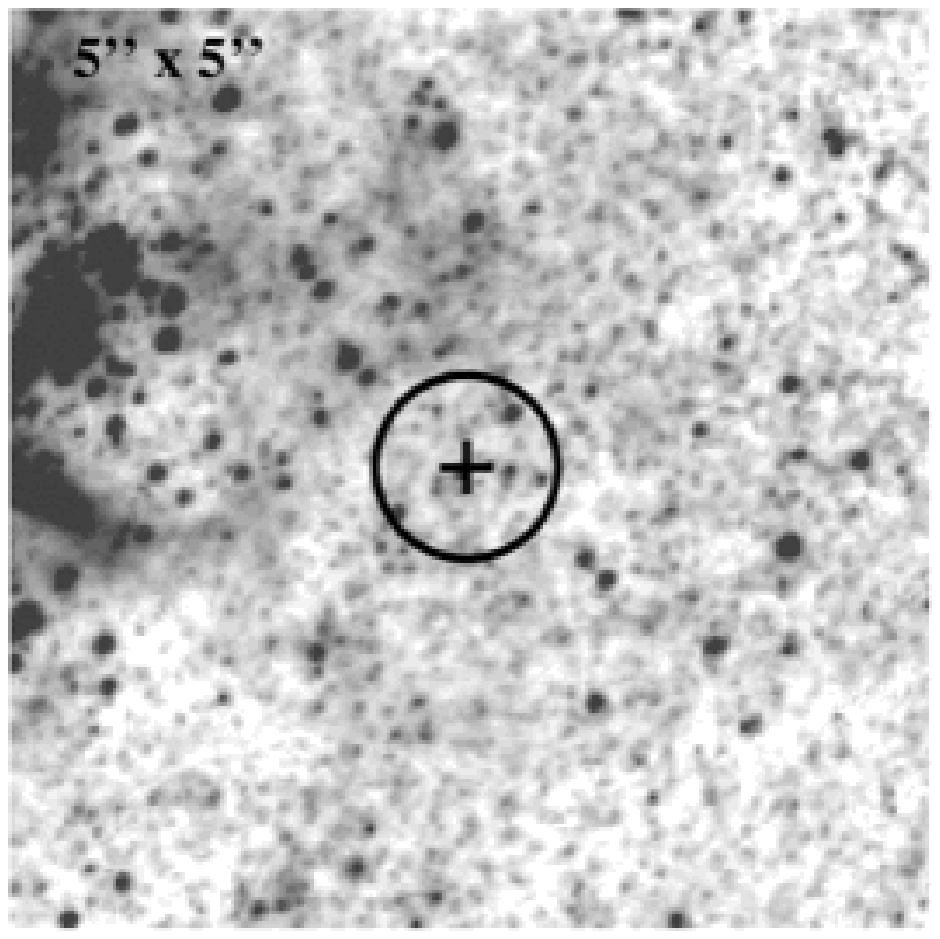} & 
\includegraphics[width=0.23\linewidth,clip=true,trim=0.5cm 3cm 0.5cm 2.5cm]{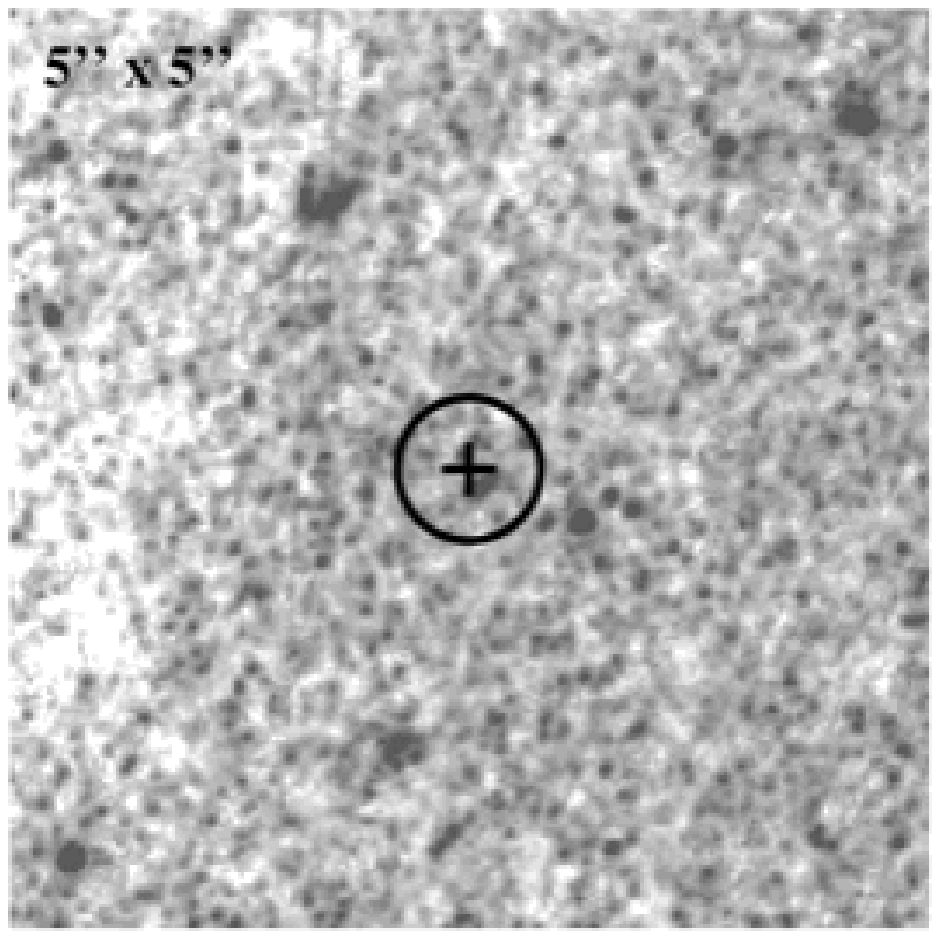} & 
\includegraphics[width=0.23\linewidth,clip=true,trim=0.5cm 3cm 0.5cm 2.5cm]{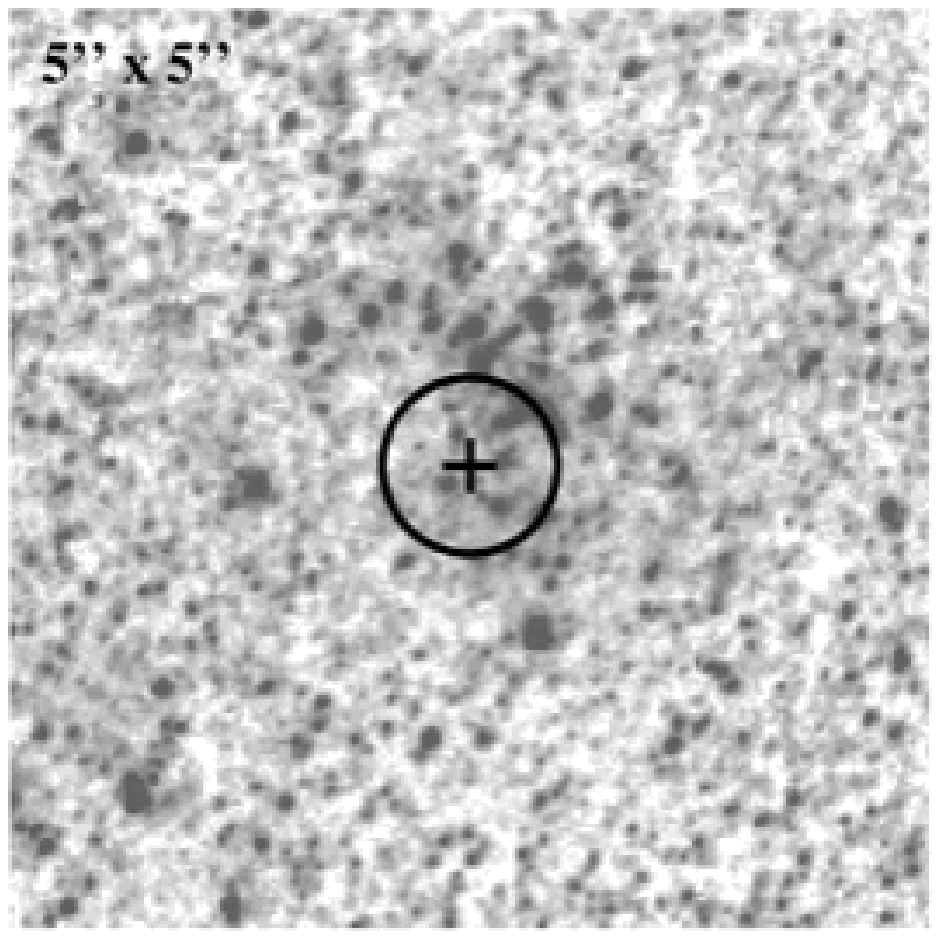} & 
\includegraphics[width=0.23\linewidth,clip=true,trim=0.5cm 3cm 0.5cm 2.5cm]{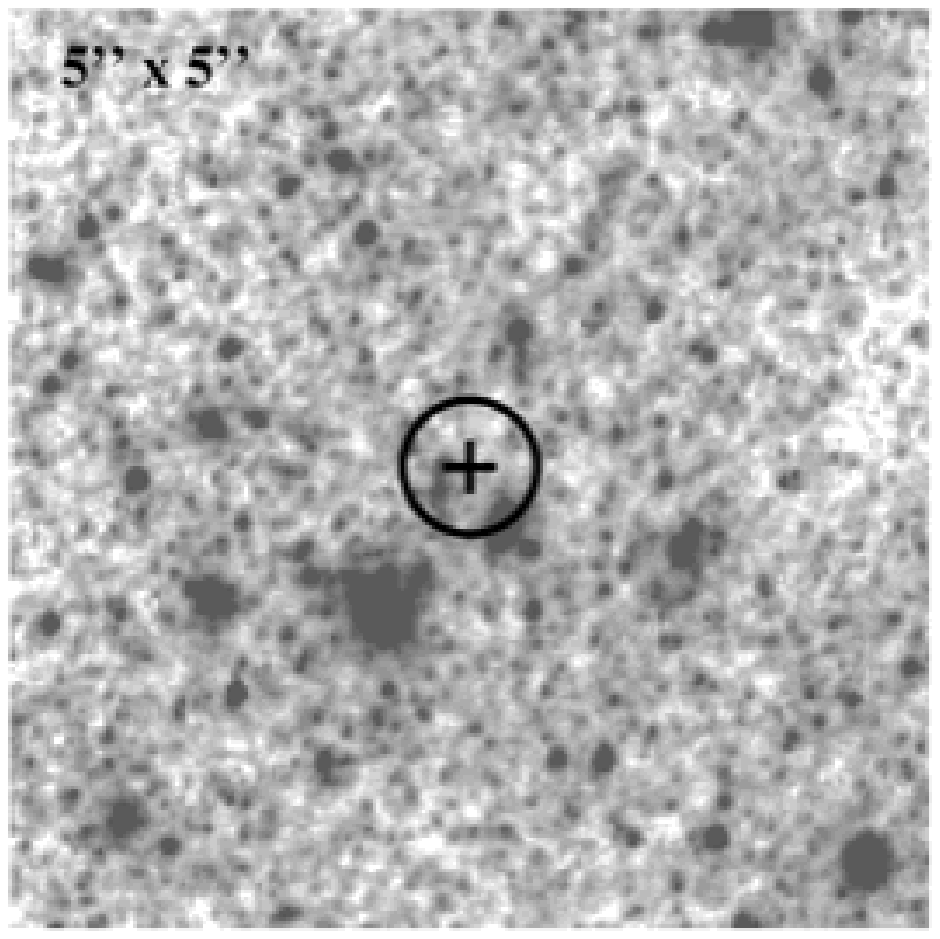} \\ 

Source 85 & Source 91 & Source 93 & Source 101 \\ 
\includegraphics[width=0.23\linewidth,clip=true,trim=0.5cm 3cm 0.5cm 2.5cm]{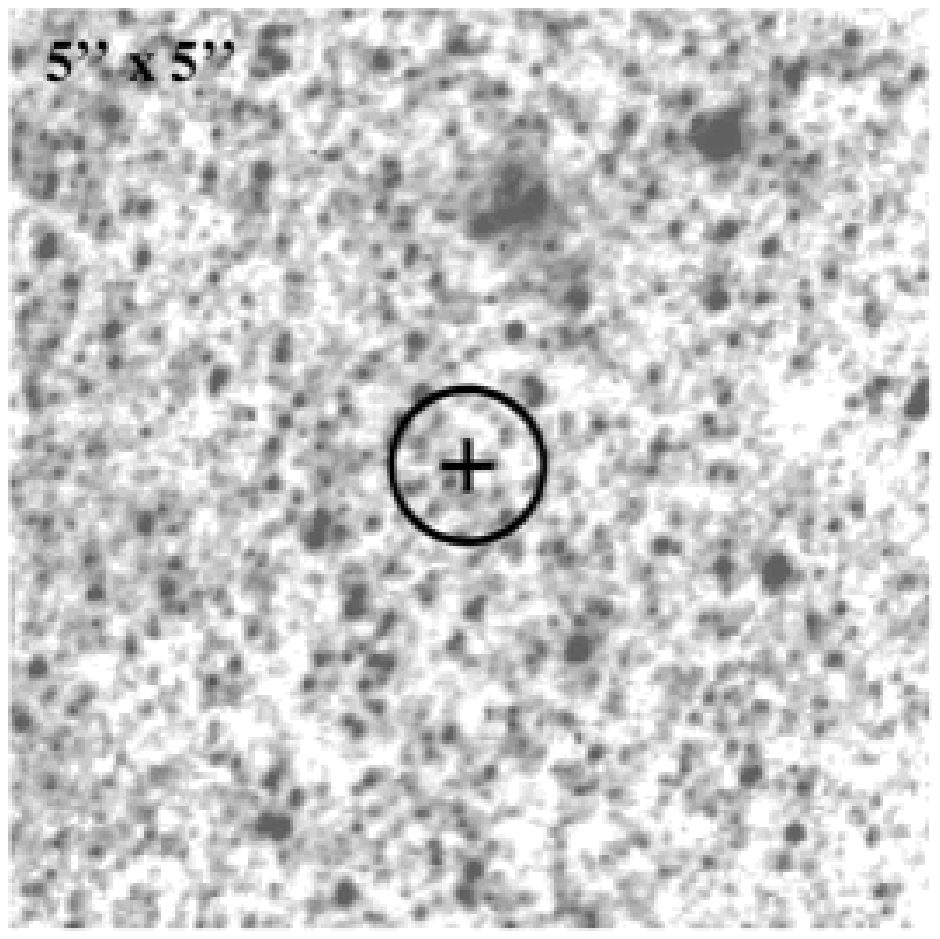} & 
\includegraphics[width=0.23\linewidth,clip=true,trim=0.5cm 3cm 0.5cm 2.5cm]{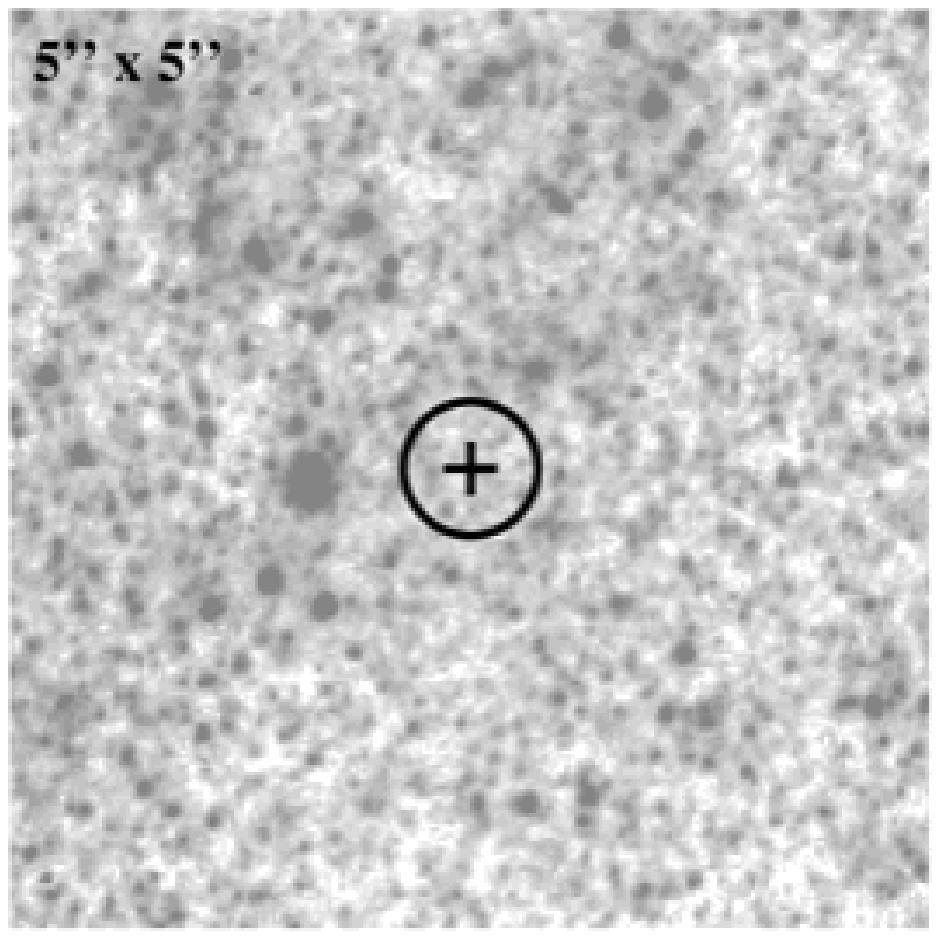} & 
\includegraphics[width=0.23\linewidth,clip=true,trim=0.5cm 3cm 0.5cm 2.5cm]{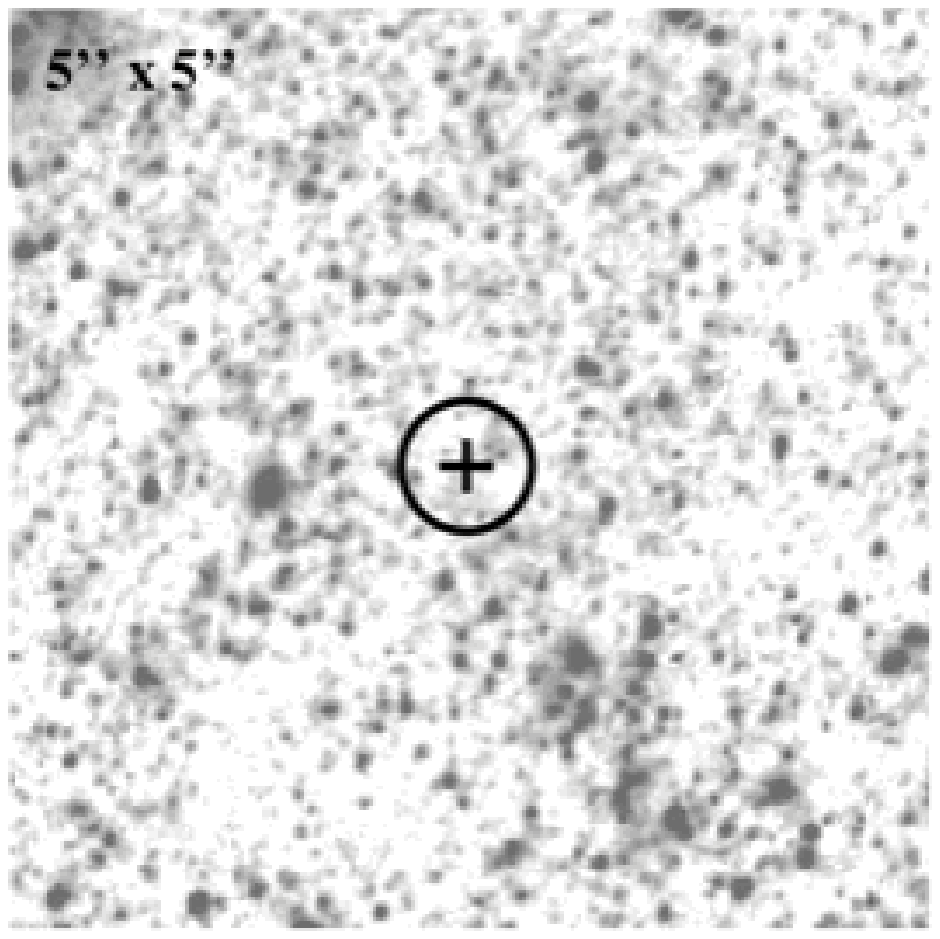} & 
\includegraphics[width=0.23\linewidth,clip=true,trim=0.5cm 3cm 0.5cm 2.5cm]{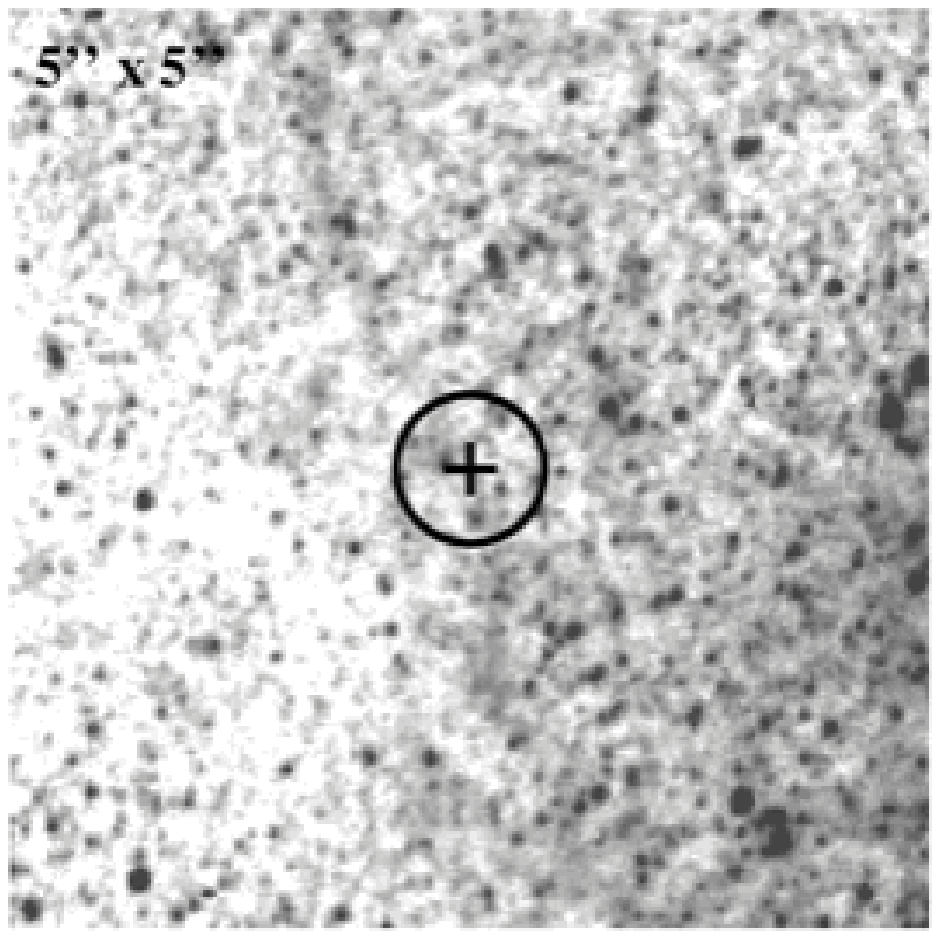} \\
\end{tabular}
\caption{{\it (Continued)} Optical \HST images for X-ray sources detected in NGC~2403. The box size (5\asn$\times$ 5\asn, 10\asn$\times$ 10\asn, or 15\asn$\times$ 15\asn) is given in the top-left corner of each image. The circle shows the \Chandra 90\% error circle, centered on the source position.} 
\label{optical_2403}
\end{figure*}

\setcounter{figure}{12}
\begin{figure*}
\centering
\begin{tabular}{cccc}
Source 107 & Source 108 & Source 109 & Source 111 \\ 
\includegraphics[width=0.23\linewidth,clip=true,trim=0.5cm 3cm 0.5cm 2.5cm]{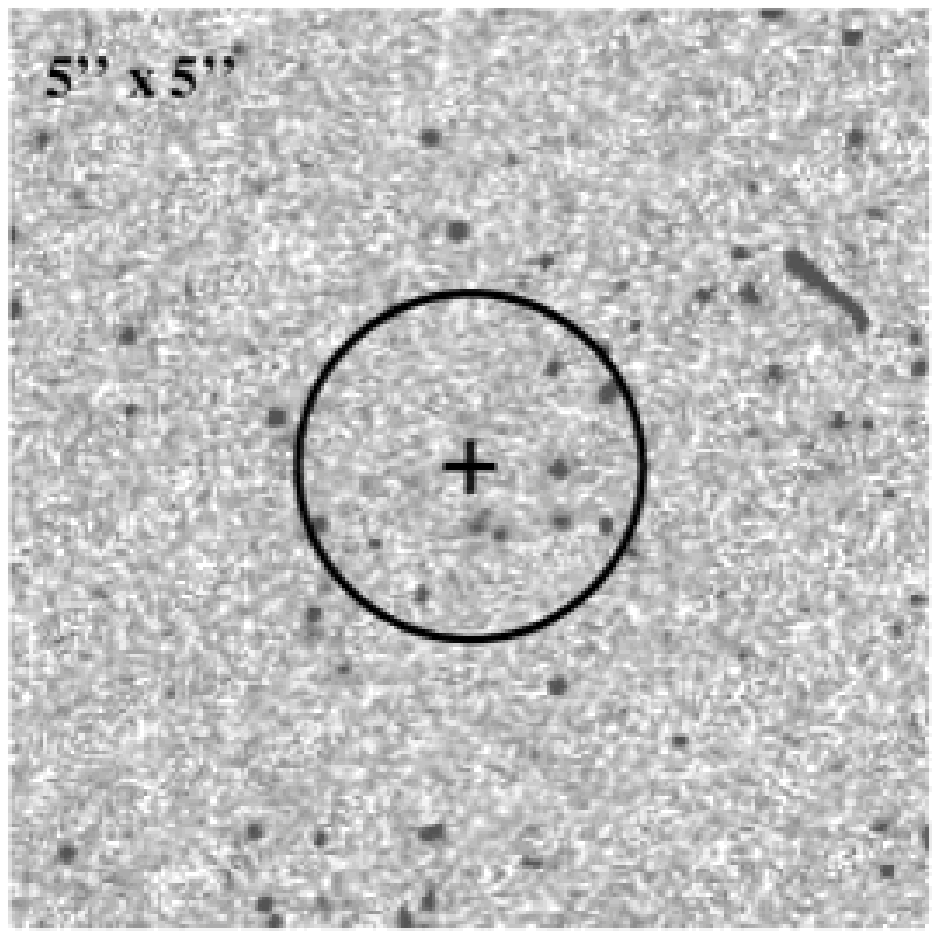} & 
\includegraphics[width=0.23\linewidth,clip=true,trim=0.5cm 3cm 0.5cm 2.5cm]{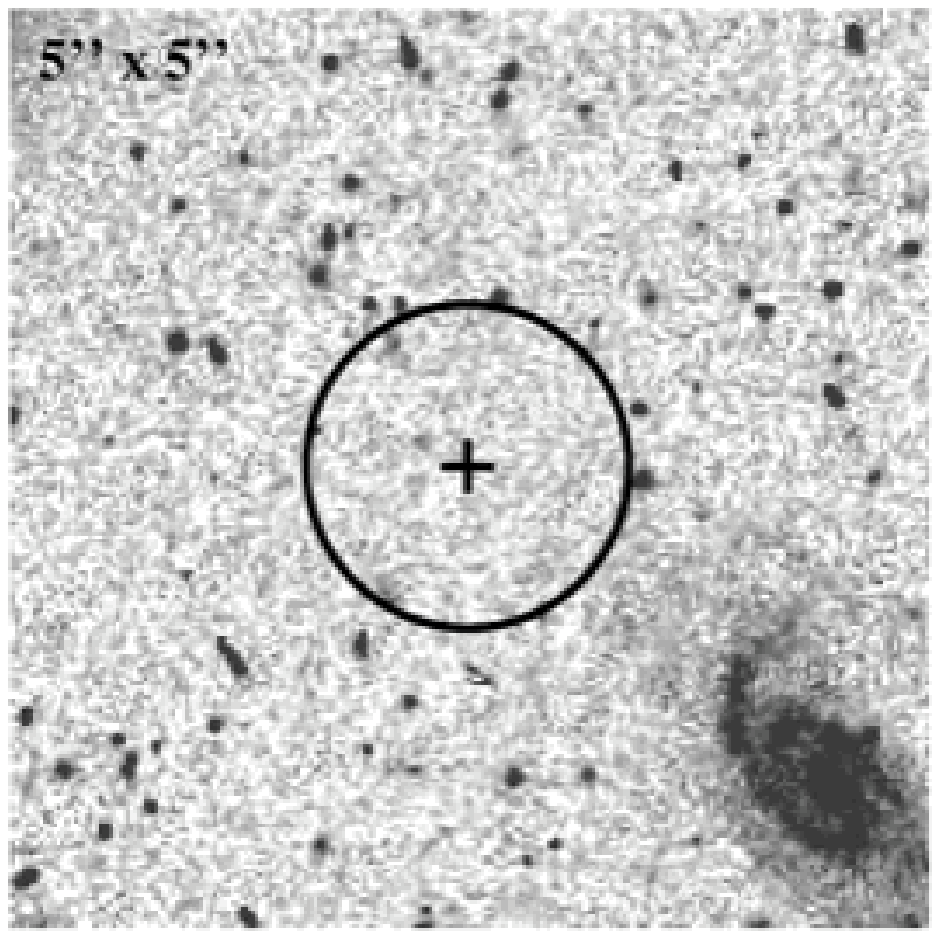} & 
\includegraphics[width=0.23\linewidth,clip=true,trim=0.5cm 3cm 0.5cm 2.5cm]{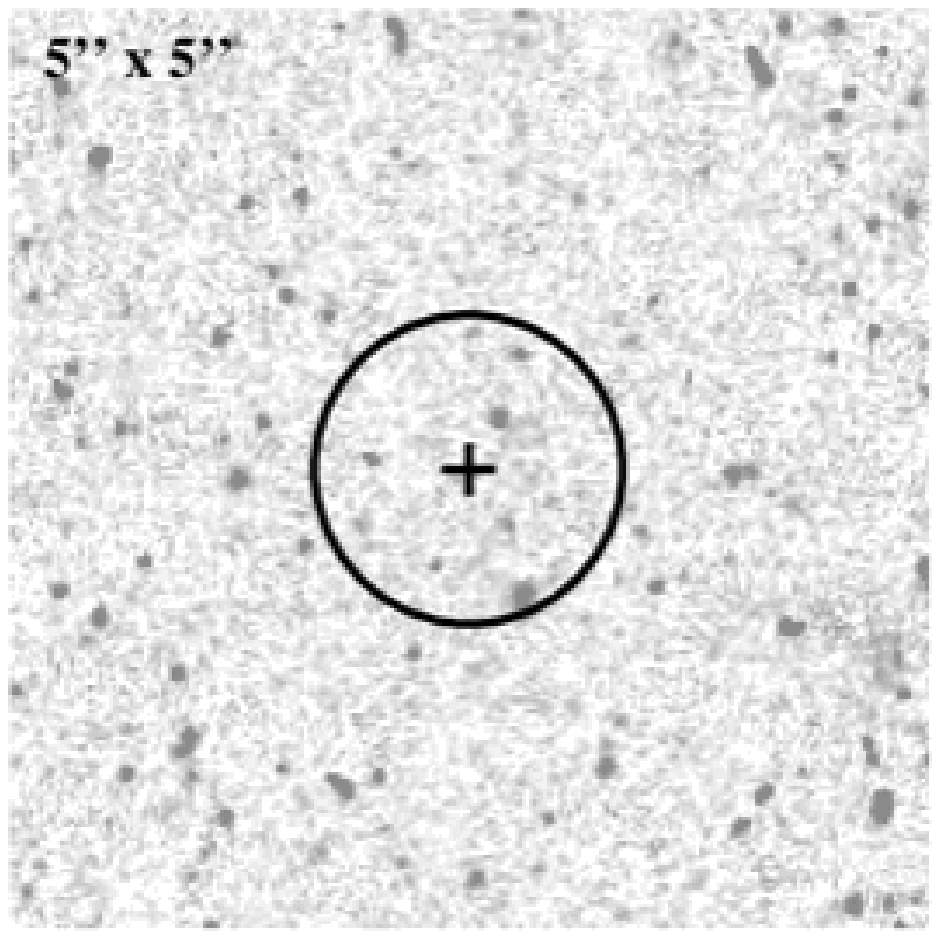} & 
\includegraphics[width=0.23\linewidth,clip=true,trim=0.5cm 3cm 0.5cm 2.5cm]{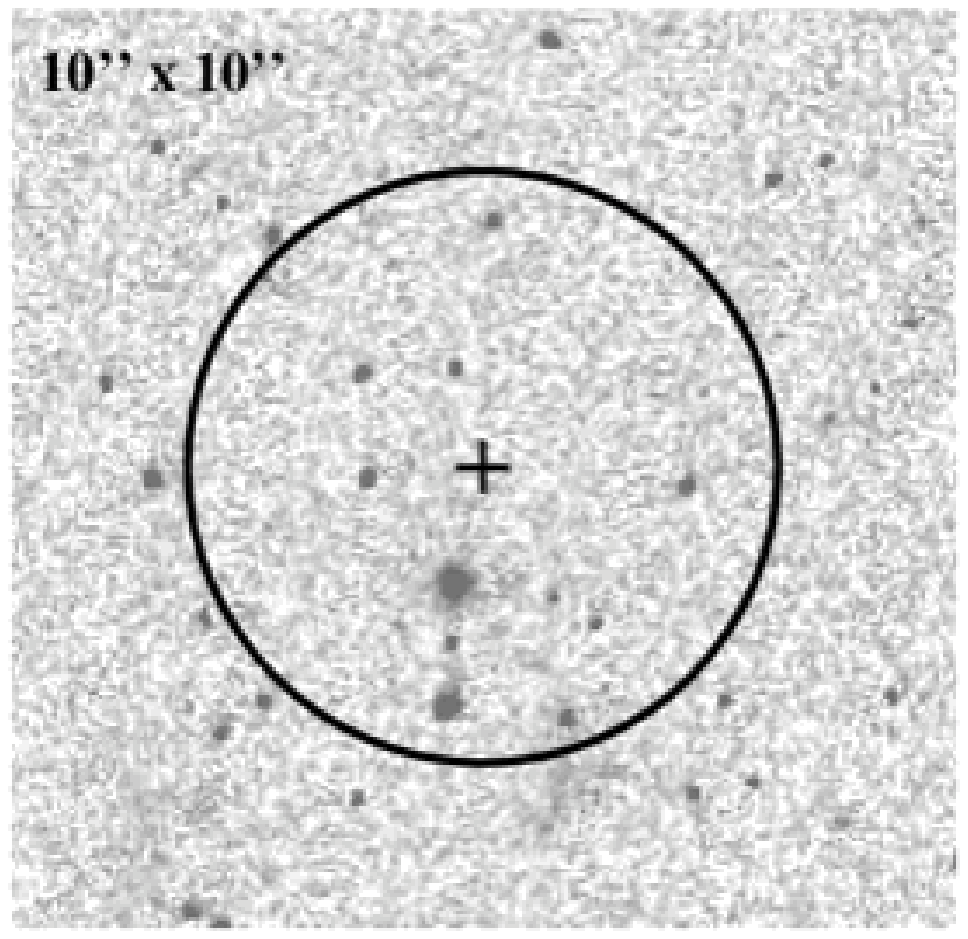} \\ 

Source 114 & Source 115 & Source 116 & Source 119 \\ 
\includegraphics[width=0.23\linewidth,clip=true,trim=0.5cm 3cm 0.5cm 2.5cm]{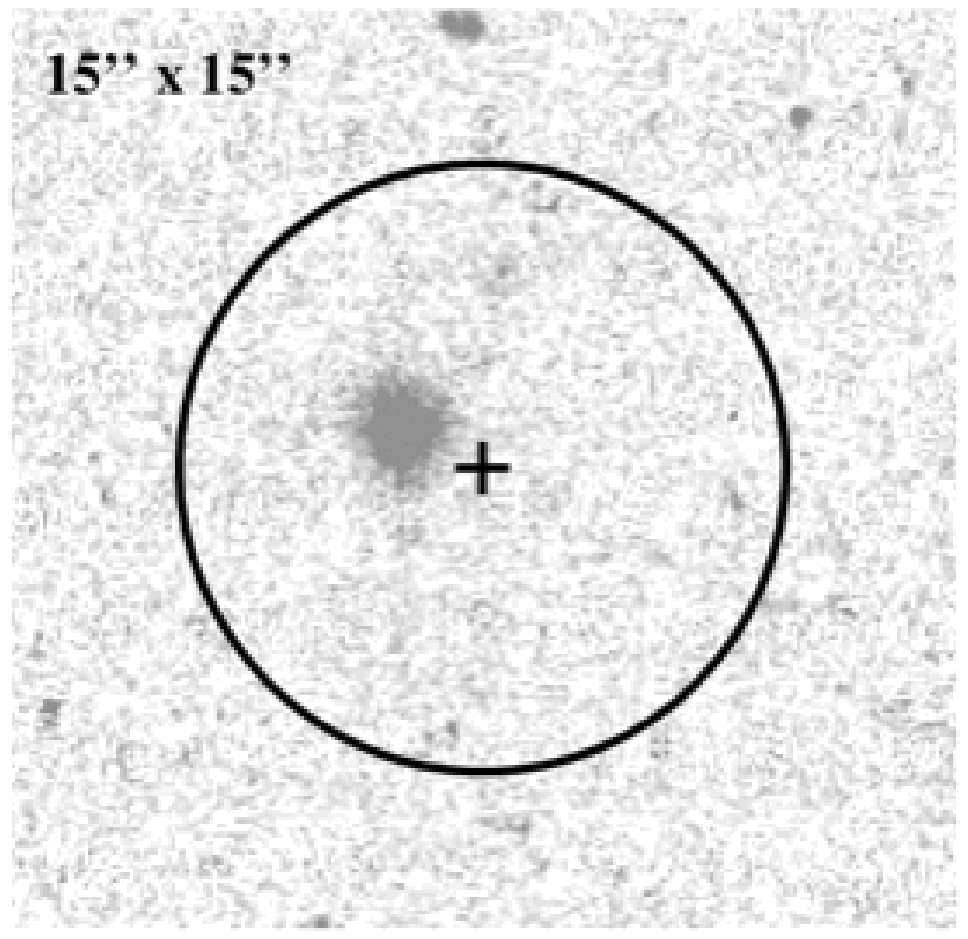} & 
\includegraphics[width=0.23\linewidth,clip=true,trim=0.5cm 3cm 0.5cm 2.5cm]{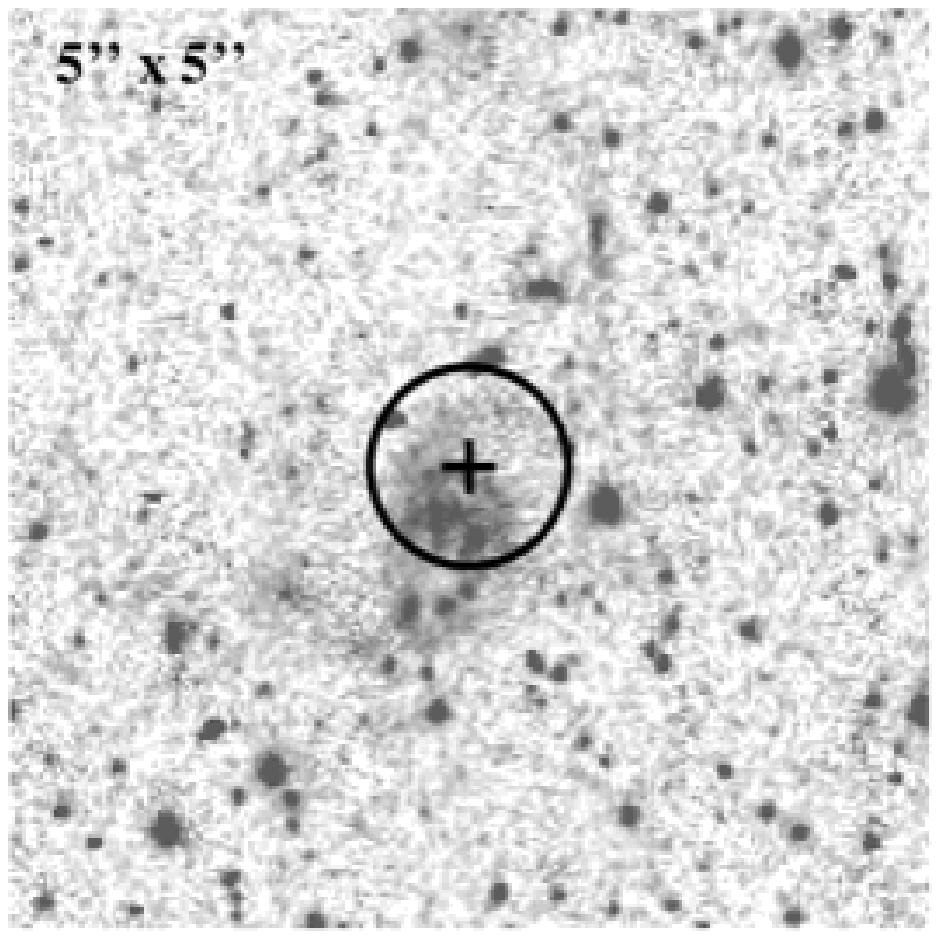} & 
\includegraphics[width=0.23\linewidth,clip=true,trim=0.5cm 3cm 0.5cm 2.5cm]{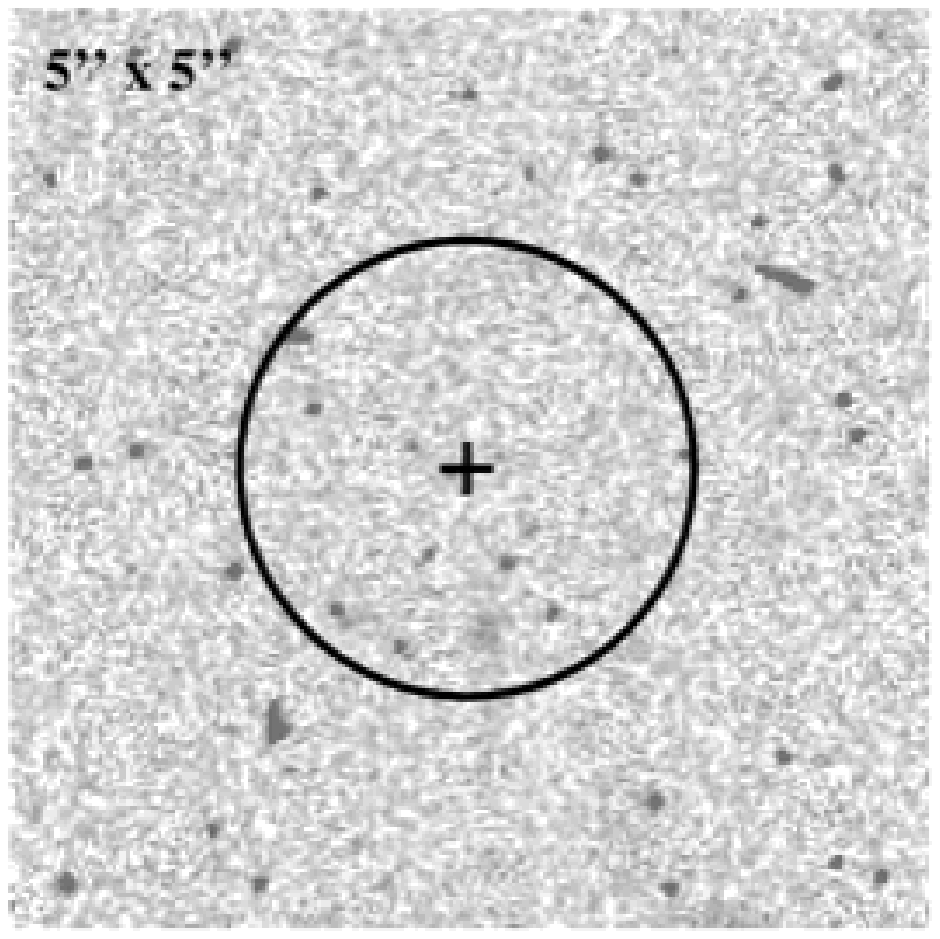} & 
\includegraphics[width=0.23\linewidth,clip=true,trim=0.5cm 3cm 0.5cm 2.5cm]{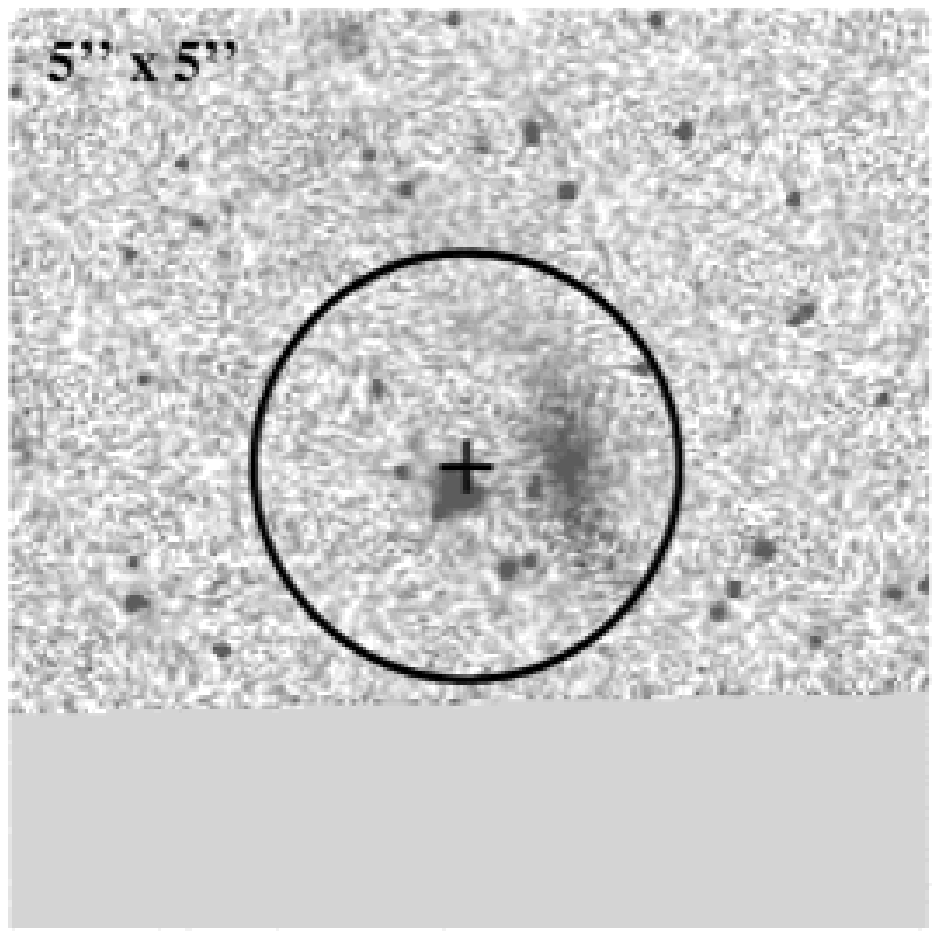} \\ 

Source 121 & Source 125 & Source 129 & Source 132 \\ 
\includegraphics[width=0.23\linewidth,clip=true,trim=0.5cm 3cm 0.5cm 2.5cm]{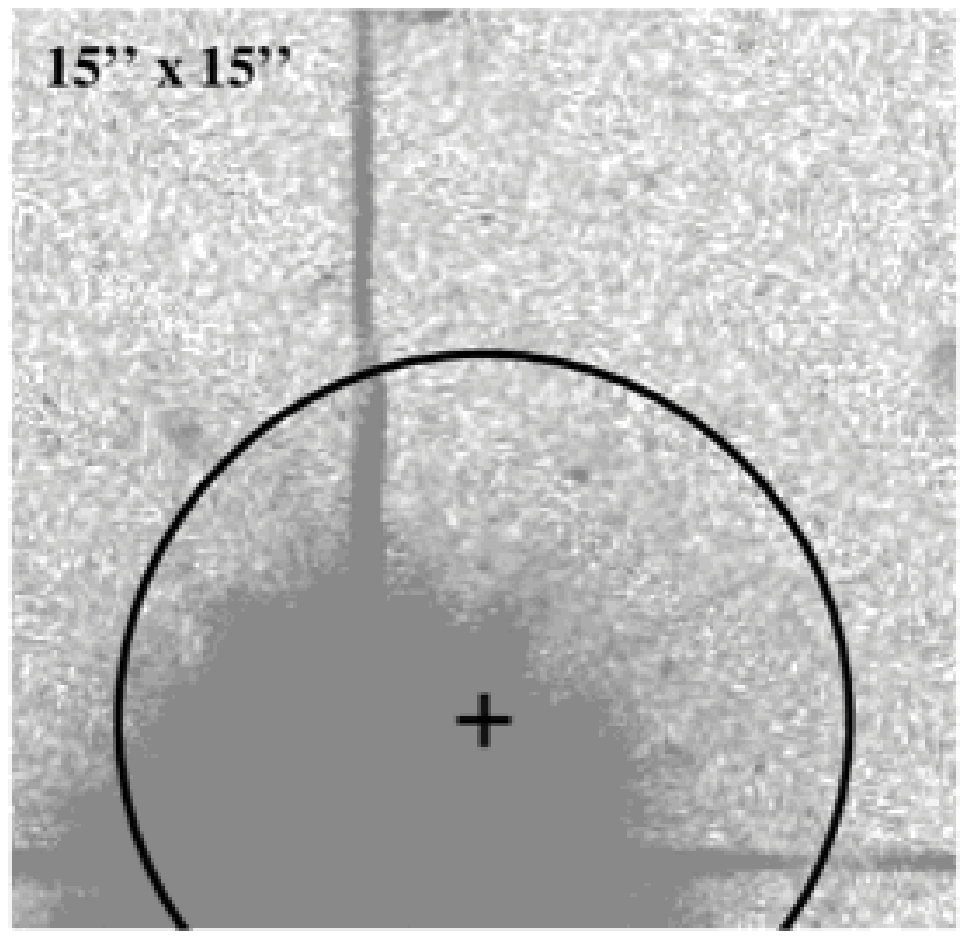} & 
\includegraphics[width=0.23\linewidth,clip=true,trim=0.5cm 3cm 0.5cm 2.5cm]{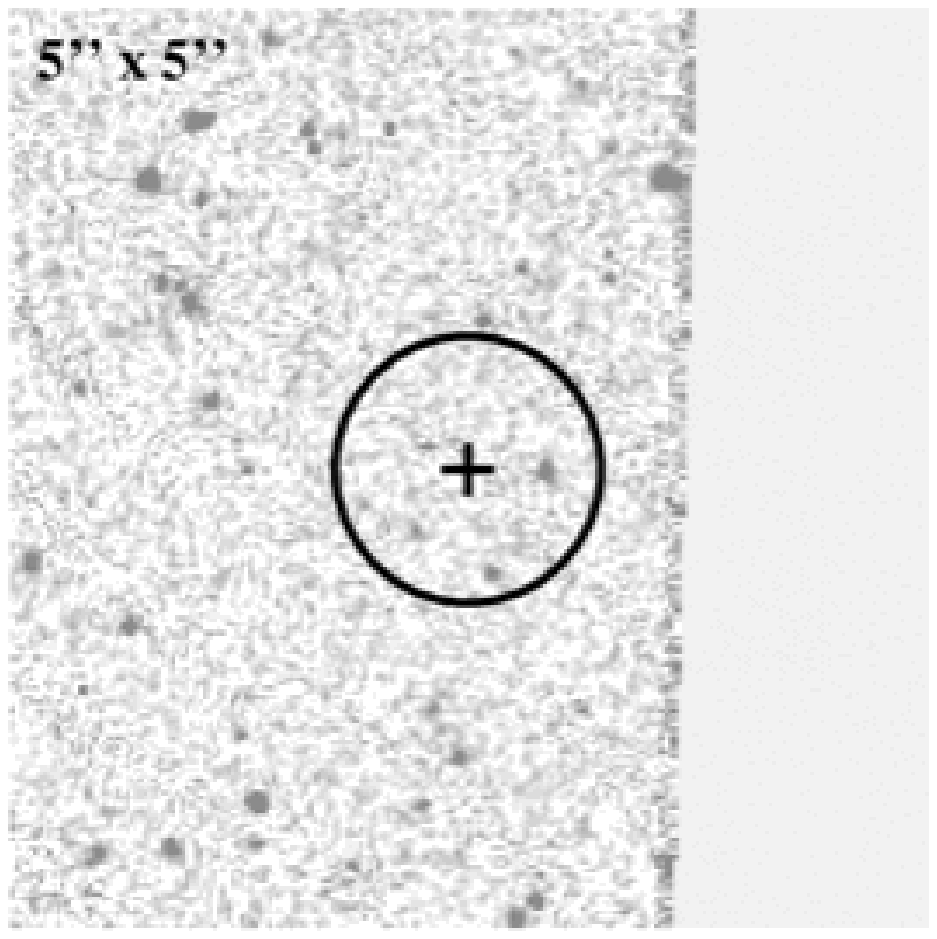} & 
\includegraphics[width=0.23\linewidth,clip=true,trim=0.5cm 3cm 0.5cm 2.5cm]{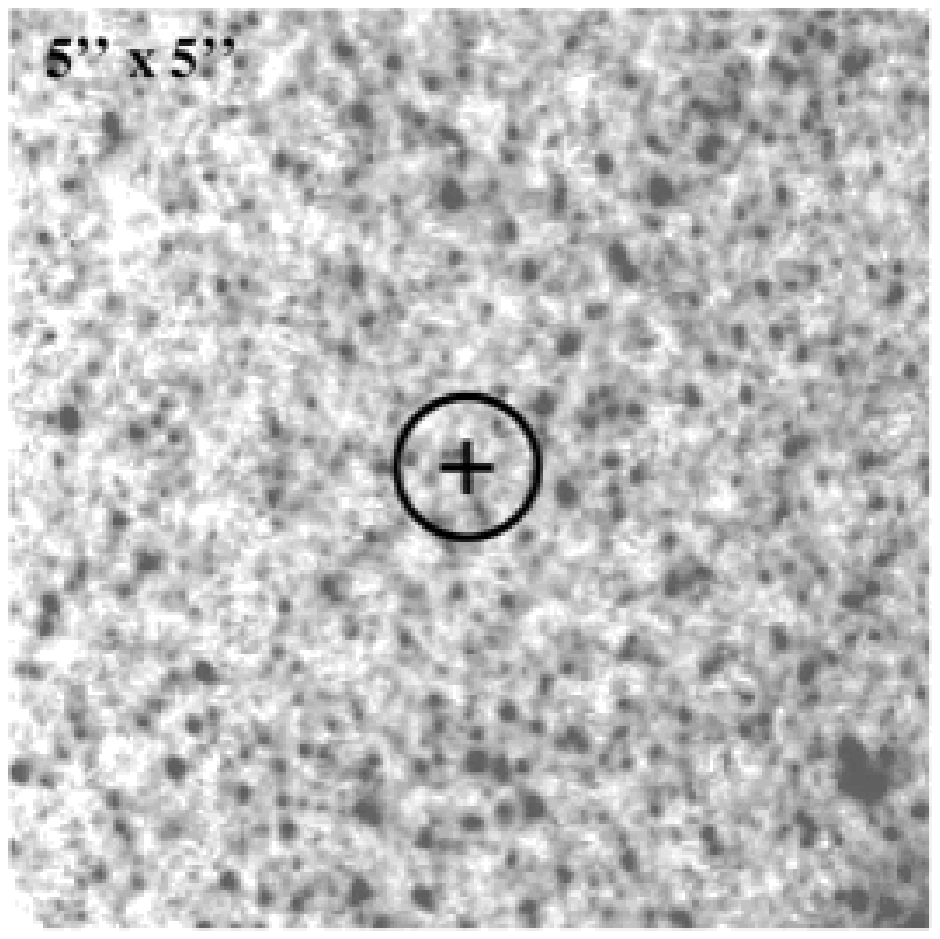} & 
\includegraphics[width=0.23\linewidth,clip=true,trim=0.5cm 3cm 0.5cm 2.5cm]{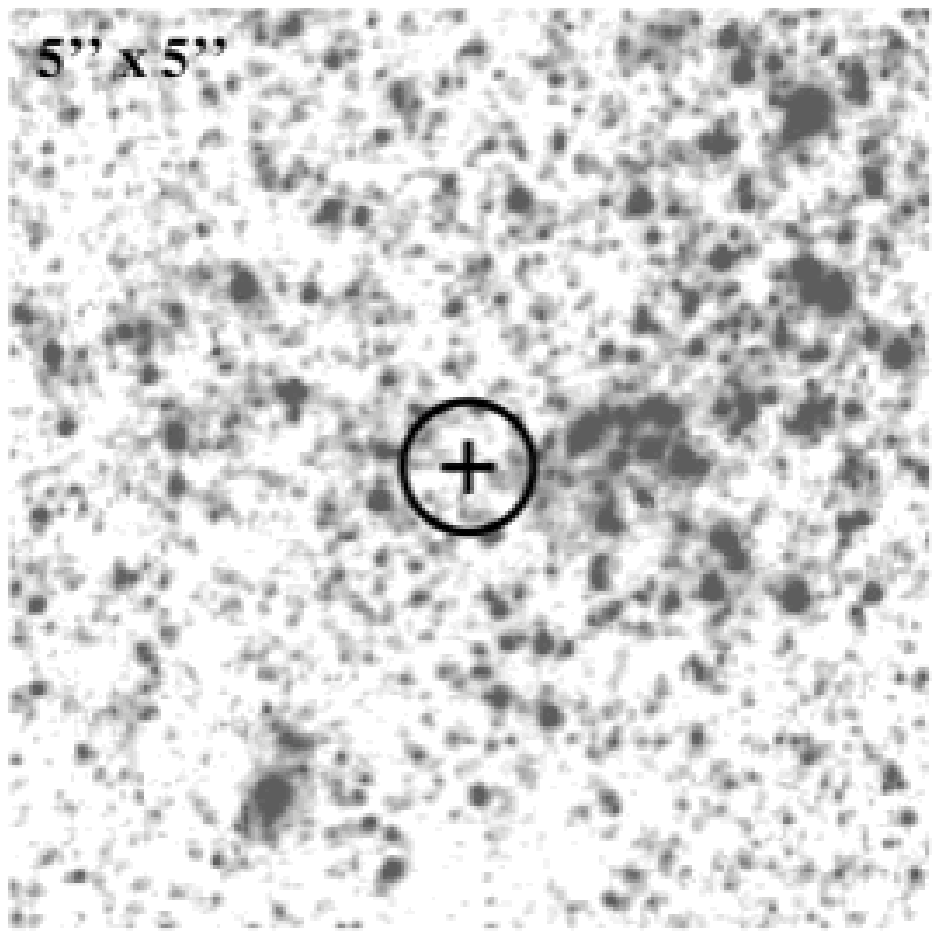} \\ 

Source 135 & Source 141 & Source 154 &   \\ 
\includegraphics[width=0.23\linewidth,clip=true,trim=0.5cm 3cm 0.5cm 2.5cm]{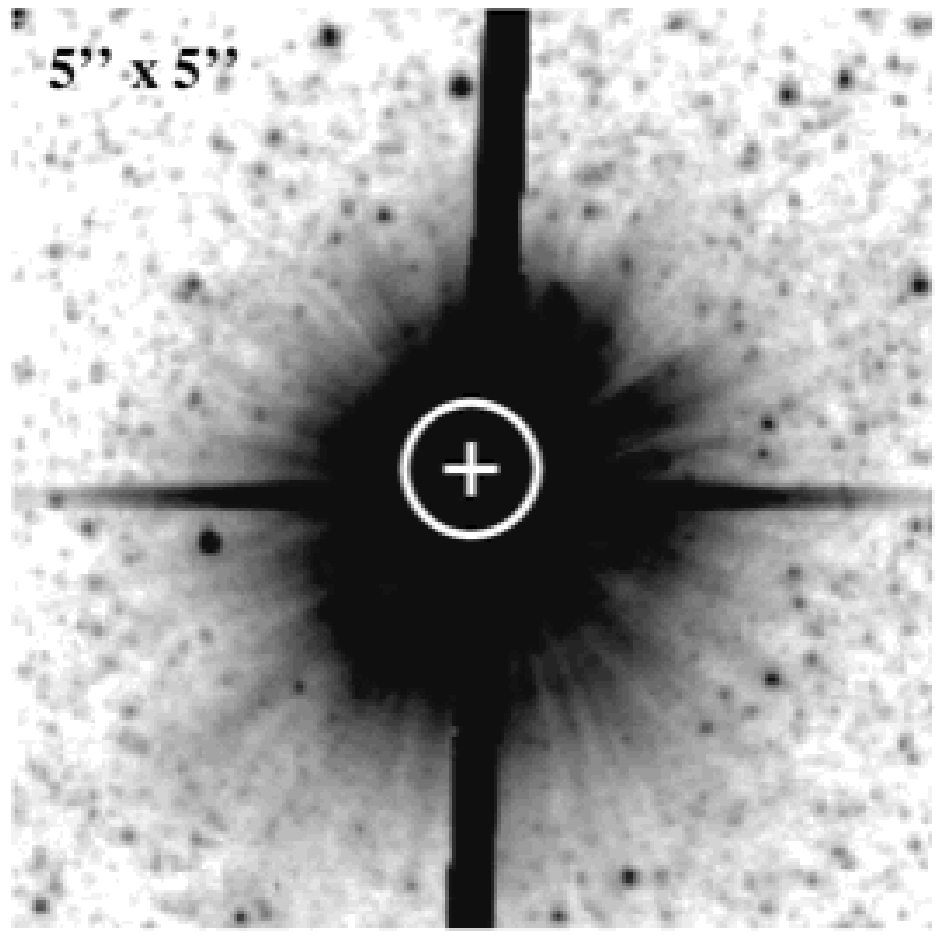} & 
\includegraphics[width=0.23\linewidth,clip=true,trim=0.5cm 3cm 0.5cm 2.5cm]{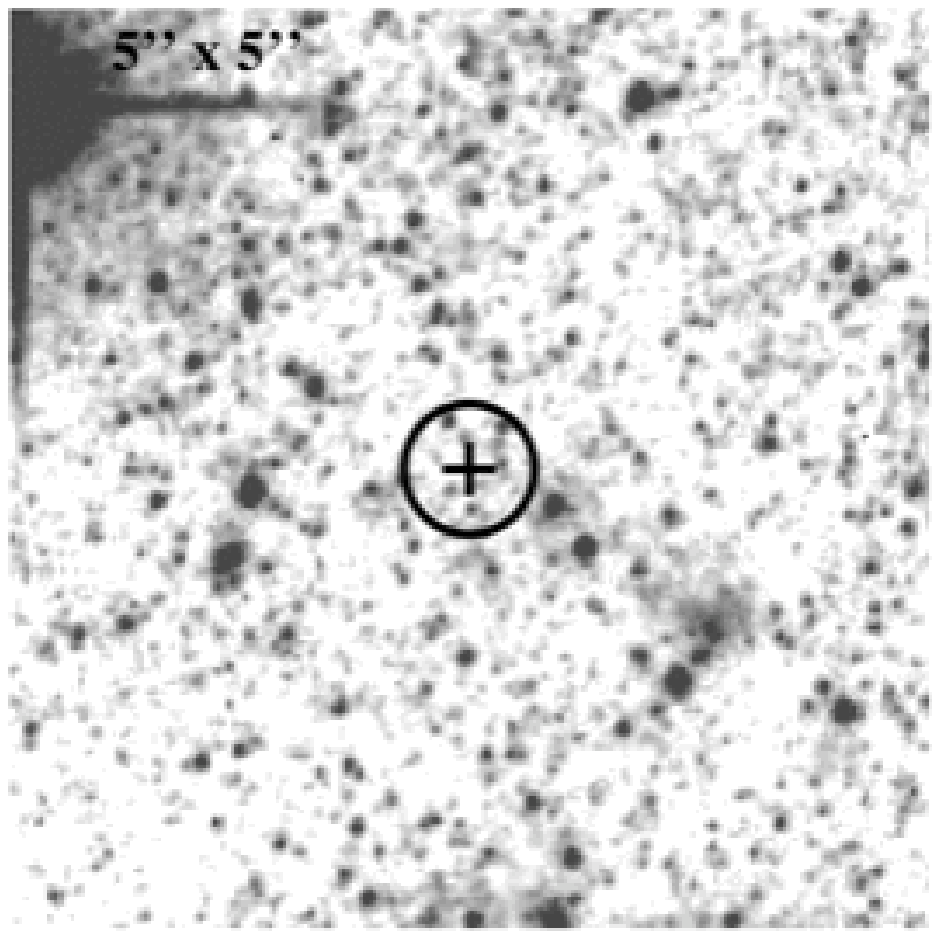} & 
\includegraphics[width=0.23\linewidth,clip=true,trim=0.5cm 3cm 0.5cm 2.5cm]{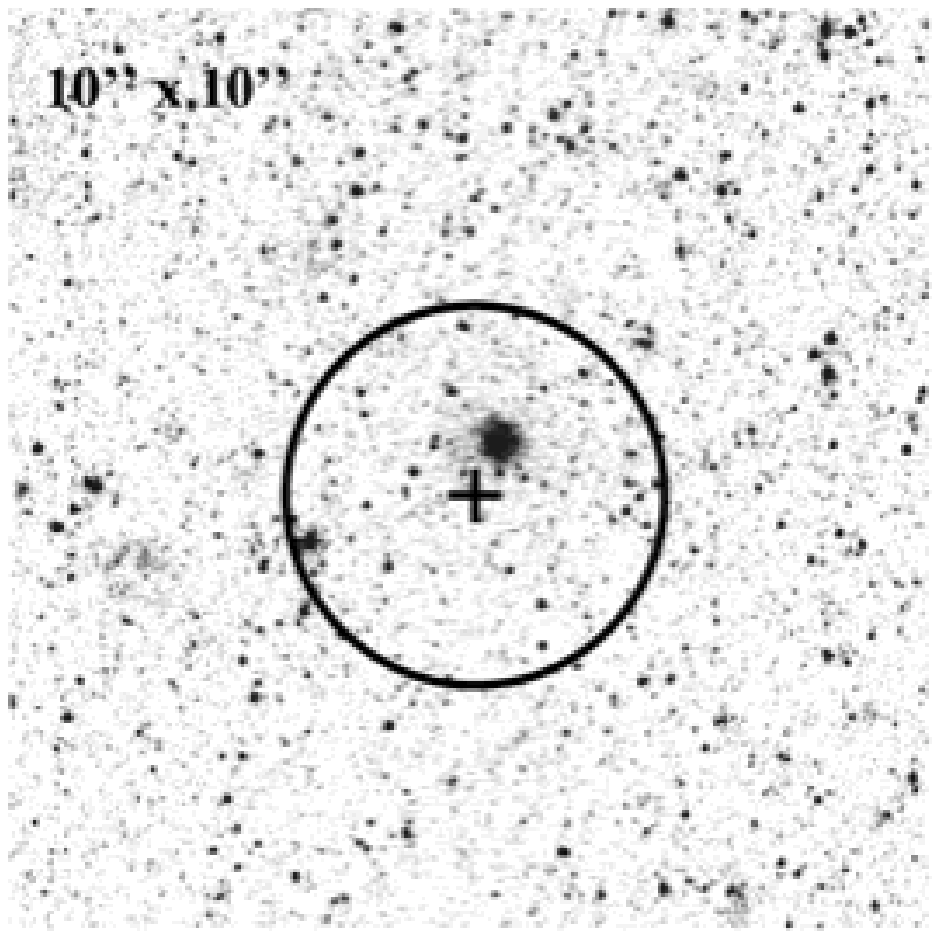} & \\ 
\end{tabular}
\caption{{\it (Continued)} Optical \HST images for X-ray sources detected in NGC~2403. The box size (5\asn$\times$ 5\asn, 10\asn$\times$ 10\asn, or 15\asn$\times$ 15\asn) is given in the top-left corner of each image. The circle shows the \Chandra 90\% error circle, centered on the source position.}
\label{optical_2403}
\end{figure*}

\begin{figure*}
\centering
\begin{tabular}{cccc}
Source 44 & Source 46 & Source 47 & Source 48 \\ 
\includegraphics[width=0.23\linewidth,clip=true,trim=0.5cm 3cm 0.5cm 2.5cm]{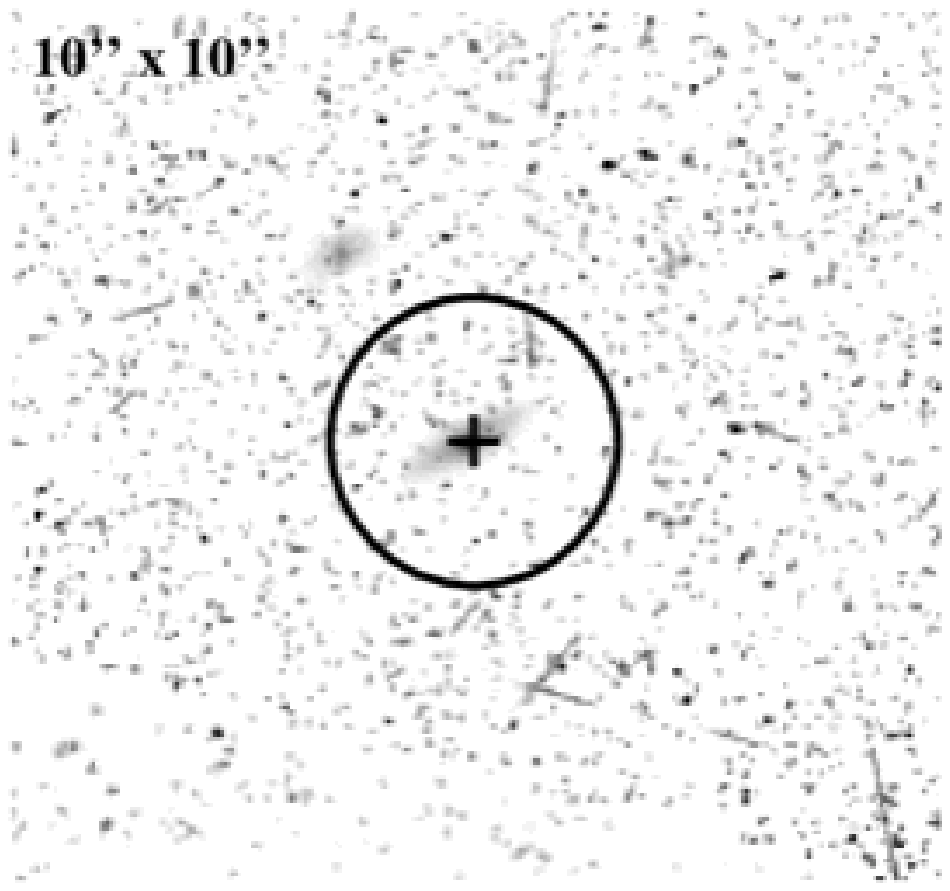} & 
\includegraphics[width=0.23\linewidth,clip=true,trim=0.5cm 3cm 0.5cm 2.5cm]{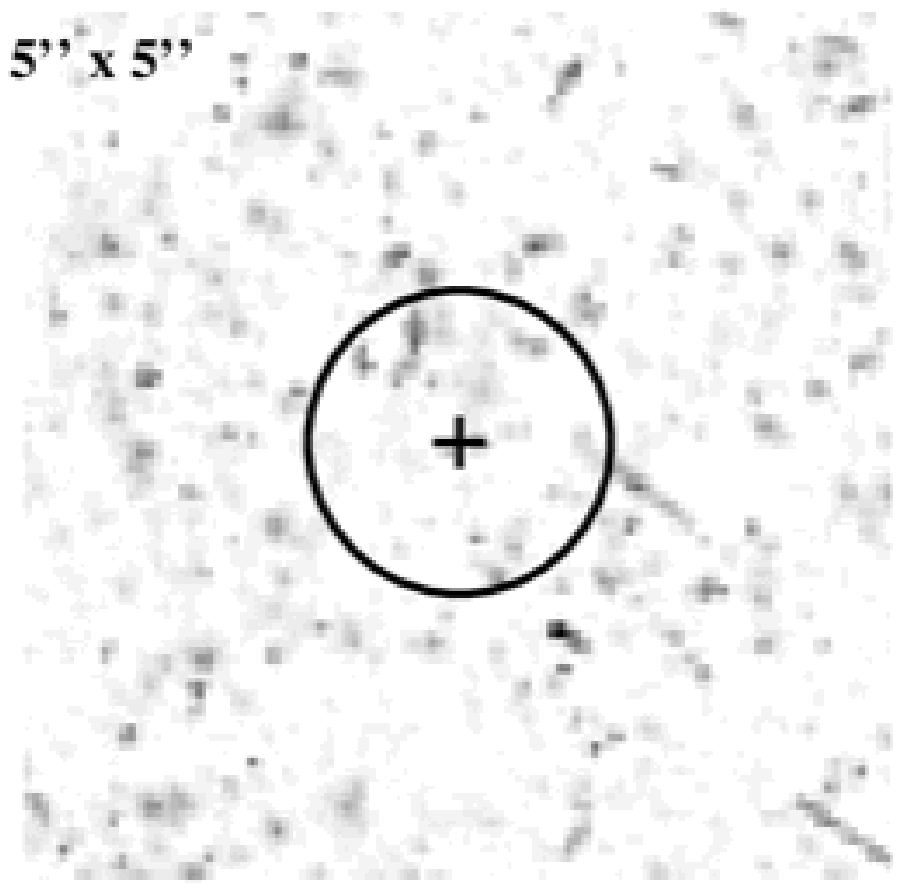} & 
\includegraphics[width=0.23\linewidth,clip=true,trim=0.5cm 3cm 0.5cm 2.5cm]{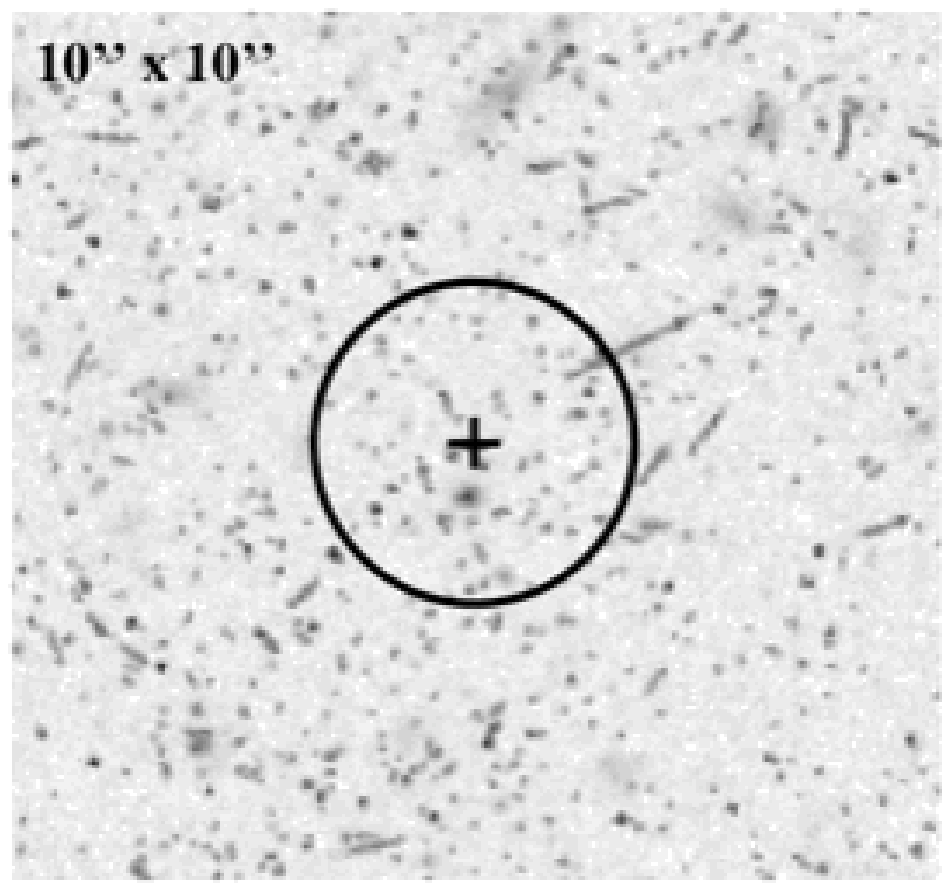} & 
\includegraphics[width=0.23\linewidth,clip=true,trim=0.5cm 3cm 0.5cm 2.5cm]{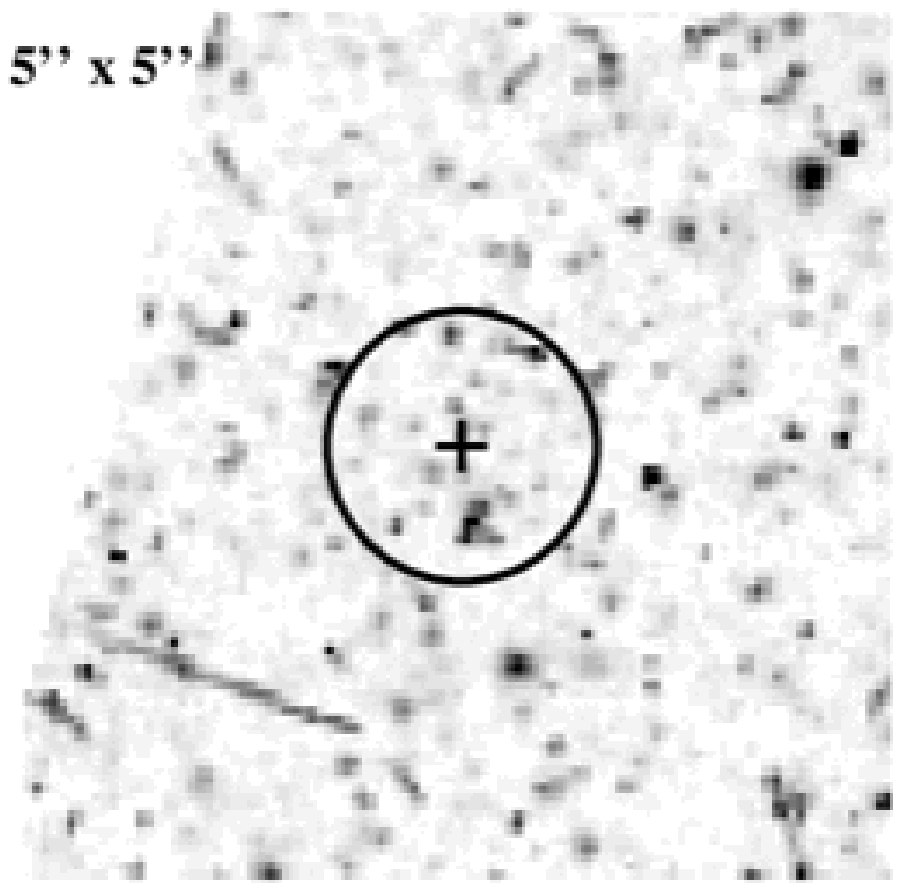} \\ 

Source 49 & Source 50 & Source 53 & Source 57 \\ 
\includegraphics[width=0.23\linewidth,clip=true,trim=0.5cm 3cm 0.5cm 2.5cm]{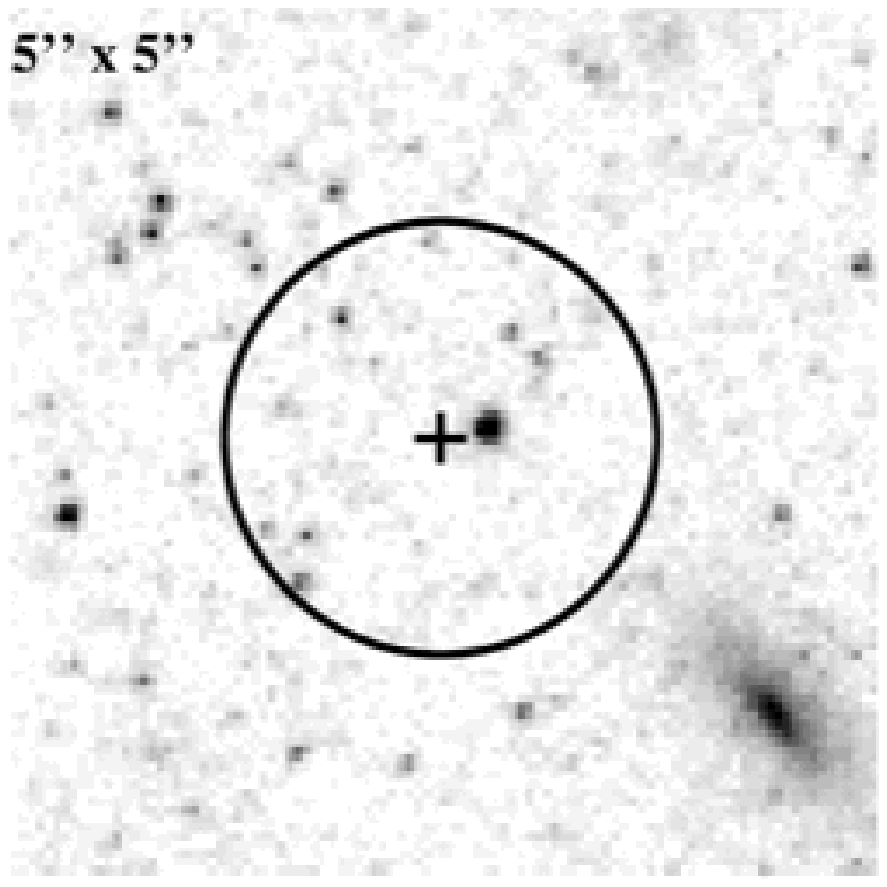} & 
\includegraphics[width=0.23\linewidth,clip=true,trim=0.5cm 3cm 0.5cm 2.5cm]{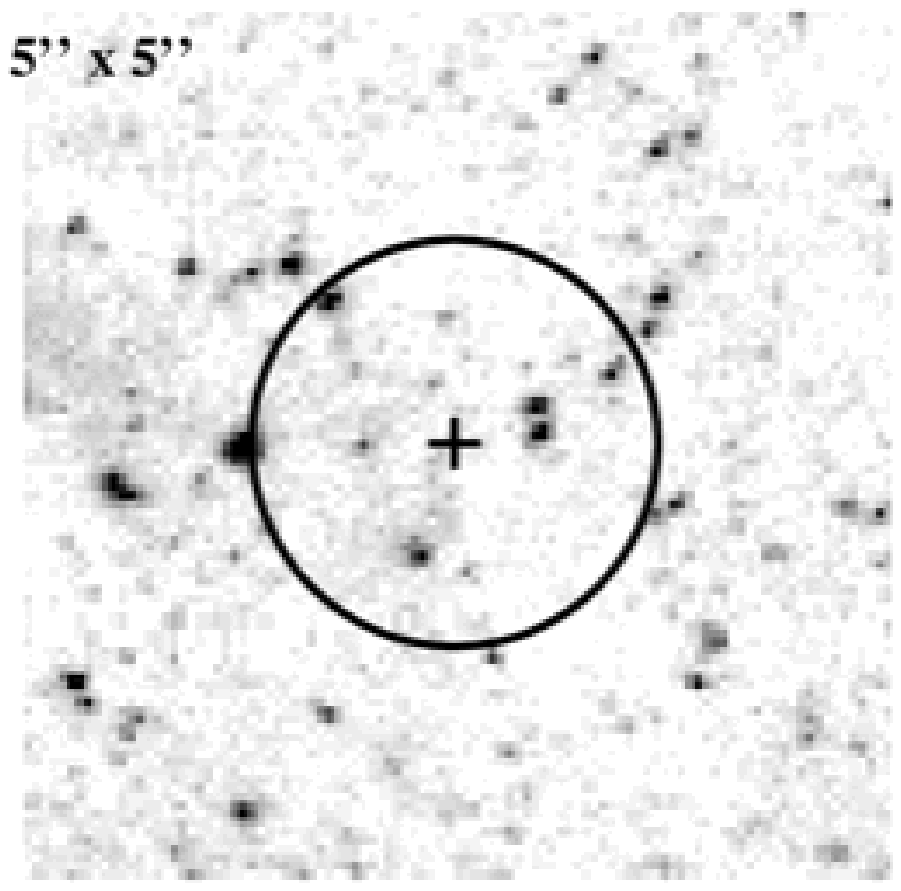} & 
\includegraphics[width=0.23\linewidth,clip=true,trim=0.5cm 3cm 0.5cm 2.5cm]{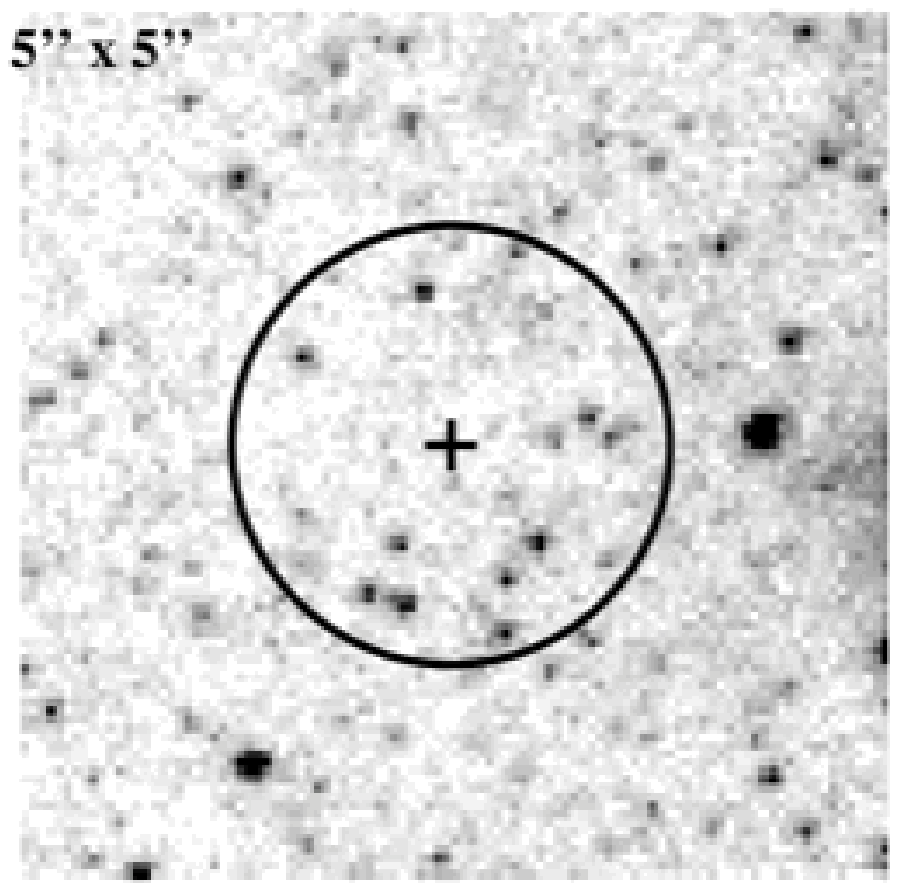} & 
\includegraphics[width=0.23\linewidth,clip=true,trim=0.5cm 3cm 0.5cm 2.5cm]{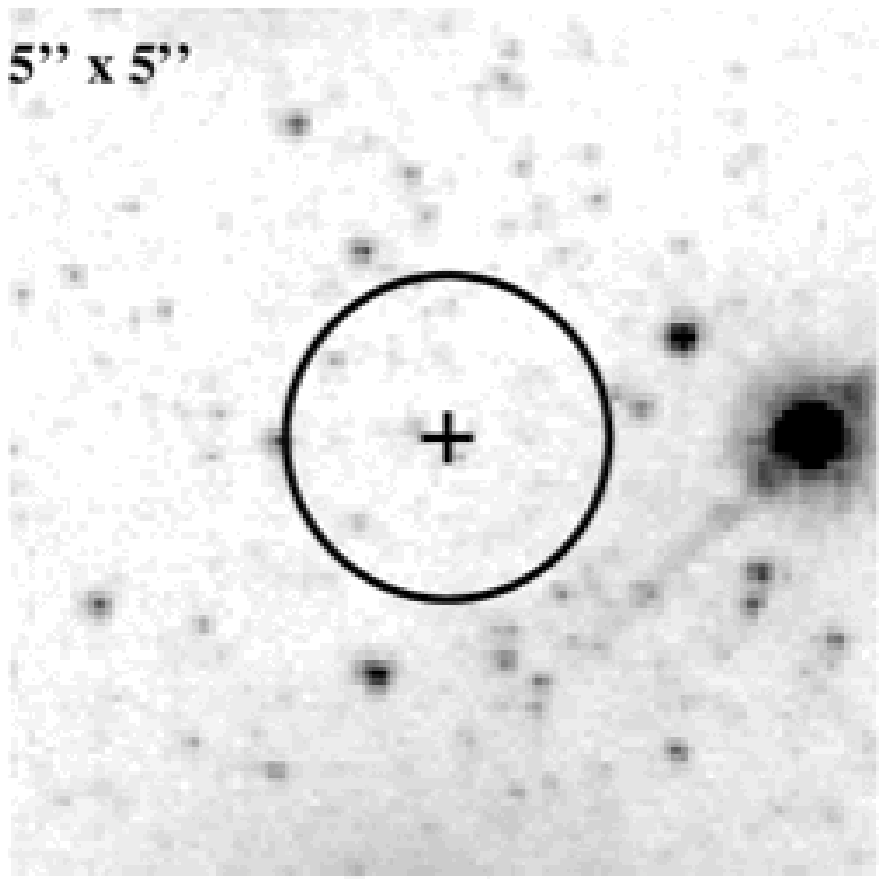} \\ 
\end{tabular}
\caption{Optical \HST images for X-ray sources detected in NGC~4214. The box size (5\asn$\times$ 5\asn, 10\asn$\times$ 10\asn, or 15\asn$\times$ 15\asn) is given in the top-left corner of each image. The circle shows the \Chandra 90\% error circle, centered on the source position.}
\label{optical_4214}
\end{figure*}

\begin{figure*}
\centering
\includegraphics[width=0.6\linewidth,clip=true]{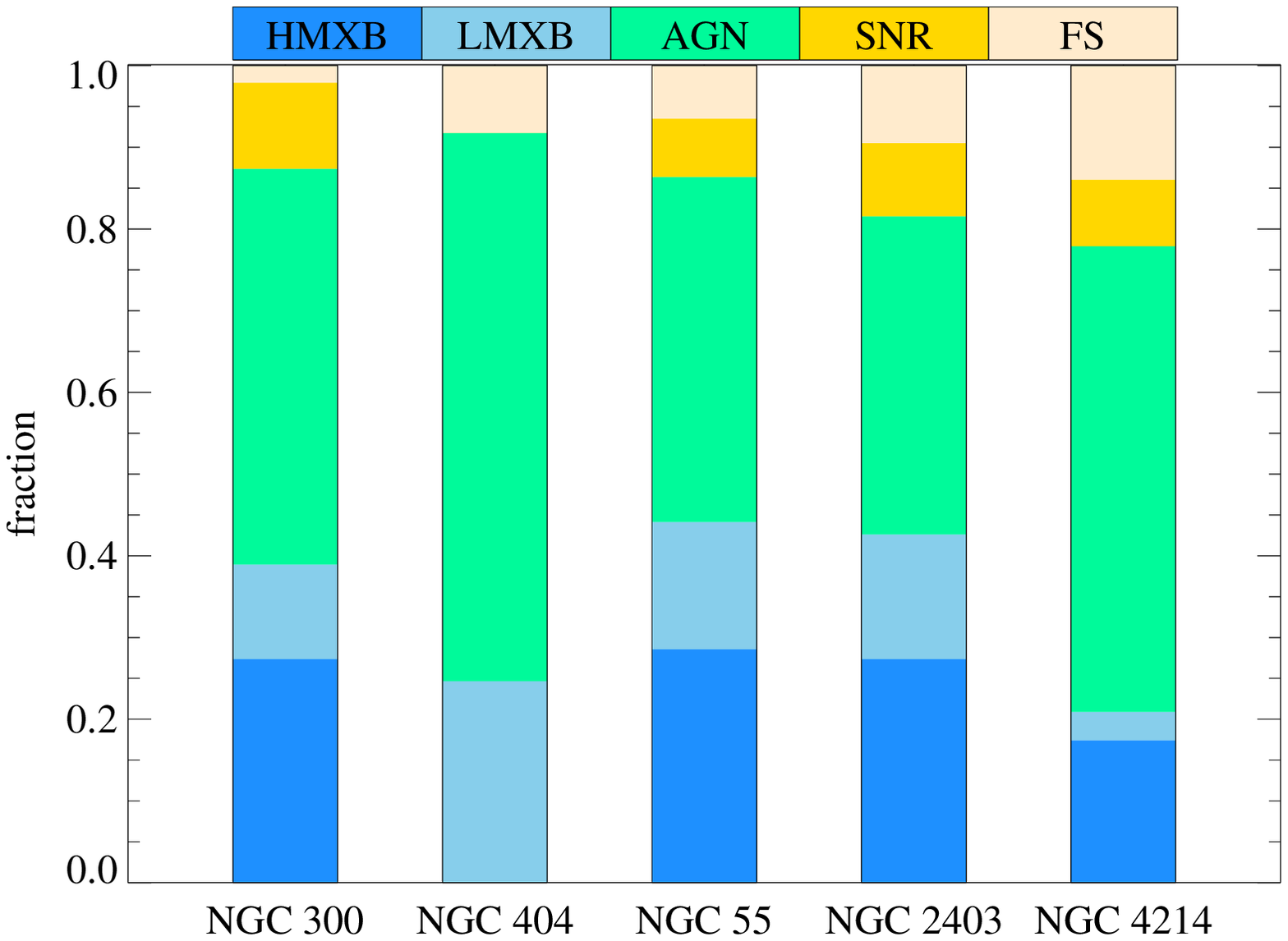} 
\caption{The fraction of X-ray sources classified as HMXBs, LMXBs, background AGN, foreground stars (`FS'), and SNRs in each galaxy.}
\label{final_classes}
\end{figure*}
                                                                                             
\end{document}